\documentclass[12pt]{article}
\usepackage{geometry,setspace}
\usepackage{lineno}
\usepackage{lipsum}
\usepackage[longnamesfirst,authoryear]{natbib}
\usepackage[T1]{fontenc}
\usepackage{geometry}
\usepackage{amssymb}
\usepackage{amsfonts,bm}
\usepackage{tabularx}
\usepackage{makeidx}
\usepackage{graphicx,eepic}
\usepackage{amsmath}
\usepackage{amsthm,bbm}
\usepackage{enumitem}
\DeclareMathOperator*{\argmax}{arg\,max}
\DeclareMathOperator*{\argmin}{arg\,min}
\DeclareMathOperator{\plim}{plim}
\usepackage{caption}
\usepackage{subcaption}
\usepackage[hyphens]{url}
\usepackage{here}
\usepackage{rotating}
\usepackage{pdfpages}
\usepackage{comment,ulem,verbatim}
\usepackage{enumitem}  
\usepackage{multirow}
\usepackage{array,booktabs}
\usepackage{adjustbox}
\usepackage{siunitx}
\usepackage{refcount}
\usepackage{hyperref}
\hypersetup{hidelinks}

\usepackage{multibib}
\newcites{appendix}{References}
\newcites{supplement}{References}

\let\counterwithin\relax
\usepackage{chngcntr}
\usepackage{apptools}
\AtAppendix{\counterwithin{lemma}{section}}
\AtAppendix{\counterwithin{assumption}{section}}
\AtAppendix{\counterwithin{step-subsection}{subsection}}
\AtAppendix{\counterwithin{step}{subsubsection}}
\AtAppendix{\counterwithin{proposition}{section}}
\AtAppendix{\counterwithin{corollary}{section}}
\AtAppendix{\counterwithin{theorem}{section}}
\AtAppendix{\counterwithin{example}{section}}

\providecommand{\U}[1]{\protect\rule{.1in}{.1in}}
\providecommand{\U}[1]{\protect\rule{.1in}{.1in}}
\providecommand{\U}[1]{\protect\rule{.1in}{.1in}}

\newtheorem{theorem}{Theorem}
\newtheorem{assumption}{Assumption}
\newtheorem{property}{Property}

\newtheorem{step-subsection}{Step}
\newtheorem{step}{Step}

\newtheorem{corollary}{Corollary}

\newtheorem{lemma}{Lemma}

\newtheorem{proposition}{Proposition}

\theoremstyle{definition}
\newtheorem{example}{Example}
\newtheorem{method}{Method}

\DeclareRobustCommand{\Var}{\mathrm{Var}}

\DeclareRobustCommand{\indep}{\mathop{\perp\!\!\!\!\perp}}
\begin{document}

\title{\vspace{-1.5cm}\LARGE{Algorithm as Experiment:}\\
Machine Learning, Market Design, and \\ Policy Eligibility Rules\footnote{{\it Keywords:} Algorithmic decision making, instrumental variables, propensity score, regression discontinuity design, COVID-19 hospital relief funding}}
\author{Yusuke Narita
\and Kohei Yata\thanks{Narita: Department of Economics, Yale University, email: \textsf{yusuke.narita@yale.edu}. Yata: Department of Economics, University of Wisconsin–Madison, email: \textsf{yata@wisc.edu}.}
\footnote{
For their suggestions, we are grateful to Joseph Altonji, Josh Angrist, Tim Armstrong, Yingying Dong, Ivan Fernandez-Val, Pat Kline, Michal Kolesár, Chris Walters, and seminar participants at the ACM Conference on Equity and Access in Algorithms, Mechanisms, and Optimization (EAAMO ’21), American Economic Association, Berkeley, BU,  Caltech, Columbia, CEMFI, Counterfactual Machine Learning Workshop, Econometric Society, European Economic Association, Hitotsubashi, JSAI, Michigan, Stanford, UC Irvine, the University of Tokyo, Virtual Market Design Seminar, and Yale. We are especially indebted to Aneesha Parvathaneni, Richard Liu, Richard Gong, and several others for expert research assistance.}}
\date{\today}
\maketitle

\begin{abstract}
Algorithms make a growing portion of policy and business decisions. We develop a treatment-effect estimator using algorithmic decisions as instruments for a class of stochastic and deterministic algorithms. Our estimator is consistent and asymptotically normal for well-defined causal effects. A special case of our setup is multidimensional regression discontinuity designs with complex boundaries. We apply our estimator to evaluate the Coronavirus Aid, Relief, and Economic Security Act, which allocated many billions of dollars worth of relief funding to hospitals via an algorithmic rule. The funding is shown to have little effect on COVID-19-related hospital activities. Naive estimates exhibit selection bias.
\end{abstract}


\newpage

\section{Introduction}
Today's society increasingly resorts to algorithms for decision making and resource allocation. 
For example, judges in the US make legal decisions aided by predictions from supervised machine learning algorithms. 
Supervised learning is also used by governments to detect potential criminals and terrorists, and by banks and insurance companies to screen potential customers. 
Tech companies like Facebook, Microsoft, and Netflix allocate digital content by reinforcement learning and bandit algorithms.
Retailers and e-commerce platforms engage in algorithmic pricing. 
Similar algorithms are encroaching on high-stakes settings, such as in education, healthcare, and the military.  

Other types of algorithms also loom large. 
School districts, college admissions systems, and labor markets use matching algorithms for position and seat allocations. 
Objects worth astronomical sums of money change hands every day in 
algorithmically run auctions.
Many public policy domains like Medicaid often use algorithmic rules to decide who is eligible. 

All of the above examples share a common trait: a decision-making algorithm makes decisions based only on its observable input variables. 
Thus conditional on the observable variables, algorithmic treatment decisions are assigned independently of any potential outcome. 
This property turns algorithm-based treatment decisions into instrumental variables (IVs) that can be used for measuring the causal effect of the final treatment assignment. 
The algorithm-based IV may produce stratified randomization, regression-discontinuity-style local variation, or some combination of the two. 

This paper shows how to use data obtained from algorithmic decision making to identify and estimate causal effects.
In our framework, the analyst observes a random iid sample $\{(Y_i,X_i,D_i,Z_i)\}_{i=1}^n$, where $Y_i$ is the outcome of interest, $X_i\in\mathbb{R}^p$ is a vector of pre-treatment covariates used as the algorithm's input variables, $D_i$ is the binary treatment assignment, possibly made by humans, and $Z_i$ is the binary treatment recommendation made by a known algorithm.
The algorithm takes $X_i$ as input and computes the probability of the treatment recommendation $A(X_i)=\Pr(Z_i=1|X_i)$.
$Z_i$ is then randomly determined based on the known probability $A(X_i)$ independently of everything else conditional on $X_i$. 
The algorithm's recommendation $Z_i$ may influence the final treatment assignment $D_i$, determined as $D_i=Z_iD_i(1)+(1-Z_i)D_i(0)$, where $D_i(z)$ is the potential treatment assignment that would be realized if $Z_i=z$.
Finally, the observed outcome $Y_i$ is determined as $Y_i=D_iY_i(1)+(1-D_i)Y_i(0)$, where $Y_i(1)$ and $Y_i(0)$ are potential outcomes that would be realized if the individual were treated and not treated, respectively. 
This setup is an IV model where the IV satisfies the conditional independence condition but may not satisfy the overlap (full-support) condition. 
This setup nests the classic propensity-score and regression-discontinuity-design (RDD) setups.

Within this framework, we first characterize the sources of causal-effect identification for a class of data-generating algorithms. This class includes all of the aforementioned examples, nesting both stochastic and deterministic algorithms. The sources of causal-effect identification turn out to be summarized by a suitable modification of the Propensity Score.  
We call it the \textit{Approximate Propensity Score} (APS). 
For each covariate value $x$, the Approximate Propensity Score is the average probability of a treatment recommendation in a shrinking neighborhood around $x$, defined as
$$
p^{A}(x)\equiv \lim_{\delta\rightarrow 0}\frac{\int_{B(x,\delta)}A(x^*)dx^*}{\int_{B(x,\delta)}dx^*},
$$
where $B(x,\delta)$ is a $p$-dimensional ball with radius $\delta$ centered at $x$.
The Approximate Propensity Score provides an easy-to-check condition for what causal effects the data from an algorithm allow us to identify.
In particular, we show that the conditional local average treatment effect (LATE; \citealp{imbens1994}) at covariate value $x$ is identified if and only if the Approximate Propensity Score is nondegenerate, i.e., $p^{A}(x)\in (0, 1)$.

The identification analysis suggests an estimator. 
The treatment effects can be estimated by two-stage least squares (2SLS) where we regress the outcome on the treatment with the algorithm's recommendation as an IV. 
To make the algorithmic recommendation a conditionally independent IV, we propose to control for the Approximate Propensity Score, as formalized below.\footnote{Code implementing this procedure in Python, R, and Stata is available at \url{https://github.com/rfgong/IVaps}.}

\begin{enumerate}
	\item \label{procedure:APS} For small bandwidth $\delta>0$ and a large number of simulation draws $S$, compute
	$$
	p^s(X_i;\delta)=\frac{1}{S}\sum_{s=1}^{S}A(X_{i,s}^*),
	$$
	where $X_{i,1}^*,...,X_{i,S}^*$ are $S$ independent simulation draws from the uniform distribution on $B(X_i,\delta)$.\footnote{To make the common $\delta$ for all dimensions reasonable, we standardize each characteristic $X_{ij}$ ($j=1,...,p$) to have variance one, where $p$ is the number of input characteristics.
	For the bandwidth $\delta$, we suggest that the analyst consider several different values and check if the 2SLS estimates are robust to bandwidth changes, as we often do in RDD applications.}
	This $p^s(X_i;\delta)$ is a simulation-based approximation to the Approximate Propensity Score $p^{A}(X_i)$.
	\item Run this 2SLS regression for observations with $p^s(X_i;\delta)\in (0,1)$:
	\begin{align*}
	D_i&=\gamma_0+\gamma_1 Z_i+\gamma_2 p^s(X_i;\delta)+\nu_i\text{ (First Stage)}\\
	Y_i&=\beta_0+\beta_1 D_i +\beta_2 p^s(X_i;\delta)+\epsilon_i\text{ (Second Stage)}.
	\end{align*}
	Let $\hat\beta_1^s$ be the estimated coefficient on $D_i$. 
\end{enumerate}



As the main theoretical result, we prove the 2SLS estimator $\hat\beta_1^s$ is a consistent and asymptotically normal estimator of a well-defined causal effect (weighted average of conditional local average treatment effects). 
Our result clarifies how to estimate other parameters by reweighting observations.
We also show that inference based on the conventional 2SLS heteroskedasticity-robust standard errors is asymptotically valid as long as the bandwidth $\delta$ goes to zero and the number of simulation draws $S$ goes to infinity at appropriate rates.
We prove the asymptotic properties by exploiting results from differential geometry and geometric measure theory. 

Our estimator is applicable even if the algorithm is deterministic and produces multidimensional regression-discontinuity variation only.
In contrast to standard multidimensional RDD methods that define the distance to the nearest boundary as a single running variable, our estimator is straightforward to use even when the multidimensional boundary is arbitrarily complex and the distance is hard to compute.
Moreover, our method applies to more general settings with stochastic algorithms, deterministic algorithms, and combinations of the two. 
In settings that mix stochastic and deterministic algorithms, our estimator exploits both the random-assignment variation and RDD variation, producing precision and representability gains compared to RDD estimators and propensity-score estimators only using either variation.

The practical performance of our estimator is demonstrated through simulation and an original application.
We first conduct a Monte Carlo simulation mimicking real-world decision making based on machine learning algorithms. We consider a data-generating process combining stochastic and deterministic algorithms. Treatment recommendations are randomly assigned for a small experimental segment of the population and are determined by a high-dimensional, deterministic machine learning algorithm for the rest of the population.
Such combinations of small experiments and deterministic treatment allocations arise in the real world when large-scale experiments are prohibited due to ethical, budget, or legislative constraints. 
Our estimator is shown to be feasible in this high-dimensional setting and has smaller median absolute errors relative to alternative estimators.
In particular, our method produces more efficient and representative estimates than a conventional propensity-score approach. 
The simulation also illustrates a setting where the RDD boundary is complex and no prior multidimensional RDD method is applicable.

Our empirical application is an analysis of COVID-19 hospital relief funding. 
The Coronavirus Aid, Relief, and Economic Security (CARES) Act designated \$175 billion for COVID-19 response efforts and reimbursement to health care entities for expenses or lost revenues \citep{kakani2020allocation}. This policy intended to help hospitals hit hard by the pandemic, as ``\textit{financially insecure hospitals may be less capable of investing in COVID-19 response efforts}'' \citep{khullar2020COVID}. 
We ask whether this problem is alleviated by the relief funding for hospitals. 

We identify the causal effects of the relief funding by exploiting the funding eligibility rule. The government runs an algorithmic rule on hospital characteristics to decide which hospitals are eligible for funding. This fact allows us to apply our method to estimate the effect of relief funding. 
Specifically, our 2SLS estimators use funding eligibility status as an IV for funding amounts, while controlling for the Approximate Propensity Score induced by the eligibility-determining algorithm. 
The funding eligibility IV boosts the funding amount by about \$15 million on average.

The resulting 2SLS estimates with Approximate Propensity Score controls suggest that COVID-19 relief funding has little to no effect on outcomes, such as the number of COVID-19 patients hospitalized at each hospital. 
The estimated causal effects of relief funding are much smaller and less significant than the naive ordinary least squares (OLS) (with and without controlling for hospital characteristics) or 2SLS estimates with no controls. 
The OLS estimates, for example, imply that a \$1 million increase in funding allows hospitals to accommodate 4.53 more COVID-19 patients. The uncontrolled 2SLS estimates produce similar, slightly smaller effects (2.44 more patients per \$1 million of funding). 
In contrast, the 2SLS estimates with Approximate Propensity Score controls show no or even negative effects (up to 2.21 \textit{fewer} patients for every \$1 million of funding).

The null effect of funding persists several months after the distribution of funding. 
We also find no clear heterogeneity in the null funding effect across different subgroups of hospitals. 
Our finding provides causal evidence for the concern that funding in the CARES Act might not have been well targeted to the clinics and hospitals with the greatest needs.\footnote{See, for example, \citet{kakani2020allocation} as well as {\it Forbes}'s article, ``Hospital Giant HCA To Return \$6 Billion in CARES Act Money,'' at \url{https://www.forbes.com/sites/brucejapsen/2020/10/08/hospital-giant-hca-to-return-6-billion-in-cares-act-money}.}

\subsection*{Related Literature} 

Our framework integrates the classic propensity-score (selection-on-observables) scenario with a multidimensional extension of the fuzzy RDD. 
We analyze this integrated setup in the IV world with noncompliance. 
This general setting appears to have no prior established estimator. \cite{armstrong2020finitesample} provide an estimator for a related setting with perfect compliance.

When we specialize our estimator to the multidimensional RDD case, our estimator has three features. First, it is a consistent and asymptotically normal estimator of a well-interpreted causal effect (average of conditional treatment effects along the RDD boundary) even if treatment effects are heterogeneous.
Second, it uses observations near all the boundary points as opposed to using only observations near one specific boundary point, thus avoiding variance explosion even when $X_i$ has many elements.
Third, it can be easily implemented even in cases with many covariates 
and complex algorithms (RDD boundaries). 
No existing estimator appears to have all of these properties \citep{papay2011extending,zajonc2012regression,keele2015geographic,cattaneo2016interpreting,imbens2019optimized}.

A popular approach to the two-dimensional RDD is to use the shortest (Euclidean) distance from each individual to the boundary as a univariate running variable and apply a univariate RDD method \citep{black1999border}.\footnote{Another common approach to the two-dimensional RDD is to first estimate the conditional average treatment effect $E[Y_i(1)- Y_i(0)|X_i = x]$ for a large number of boundary points $x$ (either by the univariate local polynomial regression using the distance to the point $x$ as a univariate covariate or by the bivariate local polynomial regression). It then computes a weighted average of the estimated conditional average treatment effects over the boundary \citep{zajonc2012regression,keele2015geographic}. However, identifying boundary points from a general decision algorithm is hard unless it has a known analytical form. Even if we can trace out the boundary, it is not straightforward to select a grid of points along the boundary.}
In general multidimensional RDDs where the boundary is complex or its analytical form is unknown, the distance-based approach requires approximation or estimation of the shortest distance to the boundary.
In such cases, no existing estimator has been proven to have properties such as consistency and asymptotic normality. 
We provide the asymptotic properties of our simulation-based estimator for a class of multidimensional RDDs, taking into account the simulation errors.


Our estimator is applicable to a class of data-generating algorithms that includes stochastic and deterministic algorithms used in practice. 
Our results thus nest existing insights on quasi-experimental variation in particular algorithms, such as supervised learning \citep{cowgill2018impact, bundorf2019humans}, 
bandit, reinforcement learning, and market-design algorithms \citep{abdulkadirouglu2017research, abdulkadiroglu2019breaking, atila2013, kawai2020using, narita2021theory, narita2018toward}. 
Our framework also reveals new sources of identification for algorithms that, at first sight, do not appear to produce a natural experiment.

The Approximate Propensity Score in this paper is related to the local random assignment interpretation of the RDD, discussed by \cite{cattaneo2015randomization}, 
\cite{frandsen2017party}, \cite{sekhon2017interpreting}, \cite{frolich/huber2018}, and  \cite{abdulkadiroglu2019breaking}. 
These papers consider special cases of this paper's framework.

Our empirical application uses the proposed method to study hospitals receiving CARES Act relief funding. Our empirical finding contributes to emerging work on how health care providers respond to financial shocks \citep{duggan2000hospital, 
dranove2017nonprofits, adelino2022hospital}. Our empirical setting is a healthcare crisis, so our work complements prior work on more normal situations. 
Our analysis also exploits rule-based locally random assignment of cash flows to hospitals. This feature provides our estimates with additional confidence in their causal interpretation.

\section{Framework}\label{framework}

Our framework is a mix of the conditional independence, multidimensional RDD, and instrumental variable scenarios. 
In the setup in the introduction, we are interested in the effect of some binary treatment $D_i\in \{0,1\}$ on some outcome of interest $Y_i\in \mathbb{R}$.
As is standard in the literature, we impose the exclusion restriction that the treatment recommendation $Z_i\in \{0,1\}$ does not affect the observed outcome other than through the treatment assignment $D_i$.
This allows us to define the potential outcomes indexed against the treatment assignment $D_i$ alone. 
$Y_i(1)$ and $Y_i(0)$ denote potential outcomes when the individual is treated and not treated, respectively.

We consider algorithms that make treatment recommendations based solely on individual $i$'s predetermined, observable covariates $X_i=(X_{i1},...,X_{ip})'\in \mathbb{R}^p$.
Let the function $A:\mathbb{R}^p\rightarrow [0,1]$ represent the decision algorithm, where $A(X_i)=\Pr(Z_i=1|X_i)$ is the probability that the treatment is recommended for individual $i$ with covariates $X_i$. 
The central assumption is that the analyst knows function $A$ and is able to simulate it. 
That is, the analyst is able to compute the recommendation probability $A(x)$ given any input value $x\in\mathbb{R}^p$. 
The treatment recommendation $Z_i$ for individual $i$ is then randomly determined with probability $A(X_i)$ independently of everything else.  
Consequently, the following conditional independence holds.

\begin{property}[Conditional Independence]\label{conditional-independence}
$Z_i\indep (Y_i(1),Y_i(0),D_i(1),D_i(0))|X_i$.  
\end{property}

The codomain of $A$ contains $0$ and $1$, allowing for deterministic treatment assignments conditional on $X_i$. 
Our framework therefore nests the RDD as a special case. 
	Another special case is the classic conditional independence scenario with the common support condition ($A(X_i)\in (0,1)$ almost surely). 
In addition to these simple settings, this framework nests many other situations, such as multidimensional RDDs and complex machine learning and market-design algorithms, as illustrated in Sections \ref{section:sim}-\ref{section:empirical} and Appendix \ref{section:examples}. 

In typical machine-learning scenarios, an algorithm first applies machine learning on $X_i$ to make some prediction and then uses the prediction to output the recommendation probability $A(X_i)$, as in the following example.\\

\textbf{Example.}
Automated disease detection algorithms use machine learning, in particular deep learning, to detect various diseases and to identify patients at risk \citep{Gulshan2016detection}.
A detection algorithm predicts whether an individual $i$ has a certain disease ($Z_i=1$) or not ($Z_i=0$) based on a digital image $X_i\in\mathbb{R}^p$ of a part of the individual's body, where each $X_{ij}\in \mathbb{R}$ denotes the intensity value of a pixel in the image. 
The algorithm uses training data to construct a binary classifier $A:\mathbb{R}^p\rightarrow \{0,1\}$.
The classifier takes an image of individual $i$ as input and makes a binary prediction of whether the individual has the disease: $Z_i\equiv A(X_i).$
The algorithm's diagnosis $Z_i$ may influence the doctor's treatment decision for the individual, denoted by $D_i\in\{0,1\}$.
We are interested in how the treatment decision $D_i$ affects the individual's health outcome $Y_i$.\\

Let $Y_{zi}$ be defined as $Y_{zi}\equiv D_i(z)Y_i(1)+(1-D_i(z))Y_i(0)$ for $z\in\{0,1\}$.
$Y_{zi}$ is the potential outcome when the treatment recommendation is $Z_i=z$.
It follows from Property \ref{conditional-independence} that $Z_i\indep (Y_{1i},Y_{0i})|X_i$.


We put assumptions on the covariates $X_i$ and the algorithm $A$. 
To simplify the exposition, the main text assumes that 
	the distribution of $X_i$ is absolutely continuous with respect to the Lebesgue measure.
Appendix \ref{section:discrete-X} extends the analysis to the case where some covariates in $X_i$ are discrete.
Let ${\cal X}$ be the support of $X_i$, ${\cal X}_0=\{x\in{\cal X}:A(x)=0\}$, ${\cal X}_1=\{x\in{\cal X}:A(x)=1\}$, ${\cal L}^p$ be the Lebesgue measure on $\mathbb{R}^p$, and ${\rm int}(S)$ be the interior of a set $S\subset\mathbb{R}^p$.

\begin{assumption}
	\label{assumption:A}~
	\begin{enumerate}[label=(\alph*)]
		\item{\rm (Almost Everywhere Continuity of $A$)} \label{assumption:A-continuity} $A$ is continuous almost everywhere with respect to the Lebesgue measure.
		\item{\rm (Measure Zero Boundaries of ${\cal X}_0$ and ${\cal X}_1$)} \label{assumption:A-measure-zero-boundary} ${\cal L}^p({\cal X}_k)={\cal L}^p({\rm int}({\cal X}_k))$ for $k=0,1$.
	\end{enumerate}
\end{assumption}

Assumption \ref{assumption:A} \ref{assumption:A-continuity} allows the function $A$ to be discontinuous on a set of points with the Lebesgue measure zero.
For example, $A$ is allowed to be a discontinuous step function as long as it is continuous almost everywhere.
Assumption \ref{assumption:A} \ref{assumption:A-measure-zero-boundary} holds if the Lebesgue measures of the boundaries of ${\cal X}_0$ and ${\cal X}_1$ are zero. Assumption \ref{assumption:A} \ref{assumption:A-measure-zero-boundary} is only for ruling out perverse cases such as the case where $A(x)=1$ if $x\in\mathbb{R}$ is an irrational number and $A(x)\neq 1$ otherwise.

\section{Identification}

What causal effects can be learned from data $(Y_i, X_i, D_i, Z_i)$ generated by the algorithm $A$? 
A key step toward answering this question is what we call the \textit{Approximate Propensity Score} (APS). To define it, we first define the \textit{fixed-bandwidth Approximate Propensity Score} as follows:
\begin{align*}
	p^{A}(x;\delta)&\equiv\frac{\int_{B(x,\delta)}A(x^*)dx^*}{\int_{B(x,\delta)}dx^*},
\end{align*}
where $B(x,\delta)=\{x^*\in\mathbb{R}^p:\|x-x^*\|<\delta\}$ is the (open) $\delta$-ball around $x\in {\cal X}$.\footnote{
We use a ball for simplicity. When we instead use a rectangle, ellipsoid, or any standard kernel function to define $p^{A}(x;\delta)$, the limit $\lim_{\delta\rightarrow 0}p^{A}(x;\delta)$ may be different at some points (e.g., at discontinuity points of $A$), but the same identification results hold under suitable conditions.}
Here, $\|\cdot\|$ denotes the Euclidean norm on $\mathbb{R}^p$. 
To make a common bandwidth $\delta$ for all dimensions reasonable, we standardize $X_{ij}$ to have variance one for each $j=1,...,p$.
We assume that $A$ is a ${\cal L}^p$-measurable function so that the integrals exist.
We then define APS as follows:
\begin{align*}
	p^{A}(x)&\equiv \lim_{\delta\rightarrow 0}p^{A}(x;\delta). 
\end{align*}
APS at $x$ is the average probability of a treatment recommendation in a shrinking ball around $x$. 
We call this the {\it Approximate} Propensity Score, since this score modifies the standard propensity score $A(X_i)$ to incorporate local variation in the score. APS exists for most covariate points and algorithms (see Appendix \ref{section:existence}).
\begin{figure}[t]
\caption{Example of the Approximate Propensity Score}\label{figure:APS-example}
	\begin{center}
\includegraphics[width=0.45\textwidth]{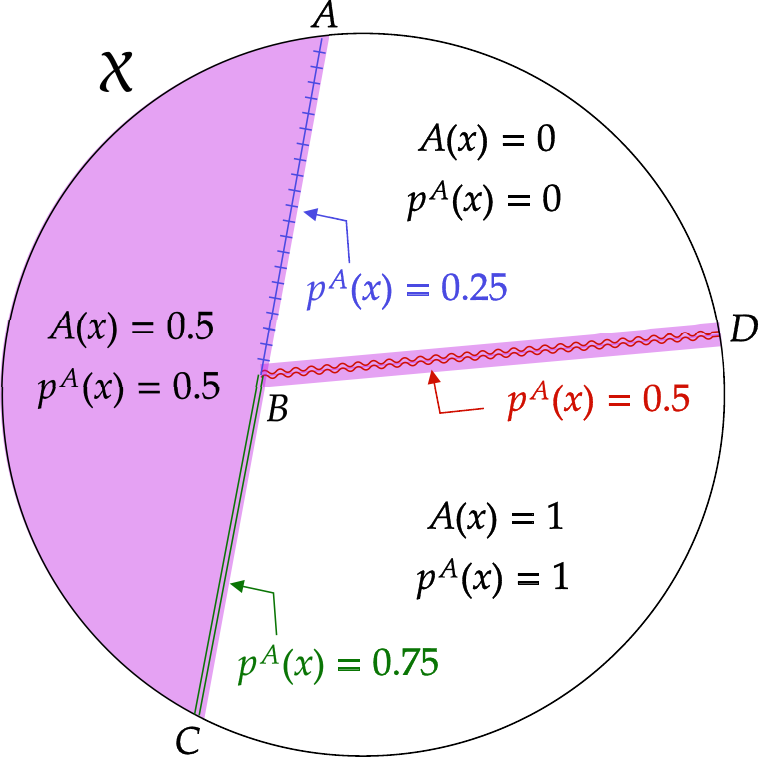}
\end{center}
\end{figure}
Figure \ref{figure:APS-example} illustrates APS.
In the example, $X_i$ is two dimensional, and the support of $X_i$ is divided into three sets depending on the value of $A$.
For the interior points of each set, APS is equal to $A$.
On the border of any two sets, APS is the average of the $A$ values in the two sets.
Thus, $p^{A}(x)=\frac{1}{2}(0+0.5)=0.25$ for any $x$ in the open line segment $AB$, $p^{A}(x)=\frac{1}{2}(0.5+1)=0.75$ for any $x$ in the open line segment $BC$, and $p^{A}(x)=\frac{1}{2}(0+1)=0.5$ for any $x$ in the open line segment $BD$.

Our identification analysis uses the following continuity condition.

\begin{assumption}[Local Mean Continuity]\label{continuity}
	For $z\in\{0,1\}$, the conditional expectation functions $E[Y_{zi}|X_i]$ and $E[D_i(z)|X_i]$ are continuous at any point $x\in {\cal X}$ such that $p^{A}(x)\in (0,1)$ and $A(x)\in \{0,1\}$. 
\end{assumption}

Assumption \ref{continuity} is a multivariate extension of the local mean continuity condition frequently assumed in the RDD. In the RDD with a single running variable, the point $x$ for which $p^{A}(x)\in (0,1)$ and $A(x)\in \{0,1\}$ is the cutoff at which the treatment probability discontinuously changes. 
$A(x)\in\{0,1\}$ means that the treatment recommendation $Z_i$ is deterministic conditional on $X_i=x$.
If APS at the point $x$ is nondegenerate ($p^{A}(x)\in (0,1)$), however, there exists a point close to $x$ that has a different value of $A$ from $x$'s, which creates variation in the treatment recommendation near $x$.
For any such point $x$, Assumption \ref{continuity} requires that the points close to $x$ have similar conditional means of the outcome $Y_{zi}$ and treatment assignment $D_i(z)$.\footnote{
In the context of the RDD with a single running variable, one sufficient condition for continuity of $E[Y_{zi}|X_i]$ is a local independence condition in the spirit of \cite{hahn2001RDD}: $(Y_i(1), Y_i(0), D_i(1), D_i(0))$ is independent of $X_i$ near $x$.
A weaker sufficient condition, which allows such dependence, is that $E[Y_i(d)|D_i(1)=d_1, D_i(0)=d_0, X_i]$ and $\Pr(D_i(1)=d_1, D_i(0)=d_0|X_i)$ are continuous at $x$ for every $d\in\{0,1\}$ and $(d_1,d_0)\in\{0,1\}^2$ \citep{Dong2018}.
This assumes that the conditional means of the potential outcomes for each of the four types determined based on the potential treatment assignment $D_i(z)$ and the conditional probabilities of those types are continuous at the cutoff.
These two sets of conditions are sufficient for continuity of $E[Y_{zi}|X_i]$ regardless of the dimension of $X_i$, accommodating multidimensional RDDs.}
Note that Assumption \ref{continuity} does not require continuity of the conditional means at $x$ for which $A(x)\in(0,1)$, since the identification of the conditional means at such points follows from Property \ref{conditional-independence} without continuity.

Under the above assumptions, APS provides an easy-to-check condition for whether an algorithm allows us to identify causal effects. We say that a causal effect is \textit{identified} if it is uniquely determined by the joint distribution of $(Y_i,X_i,D_i,Z_i)$. 

\begin{proposition}[Identification]\label{proposition:id}
	Under Assumptions \ref{assumption:A} and \ref{continuity}:
	\begin{enumerate}[label=(\alph*)]
		\item $E[Y_{1i}-Y_{0i}| X_i=x]$ and $E[D_i(1)-D_i(0)| X_i=x]$ are identified for every $x\in {\rm int}({\cal X})$ such that $p^{A}(x)\in (0,1)$.\footnote{The causal effects may not be identified at a boundary point $x$ of ${\cal X}$ for which $p^{A}(x)\in (0,1)$. For example, if $A(x^*)=1$ for all $x^*\in B(x,\delta)\cap {\cal X}$ and $A(x^*)=0$ for all $x^*\in B(x,\delta)\setminus {\cal X}$ for any sufficiently small $\delta>0$, $p^{A}(x)\in (0,1)$ but the causal effects are not identified at $x$ since $\Pr(Z_i=0|X_i\in B(x,\delta))=0$.}
		\item Let $S$ be any open subset of ${\cal X}$ such that $p^{A}(x)$ exists for all $x\in S$.
		Then either $E[Y_{1i}-Y_{0i}| X_i \in S]$ or $E[D_i(1)-D_i(0)| X_i \in S]$ or both are identified only if $p^{A}(x)\in (0,1)$ for almost every $x\in S$ (with respect to the Lebesgue measure).\footnote{We assume that $p^{A}$ is a ${\cal L}^p$-measurable function so that $\{x\in S:p^{A}(x)=0\}$ and $\{x\in S:p^{A}(x)=1\}$ are ${\cal L}^p$-measurable.}
	\end{enumerate}
\end{proposition}
\begin{proof}
	See Appendix \ref{proof:id}.
\end{proof}

Proposition \ref{proposition:id} characterizes a necessary and sufficient condition for identification.
Part (a) says that the average effects of the treatment recommendation $Z_i$ on the outcome $Y_i$ and on the treatment assignment $D_i$ conditional on $X_i=x$ are both identified if APS at $x$ is neither 0 nor 1.
Nondegeneracy of APS at $x$ implies that there are both types of individuals who receive $Z_i=1$ and $Z_i=0$ among those whose $X_i$ is close to $x$.
Assumption \ref{continuity} ensures that these individuals are similar in terms of average potential outcomes and treatment assignments.
We can therefore identify the average effects conditional on $X_i=x$. 
In Figure \ref{figure:APS-example}, $p^{A}(x)\in (0,1)$ holds for any $x$ in the shaded region. 

Part (b) provides a necessary condition for identification.
It says that if the average effect of the treatment recommendation conditional on $X_i$ being in some open set $S$ is identified, then we must have $p^{A}(x)\in (0,1)$ for almost every $x\in S$.
If, to the contrary, there is a subset of $S$ of nonzero measure for which $p^{A}(x)=1$ (or $p^{A}(x)=0$), then $Z_i$ has no variation in the subset, which makes it impossible to identify the average effect for the subset.

Proposition \ref{proposition:id} concerns causal effects of treatment {\it recommendation}, not of treatment {\it assignment}.
The proposition implies that the conditional average treatment effects and the conditional local average treatment effects (LATEs) are identified under additional assumptions.

\begin{corollary}\label{corollary:id-compliance}
	Under Assumptions \ref{assumption:A} and \ref{continuity}:
	\begin{enumerate}[label=(\alph*)]
		\item The average treatment effect conditional on $X_i=x$, $E[Y_i(1)-Y_i(0)| X_i = x]$, is identified for every $x\in {\rm int}({\cal X})$ such that $p^{A}(x)\in (0,1)$ and $\Pr(D_i(1)>D_i(0)|X_i=x)=1$ (perfect compliance).
		\item The local average treatment effect conditional on $X_i=x$, $E[Y_i(1)-Y_i(0)| D_i(1)\neq D_i(0), X_i =x]$, is identified for every $x\in {\rm int}({\cal X})$ such that $p^{A}(x)\in (0,1)$, $\Pr(D_i(1)\ge D_i(0)|X_i=x)=1$ (monotonicity), and $\Pr(D_i(1)\neq D_i(0)|X_i=x)>0$ (existence of compliers).
	\end{enumerate}
\end{corollary}

Nondegeneracy of APS $p^{A}(x)$ therefore summarizes what causal effects the data from $A$ identify. 
Note that the key condition ($p^{A}(x)\in(0,1)$) holds for some points $x$ for every standard algorithm except trivial algorithms that always recommend a treatment with probability 0 or 1. Therefore, the data from every nondegenerate algorithm identify some causal effect.

\section{Estimation}

The sources of quasi-random assignment characterized in Proposition \ref{proposition:id} suggest a way of estimating causal effects of the treatment. 
In view of Proposition \ref{proposition:id}, it is possible to nonparametrically estimate conditional average causal effects $E[Y_{1i}-Y_{0i}|X_i=x]$ and $E[D_i(1)-D_i(0)|X_i=x]$ for points $x$ such that $p^{A}(x)\in(0,1)$.
This approach is hard to use in practice, however, when $X_i$ has many elements.
	We instead seek an estimator that aggregates conditional effects at different points into a single average causal effect. 
Proposition \ref{proposition:id} suggests that conditioning on APS makes algorithm-based treatment recommendation quasi-randomly assigned. This motivates the use of an algorithm's recommendation as an instrument conditional on APS, which we operationalize as follows. 

\subsection{Two-Stage Least Squares Meets APS}\label{subsection:2sls}
Suppose that we observe a random iid sample $\{(Y_i,X_i,D_i,Z_i)\}_{i=1}^n$ of size $n$ from the population whose data-generating process is as described in the introduction and Section \ref{framework}.
Consider the following 2SLS regression using the observations with $p^{A}(X_i;\delta_n)\in (0,1)$:
\begin{align}
	D_i&=\gamma_0+\gamma_1 Z_i+\gamma_2 p^{A}(X_i;\delta_n)+\nu_i\label{2SLSestimand1}\\
	Y_i&=\beta_0+\beta_1 D_i +\beta_2 p^{A}(X_i;\delta_n)+\epsilon_i,\label{2SLSestimand2}
\end{align}
where bandwidth $\delta_n$ shrinks toward zero as the sample size $n$ increases. We drop the constant term if $A(X_i)$ takes only one nondegenerate value in the sample.
Let $I_{i,n}=1\{p^{A}(X_i;\delta_n)\in(0,1)\}$,
$\mathbf{D}_{i,n}=(1, D_i, p^{A}(X_i;\delta_n))'$, and $\mathbf{Z}_{i,n}=(1, Z_i, p^{A}(X_i;\delta_n))'$.
The 2SLS estimator $\hat\beta$ is then given by
\begin{align*}
	\hat\beta =
		(\sum_{i=1}^n \mathbf{Z}_{i,n}\mathbf{D}_{i,n}'I_{i,n})^{-1}\sum_{i=1}^n \mathbf{Z}_{i,n}Y_iI_{i,n}.
\end{align*}
Let $\hat\beta_1$ denote the 2SLS estimator of $\beta_1$ in the above regression.
The above regression uses true fixed-bandwidth APS $p^{A}(X_i;\delta_n)$, but it may be difficult to analytically compute if $A$ is complex. 
In such a case, we propose to approximate $p^{A}(X_i;\delta_n)$ using brute force simulation. 
We draw a value of $x$ from the uniform distribution on $B(X_i,\delta_n)$ a number of times, compute $A(x)$ for each draw, and take the average of $A(x)$ over the draws. 
Formally, let $X_{i,1}^*,...,X_{i,S_n}^*$ be $S_n$ independent draws from the uniform distribution on $B(X_i,\delta_n)$, and calculate
	$$
	p^s(X_i;\delta_n)=\frac{1}{S_n}\sum_{s=1}^{S_n}A(X_{i,s}^*).
	$$
	\sloppy We compute $p^s(X_i;\delta_n)$ for each $i=1,...,n$ independently across $i$ so that $p^s(X_1;\delta_n),...,p^s(X_n;\delta_n)$ are independent of each other. 
	For fixed $n$ and $X_i$, the approximation error relative to true $p^{A}(X_i;\delta_n)$ has a $1/\sqrt{S_n}$ rate of convergence.\footnote{More precisely, $|p^s(X_i;\delta_n)-p^{A}(X_i;\delta_n)|=O_{p^s}(1/\sqrt{S_n})$, where $O_{p^s}$ indicates the stochastic boundedness in terms of the probability distribution of the $S_n$ simulation draws.} 
    This rate does not depend on the dimension of $X_i$, so the simulation error can be made negligible even when $X_i$ has many elements.\footnote{When $A(X_i)\in\{0,1\}$ and $p^A(X_i;\delta_n)\in (0,1)$, the approximation error of $1\{p^s(X_i;\delta_n)\in (0,1)\}$ relative to $1\{p^A(X_i;\delta_n)\in (0,1)\}$, $|1\{p^s(X_i;\delta_n)\in (0,1)\}-1\{p^A(X_i;\delta_n)\in (0,1)\}|$, is $O_{p^s}(\sqrt{\max\{p^A(X_i;\delta_n),1-p^A(X_i;\delta_n)\}^{S_n}})$. This is because $\Var(1\{p^s(X_i;\delta_n)\in (0,1)\})=\Pr(p^s(X_i;\delta_n)\in (0,1))(1-\Pr(p^s(X_i;\delta_n)\in (0,1)))$, where $\Pr(p^s(X_i;\delta_n)\in (0,1))=1-p^A(X_i;\delta_n)^{S_n}-(1-p^A(X_i;\delta_n))^{S_n}.$
    This convergence rate may become slower with the dimension of $X_i$ since $\max\{p^A(X_i;\delta_n),1-p^A(X_i;\delta_n)\}$ tends to be close to one when $X_i$ has many elements. The rate is, however, faster than $1/\sqrt{S_n}$ regardless of the dimension.
    Furthermore, the simulation exercise in Section \ref{section:sim} shows that our method works with a moderate number of simulations ($S_n=400$) even in a high-dimensional setting ($p=100$).}
We consider the simulation version of the 2SLS regression (\ref{2SLSestimand1}) and (\ref{2SLSestimand2}), where we use the simulated fixed-bandwidth APS $p^s(X_i;\delta_n)$ in place of $p^{A}(X_i;\delta_n)$.
    Let $\hat\beta_1^s$ denote the 2SLS estimator of $\beta_1$ in the simulation-based regression.

An alternative 2SLS regression is the one controlling for the variable that equals $A(X_i)$ for observations with $A(X_i)\in (0,1)$ and $p^s(X_i;\delta_n)$ for those with $A(X_i)\in \{0,1\}$. This allows us to avoid any simulation for those with $A(X_i)\in (0,1)$. Moreover, this prevents the finite-sample bias due to using fixed-bandwidth APS instead of the standard propensity score for those with $A(X_i)\in (0,1)$.
This modification does not affect the asymptotic results below.

	
 
\subsection{Consistency and Asymptotic Normality}\label{section:consistency}
	We establish the consistency and asymptotic normality of the 2SLS estimators $\hat\beta_1$ and $\hat\beta_1^s$.
We use the following assumptions.

\begin{assumption}\label{assumption:consistency}~
	\begin{enumerate}
		\renewcommand{\theenumi}{(\alph{enumi})}
		\renewcommand{\labelenumi}{(\alph{enumi})}
		\item{\rm (Finite Moment)}\label{assumption:moment} $E[Y_i^{4}]<\infty$.
		
		Let $f_X$ denote the probability density function of $X_i$ and let ${\cal H}^{k}$ denote the $k$-dimensional Hausdorff measure on $\mathbb{R}^{p}$.\footnote{The $k$-dimensional Hausdorff measure on $\mathbb{R}^{p}$ is defined as follows.
	Let $\Sigma$ be the Lebesgue $\sigma$-algebra on $\mathbb{R}^{p}$ (the set of all Lebesgue measurable sets on $\mathbb{R}^{p}$).
		For $S\in \Sigma$ and $\delta>0$, let
		$
		{\cal H}_\delta^k(S)=\inf\{\sum_{j=1}^\infty d(E_j)^k: S\subset \cup_{j=1}^\infty E_j, d(E_j)<\delta, E_j\subset \mathbb{R}^p$ for all $j$\},
		where $d(E)=\sup\{\|x-y\|: x,y\in E\}$.
		The $k$-dimensional Hausdorff measure of $S$ on $\mathbb{R}^{p}$ is ${\cal H}^k(S)=\lim_{\delta\rightarrow 0}{\cal H}_\delta^k(S)$.}
		\item{\rm (Nonzero First Stage)}\label{assumption:relevance}
		$\int_{{\cal X}}p^A(x)(1-p^A(x))E[D_i(1)-D_i(0)|X_i=x]f_X(x)d\mu(x)\neq 0$, where $\mu$ is the Lebesgue measure ${\cal L}^p$ when $\Pr(A(X_i)\in (0,1))>0$ and is the $(p-1)$-dimensional Hausdorff measure ${\cal H}^{p-1}$ when $\Pr(A(X_i)\in (0,1))=0$.
	\end{enumerate}
	\vspace{.5em}
	\noindent
	If $\Pr(A(X_i)\in (0,1))=0$, then the following conditions \ref{assumption:nondegeneracy}--\ref{assumption:boundary-continuity} hold.
	\begin{enumerate}
		\renewcommand{\theenumi}{(\alph{enumi})}
		\renewcommand{\labelenumi}{(\alph{enumi})}
		\setcounter{enumi}{2}
		\item {\rm (Nonzero Variance)}\label{assumption:nondegeneracy}
		$\Var(A(X_i))>0$.
		\item[] For a set $S\subset\mathbb{R}^p$, let ${\rm cl}(S)$ denote the closure of $S$ and let $\partial S$ denote the boundary of $S$, i.e., $\partial S={\rm cl}(S)\setminus{\rm int}(S)$.
		\item {\rm ($C^2$ Boundary of $\Omega^*$)}\label{assumption:boundary-partition}
		There exists a partition $\{\Omega^*_{1},...,\Omega^*_{M}\}$ of $\Omega^*=\{x\in \mathbb{R}^p: A(x)=1\}$ 
  such that
		\begin{enumerate}[label=(\roman*)] 
			\item ${\rm dist}(\Omega^*_m,\Omega^*_{m'})>0$ for any $m,m'\in\{1,...,M\}$ such that $m\neq m'$. Here ${\rm dist}(S,T)=\inf_{x\in S, y\in T}\|x-y\|$ is the distance between two sets $S$ and $T\subset \mathbb{R}^p$;
			\item $\Omega^*_m$ is nonempty, bounded, open, connected, and twice continuously differentiable for each $m\in \{1,...,M\}$. Here we say that a bounded open set $S\subset \mathbb{R}^p$ is {\it twice continuously differentiable} if for every $x\in S$, there exists a ball $B(x,\epsilon)$ and a one-to-one mapping $\psi$ from $B(x,\epsilon)$ onto an open set $D\subset \mathbb{R}^{p}$ such that $\psi$ and $\psi^{-1}$ are twice continuously differentiable, $\psi(B(x,\epsilon)\cap S)\subset \{(x_1,...,x_p)\in\mathbb{R}^p:x_p>0\}$, and $\psi(B(x,\epsilon)\cap \partial S)\subset \{(x_1,...,x_p)\in\mathbb{R}^p:x_p=0\}$.
		\end{enumerate}
				 
		\item {\rm (Regularity of Deterministic $A$)} \label{assumption:boundary-measure}
		\begin{enumerate}[label=(\roman*)]
		\item \label{assumption:p-1dim} ${\cal H}^{p-1}(\partial\Omega^*)<\infty$, and $\int_{\partial\Omega^*} f_X(x) d{\cal H}^{p-1}(x)>0$;
		\item \label{assumption:zero-set} There exists $\delta>0$ such that $A(x)=0$ for almost every $x\in N({\cal X}, \delta)\setminus \Omega^*$, where $N(S,\delta)=\{x\in \mathbb{R}^p: \|x-y\|< \delta \text{ for some $y\in S$}\}$ for a set $S\subset\mathbb{R}^p$ and $\delta>0$.
		\end{enumerate}
		\item {\rm (Conditional Moments and Density Near $\partial\Omega^*$)} \label{assumption:boundary-continuity}
		There exists $\delta>0$ such that
		\begin{enumerate}[label=(\roman*)]
			\item \label{assumption:continuous-differentiable} $E[Y_{1i}|X_i]$, $E[Y_{0i}|X_i]$, $E[D_i(1)|X_i]$, $E[D_i(0)|X_i]$ and $f_X$ are continuously differentiable and have bounded partial derivatives on $N(\partial\Omega^*,\delta)$;
			\item $E[Y_{1i}^2|X_i]$, $E[Y_{0i}^2|X_i]$, $E[Y_{1i}D_i(1)|X_i]$ and $E[Y_{0i}D_i(0)|X_i]$ are continuous on $N(\partial\Omega^*,\delta)$;
			\item $E[Y_i^{4}|X_i]$ is bounded on $N(\partial\Omega^*,\delta)$.
		\end{enumerate}
	\end{enumerate}
\end{assumption}

	Assumption \ref{assumption:consistency} is a set of conditions for establishing consistency.
	Assumption \ref{assumption:consistency} \ref{assumption:relevance} assumes that the weighted average effect of the algorithm's recommendation on the treatment assignment is nonzero.\footnote{The average is taken over the covariate values for which $p^A(x)$ is nondegenerate (i.e., $p^A(x)(1-p^A(x))\in (0,1)$).
	 There is a positive mass of such covariate values when $A$ is stochastic ($\Pr(A(X_i)\in (0,1))>0$). When $A$ is deterministic ($\Pr(A(X_i)\in (0,1))=0$), APS is nondegenerate only for boundary points at which the treatment recommendation changes from one to the other. Typically, the Lebesgue measure of the boundary is zero, so we compute the integral with respect to the $(p-1)$-dimensional Hausdorff measure instead.
	 }
	 	 Under this assumption, the estimated first-stage coefficient on $Z_i$ converges to a nonzero quantity.
	 
	 
	Assumptions \ref{assumption:consistency} \ref{assumption:nondegeneracy}--\ref{assumption:boundary-continuity} are a set of conditions we require  for proving consistency and asymptotic normality of $\hat\beta_1$ when $A$ is deterministic and produces only multidimensional regression-discontinuity variation. 
	Assumption \ref{assumption:consistency} \ref{assumption:nondegeneracy} says that $A$ produces variation in the treatment recommendation.
	
	Assumption \ref{assumption:consistency} \ref{assumption:boundary-partition} imposes the differentiability of the boundary of $\Omega^*=\{x\in \mathbb{R}^p: A(x)=1\}$.
	The conditions are satisfied if, for example, $\Omega^*=\{x\in \mathbb{R}^p:f(x)\ge 0\}$
	for some twice continuously differentiable function $f:\mathbb{R}^p\rightarrow \mathbb{R}$ such that $\nabla f(x)=(\frac{\partial f(x)}{\partial x_1},...,\frac{\partial f(x)}{\partial x_p})'\neq \bm{0}$ for all $x\in \mathbb{R}^p$ with $f(x)=0$.
	$\Omega^*$ takes this form when supervised learning based on smooth models such as lasso is used to construct a binary classifier $A$ such that $A(x)=1\{f(x)\ge 0\}$, $x\in\mathbb{R}^p$.
	
	In general, the differentiability of $\Omega^*$ may not hold.
	For example, if tree-based algorithms such as Classification And Regression Tree (CART) and random forests are used to construct a classifier $A(x)=1\{f(x)\ge 0\}$, then the function $f$ is not differentiable at some points. 
	Yet, the assumptions approximately hold in that $\Omega^*$ is arbitrarily well approximated by a set that satisfies the differentiability condition.
	
	Part \ref{assumption:p-1dim} of Assumption \ref{assumption:consistency} \ref{assumption:boundary-measure} says that the boundary of $\Omega^*$ is $(p-1)$ dimensional and that the boundary has nonzero density.\footnote{The boundary of $\Omega^*$ may fail to be $(p-1)$ dimensional in trivial cases where the Lebesgue measure of $\Omega^*$ is zero and hence $A(X_i)=0$ with probability one. For example, when the covariate space is three dimensional ($p=3$) and $\Omega^*$ is a straight line, not a set with nonzero volume nor even a plane, the boundary of $\Omega^*$ is the same as $\Omega^*$, and its two-dimensional Hausdorff measure is zero.}
	Part \ref{assumption:zero-set} puts a weak restriction on the values $A$ takes on outside the support of $X_i$.
	It requires that $A(x)=0$ for almost every $x\notin\Omega^*$ outside ${\cal X}$ but in the neighborhood of ${\cal X}$.
	$A(x)$ may take on any value if $x$ is not close to ${\cal X}$.
	Without this assumption, it is possible that $p^A(X_i;\delta_n)\in (0,1)$ and observation $i$ is included in the regression even if $X_i$ is not near the boundary of $\Omega^*$.\footnote{For example, suppose that ${\cal X}=[-1,1]$, $A(x)=1$ if $x>0$, $A(x)=0$ if $x\in [-1,0]$, and $A(x)=\frac{1}{2}$ if $x<-1$ (note that $\Pr(A(X_i)\in (0,1))=0$ in this case).
	In this case, for any sufficiently small $\delta>0$, $p^A(X_i;\delta)\in (0,1)$ if $X_i\in [-1,-1+\delta)$, but these observations should not be included in the regression as they are not close to the boundary of $\Omega^*=(0,\infty)$. Such a case is ruled out by Part \ref{assumption:zero-set} of Assumption \ref{assumption:consistency} \ref{assumption:boundary-measure}.}
	These conditions hold in practice. 
		Assumption \ref{assumption:consistency} \ref{assumption:boundary-continuity} imposes continuity, continuous differentiability, and boundedness on the conditional moments of potential outcomes and the probability density near the boundary of $\Omega^*$.
	Note that Part \ref{assumption:continuous-differentiable} of Assumption \ref{assumption:consistency} \ref{assumption:boundary-continuity} implies Assumption \ref{continuity}.

Under the above conditions and technical regularity Assumptions \ref{assumption:asymptotic-normality} and \ref{assumption:simulation} in Appendix \ref{appendix:assumptions}, the 2SLS estimators $\hat\beta_1$ and $\hat\beta_1^s$ are consistent and asymptotically normal estimators of a weighted average treatment effect.\footnote{Assumption \ref{assumption:asymptotic-normality} is a set of additional regularity conditions (such as the smoothness of $A$) for proving asymptotic normality of $\hat\beta_1$ when $A$ is stochastic ($\Pr(A(X_i)\in (0,1))>0$). Assumption \ref{assumption:simulation} imposes the condition on the growth rate of the number of simulation draws $S_n$, which we require for proving asymptotic normality of the simulation-based estimator $\hat\beta_1^s$.}

\begin{theorem}[Consistency and Asymptotic Normality]\label{theorem:2sls-consistency}
	Suppose that Assumptions \ref{assumption:A} and \ref{assumption:consistency} hold and  $\delta_n\rightarrow 0$, $n\delta_n\rightarrow \infty$, and $S_n\rightarrow \infty$ as $n\rightarrow \infty$.
	Then the 2SLS estimators $\hat \beta_1$ and $\hat\beta_1^s$ converge in probability to
	$$
	\beta_1\equiv \lim_{\delta \rightarrow 0} E[\omega_i(\delta)(Y_i(1)-Y_i(0))],
	$$
	where
	$$
	\omega_i(\delta)=\frac{p^{A}(X_i;\delta)(1-p^{A}(X_i;\delta))(D_i(1)-D_i(0))}{E[p^{A}(X_i;\delta)(1-p^{A}(X_i;\delta))(D_i(1)-D_i(0))]}.
	$$
	Suppose, in addition, that Assumptions \ref{assumption:asymptotic-normality} and \ref{assumption:simulation} in Appendix \ref{appendix:assumptions} hold and $n\delta_n^2\rightarrow 0$ as $n\rightarrow \infty$.
	Then
	\begin{align*}
	\hat\sigma^{-1}_n(\hat\beta_1-\beta_1)&\stackrel{d}{\longrightarrow} {\cal N}(0,1),\\
	(\hat\sigma^{s}_n)^{-1}(\hat\beta_1^s-\beta_1)&\stackrel{d}{\longrightarrow} {\cal N}(0,1),
	\end{align*}
	
		\noindent where we define $\hat\sigma^{-1}_n$ and $(\hat\sigma_n^s)^{-1}$ as follows. Let
	$$
	\hat{\mathbf{\Sigma}}_n = (\sum_{i=1}^n \mathbf{Z}_{i,n}\mathbf{D}_{i,n}'I_{i,n})^{-1}(\sum_{i=1}^n \hat \epsilon_{i,n}^2\mathbf{Z}_{i,n}\mathbf{Z}_{i,n}'I_{i,n})(\sum_{i=1}^n \mathbf{D}_{i,n}\mathbf{Z}_{i,n}'I_{i,n})^{-1},
	$$
	where
	$$
	\hat\epsilon_{i,n}=Y_i-\mathbf{D}_{i,n}'\hat\beta.
	$$
	$\hat{\mathbf{\Sigma}}_n$ is the conventional heteroskedasticity-robust estimator for the variance of the 2SLS estimator. 
	$\hat\sigma_n^2$ is the second diagonal element of $\hat{\mathbf{\Sigma}}_n$.
	$(\hat\sigma_n^s)^2$ is the analogously-defined estimator for the variance of $\hat\beta_1^s$ from the simulation-based regression.
\end{theorem}

\begin{proof}
	See Appendix \ref{proof:2sls-consistency}.
\end{proof}

Theorem \ref{theorem:2sls-consistency} says that the 2SLS estimators converge to the limit of a weighted average of causal effects for the subpopulation whose fixed-bandwidth APS is nondegenerate ($p^{A}(X_i;\delta)\in (0,1)$) and who would switch their treatment status in response to the treatment recommendation ($D_i(1)\neq D_i(0)$). The limit $\lim_{\delta \rightarrow 0} E[\omega_i(\delta)(Y_i(1)-Y_i(0))]$ always exists under the assumptions of Theorem \ref{theorem:2sls-consistency}. It is possible to estimate other weighted averages and the unweighted average by reweighting different observations appropriately. For example, we can estimate the unweighted average treatment effect by weighting observations by the inverse of fixed-bandwidth APS. Under monotonicity ($\Pr(D_i(1)\geq D_i(0)|X_i)=1$), we could also apply \cite{abadie2003semiparametric}’s Kappa weighting method using fixed-bandwidth APS instead of the standard propensity score to estimate other weighted averages of treatment effects for compliers (see also \cite{sloczynski2020should}).

Theorem \ref{theorem:2sls-consistency} also shows that inference based on the conventional 2SLS heteroskedasticity-robust standard errors is asymptotically valid if $\delta_n$ goes to zero at an appropriate rate. 
	The convergence rate of $\hat\beta_1$ is $O_p(1/\sqrt{n})$ if $\Pr(A(X_i)\in (0,1))>0$ and is $O_p(1/\sqrt{n\delta_n})$ if $\Pr(A(X_i)\in (0,1))=0$.

Our consistency result requires that $\delta_n$ go to zero slower than $n^{-1}$. 
The rate condition ensures that, when $\Pr(A(X_i)\in (0,1))=0$, we have sufficiently many observations in the $\delta_n$-neighborhood of the boundary of $\Omega^*$.
Importantly, the rate condition does not depend on the dimension of $X_i$, unlike other bandwidth-based estimation methods such as kernel methods.
This is because we use all the observations in the $\delta_n$-neighborhood of the boundary, and the number of those observations is of order $n\delta_n$ regardless of the dimension of $X_i$ if the dimension of the boundary is $p-1$, i.e., the dimension of $X_i$ minus one.
When $\Pr(A(X_i)\in (0,1))>0$, this rate condition is not necessary since the effective sample size always goes to infinity at the rate $n$ regardless of the value of $\delta_n$.

The asymptotic normality requires that $\delta_n$ go to zero sufficiently quickly so that $n\delta_n^2\rightarrow 0$.
When $\Pr(A(X_i)\in (0,1))>0$, we need to use a small enough $\delta_n$ so that $p^{A}(X_i;\delta_n)$ converges to $p^{A}(X_i)$ fast enough. 
When $\Pr(A(X_i)\in (0,1))=0$, the asymptotic normality is based on undersmoothing, which eliminates the asymptotic bias by using the observations sufficiently close to the boundary of $\Omega^*$.
In both cases, the bias of our estimator is $O(\delta_n)$. The standard deviation is $O(1/\sqrt{n})$ when $\Pr(A(X_i)\in (0,1))>0$ and is $O(1/\sqrt{n\delta_n})$ when $\Pr(A(X_i)\in (0,1))=0$.
The condition that $n\delta_n^2\rightarrow 0$ ensures that the bias converges to zero faster than the standard deviation in either case.
When $\Pr(A(X_i)\in (0,1))=0$, a weaker condition that $n\delta_n^3\rightarrow 0$ is sufficient for eliminating the asymptotic bias.\footnote{
When $\Pr(A(X_i)\in (0,1))=0$ and $n\delta_n^3$ goes to some nonzero constant, our estimator can have a nonzero asymptotic bias (see Appendix \ref{appendix:asymptotic-bias} for the asymptotic distribution).}
When $\Pr(A(X_i)\in (0,1))>0$, it is possible to relax the rate condition that $n\delta_n^2\rightarrow 0$ at the cost of strengthening the smoothness of $A$ in Assumption \ref{assumption:asymptotic-normality}.

Whether or not $\Pr(A(X_i)\in (0,1))=0$, when we use simulated fixed-bandwidth APS, the consistency result requires that the number of simulation draws $S_n$ go to infinity as $n$ increases.
The asymptotic normality result requires a sufficiently fast growth rate of $S_n$ (satisfying Assumption \ref{assumption:simulation}) to make the bias caused by using $p^s(X_i;\delta_n)$ negligible. 

Finally, note that the weight $\omega_i(\delta)$ given in Theorem \ref{theorem:2sls-consistency} is negative if $D_i(1)<D_i(0)$, so $E[\omega_i(\delta)(Y_i(1)-Y_i(0))]$ may not be a causally interpretable convex combination of treatment effects $Y_i(1)-Y_i(0)$.
This can happen because the treatment effect of those whose treatment assignment switches from 1 to 0 in response to the treatment recommendation (i.e., defiers) negatively contributes to $E[\omega_i(\delta)(Y_i(1)-Y_i(0))]$.
Additional assumptions prevent this problem.
If the treatment effect is constant, for example, the 2SLS estimators are consistent for the treatment effect.

\begin{corollary}\label{corollary:2sls-constant}
	Suppose that Assumptions \ref{assumption:A} and \ref{assumption:consistency} hold, that the treatment effect is constant, i.e., $Y_i(1)-Y_i(0)=b$ for some constant $b$, and that $\delta_n\rightarrow 0$, $n\delta_n\rightarrow \infty$, and $S_n\rightarrow \infty$ as $n\rightarrow \infty$.
	Then the 2SLS estimators $\hat \beta_1$ and $\hat\beta_1^s$ converge in probability to $b$.
\end{corollary}

Another approach is to impose monotonicity \citep{imbens1994}.
Let $LATE(x)=E[Y_i(1)-Y_i(0)|D_i(1)\neq D_i(0),X_i=x]$ be the local average treatment effect (LATE) conditional on $X_i=x$.

\begin{corollary}\label{corollary:2sls-late}
	Suppose that Assumptions \ref{assumption:A} and \ref{assumption:consistency} hold, that $\Pr(D_i(1)\ge D_i(0)|X_i=x)=1$ for any $x\in {\cal X}$ with $p^{A}(x)\in (0,1)$ (monotonicity), and that $\delta_n\rightarrow 0$, $n\delta_n\rightarrow \infty$, and $S_n\rightarrow \infty$ as $n\rightarrow \infty$.
	Then the 2SLS estimators $\hat \beta_1$ and $\hat\beta_1^s$ converge in probability to
	$$
	\lim_{\delta \rightarrow 0}E[\omega(X_i;\delta)LATE(X_i)],
	$$
	where
	$$
	\omega(x;\delta)=\frac{p^{A}(x;\delta)(1-p^{A}(x;\delta))E[D_i(1)-D_i(0)|X_i=x]}{E[p^{A}(X_i;\delta)(1-p^{A}(X_i;\delta))(D_i(1)-D_i(0))]}.
	$$
\end{corollary}


\subsection{Special Cases}\label{section:consistency-special-cases}

Theorem \ref{theorem:2sls-consistency} holds whether $A$ is stochastic ($\Pr(A(X_i)\in (0,1))>0$) or deterministic ($\Pr(A(X_i)\in (0,1))=0$). 
If we consider these two underlying cases separately, the probability limit of the 2SLS estimators has a more specific expression, as shown in the proof of Theorem \ref{theorem:2sls-consistency} in Appendix \ref{proof:2sls-consistency}.
If $\Pr(A(X_i)\in (0,1))>0$,
\begin{align}
\plim\hat\beta_1 =\plim\hat\beta_1^s=\frac{E[A(X_i)(1-A(X_i))(D_i(1)-D_i(0))(Y_i(1)-Y_i(0))]}{E[A(X_i)(1-A(X_i))(D_i(1)-D_i(0))]}.\label{eq:plim-stochastic}
\end{align}
The 2SLS estimators converge to a weighted average of treatment effects for the subpopulation with nondegenerate $A(X_i)$.

To relate this result to existing work, consider the following 2SLS regression with the (standard) propensity score $A(X_i)$ control:
\begin{align}
	D_i&=\gamma_0+\gamma_1 Z_i+\gamma_2 A(X_i)+\nu_i \label{eq:pscore-first-stage}\\
	Y_i&=\beta_0+\beta_1 D_i+\beta_2 A(X_i)+\epsilon_i \label{eq:pscore-main}.
\end{align}
Under conditional independence, the 2SLS estimator from this regression converges in probability to the treatment-variance weighted average of treatment effects in (\ref{eq:plim-stochastic}) \citep{hull}. 
Not surprisingly, for this selection-on-observables case, our result shows that the 2SLS estimator is consistent for the same treatment effect whether we control for the propensity score, fixed-bandwidth APS, or simulated fixed-bandwidth APS.

Importantly, using fixed-bandwidth APS as a control allows us to consistently estimate a causal effect even if $A$ is deterministic and produces multidimensional regression-discontinuity variation. 
If $\Pr(A(X_i)\in (0,1))=0$,
\begin{align}
\plim\hat\beta_1 =\plim\hat\beta_1^s 
=\frac{\int_{\partial\Omega^*}E[(D_i(1)-D_i(0))(Y_i(1)-Y_i(0))|X_i=x]f_X(x)d{\cal H}^{p-1}(x)}{\int_{\partial\Omega^*}E[D_i(1)-D_i(0)|X_i=x]f_X(x)d{\cal H}^{p-1}(x)}.\label{eq:plim-beta-boundary}
\end{align}
The 2SLS estimators converge to a weighted average of treatment effects for the subpopulation who are on the boundary of the treated region.\footnote{We prove this result using techniques from differential geometry and geometric measure theory. Our approach using geometric theory shows that $\hat\beta_1$ converges to an integral of the conditional treatment effect over boundary points with respect to the Hausdorff measure.
In contrast, prior studies on multidimensional RDDs express treatment effect estimands in terms of expectations conditional on $X_i$ being in the boundary like $E[Y_{1i}-Y_{0i}|X_i\in \partial\Omega^*]$ \citep{zajonc2012regression}.
However, those conditional expectations are, formally, not well-defined, since ${\cal L}^p(\partial\Omega^*)=0$ and hence $\Pr(X_i\in \partial\Omega^*)=0$.
We therefore prefer our expression in terms of an integral with respect to the Hausdorff measure.} 

The estimand in (\ref{eq:plim-beta-boundary}) nests parameters considered in the Regression Discontinuity (RD) literature.
Under the monotonicity condition in Corollary \ref{corollary:2sls-late}, the estimand in (\ref{eq:plim-beta-boundary}) equals
    \begin{align*}
        \frac{\int_{\partial\Omega^*}LATE(x)E[D_i(1)-D_i(0)|X_i=x]f_X(x)d{\cal H}^{p-1}(x)}{\int_{\partial\Omega^*}E[D_i(1)-D_i(0)|X_i=x]f_X(x)d{\cal H}^{p-1}(x)}.
    \end{align*}
    This estimand can be interpreted as the average treatment effect for the subpopulation of compliers who are on the boundary \citep{zajonc2012regression}.
    In the univariate RDD with a single cutoff, this estimand further reduces to the average treatment effect for the compliers at the cutoff, the standard parameter in the fuzzy RDD \citep{hahn2001RDD}.
If we instead assume $\Pr(D_i(1)>D_i(0)|X_i=x)=1$ (perfect compliance) for all $x\in \partial\Omega^*$, the estimand in (\ref{eq:plim-beta-boundary}) equals
    \begin{align*}
        \frac{\int_{\partial\Omega^*}E[Y_i(1)-Y_i(0)|X_i=x]f_X(x)d{\cal H}^{p-1}(x)}{\int_{\partial\Omega^*}f_X(x)d{\cal H}^{p-1}(x)}.
    \end{align*}
    This estimand represents the average treatment effect for the subpopulation who are on the boundary \citep{zajonc2012regression,keele2015geographic}.

\subsection{Comparison with Existing Approaches}
    To compare our method with existing ones, first consider the univariate RDD with a single cutoff $c$. In this special case, $p^{A}(X_i;\delta_n)=\frac{X_i-c}{2\delta_n}+\frac{1}{2}$ if $X_i\in[c-\delta_n,c+\delta_n]$ and $p^{A}(X_i;\delta_n)\in\{0,1\}$ otherwise.
    Therefore, the estimator $\hat\beta_1$ from the 2SLS regression (\ref{2SLSestimand1}) and (\ref{2SLSestimand2}) is numerically equivalent to a version of the regression discontinuity (RD) local linear estimator \citep{hahn2001RDD} that uses a box kernel and places the same slope coefficient of $X_i$ on both sides of the cutoff.\footnote{It is possible to allow for slope changes at the cutoff by viewing $p^{A}(X_i;\delta_n)$ as a running variable with cutoff $\frac{1}{2}$ and applying standard RD local linear estimators (i.e., adding interaction terms $Z_i(p^{A}(X_i;\delta_n)-\frac{1}{2})$ and $D_i(p^{A}(X_i;\delta_n)-\frac{1}{2})$ to (\ref{2SLSestimand1}) and (\ref{2SLSestimand2}), respectively).
    Under twice continuous differentiability of $E[Y_{1i}|X_i]$ and $E[Y_{0i}|X_i]$ near the cutoff (which is stronger than our smoothness assumption), this estimator achieves the optimal rate of $n^{-2/5}$.
	For the multidimensional RDD with a nonlinear boundary, $Z_i$ is not a deterministic function of $p^A(X_i;\delta_n)$. In this case, it is not straightforward to use $p^A(X_i;\delta_n)$ as a single running variable, since no appropriate cutoff value exists. 
	We leave to future research how to allow for more flexible 2SLS specifications in the general multidimensional setting.}
    When we use the bandwidth $\delta_n$ that converges to zero at the $n^{-1/3}$ rate instead of undersmoothing, our estimator achieves a convergence rate of $n^{-1/3}$. This rate is optimal for the estimation of the conditional LATE at the cutoff under our smoothness condition (continuous differentiability of $E[Y_{1i}|X_i]$ and $E[Y_{0i}|X_i]$ near the cutoff) in Assumption \ref{assumption:consistency} \ref{assumption:boundary-continuity}.\footnote{Note that the continuous differentiability of $E[Y_{1i}|X_i]$ and $E[Y_{0i}|X_i]$ near the cutoff implies that they have bounded derivatives near the cutoff, so the latter condition can be omitted from Assumption \ref{assumption:consistency} \ref{assumption:boundary-continuity} for the univariate case.}

    Our approach is particularly useful in more general scenarios.
    One such scenario is the multidimensional sharp or fuzzy RDD when the boundary is complex or its analytical form is unknown. Our estimator is computationally feasible for any decision boundary as long as we can simulate the underlying algorithm. More importantly, our estimator is shown to be consistent and asymptotically normal.
    As far as we know, there appear to be no existing estimators that are computationally feasible and have theoretical validity for a general class of multidimensional RDDs.
    Moreover, our method is applicable to a more general setting that mixes stochastic and deterministic algorithms. We illustrate such a case in the next section.    

\section{Monte Carlo Simulation}\label{section:sim}

This section assesses the feasibility and performance of our method. We do so through a Monte Carlo experiment motivated by decision making by machine learning with high-dimensional data. 
Consider a government or tech company that applies a machine-learning-based deterministic decision algorithm to a large segment of the population. At the same time, they conduct a randomized controlled trial (RCT) using the rest of the population. They are interested in estimating treatment effects using data from both segments.
Our approach offers a way of exploiting not only the RCT segment but also the deterministic algorithm segment.
Even if we focus on the deterministic algorithm segment, it is not straightforward to apply existing RDD methods to our simulation setup since the decision boundary is high dimensional and complex.
We demonstrate the applicability of our approach in such a setup.

We simulate $1,000$ hypothetical samples from the following data-generating process.
Each sample $\{(Y_i,X_i,D_i,Z_i)\}_{i=1}^n$ is of size $n=10,000$. 
There are $100$ covariates ($p=100$), and $X_i\sim {\cal N}(\bm{0},\mathbf{\Sigma})$. 
$Y_i(0)$ is generated as $Y_i(0)=0.75X_i'\alpha_0+0.25 \epsilon_{0i}$, where $\alpha_0\in\mathbb{R}^{100}$,
and $\epsilon_{0i}\sim {\cal N}(0,1)$.
We consider two models for $Y_i(1)$, one in which the treatment effect $Y_i(1)-Y_i(0)$ does not depend on $X_i$ and one in which the treatment effect depends on $X_i$.
\begin{enumerate}[label={}]
	\item Model A. $Y_i(1)=Y_i(0)+\epsilon_{1i}$, where $\epsilon_{1i}\sim {\cal N}(0,1)$.
	\item Model B. $Y_i(1)=Y_i(0)+X_i'\alpha_1$, where $\alpha_1\in\mathbb{R}^{100}$.
\end{enumerate}
The choice of parameters $\mathbf{\Sigma}$, $\alpha_0$ and $\alpha_1$ is explained in Appendix \ref{appendix:simulation}.
$D_i(0)$ and $D_i(1)$ are generated as $D_i(0)=0$ and $D_i(1)=1\{Y_i(1)-Y_i(0)>u_i\}$, where $u_i\sim {\cal N}(0,1)$.

To generate $Z_i$, let $q_{0.495}$ and $q_{0.505}$ be the $49.5$th and $50.5$th (empirical) quantiles of the first covariate $X_{i1}$. 
Let $\tau_{pred}(X_i)$ be a real-valued function of $X_i$, 
which is constructed by random forests using an independent sample (see Appendix \ref{appendix:simulation} for the details).
$Z_i$ is then generated as
\begin{align*}
Z_i=\begin{cases}
Z_i^*\sim {\rm Bernoulli}(0.5) & \text{if $X_{i1}\in [q_{0.495}, q_{0.505}]$}\\
1 & \text{if $X_{i1}\notin [q_{0.495}, q_{0.505}]$ and $\tau_{pred}(X_i)\ge 0$}\\
0 & \text{if $X_{i1}\notin [q_{0.495}, q_{0.505}]$ and $\tau_{pred}(X_i)< 0$}.
\end{cases}
\end{align*} 
The first case corresponds to the RCT segment while the latter two cases to the deterministic algorithm segment. 
The algorithm function $A$ is given by
\begin{align*}
A(x)=\begin{cases}
0.5 & \text{if $x_1\in [q_{0.495}, q_{0.505}]$}\\
1 & \text{if $x_1\notin [q_{0.495}, q_{0.505}]$ and $\tau_{pred}(x)\ge 0$}\\
0 & \text{if $x_1\notin [q_{0.495}, q_{0.505}]$ and $\tau_{pred}(x)< 0$}.
\end{cases}
\end{align*}
Finally, $D_i$ and $Y_i$ are generated as $D_i=Z_iD_i(1)+(1-Z_i)D_i(0)$ and $Y_i=D_iY_i(1)+(1-D_i)Y_i(0)$, respectively.

{\bf Estimands and Estimators.}
We consider four parameters as target estimands: ${\rm ATE}\equiv E[Y_i(1)-Y_i(0)]$; ${\rm ATE(RCT)}\equiv E[Y_i(1)-Y_i(0)|X_{i1}\in [q_{0.495}, q_{0.505}]]$; ${\rm LATE}\equiv E[Y_i(1)-Y_i(0)|D_i(1)\neq D_i(0)]$; and 
${\rm LATE(RCT)}\equiv E[Y_i(1)-Y_i(0)|D_i(1)\neq D_i(0),X_{i1}\in [q_{0.495}, q_{0.505}]]$.
In the case where the treatment effect does not depend on $X_i$ (Model A), ATE and LATE are the same as ATE(RCT) and LATE(RCT), respectively.
In the case where the treatment effect depends on $X_i$ (Model B), the conditional effects are heterogeneous.
However, since the RCT segment consists of those in the middle of the distribution of $X_{i1}$, the average effect for the RCT segment is close to the unconditional average effect.
As a result, ATE is similar to ATE(RCT), and LATE is similar to LATE(RCT).


We use the data $\{(Y_i,X_i,D_i,Z_i)\}_{i=1}^n$ to estimate the treatment effect parameters.
Our main approach is 2SLS with fixed-bandwidth APS controls in Theorem \ref{theorem:2sls-consistency}. To compute fixed-bandwidth APS, we use $S=400$ simulation draws for each observation. 

We compare our approach with two naive alternatives. 
The first alternative is OLS of $Y_i$ on a constant and $D_i$ (i.e., the difference in the sample mean of $Y_i$ between the treated group and untreated group) using all observations. 
The second alternative is 2SLS with $A(X_i)$ controls. 
This method uses the observations with $A(X_i)\in (0,1)$ to run the 2SLS regression of $Y_i$ on $D_i$ and $A(X_i)$ using $Z_i$ as an instrument for $D_i$ (see (\ref{eq:pscore-first-stage}) and (\ref{eq:pscore-main}) in Section \ref{section:consistency-special-cases}) and reports the coefficient on $D_i$.

For both models, the 2SLS estimator converges in probability to ${\rm LATE(RCT)}$ (equivalently, the right-hand side of equation (\ref{eq:plim-stochastic})) whether we control for fixed-bandwidth APS or $A(X_i)$. 
However, 2SLS with $A(X_i)$ controls uses only the RCT segment while 2SLS with fixed-bandwidth APS controls additionally uses the individuals near the decision boundary of the deterministic algorithm (i.e., the boundary of the region for which $\tau_{pred}(x)\ge 0$).
Therefore, 2SLS with fixed-bandwidth APS controls is expected to produce a more precise estimate than 2SLS with $A(X_i)$ controls if the conditional effects for those near the boundary are not far from the target estimand.

We do not apply any multidimensional RD estimators as alternatives, since there appear to be no existing RD estimators applicable to this setup. It is hard to apply distance-based RD methods since it is difficult to compute the distance from each $X_i$ to the high-dimensional random-forests decision boundary.
An alternative is to use the individual's predicted effect $\tau_{pred}(X_i)$ as a univariate running variable. However, $\tau_{pred}(X_i)$ may not be a continuous variable since $\tau_{pred}$ is constructed by tree-based methods.

{\bf Performance Measures.}
2SLS with a single instrument has no moments, so we cannot consider the bias, standard deviation, or mean squared error. As an alternative, we calculate the median bias, ${\rm med}(\hat\theta)-\theta$, median absolute deviation from the median, ${\rm med}(|\hat\theta-{\rm med}(\hat\theta)|)$, and median absolute error, ${\rm med}(|\hat\theta-\theta|)$, where $\theta$ and $\hat\theta$ denote the estimand and estimator, respectively. 

{\bf Results.}
Table \ref{table:simulation_rf} reports the median bias, median absolute deviation from the median (a measure of dispersion), and median absolute error (an overall performance measure). 
Panels A and B present the results for the cases where the conditional effects are homogeneous and heterogeneous, respectively.
OLS with no controls is significantly biased, showing the importance of correcting for omitted variable bias.
2SLS with fixed-bandwidth APS controls achieves this goal, as demonstrated by its smaller biases across models, target parameters, and small values of the bandwidth $\delta$.

2SLS with fixed-bandwidth APS controls shows a consistent pattern; as the bandwidth $\delta$ grows, the bias increases while the absolute deviation from the median declines. 
For several values of $\delta$, 2SLS with fixed-bandwidth APS controls outperforms 2SLS with $A(X_i)$ controls in terms of the median absolute error. 
This finding implies that exploiting individuals near the multidimensional decision boundary of the deterministic algorithm can lead to better performance than using only the RCT segment.

We also evaluate our inference procedure based on Theorem \ref{theorem:2sls-consistency}. Table \ref{table:simulation_rf} reports the coverage probabilities of the 95\% confidence intervals for LATE(RCT) constructed from the estimates and their heteroskedasticity-robust standard errors. 
The confidence intervals for 2SLS offer nearly correct coverage when $\delta$ is small, which supports the implication of Theorem \ref{theorem:2sls-consistency} that the inference procedure is valid when we use a sufficiently small $\delta$. Overall, Table \ref{table:simulation_rf} shows that our estimator works well in this high-dimensional setting and performs better than alternative estimators. 


\begin{table}[bp] \centering
\newcolumntype{C}{>{\centering\arraybackslash}X}

\caption{Comparison of Estimators and Coverage of Confidence Intervals}
{\scriptsize
\begin{tabularx}{\textwidth}{lCCCCCCCC}

\toprule
{}&{}&{}&\multicolumn{6}{c}{Our Method: 2SLS with Approximate Propensity Score Controls} \tabularnewline {}&{OLS}&{2SLS}&\tabularnewline[-2.5ex] \cline{4-9} \tabularnewline
{}&{with No Controls}&{with $A(X_i)$ Controls}&{$\delta = 0.01$}&{$\delta = 0.05$}&{$\delta = 0.1$}&{$\delta = 0.25$}&{$\delta = 0.5$}&{$\delta = 1$} \tabularnewline
\midrule\addlinespace[1.5ex]
\multicolumn{9}{c}{Panel A: Homogeneous Conditional Effects (Model A)} \tabularnewline
\addlinespace[1.5ex]
\multicolumn{9}{l}{Estimand: ATE = 0} \tabularnewline
\addlinespace[0.75ex]
~~Median Bias&0.663&0.562&0.558&0.618&0.652&0.716&0.811&0.965 \tabularnewline
~~Median Absolute Error&0.663&0.562&0.558&0.618&0.652&0.716&0.811&0.965 \tabularnewline
\addlinespace[0.75ex]
\multicolumn{9}{l}{Estimand: ATE(RCT) = $-$0.001} \tabularnewline
\addlinespace[0.75ex]
~~Median Bias&0.663&0.562&0.559&0.618&0.653&0.717&0.811&0.966 \tabularnewline
~~Median Absolute Error&0.663&0.563&0.559&0.618&0.653&0.717&0.811&0.966 \tabularnewline
\addlinespace[0.75ex]
\multicolumn{9}{l}{Estimand: LATE = 0.564} \tabularnewline
\addlinespace[0.75ex]
~~Median Bias&0.098&$-$0.002&$-$0.006&0.053&0.088&0.152&0.246&0.401 \tabularnewline
~~Median Absolute Error&0.098&0.222&0.129&0.084&0.091&0.152&0.246&0.401 \tabularnewline
\addlinespace[0.75ex]
\multicolumn{9}{l}{Estimand: LATE(RCT) = 0.566} \tabularnewline
\addlinespace[0.75ex]
~~Median Bias&0.096&$-$0.005&$-$0.008&0.051&0.086&0.150&0.244&0.399 \tabularnewline
~~Median Absolute Error&0.096&0.223&0.129&0.083&0.089&0.150&0.244&0.399 \tabularnewline
\addlinespace[0.75ex]
Med. Abs. Deviation from Median&0.014&0.224&0.128&0.075&0.062&0.045&0.041&0.041 \tabularnewline
\addlinespace[0.75ex]
Coverage&0.4\%&95.2\%&94.4\%&92.7\%&84.0\%&46.0\%&3.1\%&0.0\% \tabularnewline
\addlinespace[0.75ex]
Avg N&10000&100&397&1175&1722&2613&3349&3994 \tabularnewline\tabularnewline
\multicolumn{9}{c}{Panel B: Heterogeneous Conditional Effects (Model B)} \tabularnewline
\addlinespace[1.5ex]
\multicolumn{9}{l}{Estimand: ATE = 0} \tabularnewline
\addlinespace[0.75ex]
~~Median Bias&1.010&0.561&0.471&0.496&0.525&0.590&0.698&0.882 \tabularnewline
~~Median Absolute Error&1.010&0.566&0.471&0.496&0.525&0.590&0.698&0.882 \tabularnewline
\addlinespace[0.75ex]
\multicolumn{9}{l}{Estimand: ATE(RCT) = $-$0.004} \tabularnewline
\addlinespace[0.75ex]
~~Median Bias&1.015&0.566&0.475&0.500&0.529&0.595&0.702&0.887 \tabularnewline
~~Median Absolute Error&1.015&0.571&0.475&0.500&0.529&0.595&0.702&0.887 \tabularnewline
\addlinespace[0.75ex]
\multicolumn{9}{l}{Estimand: LATE = 0.564} \tabularnewline
\addlinespace[0.75ex]
~~Median Bias&0.446&$-$0.002&$-$0.093&$-$0.068&$-$0.039&0.026&0.134&0.318 \tabularnewline
~~Median Absolute Error&0.446&0.262&0.163&0.099&0.073&0.057&0.134&0.318 \tabularnewline
\addlinespace[0.75ex]
\multicolumn{9}{l}{Estimand: LATE(RCT) = 0.559} \tabularnewline
\addlinespace[0.75ex]
~~Median Bias&0.451&0.002&$-$0.089&$-$0.063&$-$0.035&0.031&0.138&0.323 \tabularnewline
~~Median Absolute Error&0.451&0.265&0.162&0.097&0.074&0.057&0.138&0.323 \tabularnewline
\addlinespace[0.75ex]
Med. Abs. Deviation from Median&0.012&0.264&0.153&0.084&0.071&0.052&0.045&0.044 \tabularnewline
\addlinespace[0.75ex]
Coverage&0.0\%&94.6\%&92.4\%&91.7\%&94.1\%&93.5\%&51.8\%&0.3\% \tabularnewline
\addlinespace[0.75ex]
Avg N&10000&100&397&1175&1722&2613&3349&3994 \tabularnewline
\bottomrule 

\end{tabularx}
\caption*{
\scriptsize {\it Note}: This table shows the median bias, median of absolute errors, and median of absolute deviations from the median of OLS with no controls, 2SLS with $A(X_i)$ controls, and 2SLS with Approximate Propensity Score controls. These statistics are computed with the estimand set to ATE, ATE(RCT), LATE, or LATE(RCT). The ``Coverage'' row  in each panel shows the probabilities that the 95\% confidence intervals of the form $[\hat\beta_1^{s}-1.96\hat\sigma_n^{s},\hat\beta_1^{s}+1.96\hat\sigma_n^{s}]$ contains LATE(RCT), where $\hat\beta_1^{s}$ is the estimate and $\hat\sigma_n^{s}$ is its heteroskedasticity-robust standard error. We use 1,000 replications of a size 10,000 simulated sample to compute these statistics. We use several possible values of $\delta$ to compute the Approximate Propensity Score. All Approximate Propensity Scores are computed by averaging 400 simulation draws of $A(X_i)$. Panel A reports the results under the model in which the treatment effect does not depend on $X_i$ (Model A). Panel B reports the results under the model in which the treatment effect depends on $X_i$ (Model B). The bottom row  in each panel shows the average number of observations used for estimation (i.e., the average number of observations for which the Approximate Propensity Score or $A(X_i)$ is strictly between 0 and 1).}
}
\label{table:simulation_rf}
\end{table}


\section{Empirical Policy Application}\label{section:empirical}

\subsection{Hospital Relief Funding during the Pandemic}

The COVID-19 pandemic has afflicted millions of people across the country and has imposed historic challenges for hospitals and the health system. The pandemic led to revenue losses coupled with skyrocketing expenses, pushing many already overburdened hospitals further to their financial brink.

To deal with this crisis, as part of the 3-phase Coronavirus Aid, Relief, and Economic Security (CARES) Act, the US government distributed tens of billions of dollars of relief funding to hospitals since April 2020. This funding intended to help health care providers hit hardest by the COVID-19 outbreak. 
The bill specified that providers may (but are not required to) use the funds for COVID-19-related expenses such as construction of temporary structures, purchasing medical supplies and equipment (including personal protective equipment and testing supplies), increased workforce utilization and training, establishing emergency operation centers, retrofitting facilities, and managing the surge in capacity. 


We ask whether this funding had a causal impact on hospital operation and activities in dealing with COVID-19 patients. 
Answering this question would help the government design better funding policies to respond to future healthcare crises.
We focus on an initial portion of this funding (\$10 billion). This portion was allocated to hospitals that qualified as ``safety net hospitals'' according to a specific eligibility criterion. This eligibility criterion intends to direct funding towards hospitals that ``\textit{disproportionately provide care to the most vulnerable, and operate on thin margins}.'' Specifically, an acute care hospital is deemed eligible for funding if the following conditions hold:
\begin{itemize}
    \item Medicare Disproportionate Patient Percentage (DPP) of 20.2\% or greater. DPP is equal to the sum of (1) the percentage of Medicare inpatient days attributable to patients eligible for both Medicare Part A and Supplemental Security Income (SSI), and (2) the percentage of total inpatient days attributable to patients eligible for Medicaid but not Medicare Part A.
    \item Annual Uncompensated Care (UCC) of at least $\$25,000$ per bed. UCC is a measure of hospital care provided for which no payment was received from the patient or insurer. It is the sum of a hospital's bad debt and the financial assistance it provides.
    \item Profit Margin (net income$/$(net patient revenue $+$ total other income)) of 3.0\% or less. 
\end{itemize}

\begin{figure}[bp]
    \centering
    \caption{Regression Discontinuity in Hospital Funding Eligibility}
    \includegraphics[height=7in]{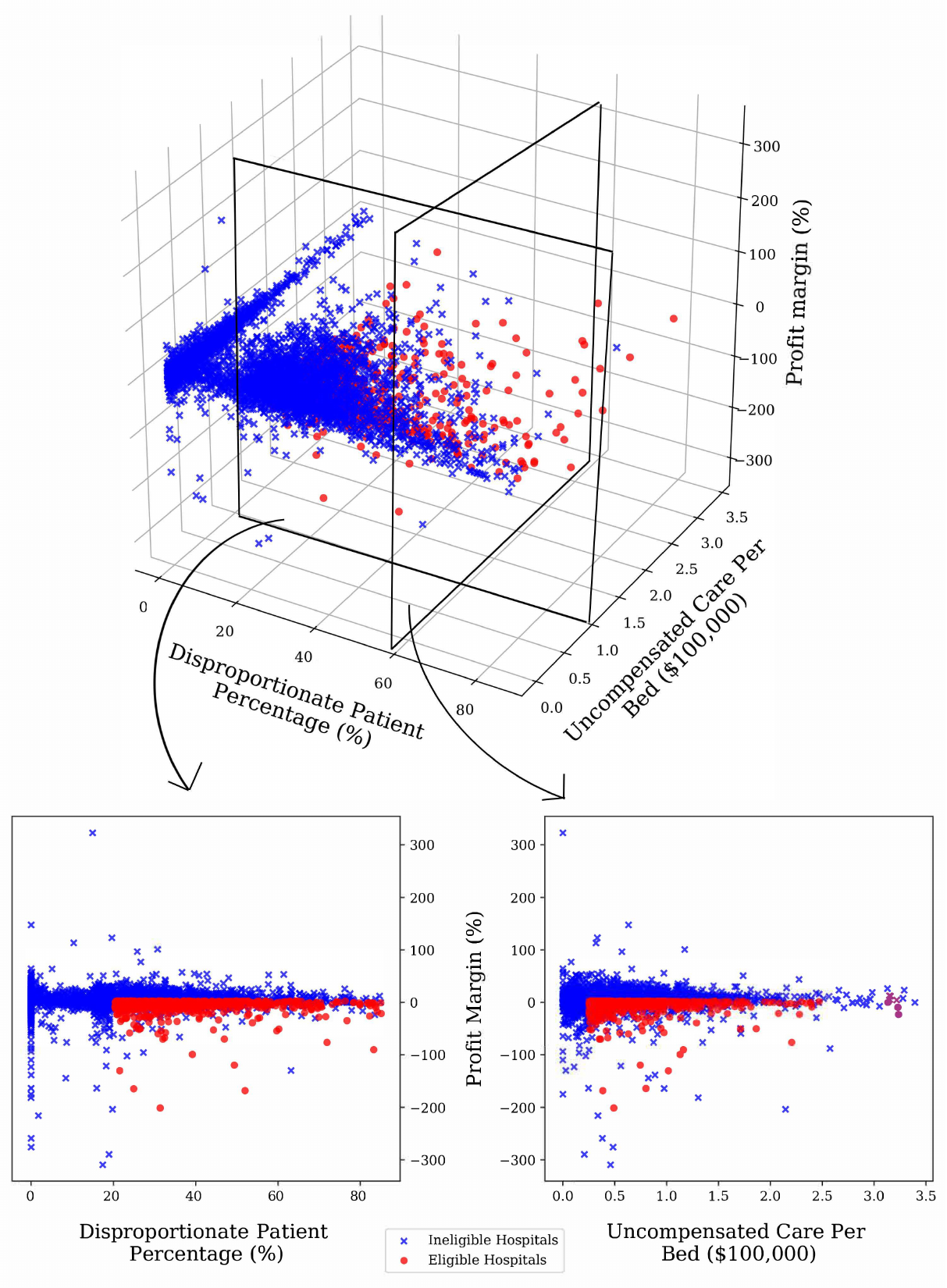}
     \caption*{\scriptsize{\it Note}: The top figure visualizes the three hospital characteristics that determine funding eligibility. The bottom figures show the data points plotted along 2 out of 3 dimensions. The bottom left panel plots disproportionate patient percentage against profit margin, while the bottom right panel plots uncompensated care per bed against profit margin. We remove hospitals above the 99th percentile of disproportionate patient percentage and uncompensated care per bed, for visibility purposes.}
     \label{fig:rdd}
\end{figure}

Hospitals that do not qualify on any of the three dimensions are funding ineligible.  Figure \ref{fig:rdd} visualizes how the three dimensions determine funding eligibility. 
From the original space of the three eligibility determinants, we extract two-dimensional planes to better visualize the structure of quasi-experimental variation. 
As the bottom two-dimensional planes show, eligibility discontinuously changes as hospitals cross the eligibility boundary in the characteristic space. 
This setting is a three-dimensional RDD, falling under our framework. 

Our treatment is the funding amount, which is calculated as follows. Each eligible hospital is assigned a facility score, which is calculated as the product of DPP and the number of beds in that hospital. This facility score determines the share of funding allocated to the hospital, out of the total \$10 billion. The share received by each hospital is determined by the ratio of the hospital's facility score to the sum of facility scores across all eligible hospitals. The amount of funding that can be received by a hospital is bounded below at \$5 million and capped above at \$50 million.
We compute the funding eligibility status as well as the amount of funding received, by using data from the Healthcare Cost Report Information System (HCRIS) for the 2018 financial year.\footnote{We use the methodology detailed in the \href{https://www.hhs.gov/coronavirus/cares-act-provider-relief-fund/general-information/index.html}{CARES Act website} to project funding based on 2018 financial year cost reports. We use the RAND cleaned version of the dataset \citep{rand2018}.
See Appendix \ref{appendix:empirics} for details on the construction of our dataset.} 

A majority of eligible hospitals receive the minimum amount of \$5 million. A small mass of hospitals receive amounts close to the maximum of \$50 million. Figure \ref{fig:dist_funding} in Appendix \ref{appendix:additional_empirics} shows the distribution of funding amounts received by eligible hospitals.

\begin{table}[!t] \centering
\newcolumntype{C}{>{\centering\arraybackslash}X}
\caption{Hospital Characteristics and Outcomes}
{\scriptsize
\begin{tabularx}{\textwidth}{lCCCC}

\hline\tabularnewline[-1.8ex]
{}&{All}&{Ineligible}&{Eligible}&{Hospitals w/} \\
{}&{}&{Hospitals}&{Hospitals}&{APS $\in$ (0,1)} \\
\midrule\addlinespace[1.5ex]
\multicolumn{5}{c}{Panel A: Outcome Variable Means} \\ \tabularnewline[-2.0ex]
\# Confirmed/Suspected COVID Patients&105.59&98.41&136.61&123.83 \tabularnewline
\# Confirmed COVID Patients&80.10&73.86&107.83&85.53 \tabularnewline
\# Confirmed/Suspected COVID Patients in ICU&31.37&28.92&42.10&36.33 \tabularnewline
\# Confirmed COVID Patients in ICU&26.62&24.41&36.56&30.89 \tabularnewline
N&4,008&3,293&715&438 \tabularnewline\tabularnewline
\multicolumn{5}{c}{Panel B: Hospital Characteristics Means} \tabularnewline\tabularnewline[-2.0ex]
Beds&143.66&134.60&188.35&205.30 \tabularnewline
Interns and residents (full-time equivalents) per bed&.06&.05&.11&.09 \tabularnewline
Adult and pediatric hospital beds&120.26&113.29&154.66&169.64 \tabularnewline
Ownership: Proprietary (for-profit)&.19&.20&.18&.16 \tabularnewline
Ownership: Governmental&.22&.22&.23&.16 \tabularnewline
Ownership: Voluntary (non-profit)&.58&.58&.59&.68 \tabularnewline
Inpatient length of stay&9.21&10.14&4.66&4.37 \tabularnewline
Employees on payroll (full-time equivalents)&973.90&897.31&1351.57&1511.87 \tabularnewline
Disproportionate patient percentage&.21&.18&.38&.36 \tabularnewline
Uncompensated care per bed (\$) &59,850.00&56,556.03&76,096.31&45,575.28 \tabularnewline
Profit margin&.02&.04&$-$.07&$-$.03 \tabularnewline
N&4,633&3,852&781&494 \tabularnewline
\bottomrule 
\end{tabularx}
\caption*{
\scriptsize {\it Note}: This table reports averages of outcome variables and hospital characteristics by funding eligibility. 
Panel A reports the outcome variable means. Outcome variable estimates are 7 day sums for the week spanning July 31st 2020 to August 6th 2020.  Confirmed or Suspected COVID patients refer to the sum of patients in inpatient beds with lab-confirmed/suspected COVID. Confirmed COVID patients refer to the sum of patients in inpatient beds with lab-confirmed COVID, including those with both lab-confirmed COVID and influenza. Inpatient bed totals also include observation beds. Similarly, Confirmed/Suspected COVID patients in ICU refer to the sum of patients in ICU beds with lab-confirmed or suspected COVID. Confirmed COVID patients in ICU refers to the sum of patients in ICU beds with lab-confirmed COVID, including those with both lab-confirmed COVID and influenza. Panel B reports the means for hospital characteristics for the financial year 2018. Column 1 shows the means for all hospitals. Columns 2 and 3 show the means for hospitals that are ineligible and eligible to receive funding, respectively. Column 4 shows the means for the hospitals with nondegenerate Approximate Propensity Score with bandwidth $\delta=0.05$. Approximate Propensity Score is computed by averaging 10,000 simulation draws. 
}
}
\label{tbl:descriptives}
\end{table}

Our outcomes are a few different versions of the number of COVID patients hospitalized at each hospital. 
To obtain these outcomes, we use the publicly available COVID-19 Reported Patient Impact and Hospital Capacity by Facility dataset \citep{hhs2021}. This provides facility-level data on hospital utilization aggregated on a weekly basis, from July 31st 2020 onwards. 
Summary statistics about hospital outcomes and  characteristics are documented in Table \ref{tbl:descriptives}. Eligible hospitals have larger numbers of inpatient and ICU beds occupied by COVID-19 patients. Eligible hospitals also have a higher disproportionate patient percentage, higher uncompensated care per bed, lower profit margins, more   employees and beds, and shorter lengths of inpatient stay. These patterns are consistent with the funding's goal of helping struggling hospitals. 


\subsection{Covariate Balance Estimates}

We first validate our method by evaluating the balancing property of fixed-bandwidth APS conditioning. We calculate fixed-bandwidth-APS-controlled differences in covariate means for hospitals who are and are not eligible for funding.
We run the following OLS regression of hospital-level characteristics on the eligibility status using observations with $p^s(X_i;\delta)\in (0,1)$:
$$
W_i = \gamma_0+\gamma_1Z_i+\gamma_2p^s(X_i;\delta)  + \eta_i,
$$
where $W_i$ is one of the predetermined characteristics of the hospital, $Z_i$ is a funding eligibility dummy, $X_i$ is a vector of the three input variables (DPP, UCC, and profit margin) that determine the funding eligibility, and $p^s(X_i; \delta)$ is the simulated fixed-bandwidth APS.
We compute fixed-bandwidth APS using $S=10,000$ simulation draws for different bandwidth values.\footnote{We standardize the three input variables when computing fixed-bandwidth APS. Figure \ref{fig:aps_sim} in Appendix \ref{appendix:additional_empirics} reports fixed-bandwidth APS for several hospitals with varying numbers of simulation draws. We find that $S=10,000$ is sufficient for well stabilizing fixed-bandwidth APS simulation.}
The estimated coefficient on $Z_i$ is the fixed-bandwidth-APS-controlled difference in the mean of the covariate between eligible and ineligible hospitals.
For comparison, we also run the OLS regression of hospital characteristics on the eligibility status with no controls using the whole sample.

\begin{table}[!t] \centering
\newcolumntype{C}{>{\centering\arraybackslash}X}

\caption{Covariate Balance Regressions}
{\scriptsize
\begin{tabularx}{\textwidth}{lCCCCCCCCC}

\toprule
{}&{}&{}&\multicolumn{7}{c}{Our Method: OLS with Approximate Propensity Score Controls} \tabularnewline[-2ex] {}&{Mean (Ineli-}&{OLS with}&\tabularnewline[-1ex] \cline{4-10} \tabularnewline[-2ex]
{}&{gible Hospitals)}&{No Controls}&{ $\delta = 0.01$}&{ $\delta = 0.025$}&{ $\delta = 0.05$}&{ $\delta = 0.075$}&{ $\delta = 0.1$}&{ $\delta = 0.25$}&{ $\delta = 0.5$} \tabularnewline
&{(1)}&{(2)}&{(3)}&{(4)}&{(5)}&{(6)}&{(7)}&{(8)}&{(9)} \\
\midrule\addlinespace[1.5ex]
\multicolumn{10}{c}{Panel A: Determinants of Funding Eligibility} \\ \tabularnewline[-2ex]
Profit margin&0.04&$-$0.11&$-$0.01&$-$0.00&0.02&0.01&0.03&0.05&0.04 \tabularnewline
&&(0.01)&(0.05)&(0.04)&(0.03)&(0.03)&(0.02)&(0.01)&(0.01) \tabularnewline
&&N=4633&N=91&N=239&N=494&N=684&N=905&N=1751&N=2397 \tabularnewline
Uncompensated&56,556&19,540&2,941&6,235&$-$8,408&$-$10,882&$-$9,432&$-$7,232&$-$8,071 \tabularnewline
care per bed (\$)&&(3,827)&(10,419)&(9,375)&(7,741)&(6,634)&(6,181)&(3,924)&(3,450) \tabularnewline
&&N=4633&N=91&N=239&N=494&N=684&N=905&N=1751&N=2397 \tabularnewline
Disproportionate&0.18&0.21&$-$0.06&$-$0.09&$-$0.09&$-$0.07&$-$0.07&$-$0.07&$-$0.07 \tabularnewline
patient percentage&&(0.01)&(0.07)&(0.07)&(0.06)&(0.05)&(0.04)&(0.02)&(0.01) \tabularnewline
&&N=4633&N=91&N=239&N=494&N=684&N=905&N=1751&N=2397 \tabularnewline\tabularnewline
\multicolumn{10}{c}{Panel B: Other Hospital Characteristics} \\ \tabularnewline[-2ex]
Full time employees&897.32&454.26&1,963.70&155.03&$-$58.52&65.12&39.85&192.77&$-$3.40 \tabularnewline
&&(69.23)&(1,382.13)&(897.52)&(561.57)&(432.30)&(354.30)&(178.37)&(124.68) \tabularnewline
&&N=4626&N=91&N=238&N=493&N=683&N=904&N=1748&N=2394 \tabularnewline
Medicare net revenue&20.04&18.36&25.85&$-$8.58&$-$6.49&$-$1.53&2.67&2.55&$-$4.40 \tabularnewline
(in millions \$)&&(2.39)&(25.85)&(16.60)&(12.46)&(10.61)&(9.49)&(5.66)&(4.12) \tabularnewline
&&N=4511&N=90&N=238&N=492&N=680&N=901&N=1709&N=2351 \tabularnewline
Occupancy&0.44&0.07&0.17&0.04&$-$0.01&0.01&0.02&0.04&0.03 \tabularnewline
&&(0.01)&(0.08)&(0.05)&(0.04)&(0.03)&(0.03)&(0.02)&(0.01) \tabularnewline
&&N=4624&N=91&N=239&N=494&N=684&N=905&N=1751&N=2397 \tabularnewline
Operating margin&0.02&$-$0.11&$-$0.01&0.01&0.03&0.02&0.04&0.06&0.06 \tabularnewline
&&(0.01)&(0.05)&(0.04)&(0.03)&(0.03)&(0.02)&(0.02)&(0.01) \tabularnewline
&&N=4541&N=90&N=238&N=486&N=674&N=894&N=1701&N=2343 \tabularnewline
Beds&134.60&53.75&154.56&17.33&$-$3.93&8.62&12.90&13.92&$-$5.80 \tabularnewline
&&(7.05)&(91.55)&(60.39)&(41.41)&(33.85)&(28.77)&(17.25)&(12.99) \tabularnewline
&&N=4633&N=91&N=239&N=494&N=684&N=905&N=1751&N=2397 \tabularnewline
Costs per discharge&66.28&$-$49.95&3.78&3.00&1.21&$-$5.41&1.46&5.77&$-$5.80 \tabularnewline
(in thousands \$)&&(17.93)&(1.91)&(1.38)&(1.08)&(7.00)&(0.90)&(5.09)&(10.28) \tabularnewline
&&N=3539&N=91&N=239&N=494&N=684&N=905&N=1751&N=2397 \tabularnewline
p-value joint significance&&0&.717&.494&.713&.517&.134&0&0 \tabularnewline
\bottomrule 

\end{tabularx}
\caption*{
\scriptsize{\it Note}: This table shows the results of the covariate balance regressions at the hospital level. The dependent variables for these regressions are drawn from the
Healthcare Cost Report Information System for the financial year 2018. Disproportionate patient percentage, profit margin and uncompensated care per bed
are used to determine the hospital’s funding eligibility. Other dependent variables shown indicate the financial health and utilization of the hospitals.
In column 2, we regress the dependent variables on the eligibility of the hospital with no controls. In columns 3–9, we regress the dependent variables on funding eligibility controlling for the Approximate Propensity Score with different values of bandwidth $\delta$. All Approximate Propensity Scores are
computed by averaging 10,000 simulation draws. Column 1 shows the mean of dependent variables for hospitals that are ineligible to receive safety net funding.
Robust standard errors are reported in the parenthesis and the number of observations is reported separately for each regression. The last row reports the $p$-value
of the joint significance test.
}
}
\label{table:CBR}
\end{table}

Table \ref{table:CBR} reports the covariate balance estimates.
Column 2 shows that, without controlling for fixed-bandwidth APS, eligible hospitals are significantly different from ineligible hospitals. All the relevant hospital eligibility characteristics are strongly associated with eligibility. 
Once we control for fixed-bandwidth APS with small enough bandwidth $\delta$, eligible and ineligible hospitals have similar financial and utilization characteristics, as reported in columns 3--7 of Table \ref{table:CBR}. These estimates are consistent with our theoretical results, establishing the empirical ability of fixed-bandwidth APS controls to eliminate selection bias.

\subsection{Effects of Funding: OLS and 2SLS Estimates}

The balancing performance of fixed-bandwidth APS motivates us to estimate the causal effects of funding by using algorithmically-determined funding eligibility as an IV.  
We study the effect of funding on relevant hospital outcomes, such as the number of inpatient beds occupied by adult COVID patients between July 31st 2020 and August 6th 2020. 

We first run the following OLS (reduced-form) regression of each outcome on the binary funding eligibility, while controlling for fixed-bandwidth APS: 
\begin{align*}
Y_i&=\alpha_0+\alpha_1 Z_i+\alpha_2 p^s(X_i;\delta)+\nu_i,
\end{align*}
\noindent where $Y_i$ is a hospital-level outcome and $Z_i$ is the binary indicator for funding eligibility. This OLS (reduced-form) specification is a special case of the 2SLS specification treated in the theoretical analysis.


We then estimate the following 2SLS regression using the funding amount as the treatment and funding eligibility as an instrument.
We run the regression on two different hospital-level outcome variables, using hospitals with $p^s(X_i;\delta)\in (0,1)$:
\begin{align*}
    D_i &= \gamma_0 + \gamma_1 Z_i + \gamma_2 p^s(X_i; \delta) + v_i \\ 
    Y_i &= \beta_0 + \beta_1 D_i + \beta_2 p^s(X_i; \delta) + \epsilon_i, 
\end{align*}
where $D_i$ is the funding amount. 
We also run the OLS and 2SLS regressions with no controls, as well as OLS regression controlling for the three eligibility determinants. 
These alternative regressions are computed using the sample of all hospitals, as benchmark estimators. 

\begin{table}[!t] \centering
\newcolumntype{C}{>{\centering\arraybackslash}X}

\caption{Effects of Funding on Hospital Behavior}
{\scriptsize
\begin{tabularx}{\textwidth}{lCCCCCCCCCC}

\hline\hline\tabularnewline[-1.8ex]
{}&{OLS with}&{OLS with}&{2SLS with}&\multicolumn{7}{c}{Our Method: 2SLS with Approximate Propensity Score Controls} \tabularnewline
{}&{No}&{Covariate}&{No} \tabularnewline[-1.8ex]\cline{5-11} \tabularnewline[-1.8ex]
{}&{Controls}&{Controls}&{Controls}&{ $\delta = 0.01$}&{ $\delta = 0.025$}&{ $\delta = 0.05$}&{ $\delta = 0.075$}&{ $\delta = 0.1$}&{ $\delta = 0.25$}&{ $\delta = 0.5$} \tabularnewline
&(1)&(2)&(3)&(4)&(5)&(6)&(7)&(8)&(9)&(10)\\
\midrule\addlinespace[1.5ex]
\multicolumn{11}{l}{\# Confirmed COVID Patients}\tabularnewline
\midrule\addlinespace[1.5ex]
Reduced Form&&&33.97***&8.82&$-$31.57&25.21&15.28&10.63&9.17&5.23 \tabularnewline
&&&(7.44)&(66.76)&(54.25)&(34.19)&(30.38)&(25.91)&(15.96)&(12.41) \tabularnewline
First stage&&&13.90***&16.55***&14.37***&15.05***&14.81***&14.42***&14.10***&13.19*** \tabularnewline
(in millions \$)&&&(0.50)&(6.11)&(3.66)&(2.33)&(1.91)&(1.64)&(1.04)&(0.75) \tabularnewline
\$1mm of funding&4.53***&2.50***&2.44***&0.05&$-$2.14&1.42&0.13&$-$0.03&$-$0.09&$-$0.63 \tabularnewline
&(0.63)&(0.79)&(0.50)&(4.33)&(3.97)&(2.17)&(1.97)&(1.74)&(1.12)&(0.96) \tabularnewline
N&3558&3558&3558&70&191&385&539&709&1366&1923 \tabularnewline
\midrule\addlinespace[1.5ex]
\multicolumn{10}{l}{\# Confirmed COVID Patients in ICU}\tabularnewline
\midrule\addlinespace[1.5ex]
Reduced Form&&&12.16***&10.75&$-$0.22&4.32&4.45&4.52&0.55&$-$1.58 \tabularnewline
&&&(2.58)&(23.34)&(18.35)&(11.38)&(10.13)&(8.71)&(5.64)&(4.39) \tabularnewline
First stage&&&13.89***&15.80**&13.79***&15.78***&15.53***&15.08***&14.43***&13.40*** \tabularnewline
(in millions \$)&&&(0.50)&(6.15)&(3.73)&(2.41)&(2.02)&(1.73)&(1.09)&(0.77) \tabularnewline
\$1mm of funding&1.51***&0.82***&0.88***&0.50&$-$0.11&0.18&0.04&0.12&$-$0.13&$-$0.35 \tabularnewline
&(0.21)&(0.27)&(0.17)&(1.54)&(1.37)&(0.70)&(0.64)&(0.56)&(0.39)&(0.34) \tabularnewline
N&3503&3503&3503&67&181&370&514&671&1321&1868 \tabularnewline
\bottomrule 

\end{tabularx}
\caption*{
\scriptsize{\it Notes}: In this table, we estimate the effects of the funding amount on hospital-level outcomes. Column 1 presents the results of OLS regression of the outcome variables on funding without any controls. Column 2 presents the results of OLS regression of the outcome variables on funding controlling for disproportionate patient percentage, uncompensated care per bed and profit margin.  In columns 3--10, we instrument the amount of funding with eligibility to receive this funding and present the results of 2SLS regressions. The reduced form shows the effect of funding eligibility on the outcome. The first stage shows the effect of being deemed eligible on the amount of relief funding received by hospitals, in millions of dollars. Column 3 shows the results of a 2SLS regression with no controls.  In columns 4--10, we run this regression controlling for the Approximate Propensity Score with different values of bandwidth $\delta$ on the sample with nondegenerate Approximate Propensity Scores.  All Approximate Propensity Scores are computed by averaging 10,000 simulation draws. The outcome variables are the 7 day totals for the week spanning July 31st, 2020 to August 6th, 2020. Confirmed COVID patients refer to the sum of patients in inpatient beds with lab-confirmed COVID-19, including those with both lab-confirmed COVID-19 and influenza. Inpatient bed totals also include observation beds. Confirmed COVID patients in ICU refers to the sum of patients in ICU beds with lab-confirmed COVID-19, including those with both lab-confirmed COVID-19 and influenza. Robust standard errors are reported in parentheses. */**/*** indicate $p <$ 0.10/0.05/0.01.
}
\label{table:2sls_safetynet}
}
\end{table}

The first-stage effects of funding eligibility on the funding amount are substantial (columns 4--10 of Table \ref{table:2sls_safetynet}).
In column 4 of Table \ref{table:2sls_safetynet}, funding eligibility increases the amount of funding received by \$15.85 million on average.

OLS estimates of funding effects, reported as the benchmark in column 1 of Table \ref{table:2sls_safetynet}, indicate that funding is associated with a higher number of adult inpatient beds and higher number of staffed ICU beds utilized by COVID patients. For example, the estimates indicate that receiving an additional \$1 million in funding is associated with 4.53 more beds occupied by patients. These uncontrolled OLS estimates show a similar picture as the descriptive statistics in Table \ref{tbl:descriptives}. 
Naive 2SLS estimates with no controls and OLS with covariate controls produce similar significantly positive associations of funding with outcomes.

However, the OLS or uncontrolled 2SLS estimates turn out to be an artifact of selection bias. In contrast with them, our preferred reduced-form and 2SLS estimates with fixed-bandwidth APS controls show a different picture (columns 4--10). The gains in the number of inpatient beds and staffed ICU beds occupied by COVID patients become much smaller and lose significance across all bandwidth specifications. In fact, even the sign of the estimated funding effect is reversed for several combinations of the outcome and bandwidth. Once we control for fixed-bandwidth APS to eliminate the bias, therefore, funding has little to no effect on the hospital utilization level by COVID-19 patients. These results suggest that fixed-bandwidth APS reveals important selection bias in the naively estimated effects of funding.\footnote{The 2SLS estimates in Table \ref{table:2sls_safetynet} are unlikely to be compromised by differential attrition. Estimates reported in Table \ref{table:differential_attrition} in Appendix \ref{appendix:additional_empirics} show little difference in outcome availability rates between eligible and ineligible hospitals once we control for fixed-bandwidth APS.}

We also estimate the evolving effects of funding for each week from July 31st, 2020 to April 2nd, 2021 (Appendix \ref{appendix:additional_empirics}). 
The estimated dynamic effects are similar to the initial null effects, suggesting that funding has no substantial effect even in the long run.
Furthermore, we estimate dynamic effects for different groups of hospitals defined by hospital size and ownership type.
We do not find any strong evidence of heterogeneity at any point in time.

The overall insignificance of the estimates suggests that funding by the CARES Act had little effect on hospital utilization during the pandemic. The null effect is widely observed for subgroups of hospitals at different points in time. This finding is consistent with policy and media arguments that CARES Act funding was not well targeted toward needy providers. Unlike the previous media arguments and descriptive analyses, the analysis here provides causal evidence supporting the concern. 




\section{Conclusion}

As algorithmic decisions become the new norm, the world becomes a mountain of natural experiments. 
We develop a general method to use these algorithm-produced instruments to identify and estimate causal treatment effects. 
Our analysis of the CARES Act hospital relief funding uses the proposed method to find that relief funding has little effect on COVID-19-related hospital activities. OLS or uncontrolled 2SLS estimates, by contrast, show considerably
larger and more significant effects. The large estimates appear to be an artifact of selection bias; relief funding just went to hospitals with more COVID-19 patients, without helping hospitals accommodate additional patients.

Our analysis provides a few implications for policy and management practices of decision-making algorithms. It is important to record the implementation of algorithms in a replicable way, including what input variables $X_i$ are used to make algorithmic recommendation $Z_i$. 
Another key lesson is the importance of recording an algorithm's recommendation $Z_i$ even if they are superseded by a human decision $D_i$. 
These data retention efforts go a long way to exploit the full potential of algorithms as natural experiments. 

An important topic for future research is estimation and inference details, such as data-driven bandwidth selection. This work needs to extend \cite{imbens2012bandwidth} and \cite{Calonico2014robustRD}'s bandwidth selection methods in the univariate RDD to our setting.\footnote{For univariate RDDs, \cite{imbens2012bandwidth} and \cite{Calonico2014robustRD} estimate the bandwidth that minimizes the asymptotic mean squared error (AMSE). It is not straightforward to estimate the AMSE-optimal bandwidth in our setting with many running variables and complex IV assignment, since it requires nonparametric estimation of functions on the multidimensional covariate space such as conditional mean functions, their derivatives, the curvature of the RDD boundary, etc.} Inference on treatment effects in our framework relies on large sample reasoning. It seems natural to additionally consider permutation or randomization inference. It will also be challenging but interesting to develop finite-sample optimal estimation and inference strategies such as those recently introduced by \cite{Armstrong2018inference, armstrong2020finitesample} and \cite{imbens2019optimized}. Finally, we look forward to empirical applications of our method in a variety of business, policy, and scientific domains.


\bibliographystyle{ecca}
\bibliography{reference}

\newpage

\clearpage

\appendix
\pagenumbering{arabic}
\renewcommand*{\thepage}{A-\arabic{page}}
\setcounter{table}{0}
\renewcommand{\thetable}{A.\arabic{table}}
\setcounter{figure}{0}
\renewcommand{\thefigure}{A.\arabic{figure}}
\renewcommand{\theequation}{A.\arabic{equation}}
\setcounter{equation}{0}

\begin{center}
\begin{LARGE}
\textbf{Appendix}\\
\end{LARGE}
\end{center}

\section{Assumptions for Asymptotic Normality in Theorem \ref{theorem:2sls-consistency}}\label{appendix:assumptions}

    Here we provide additional assumptions required for proving asymptotic normality of $\hat\beta_1$ and $\hat\beta_1^s$ in Theorem \ref{theorem:2sls-consistency}.
	First, when $A$ is stochastic, we assume the following. Let
	$$C^*=\{x\in \mathbb{R}^p: \text{$A$ is continuously differentiable at $x$}\},
	$$
	and let $D^*=\mathbb{R}^p\setminus C^*$ be the set of points at which $A$ is not continuously differentiable.
\begin{assumption}\label{assumption:asymptotic-normality}
	If $\Pr(A(X_i)\in (0,1))>0$, then the following conditions \ref{assumption:neighborhood-O-delta}--\ref{assumption:bounded-mean} hold.
\begin{enumerate}
		\renewcommand{\theenumi}{(\alph{enumi})}
		\renewcommand{\labelenumi}{(\alph{enumi})}
	\item {\rm (Probability of Neighborhood of $D^*$)}\label{assumption:neighborhood-O-delta}
	$\Pr(X_i\in N(D^*,\delta))=O(\delta)$.
	\item {\rm (Bounded Partial Derivatives of $A$)}\label{assumption:bounded-derivatives-A}
	The partial derivatives of $A$ are bounded on $C^*$.
	\item {\rm (Bounded Conditional Mean)}\label{assumption:bounded-mean} $E[Y_i|X_i]$ is bounded on ${\cal X}$.
\end{enumerate}
\end{assumption}

	To explain the role of Assumption \ref{assumption:asymptotic-normality} \ref{assumption:neighborhood-O-delta}, consider a path of covariate points $x_\delta\in N(D^*,\delta)\cap C^*$ indexed by $\delta>0$.
	Since $A$ is continuous at $x_\delta$, $p^{A}(x_\delta)=A(x_\delta)$ (as formally implied by Proposition \ref{proposition:LPSexistence} in Appendix \ref{section:existence}).
	However, $p^{A}(x_\delta;\delta)$ does not necessarily get sufficiently close to $A(x_\delta)$ even as $\delta\rightarrow 0$, since $x_\delta$ is in the $\delta$-neighborhood of $D^*$ and hence $A$ may discontinuously change within the $\delta$-ball $B(x_\delta,\delta)$.
	Assumption \ref{assumption:asymptotic-normality} \ref{assumption:neighborhood-O-delta} requires that the probability of $X_i$ being in the $\delta$-neighborhood of $D^*$ shrink to zero at the rate of $\delta$, which makes the points in the neighborhood negligible.
	
	Assumption \ref{assumption:asymptotic-normality} \ref{assumption:neighborhood-O-delta} often holds in practice. 
	If $A$ is continuously differentiable on ${\cal X}$, then $D^*\cap {\cal X}=\emptyset$, so this condition holds.
	If, for example, the treatment recommendation is randomly assigned based on a stratified randomized experiment, 
	$D^*$ is the boundary at which the recommendation probability changes discontinuously.
	For any boundary of standard shape, 
    $\Pr(X_i\in N(D^*,\delta))$ vanishes at the rate of $\delta$, and the required condition is satisfied.
	Assumption \ref{assumption:asymptotic-normality} \ref{assumption:bounded-derivatives-A} and \ref{assumption:bounded-mean} are regularity conditions, imposing the boundedness of the partial derivatives of $A$ and of the conditional mean of the outcome.
	
	The following is the key to asymptotic normality of the simulation-based estimator $\hat\beta_1^s$.
	
\begin{assumption}[The Number of Simulation Draws]\label{assumption:simulation}
	$(n\delta_n)^{-1/2}S_n\rightarrow \infty$, and $\Pr(p^{A}(X_i;\delta_n)\in (0,\gamma\frac{\log n}{S_n})\cup(1-\gamma\frac{\log n}{S_n},1))=o(n^{-1/2}\delta_n^{1/2})$ for some $\gamma>\frac{1}{2}$.
\end{assumption}


Assumption \ref{assumption:simulation} imposes the condition on the growth rate of the number of simulation draws $S_n$. This assumption ensures that the bias caused by using $p^s(X_i;\delta_n)$ instead of $p^A(X_i;\delta_n)$ is asymptotically negligible. To understand this condition, note that $p^s(X_i;\delta_n)$ enters the 2SLS first-order condition, $\sum_{i=1}^n (1, Z_i, p^s(X_i;\delta_n))' (Y_i - \beta_0-\beta_1 D_i - \beta_2 p^s(X_i;\delta_n)) 1\{p^s(X_i;\delta_n)\in (0,1)\}=0$, in two ways. First, $p^s(X_i;\delta_n)$ enters the condition in a nonlinear but smooth way through the $p^s(X_i;\delta_n)^2$ term. The asymptotic bias due to simulation errors is $O(\sqrt{n}\delta_n^2/S_n)$ under Assumption \ref{assumption:asymptotic-normality} if $\Pr(A(X_i)\in (0,1))>0$ and $n\delta_n^2\rightarrow 0$, and is $O(\sqrt{n\delta_n}/S_n)$ if $\Pr(A(X_i)\in (0,1))=0$.
The bias diminishes under the first part of Assumption \ref{assumption:simulation}. Second, $p^s(X_i;\delta_n)$ also enters the first-order condition in a nonsmooth way, since we only use observations for which $p^s(X_i;\delta_n)\in (0,1)$. If $p^A(X_i;\delta_n)$ is nondegenerate but close to zero or one, $p^s(X_i;\delta_n)$ may be degenerate (i.e., $A(X_{i,s}^*)=0$ for all $s$ or $A(X_{i,s}^*)=1$ for all $s$) with a large probability. The second part of Assumption \ref{assumption:simulation} ensures that the fraction of such observations goes to zero sufficiently fast, which eliminates the asymptotic bias caused by not using observations with $p^A(X_i;\delta_n)\in (0,1)$.

To illustrate how this assumption restricts the rate at which $S_n$ goes to infinity, consider an example where $\Pr(p^{A}(X_i;\delta_n)\in (0,1))=O(\delta_n)$, and $p^{A}(X_i;\delta_n)$ is approximately uniformly distributed on the tails $(0,\gamma\frac{\log n}{S_n})\cup(1-\gamma\frac{\log n}{S_n},1)$.
In this case, $\Pr(p^{A}(X_i;\delta_n)\in (0,\gamma\frac{\log n}{S_n})\cup(1-\gamma\frac{\log n}{S_n},1))=O(\delta_n\frac{\log n}{S_n})$, and the second part of Assumption \ref{assumption:simulation} requires that $S_n$ grow sufficiently fast so that $\frac{n^{1/2}\delta_n^{1/2}\log n}{S_n}=o(1)$.
One choice of $(\delta_n, S_n)$ that satisfies both parts of Assumption \ref{assumption:simulation} is $\delta_n=\alpha_1n^{-\kappa_1}$ and $S_n=\alpha_2n^{\kappa_2}$ for some $\alpha_1,\alpha_2>0$, $\kappa_1\in (\frac{1}{2},1)$ and $\kappa_2>\frac{1}{2}(1-\kappa_1)$.

\section{Notation and Lemmas}\label{appendix:notations-lemmas}


For a scalar-valued differentiable function $f:S\subset\mathbb{R}^n\rightarrow \mathbb{R}$, let $\nabla f:S\rightarrow \mathbb{R}^n$ be a gradient of $f$: for every $x\in S$,
$
\nabla f(x)=\left(\frac{\partial f(x)}{\partial x_1}, \cdots, \frac{\partial f(x)}{\partial x_n}\right)'.
$
When the second-order partial derivatives of $f$ exist, let $D^2 f(x)$ be the Hessian matrix:
$$
D^2 f(x)=\begin{bmatrix}\frac{\partial^2 f(x)}{\partial x_1^2} & \cdots & \frac{\partial^2 f(x)}{\partial x_1\partial x_n}\\
\vdots&\ddots&\vdots\\
\frac{\partial^2 f(x)}{\partial x_n\partial x_1} & \cdots & \frac{\partial^2 f(x)}{\partial x_n^2}
\end{bmatrix}
$$
for each $x\in S$. 
Let $f:S\subset\mathbb{R}^m\rightarrow \mathbb{R}^n$ be a function such that its first-order partial derivatives exist.
For each $x\in S$, let $Jf(x)$ be the Jacobian matrix of $f$ at $x$:
$$
Jf(x)=\begin{bmatrix}\frac{\partial f_1(x)}{\partial x_1} & \cdots & \frac{\partial f_1(x)}{\partial x_m}\\
\vdots&\ddots&\vdots\\
\frac{\partial f_n(x)}{\partial x_1} & \cdots & \frac{\partial f_n(x)}{\partial x_m}
\end{bmatrix}.
$$

\noindent For a positive integer $n$, let $I_n$ denote the $n\times n$ identity matrix.

\subsection{Differential Geometry}\label{section:differential-geometry}

We provide some concepts and facts from differential geometry of twice continuously differentiable sets, following \citeappendix{Crasta2007SDF}.
Let $S\subset \mathbb{R}^p$ be a twice continuously differentiable set.
For each $x\in \partial S$, we denote by $\nu_S(x)\in \mathbb{R}^p$ the inward unit normal vector of $\partial S$ at $x$, that is, the unit vector orthogonal to all vectors in the tangent space of $\partial S$ at $x$ that points toward the inside of $S$.
For a set $S\subset\mathbb{R}^p$, let $d_S^s:\mathbb{R}^p\rightarrow \mathbb{R}$ be the signed distance function of $S$, defined by
\begin{align*}
	d_S^s(x) = \begin{cases}
		d(x,\partial S) & \ \ \ \text{if $x\in {\rm cl}(S)$}\\
		-d(x,\partial S) & \ \ \ \text{if $x\in \mathbb{R}^p\setminus {\rm cl}(S)$},
	\end{cases}
\end{align*}
where $d(x,B)=\inf_{y\in B} \|y-x\|$ for any $x\in \mathbb{R}^p$ for a set $B\subset \mathbb{R}^p$.
Note that we can write $N(\partial S,\delta)=\{x\in \mathbb{R}^p:-\delta<d_S^s(x)<\delta\}$ for $\delta>0$.
Lastly, let $\Pi_{\partial S}(x)=\{y\in\partial S:\|y-x\|=d(x,\partial S)\}$ be the set of projections of $x$ on $\partial S$.
\begin{lemma}[Corollary of Theorem 4.16, \citeappendix{Crasta2007SDF}]\label{lemma:Crasta2007SDF-C2}
	Let $S\subset\mathbb{R}^p$ be nonempty, bounded, open, connected and twice continuously differentiable.
	Then the function $d_S^s$ is twice continuously differentiable on $N(\partial S,\mu)$ for some $\mu>0$.
	In addition, for every $x_0\in \partial S$, $\Pi_{\partial S}(x_0+t\nu_S(x_0))=\{x_0\}$ for every $t\in(-\mu,\mu)$.
	Furthermore, for every $x\in N(\partial S,\mu)$, $\Pi_{\partial S}(x)$ is a singleton, $\nabla d_S^s(x)=\nu_S(y)$ and $x=y+d_S^s(x)\nu_S(y)$ for $y\in \Pi_{\partial S}(x)$, and $\|\nabla d_S^s(x)\|=1$.
\end{lemma}

	\begin{proof}
		See Appendix \ref{proof:lemma:Crasta2007SDF-C2}.
	\end{proof}

We say that a set $S\subset \mathbb{R}^p$ is an \textit{$m$-dimensional $C^1$ submanifold of $\mathbb{R}^{p}$} if for every point $x\in S$, there exist an open neighborhood $V\subset \mathbb{R}^p$ of $x$ and a one-to-one continuously differentiable function $\phi$ from an open set $U\subset \mathbb{R}^{m}$ to $\mathbb{R}^p$ such that the Jacobian matrix $J\phi(u)$ is of rank $m$ for all $u\in U$, and $\phi(U)=V\cap S$.

\begin{lemma}\label{lemma:C1-submanifold}
	Let $S\subset\mathbb{R}^p$ be nonempty, bounded, open, connected and twice continuously differentiable.
	Then $\partial S$ is a $(p-1)$-dimensional $C^1$ submanifold of $\mathbb{R}^{p}$.
\end{lemma}

	\begin{proof}
		See Appendix \ref{proof:lemma:C1-submanifold}.
	\end{proof}

\subsection{Geometric Measure Theory}\label{section:geometric-theory}

We provide some concepts and facts from geometric measure theory, following \citeappendix{Krantz2008book}.
Let $S$ be an $m$-dimensional $C^1$ submanifold of $\mathbb{R}^p$.
Let $x\in S$ and let $\phi: U\subset \mathbb{R}^m\rightarrow \mathbb{R}^p$ be as in the definition of $m$-dimensional $C^1$ submanifold.
We denote by $T_S(x)$ the tangent space of $S$ at $x$, $\{J\phi(u) v: v\in \mathbb{R}^m\}$, where $u=\phi^{-1}(x)$.
Also, for each $x\in\mathbb{R}^p$ at which $d_S^s$ is differentiable and for each $\lambda\in\mathbb{R}$, let $\psi_S(x,\lambda)=x+\lambda\nabla d_S^s(x)$.
Furthermore, for a Lipschitz function $f:\mathbb{R}^p\rightarrow \mathbb{R}^\nu$ with $\nu\ge m$, let 
		$J_m^Sf(x)=\frac{{\cal H}^m(\{Jf(x)y:y\in P\})}{{\cal H}^m(P)}$ for each $x\in \mathbb{R}^p$ at which $f$ is differentiable,
	where $P$ is an arbitrary $m$-dimensional parallelepiped contained in $T_S(x)$.

\begin{lemma}\label{lemma:neighborhood-integral}
	For $\Omega\subset \mathbb{R}^p$, suppose that there is a partition $\{\Omega_{1},...,\Omega_{M}\}$ of $\Omega$ with
	\begin{enumerate}[label=(\roman*)]
		\item ${\rm dist}(\Omega_m,\Omega_{m'})>0$ for any $m,m'\in\{1,...,M\}$ such that $m\neq m'$;
		\item $\Omega_m$ is nonempty, bounded, open, connected and twice continuously differentiable for each $m\in \{1,...,M\}$.
	\end{enumerate}
	Then there is $\mu>0$ such that $d_{\Omega}^s$ is twice continuously differentiable on $N(\partial\Omega,\mu)$ and
	$$
	\int_{N(\partial \Omega,\delta)}g(x)dx = \int_{-\delta}^{\delta}\int_{\partial \Omega}g(u+\lambda \nu_\Omega(u))J_{p-1}^{\partial\Omega}\psi_\Omega(u,\lambda)d{\cal H}^{p-1}(u)d\lambda
	$$
	for every $\delta\in (0,\mu)$ and every function $g:\mathbb{R}^p\rightarrow \mathbb{R}$ that is integrable on $N(\partial \Omega,\delta)$, where for each fixed $\lambda\in(-\mu,\mu)$, $J_{p-1}^{\partial\Omega}\psi_\Omega(\cdot,\lambda)$ is calculated by applying the operation $J_{p-1}^{\partial \Omega}$ to the function $\psi_\Omega(\cdot,\lambda)$.
	Furthermore, $J_{p-1}^{\partial\Omega}\psi_\Omega(x,\cdot)$ is continuously differentiable in $\lambda$ and $J_{p-1}^{\partial\Omega}\psi_\Omega(x,0)=1$ for every $x\in\partial\Omega$, and  $J_{p-1}^{\partial\Omega}\psi_\Omega(\cdot,\cdot)$ and $\tfrac{\partial J_{p-1}^{\partial\Omega}\psi_\Omega(\cdot,\cdot)}{\partial\lambda}$ are bounded on $\partial\Omega\times (-\mu,\mu)$.
\end{lemma}

	\begin{proof}
		See Appendix \ref{proof:lemma:neighborhood-integral}.
	\end{proof}

\subsection{Other Lemmas}

		\begin{lemma}\label{lemma:mean-plim}
			Let $\{V_i\}_{i=1}^\infty$ be i.i.d. random variables such that $E[V_i^2]<\infty$.
			If Assumption \ref{assumption:A} holds, then for $l\ge 0$ and $m=0,1$,
			$$
			E[V_i p^{A}(X_i;\delta)^l 1\{p^{A}(X_i;\delta)\in (0,1)\}^m]\rightarrow E[V_i A(X_i)^l 1\{A(X_i)\in (0,1)\}^m]
			$$
			as $\delta\rightarrow 0$.
			Moreover, if, in addition, $\delta_n\rightarrow 0$ as $n\rightarrow\infty$, then for $l\ge 0$, as $n\rightarrow \infty$,
			$$\frac{1}{n}\sum_{i=1}^n V_i p^{A}(X_i;\delta_n)^l I_{i,n}\stackrel{p}{\longrightarrow} E[V_i A(X_i)^l 1\{A(X_i)\in (0,1)\}].
			$$
		\end{lemma}
		
	\begin{proof}
		See Appendix \ref{proof:lemma:mean-plim}.
	\end{proof}

\begin{lemma}\label{lemma:simulation-error}
	Let $\{(\delta_n,S_n)\}_{n=1}^\infty$ be any sequence of positive numbers and positive integers.
	Fix $x\in{\cal X}$, and let $X_1^*,...,X_{S_n}^*$ be $S_n$ independent draws from the uniform distribution on $B(x,\delta_n)$ so that
	$
	p^s(x;\delta_n)=\frac{1}{S_n}\sum_{s=1}^{S_n}A(X_s^*).
	$
	Then,
	\begin{align*}
		& E[p^s(x;\delta_n)-p^{A}(x;\delta_n)] =0,~~
		E[(p^s(x;\delta_n)-p^{A}(x;\delta_n))^2]\le\frac{1}{S_n},\\
		&|E[p^s(x;\delta_n)^2-p^{A}(x;\delta_n)^2]|\le\frac{1}{S_n},~~
		E[(p^s(x;\delta_n)^2-p^{A}(x;\delta_n)^2)^2]\le\frac{4}{S_n}, \text{and}\\
		&\Pr(p^s(x;\delta_n)\in\{0,1\})
		\le (1-p^{A}(x;\delta_n))^{S_n}+p^{A}(x;\delta_n)^{S_n}.
	\end{align*}
	Moreover, for any $\epsilon>0$,
	$
	E[|p^s(x;\delta_n)-p^{A}(x;\delta_n)|]\le \frac{1}{S_n\epsilon^2}+\epsilon,
	$
	and if $S_n\rightarrow \infty$, then
		$E[|p^s(x;\delta_n)-p^{A}(x;\delta_n)|]\rightarrow 0$
	as $n\rightarrow\infty$.    
\end{lemma}

	\begin{proof}
		See Appendix \ref{proof:lemma:simulation-error}.
	\end{proof}

\begin{lemma}\label{lemma:mean-simulation-plim}
	Suppose $\Pr(A(X_i)\in (0,1))>0$.
    Let $I_{i,n}^s=1\{p^s(X_i;\delta_n)\in (0,1)\}$ and $\{V_i\}_{i=1}^\infty$ be i.i.d. random variables such that $E[V_i^2]<\infty$.
	If Assumption \ref{assumption:A} holds, $S_n\rightarrow \infty$, and $\delta_n\rightarrow 0$, then
	$$
	\frac{1}{n}\sum_{i=1}^n V_i p^s(X_i;\delta_n)^lI_{i,n}^s-\frac{1}{n}\sum_{i=1}^n V_i p^{A}(X_i;\delta_n)^l I_{i,n}=o_p(1)
	$$
	for $l=0,1,2,3,4$.
	If, in addition, Assumptions \ref{assumption:asymptotic-normality} and \ref{assumption:simulation} hold, $n\delta_n^2\rightarrow 0$, and $E[V_i|X_i]$ is bounded, then
	$$
	\frac{1}{\sqrt{n}}\sum_{i=1}^n V_i p^s(X_i;\delta_n)^lI_{i,n}^s-\frac{1}{\sqrt{n}}\sum_{i=1}^n V_i p^{A}(X_i;\delta_n)^l I_{i,n}=o_p(1)
	$$
	for $l=0,1,2$.
\end{lemma}

	\begin{proof}
		See Appendix \ref{proof:lemma:mean-simulation-plim}.
	\end{proof}

\section{Proofs of Main Results}

\subsection{Proof of Proposition \ref{proposition:id}}\label{proof:id}

	Suppose that Assumptions \ref{assumption:A} and \ref{continuity} hold.
	Here, we only show that
	\begin{enumerate}[label=(\alph*)]
		\item $E[Y_{1i}-Y_{0i}| X_i=x]$ is identified for every $x\in {\rm int}({\cal X})$ such that $p^{A}(x)\in (0,1)$.
		\item Let $S$ be any open subset of ${\cal X}$ such that $p^{A}(x)$ exists for all $x\in S$. 
		Then $E[Y_{1i}-Y_{0i}| X_i \in S]$ is identified only if $p^{A}(x)\in (0,1)$ for almost every $x\in S$. 
	\end{enumerate}
	The proofs for $E[D_i(1)-D_i(0)| X_i =x]$ and $E[D_i(1)-D_i(0)| X_i \in S]$ are similar.
	\\~\\
	{\bf Proof of Part (a).}
		Pick an $x\in {\rm int}({\cal X})$ such that $p^{A}(x)\in (0,1)$.
		If $A(x)\in (0,1)$, $E[Y_{1i}-Y_{0i}| X_i=x]$ is trivially identified by Property \ref{conditional-independence}:
		\begin{align*} E[Y_i|X_i=x,Z_i=1]-E[Y_i|X_i=x,Z_i=0]&=E[Y_{1i}-Y_{0i}|X_i=x].
		\end{align*}

		We next consider the case where $A(x)\in \{0,1\}$.
		Since $x\in {\rm int}({\cal X})$, $B(x,\delta)\subset {\cal X}$ for any sufficiently small $\delta>0$.
		Since $p^{A}(x)=\lim_{\delta\rightarrow 0} p^{A}(x;\delta) \in(0,1)$, $p^{A}(x;\delta)\in (0,1)$ for any sufficiently small $\delta>0$.
		This implies that we can find points $x_{0,\delta}, x_{1,\delta}\in B(x,\delta)(\subset{\cal X})$ such that $A(x_{0,\delta})<1$ and $A(x_{1,\delta})>0$ for any sufficiently small $\delta>0$, for otherwise $p^{A}(x;\delta)\in\{0,1\}$.
		Since $x_{0,\delta}\rightarrow x$ and $x_{1,\delta}\rightarrow x$ as $\delta\rightarrow 0$,
		\begin{align*}
		&~\lim_{\delta\rightarrow 0} (E[Y_i|X_i=x_{1,\delta},Z_i=1]-E[Y_i|X_i=x_{0,\delta},Z_i=0])\\=&~\lim_{\delta\rightarrow 0} (E[Y_{1i}|X_i=x_{1,\delta}]-E[Y_{0i}|X_i=x_{0,\delta}])
		=E[Y_{1i}-Y_{0i}|X_i=x],
		\end{align*}
		where the two equalities follow from Property \ref{conditional-independence} and Assumption \ref{continuity}, respectively.
		\qed
	\\~\\
	{\bf Proof of Part (b).}
	Suppose to the contrary that ${\cal L}^p(\{x\in S:p^{A}(x)\in \{0,1\}\})>0$. 
	Without loss of generality, assume ${\cal L}^p(\{x\in S:p^{A}(x)=1\})>0$.
	
	\begin{step-subsection}\label{step:necessary-nonzero}
	${\cal L}^p(S\cap {\cal X}_1)>0$.
	\end{step-subsection}
	\begin{proof}
	By Assumption \ref{assumption:A}, $A$ is continuous almost everywhere.
	Part 1 of Corollary \ref{corollary:LPSexistence} then implies $p^{A}(x)=A(x)$ for almost every $x\in \{x^*\in S:p^{A}(x^*)=1\}$.
	Since ${\cal L}^p(\{x\in S:p^{A}(x)=1\})>0$, ${\cal L}^p(\{x\in S:p^{A}(x)=1, p^{A}(x)=A(x)\})>0$, and hence ${\cal L}^p(S\cap{\cal X}_1)>0$.
	\end{proof}
	

	\begin{step-subsection}\label{step:necessary-nonempty}
	$S\cap {\rm int}({\cal X}_1)\neq \emptyset$.
	\end{step-subsection}
	\begin{proof}
	Suppose $S\cap {\rm int}({\cal X}_1)= \emptyset$.
	Then, we must have $S\cap {\cal X}_1\subset {\cal X}_1\setminus {\rm int}({\cal X}_1)$.
	It then follows that ${\cal L}^p(S\cap {\cal X}_1) \le {\cal L}^p({\cal X}_1\setminus {\rm int}({\cal X}_1))= {\cal L}^p({\cal X}_1)-{\cal L}^p({\rm int}({\cal X}_1))=0$, where the last equality holds by Assumption \ref{assumption:A}.
	But this is a contradiction to Step \ref{step:necessary-nonzero}.
	\end{proof}

	\begin{step-subsection}\label{step:necessary-interior}
	$p^{A}(x)=1$ for any $x\in {\rm int}({\cal X}_1)$.
	\end{step-subsection}
	\begin{proof}
	Pick any $x\in {\rm int}({\cal X}_1)$.
	By the definition of interior, $B(x,\delta)\subset {\cal X}_1$ for any sufficiently small $\delta>0$.
	Therefore, $p^{A}(x;\delta)=1$ for any sufficiently small $\delta>0$.
	\end{proof}

	\begin{step-subsection}\label{step:necessary-nonID}
	$E[Y_{1i}-Y_{0i}|X_i\in S]$ is not identified.
	\end{step-subsection}
	\begin{proof}
	We first introduce some notation.
	Let ${\bf Q}$ be the set of all distributions of $(Y_{1i},Y_{0i},X_i,Z_i)$ satisfying Property \ref{conditional-independence} and Assumptions \ref{assumption:A} and \ref{continuity}.
	Let ${\bf P}$ be the set of all distributions of $(Y_i,X_i,Z_i)$.
	Let $T:{\bf Q}\rightarrow {\bf P}$ be a function such that, for $Q\in{\bf Q}$, $T(Q)$ is the distribution of $(Z_i Y_{1i}+(1-Z_i) Y_{0i},X_i,Z_i)$, where the distribution of $(Y_{1i},Y_{0i},X_i,Z_i)$ is $Q$.
	Let $Q_0$ and $P_0$ denote the true distributions of $(Y_{1i},Y_{0i},X_i,Z_i)$ and $(Y_i,X_i,Z_i)$, respectively.
	Given $P_0$, the identified set of $E[Y_{1i}-Y_{0i}|X_i\in S]$ is given by $\{E_Q[Y_{1i}-Y_{0i}|X_i\in S]: P_0=T(Q), Q\in{\bf Q}\}$, where $E_Q[\cdot]$ is the expectation under distribution $Q$.
	We show that this set contains two distinct values.
	In what follows, $\Pr(\cdot)$ and $E[\cdot]$ without a subscript denote the probability and expectation under the true distributions $Q_0$ and $P_0$ as up until now.
	
	Now pick any $x^*\in S\cap {\rm int}({\cal X}_1)$.
	Since $S$ and ${\rm int}({\cal X}_1)$ are open, there is a $\delta>0$ such that $B(x^*,\delta)\subset S\cap {\rm int}({\cal X}_1)$.	
	Let $\epsilon=\frac{\delta}{2}$, and consider a function $f:{\cal X}\rightarrow \mathbb{R}$ such that $f(x)=E[Y_{0i}|X=x]$ for all $x\in {\cal X}\setminus B(x^*,\epsilon)$ and $f(x)=E[Y_{0i}|X=x]-1$ for all $x\in B(x^*,\epsilon)$.
	Below, we show that $f$ is continuous at any point $x\in {\cal X}$ such that $p^{A}(x)\in (0,1)$ and $A(x)\in \{0,1\}$.
	Pick any $x\in {\cal X}$ such that $p^{A}(x)\in (0,1)$ and $A(x)\in \{0,1\}$.
	Since $B(x^*,\delta)\subset {\rm int}({\cal X}_1)$ and ${\rm int}({\cal X}_1)\subset\{x'\in {\cal X}:p^{A}(x')=1\}$ by Step \ref{step:necessary-interior}, $x\notin B(x^*,\delta)$.
	Hence, $B(x,\epsilon)\subset {\cal X}\setminus B(x^*,\epsilon)$.
	By Assumption \ref{continuity} and the definition of $f$, $f$ is continuous at $x$.
	
	
	Now take any random vector $(Y_{1i}^*,Y_{0i}^*,X^*_i,Z^*_i)$ that is distributed according to the true distribution $Q_0$.
	Let $Q$ be the distribution of $(Y_{1i}^Q,Y_{0i}^Q,X^Q_i,Z^Q_i)$, where $(Y_{1i}^Q,X^Q_i,Z^Q_i)=(Y_{1i}^*,X^*_i,Z^*_i)$, and
	$$
	Y_{0i}^Q=\begin{cases}
	Y_{0i}^* & \ \ \ \text{if } X^*_i\in{\cal X}\setminus B(x^*,\epsilon)\\ 
	Y_{0i}^*-1 & \ \ \ \text{if } X^*_i\in B(x^*,\epsilon).\\ 
	\end{cases}
	$$
	Note first that $Q\in{\bf Q}$, since $E_Q[Y_{1i}^Q|X^Q_i=x]=E[Y_{1i}^*|X_i^*=x]$ and $E_Q[Y_{0i}^Q|X^Q_i=x]= f(x)$, where $E[Y_{1i}^*|X_i^*]$ and $f$ are both continuous at any point $x\in {\cal X}$ such that $p^{A}(x)\in (0,1)$ and $A(x)\in \{0,1\}$.
	Also, $Z_i^Q=Z_i^*=1$ if $X_i^*\in B(x^*,\epsilon)$.
	It then follows 
	\begin{align*}
	Y_i^Q&=Z_i^Q Y_{1i}^Q+(1-Z_i^Q)Y_{0i}^Q=\begin{cases}
	Z_i^* Y_{1i}^*+(1-Z_i^*)Y_{0i}^* & \ \ \ \text{if } X^*_i\in{\cal X}\setminus B(x^*,\epsilon)\\ 
	Z_i^* Y_{1i}^* & \ \ \ \text{if } X^*_i\in B(x^*,\epsilon)\\ 
	\end{cases}
	\end{align*}
	and
	\begin{align*}
	Y_i^*&=Z_i^* Y_{1i}^*+(1-Z_i^*)Y_{0i}^*=\begin{cases}
	Z_i^* Y_{1i}^*+(1-Z_i^*)Y_{0i}^* & \ \ \ \text{if } X^*_i\in{\cal X}\setminus B(x^*,\epsilon)\\ 
	Z_i^* Y_{1i}^* & \ \ \ \text{if } X^*_i\in B(x^*,\epsilon).
	\end{cases}
	\end{align*}
	Thus, $Y_i^Q=Y_i^*$, and hence $T(Q)=T(Q_0)=P_0$.
	
	Using $E_Q[Y_{1i}^Q|X^Q_i=x]=E[Y_{1i}^*|X_i^*=x]$ and $E_Q[Y_{0i}^Q|X^Q_i=x]= f(x)$, we have
	\begin{align*}
		&~E_Q[Y_{1i}^Q-Y_{0i}^Q|X^Q_i\in S]\\
		=&~E_Q[E_Q[Y_{1i}^Q|X^Q_i]|X^Q_i\in S]\\
		&-E_Q[E_Q[Y_{0i}^Q|X^Q_i]|X^Q_i\in S, X^Q_i\notin B(x^*,\epsilon)]{\rm Pr}_{Q}(X^Q_i\notin B(x^*,\epsilon)|X^Q_i\in S)\\
		& -E_Q[E_Q[Y_{0i}^Q|X^Q_i]|X^Q_i\in B(x^*,\epsilon)]{\rm Pr}_{Q}(X^Q_i\in B(x^*,\epsilon)|X^Q_i\in S)\\
		=&~E[E[Y_{1i}^*|X^*_i]|X^*_i\in S]-E[f(X^*_i)|X^*_i\in S, X^*_i\notin B(x^*,\epsilon)]\Pr(X^*_i\notin B(x^*,\epsilon)|X^*_i\in S)\\
		& -E[f(X^*_i)|X^*_i\in B(x^*,\epsilon)]\Pr(X^*_i\in B(x^*,\epsilon)|X^*_i\in S)\\
		=&~E[Y_{1i}^*|X^*_i\in S]-E[Y_{0i}^*|X^*_i\in S, X^*_i\notin B(x^*,\epsilon)]\Pr(X^*_i\notin B(x^*,\epsilon)|X^*_i\in S)\\
		& -E[Y_{0i}^*-1|X^*_i\in B(x^*,\epsilon)]\Pr(X^*_i\in B(x^*,\epsilon)|X^*_i\in S)\\
		=&~E[Y_{1i}^*-Y_{0i}^*|X^*_i\in S]+\Pr(X^*_i\in B(x^*,\epsilon)|X^*_i\in S).
	\end{align*}
	By the definition of support, $\Pr(X_i^*\in B(x^*,\epsilon))>0$.
	Since $T(Q)=T(Q_0)=P_0$ but $E_Q[Y_{1i}^Q-Y_{0i}^Q|X^Q_i\in S]\neq E[Y_{1i}^*-Y_{0i}^*|X^*_i\in S]$, $E[Y_{1i}-Y_{0i}|X_i\in S]$ is not identified.
	\end{proof}

\subsection{Proof of Theorem \ref{theorem:2sls-consistency}}\label{proof:2sls-consistency}

As we mention when we define our 2SLS estimators in Section \ref{subsection:2sls}, we drop the constant term if $A(X_i)$ takes on only one nondegenerate value in the sample.
Formally,
consider the following 2SLS regression using the observations with $p^{A}(X_i;\delta_n)\in (0,1)$:
\begin{align}
	D_i&=\gamma_0(1-\mathbf{I}_n)+\gamma_1 Z_i+\gamma_2 p^{A}(X_i;\delta_n)+\nu_i\label{2SLSestimand1-general}\\
	Y_i&=\beta_0(1-\mathbf{I}_n)+\beta_1 D_i +\beta_2 p^{A}(X_i;\delta_n)+\epsilon_i.\label{2SLSestimand2-general}
\end{align}
Here $\mathbf{I}_n$ is a dummy random variable which equals one if there exists a constant $q\in (0,1)$ such that $A(X_i)\in \{0,q,1\}$ for all $i\in\{1,...,n\}$. 
$\mathbf{I}_n$ is the indicator that $A(X_i)$ takes on only one nondegenerate value in the sample.
If the support of $A(X_i)$ (in the population) contains only one value in $(0,1)$, $p^{A}(X_i;\delta_n)$ is asymptotically constant conditional on $p^{A}(X_i;\delta_n)\in (0,1)$.
To avoid the multicollinearity between asymptotically constant $p^{A}(X_i;\delta_n)$ and a constant, we do not include the constant term if $\mathbf{I}_n=1$.
Let $I_{i,n}=1\{p^{A}(X_i;\delta_n)\in(0,1)\}$,
$\mathbf{D}_{i,n}=(1, D_i, p^{A}(X_i;\delta_n))'$, $\mathbf{Z}_{i,n}=(1, Z_i, p^{A}(X_i;\delta_n))'$, $\mathbf{D}_{i,n}^{nc}=(D_i, p^{A}(X_i;\delta_n))'$, and $\mathbf{Z}_{i,n}^{nc}=(Z_i, p^{A}(X_i;\delta_n))'$.
The 2SLS estimator $\hat\beta$ from this regression is then given by
\begin{align*}
	\hat\beta =\begin{cases}
		(\sum_{i=1}^n \mathbf{Z}_{i,n}\mathbf{D}_{i,n}'I_{i,n})^{-1}\sum_{i=1}^n \mathbf{Z}_{i,n}Y_iI_{i,n} ~~&\text{if $\mathbf{I}_n$=0}\\
		(\sum_{i=1}^n \mathbf{Z}_{i,n}^{nc}(\mathbf{D}_{i,n}^{nc})'I_{i,n})^{-1}\sum_{i=1}^n \mathbf{Z}_{i,n}^{nc}Y_iI_{i,n} ~~&\text{if $\mathbf{I}_n$=1}.
	\end{cases}
\end{align*}
Let $\hat\beta_1$ denote the 2SLS estimator of $\beta_1$ in the above regression.
Similarly, we consider the simulation version of the 2SLS regression (\ref{2SLSestimand1-general}) and (\ref{2SLSestimand2-general}), where we use $p^s(X_i;\delta_n)$ in place of $p^A(X_i;\delta_n)$.
Let $\hat\beta_1^s$ be the simulation-based 2SLS estimator of $\beta_1$.

Below, we prove that the statement in Theorem \ref{theorem:2sls-consistency} holds for the above modified estimators $\hat\beta_1$ and $\hat\beta_1^s$.
Throughout the proof, we omit the subscript $n$ from $I_{i,n}$, $\mathbf{D}_{i,n}$, $\mathbf{Z}_{i,n}$, $\hat\epsilon_{i,n}$, $\hat{\mathbf{\Sigma}}_n$, $\hat \sigma_n$, etc. for notational brevity.
We provide proofs separately for the two cases, the case in which $\Pr(A(X_i)\in (0,1))>0$ and the case in which $\Pr(A(X_i)\in (0,1))=0$.
For each case, we first prove consistency and asymptotic normality of $\hat\beta_1$, and then prove those of $\hat\beta^s_1$.

\subsubsection{Proof of Asymptotic Properties of \texorpdfstring{$\hat\beta_1$}{hat beta} When \texorpdfstring{$\Pr(A(X_i)\in (0,1))>0$}{A is stochastic}}
\label{subsubsection:consistency}

By Lemma \ref{lemma:mean-plim},
$$
\lim_{\delta \rightarrow 0}E[p^{A}(X_i;\delta)(1-p^{A}(X_i;\delta))(D_i(1)-D_i(0))]= E[A(X_i)(1-A(X_i))(D_i(1)-D_i(0))].
$$
When $\Pr(A(X_i)\in (0,1))>0$, $E[A(X_i)(1-A(X_i))(D_i(1)-D_i(0))]=E[p^{A}(X_i)(1-p^{A}(X_i))(D_i(1)-D_i(0))]$, since $p^{A}(x)=A(x)$ for almost every $x\in{\cal X}$ by Proposition \ref{proposition:APSexistence-a.e.}.
Note that $E[p^{A}(X_i)(1-p^{A}(X_i))(D_i(1)-D_i(0))]=\int_{{\cal X}} p^A(x)(1-p^A(x))E[D_i(1)-D_i(0)|X_i=x]f_X(x)dx$.
Hence, under Assumption \ref{assumption:consistency} \ref{assumption:relevance}, $E[p^{A}(X_i)(1-p^{A}(X_i))(D_i(1)-D_i(0))]\neq 0$. 
Again by Lemma \ref{lemma:mean-plim},
$$
\lim_{\delta \rightarrow 0}E[\omega_i(\delta)(Y_i(1)-Y_i(0))]=\frac{E[A(X_i)(1-A(X_i))(D_i(1)-D_i(0))(Y_i(1)-Y_i(0)]}{E[A(X_i)(1-A(X_i))(D_i(1)-D_i(0))]}.
$$

Let $\beta_1=\frac{E[A(X_i)(1-A(X_i))(D_i(1)-D_i(0))(Y_i(1)-Y_i(0)]}{E[A(X_i)(1-A(X_i))(D_i(1)-D_i(0))]}$,
	$\hat\beta^c = (\sum_{i=1}^n \mathbf{Z}_i\mathbf{D}_i'I_{i})^{-1}\sum_{i=1}^n \mathbf{Z}_iY_iI_{i}$,
	$\hat\beta^{nc} =  (\sum_{i=1}^n \mathbf{Z}_i^{nc}(\mathbf{D}_i^{nc})'I_{i})^{-1}\sum_{i=1}^n \mathbf{Z}_i^{nc}Y_iI_{i}$,
$\hat\beta_1^c=(0,1,0)\hat\beta^c$ and $\hat\beta_1^{nc}=(1,0)\hat\beta^{nc}$.
$\hat\beta_1$ is then given by
$
\hat\beta_1=\hat\beta_1^c (1-\mathbf{I}_n)+\hat\beta_1^{nc}\mathbf{I}_n.
$
Also, let $\tilde{\mathbf{D}}_i=(1, D_i, A(X_i))'$, $\tilde{\mathbf{Z}}_i=(1, Z_i, A(X_i))'$, $\tilde{\mathbf{D}}_i^{nc}=(D_i, A(X_i))'$, $\tilde{\mathbf{Z}}_i^{nc}=(Z_i, A(X_i))'$, and $I_i^{A}=1\{A(X_i)\in (0,1)\}$.

We claim that $\Pr(\mathbf{I}_n=1)\rightarrow0$ when $\Var(A(X_i)|I_i^{A}=1)>0$, and that $\Pr(\mathbf{I}_n=1)\rightarrow 1$ when $\Var(A(X_i)|I_i^{A}=1)=0$.
To show the first claim, observe that $\mathbf{I}_n=1$ if and only if $\hat V_n=0$, where
$
\hat V_n =\frac{1}{\sum_{i=1}^nI_i^{A}}\sum_{i=1}^n (A(X_i)-\frac{\sum_{i=1}^nA(X_i)I_i^{A}}{\sum_{i=1}^nI_i^{A}})^2I_i^{A}
$
is the sample variance of $A(X_i)$ conditional on $I_i^{A}=1$.
When $\Var(A(X_i)|I_i^{A}=1)>0$,
\begin{align*}
	\Pr(\mathbf{I}_n=1)&=\Pr(\hat V_n=0)\le \Pr(|\hat V_n-\Var(A(X_i)|I_i^{A}=1)|\ge \Var(A(X_i)|I_i^{A}=1))\rightarrow 0,
\end{align*}
where the convergence follows since $\hat V_n\stackrel{p}{\longrightarrow} \Var(A(X_i)|I_i^{A}=1)>0$. 
To show the second claim, note that, when $\Var(A(X_i)|I_i^{A}=1)=0$, there exists $q\in (0,1)$ such that $\Pr(A(X_i)=q|I_i^{A}=1)=1$.
It follows that
\begin{align*}
	\Pr(\mathbf{I}_n=0)	&=\Pr(A(X_i)\in \{0,1\} \text{ for all $i=1,...,n$})\\
	&~~~~+\Pr(\text{$A(X_i)=q'$ and $A(X_j)=q''$ for some $q',q''\in (0,1)$ with $q'\neq q''$} \\
	&\hspace{18.7em}\text{for some $i,j \in\{1,...,n\}$})\\
	&=\Pr(A(X_i)\in \{0,1\} \text{ for all $i=1,...,n$})=(1-\Pr(A(X_i)\in (0,1)))^n,
\end{align*}
which converges to zero as $n\rightarrow \infty$, since $\Pr(A(X_i)\in (0,1))>0$.

The above claims imply that,
to prove consistency and asymptotic normality of $\hat\beta_1$, it suffices to show those of $\hat\beta_1^c$ when $\Var(A(X_i)|I_i^{A}=1)>0$ and those of $\hat\beta_1^{nc}$ when $\Var(A(X_i)|I_i^{A}=1)=0$.

Below we first show that $\hat\beta_1\stackrel{p}{\longrightarrow}\beta_1$ if Assumptions \ref{assumption:A} and \ref{assumption:consistency} hold and $\delta_n\rightarrow 0$. We then show that $\hat\sigma^{-1}(\hat\beta_1-\beta_1)\stackrel{d}{\longrightarrow} {\cal N}(0,1)$ if, in addition, Assumption \ref{assumption:asymptotic-normality} holds and $n\delta_n^2\rightarrow 0$.
\\\par
\noindent
{\bf Proof of Consistency.}
We only show that $\hat\beta_1^c\stackrel{p}{\longrightarrow}\beta_1$ when $\Var(A(X_i)|I_i^{A}=1)>0$.
We can show that $\hat\beta_1^{nc}\stackrel{p}{\longrightarrow}\beta_1$ whether or not $\Var(A(X_i)|I_i^{A}=1)>0$ analogously.
Note first that, a few lines of algebra gives
\begin{align*}
	&{\rm det}(E[\tilde{\mathbf{Z}}_i\tilde{\mathbf{D}}_i'I_i^{A}])\\
	=&\Pr(I_i^{A}=1)^2\Var(A(X_i)|I_i^{A}=1)E[D_i(Z_i-A(X_i))I_i^{A}]\\
	=&\Pr(I_i^{A}=1)^2\Var(A(X_i)|I_i^{A}=1)E[(Z_iD_i(1)+(1-Z_i)D_i(0))(Z_i-A(X_i))I_i^{A}]\\
	=&\Pr(I_i^{A}=1)^2\Var(A(X_i)|I_i^{A}=1)E[((Z_i-Z_iA(X_i))D_i(1)-(1-Z_i)A(X_i)D_i(0))I_i^{A}]\\
	=&\Pr(I_i^{A}=1)^2\Var(A(X_i)|I_i^{A}=1)E[((A(X_i)-A(X_i)^2)D_i(1)-(1-A(X_i))A(X_i)D_i(0))I_i^{A}]\\
	=&\Pr(I_i^{A}=1)^2\Var(A(X_i)|I_i^{A}=1)E[A(X_i)(1-A(X_i))(D_i(1)-D_i(0))I_i^{A}]\\
	=&\Pr(I_i^{A}=1)^2\Var(A(X_i)|I_i^{A}=1)E[A(X_i)(1-A(X_i))(D_i(1)-D_i(0))],
\end{align*}
where the fourth equality follows from Property \ref{conditional-independence}.
Therefore, $E[\tilde{\mathbf{Z}}_i\tilde{\mathbf{D}}_i'I_i^{A}]$ is invertible when $\Var(A(X_i)|I_i^{A}=1)>0$.
Another few lines of algebra gives
\begin{align*}
	(E[\tilde{\mathbf{Z}}_i\tilde{\mathbf{D}}_i'I_i^{A}])^{-1}=\frac{1}{E[A(X_i)(1-A(X_i))(D_i(1)-D_i(0))]}\begin{bmatrix} *~ & * & *\\
		0~ & 1 & -1\\
		*~ & * & *
	\end{bmatrix}.
\end{align*}
Therefore, when $\Var(A(X_i)|I_i^{A}=1)>0$, by Lemma \ref{lemma:mean-plim},
\begin{align*}
	\hat\beta^c=&~(\sum_{i=1}^n \mathbf{Z}_i\mathbf{D}_i'I_{i})^{-1}\sum_{i=1}^n \mathbf{Z}_iY_iI_{i}\stackrel{p}{\longrightarrow}(E[\tilde{\mathbf{Z}}_i\tilde{\mathbf{D}}_i'I_i^{A}])^{-1}E[\tilde{\mathbf{Z}}_iY_iI_i^{A}]\\
    =&\frac{E[Z_iY_iI_i^{A}]-E[A(X_i)Y_iI_i^{A}]}{E[A(X_i)(1-A(X_i))(D_i(1)-D_i(0))]}
	=\frac{E[Z_iY_{1i}I_i^{A}]-E[A(X_i)(Z_iY_{1i}+(1-Z_i)Y_{0i})I_i^{A}]}{E[A(X_i)(1-A(X_i))(D_i(1)-D_i(0))]}\\
	=&~\frac{E[A(X_i)Y_{1i}I_i^{A}]-E[A(X_i)(A(X_i)Y_{1i}+(1-A(X_i))Y_{0i})I_i^{A}]}{E[A(X_i)(1-A(X_i))(D_i(1)-D_i(0))]}\\
	=&~\frac{E[A(X_i)(1-A(X_i))(Y_{1i}-Y_{0i})I_i^{A}]}{E[A(X_i)(1-A(X_i))(D_i(1)-D_i(0))]}\\
	=&~\frac{E[A(X_i)(1-A(X_i))(D_i(1)-D_i(0))(Y_i(1)-Y_i(0))]}{E[A(X_i)(1-A(X_i))(D_i(1)-D_i(0))]}=\beta_1,
\end{align*}
where the third line follows from Property \ref{conditional-independence}, and the second last equality follows from the definitions of $Y_{1i}$ and $Y_{0i}$.
\qed

~\par
\noindent
{\bf Proof of Asymptotic Normality.}
Let $(\hat\sigma^c)^2$ be the second diagonal element of
	$$
	\hat{\mathbf{\Sigma}}^c=(\sum_{i=1}^n \mathbf{Z}_{i}\mathbf{D}_{i}'I_{i})^{-1}(\sum_{i=1}^n \hat \epsilon_{i}^2\mathbf{Z}_{i}\mathbf{Z}_{i}'I_{i})(\sum_{i=1}^n \mathbf{D}_{i}\mathbf{Z}_{i}'I_{i})^{-1}
	$$
and $(\hat\sigma^{nc})^2$ be the first diagonal element of
	$$
	\hat{\mathbf{\Sigma}}^{nc}=(\sum_{i=1}^n \mathbf{Z}_{i,n}^{nc}(\mathbf{D}_{i,n}^{nc})'I_i)^{-1}(\sum_{i=1}^n \hat \epsilon_{i,n}^2\mathbf{Z}_{i,n}^{nc}(\mathbf{Z}_{i,n}^{nc})'I_i)(\sum_{i=1}^n \mathbf{D}_{i,n}^{nc}(\mathbf{Z}_{i,n}^{nc})'I_i)^{-1}.
	$$
We only show $(\hat\sigma^c)^{-1}(\hat\beta_1^c-\beta_1)\stackrel{d}{\longrightarrow} {\cal N}(0,1)$ when $\Var(A(X_i)|I_i^{A}=1)>0$.
We can show $(\hat\sigma^{nc})^{-1}(\hat\beta_1^{nc}-\beta_1)\stackrel{d}{\longrightarrow} {\cal N}(0,1)$ analogously.
The proof proceeds in steps.
\setcounter{step}{0}
\begin{step}\label{step:tilde-beta_1}
	Let $\tilde\beta_n= (E[\tilde{\mathbf{Z}}_i\tilde{\mathbf{D}}_i'I_i])^{-1}E[\tilde{\mathbf{Z}}_iY_iI_i]$, and let $\tilde\beta_{1,n}$ denote the second element of $\tilde\beta_n$.
	Then
	$\tilde\beta_{1,n} =\beta_1$ for any choice of $\delta_n>0$.
\end{step}
\begin{proof}
	Note first that, for every $\delta>0$, $p^{A}(x;\delta)\in (0,1)$ for almost every $x\in \{x'\in {\cal X}:A(x')\in (0,1)\}$, since by almost everywhere continuity of $A$, for almost every $x\in \{x'\in {\cal X}:A(x')\in (0,1)\}$, there exists an open ball $B\subset B(x,\delta)$ such that $A(x')\in (0,1)$ for every $x' \in B$.
	After a few lines of algebra, we have
	\begin{align*}
		{\rm det}(E[\tilde{\mathbf{Z}}_i\tilde{\mathbf{D}}_i'I_i])
		=&~\Pr(I_{i}=1)^2\Var(A(X_i)|I_{i}=1)E[D_i(Z_i-A(X_i))I_{i}]\\
		=&~\Pr(I_{i}=1)^2\Var(A(X_i)|I_{i}=1)E[A(X_i)(1-A(X_i))(D_i(1)-D_i(0))I_{i}]\\
		=&~\Pr(I_{i}=1)^2\Var(A(X_i)|I_{i}=1)E[A(X_i)(1-A(X_i))(D_i(1)-D_i(0))],
	\end{align*}
	where the last equality holds since $p^{A}(x;\delta)\in (0,1)$ for almost every $x\in \{x'\in {\cal X}:A(x')\in (0,1)\}$.
	By the law of total conditional variance,
	\begin{align*}
		&\Var(A(X_i)|I_{i}=1)
		=E[\Var(A(X_i)|I_{i}=1,I_i^{A})|I_{i}=1]+\Var(E[A(X_i)|I_{i}=1,I_i^{A}]|I_{i}=1)\\
		&\ge\sum_{t\in \{0,1\}}\Var(A(X_i)|I_{i}=1,I_i^{A}=t)\Pr(I_i^{A}=t|I_{i}=1)\\
		&\ge\Var(A(X_i)|I_{i}=1,I_i^{A}=1)\Pr(I_i^{A}=1|I_{i}=1)
		=\Var(A(X_i)|I_i^{A}=1)\Pr(I_i^{A}=1|I_{i}=1)
		>0.
	\end{align*}
	Therefore, $E[\tilde{\mathbf{Z}}_i\tilde{\mathbf{D}}_i'I_i]$ is invertible.
	Another few lines of algebra gives
	\begin{align*}
		(E[\tilde{\mathbf{Z}}_i\tilde{\mathbf{D}}_i'I_i])^{-1}=\frac{1}{E[A(X_i)(1-A(X_i))(D_i(1)-D_i(0))]}\begin{bmatrix} *~ & * & *\\
			0~ & 1 & -1\\
			*~ & * & *
		\end{bmatrix}.
	\end{align*}
	It follows that
	\begin{align*}
		\tilde\beta_{1,n}&=\frac{E[Z_iY_iI_{i}]-E[A(X_i)Y_iI_{i}]}{E[A(X_i)(1-A(X_i))(D_i(1)-D_i(0))]}\\
		&=\frac{E[A(X_i)(1-A(X_i))(D_i(1)-D_i(0))(Y_i(1)-Y_i(0))I_{i}]}{E[A(X_i)(1-A(X_i))(D_i(1)-D_i(0))]}
        =\beta_1.
		\qedhere
	\end{align*}

\end{proof}

We can write
\begin{align*}
	\sqrt{n}(\hat\beta^c -\tilde\beta_n)=&~\underbrace{(\frac{1}{n}\sum_{i=1}^n \mathbf{Z}_i\mathbf{D}_i'I_{i})^{-1}\frac{1}{\sqrt{n}}\sum_{i=1}^n \mathbf{Z}_iY_iI_{i} - (\frac{1}{n}\sum_{i=1}^n \tilde{\mathbf{Z}}_i\tilde{\mathbf{D}}_i'I_{i})^{-1}\frac{1}{\sqrt{n}}\sum_{i=1}^n \tilde{\mathbf{Z}}_iY_iI_{i}}_{=(A)}\\
	&+\underbrace{(\frac{1}{n}\sum_{i=1}^n \tilde{\mathbf{Z}}_i\tilde{\mathbf{D}}_i'I_{i})^{-1}\frac{1}{\sqrt{n}}\sum_{i=1}^n \tilde{\mathbf{Z}}_iY_iI_{i}-(E[\tilde{\mathbf{Z}}_i\tilde{\mathbf{D}}_i'I_i])^{-1}\sqrt{n}E[\tilde{\mathbf{Z}}_iY_iI_i]}_{=(B)}.
\end{align*}
We first consider $(B)$. Let $\tilde\epsilon_{i,n}=Y_i-\tilde{\mathbf{D}}_i'\tilde\beta_n$ so that
$
E[\tilde{\mathbf{Z}}_i\tilde\epsilon_{i,n}I_i]=E[\tilde{\mathbf{Z}}_i(Y_i-\tilde{\mathbf{D}}_i'\tilde\beta_n)I_i]=E[\tilde{\mathbf{Z}}_iY_iI_{i}]-E[\tilde{\mathbf{Z}}_i\tilde{\mathbf{D}}_i'I_{i}]\tilde\beta_n=0.
$
Then
\begin{align*}
	(B)
	=&~(\frac{1}{n}\sum_{i=1}^n \tilde{\mathbf{Z}}_i\tilde{\mathbf{D}}_i'I_{i})^{-1}\frac{1}{\sqrt{n}}\sum_{i=1}^n \tilde{\mathbf{Z}}_i(\tilde{\mathbf{D}}_i'\tilde\beta_n+\tilde\epsilon_{i,n})I_{i}-(E[\tilde{\mathbf{Z}}_i\tilde{\mathbf{D}}_i'I_i])^{-1}\sqrt{n}E[\tilde{\mathbf{Z}}_i(\tilde{\mathbf{D}}_i'\tilde\beta_n+\tilde\epsilon_{i,n})I_i]\\
	=&~\sqrt{n}(\tilde\beta_n-\tilde\beta_n)+(\frac{1}{n}\sum_{i=1}^n \tilde{\mathbf{Z}}_i\tilde{\mathbf{D}}_i'I_{i})^{-1}\frac{1}{\sqrt{n}}\sum_{i=1}^n \tilde{\mathbf{Z}}_i\tilde\epsilon_{i,n}I_{i}-(E[\tilde{\mathbf{Z}}_i\tilde{\mathbf{D}}_i'I_i])^{-1}\sqrt{n}E[\tilde{\mathbf{Z}}_i\tilde\epsilon_{i,n}I_i]\\
	=&~(\frac{1}{n}\sum_{i=1}^n \tilde{\mathbf{Z}}_i\tilde{\mathbf{D}}_i'I_{i})^{-1}\frac{1}{\sqrt{n}}\sum_{i=1}^n \tilde{\mathbf{Z}}_i\tilde\epsilon_{i,n}I_{i}.
\end{align*}

\begin{step}\label{step:clt}
Let $\beta=(E[\tilde{\mathbf{Z}}_i\tilde{\mathbf{D}}_i'I_i^{A}])^{-1}E[\tilde{\mathbf{Z}}_iY_iI_i^{A}]$ and $\tilde \epsilon_i=Y_i-\tilde{\mathbf{D}}_i'\beta$.
Then
	$
	\frac{1}{\sqrt{n}}\sum_{i=1}^n \tilde{\mathbf{Z}}_i\tilde\epsilon_{i,n}I_{i}\stackrel{d}{\longrightarrow}{\cal N}(0,E[\tilde\epsilon_i^2\tilde{\mathbf{Z}}_i\tilde{\mathbf{Z}}_i'I_i^{A}]).
	$
\end{step}

\begin{proof}
	We use the triangular-array Lyapunov CLT and the Cram\'{e}r-Wold device.
	Pick a nonzero $\lambda\in \mathbb{R}^p$, and let $V_{i,n}=\frac{1}{\sqrt{n}}\lambda' \tilde{\mathbf{Z}}_i\tilde\epsilon_{i,n}I_{i}$.
	First, 
	by Lemma \ref{lemma:mean-plim},
	$
		\tilde\beta_n\rightarrow (E[\tilde{\mathbf{Z}}_i\tilde{\mathbf{D}}_i'I_i^{A}])^{-1}E[\tilde{\mathbf{Z}}_iY_iI_i^{A}]=\beta
	$
	as $n\rightarrow \infty$.
	We have
	\begin{align*}
		E[\tilde\epsilon_{i,n}^2\tilde{\mathbf{Z}}_i\tilde{\mathbf{Z}}_i'I_{i}]&=E[(Y_i-\tilde{\mathbf{D}}_i'\tilde \beta_n)^2\tilde{\mathbf{Z}}_i\tilde{\mathbf{Z}}_i'I_{i}]=E[(\tilde\epsilon_i-\tilde{\mathbf{D}}_i'(\tilde \beta_n-\beta))^2\tilde{\mathbf{Z}}_i\tilde{\mathbf{Z}}_i'I_{i}]\\
		&=E[\tilde\epsilon_i^2\tilde{\mathbf{Z}}_i\tilde{\mathbf{Z}}_i'I_{i}]-2E[\tilde\epsilon_i((\tilde \beta_{0,n}-\beta_0)+D_i(\tilde \beta_{1,n}-\beta_1)+A(X_i)(\tilde \beta_{2,n}-\beta_2))\tilde{\mathbf{Z}}_i\tilde{\mathbf{Z}}_i'I_{i}]\\
		&~~~~+E[((\tilde \beta_{0,n}-\beta_0)+D_i(\tilde \beta_{1,n}-\beta_1)+A(X_i)(\tilde \beta_{2,n}-\beta_2))^2\tilde{\mathbf{Z}}_i\tilde{\mathbf{Z}}_i'I_{i}]\rightarrow E[\tilde\epsilon_i^2\tilde{\mathbf{Z}}_i\tilde{\mathbf{Z}}_i'I_i^{A}],
	\end{align*}
where the convergence follows from Lemma \ref{lemma:mean-plim} and $\tilde\beta_n\rightarrow \beta$.
	Therefore,
	$
	\sum_{i=1}^n E[V_{i,n}^2]=\lambda'E[\tilde\epsilon_{i,n}^2\tilde{\mathbf{Z}}_i\tilde{\mathbf{Z}}_i'I_{i}]\lambda\rightarrow \lambda'E[\tilde\epsilon_i^2\tilde{\mathbf{Z}}_i\tilde{\mathbf{Z}}_i'I_i^{A}]\lambda.
	$
	
	We next verify the Lyapunov condition: for some $t>0$,
	$
	\sum_{i=1}^n E[|V_{i,n}|^{2+t}]\rightarrow 0.
	$
	Consider
	$	\sum_{i=1}^n E[|V_{i,n}|^{4}]=\frac{1}{n} E[|\lambda' \tilde{\mathbf{Z}}_i\tilde\epsilon_{i,n}I_{i}|^{4}].
	$
	We use the $c_r$-inequality: $E[|X+Y|^r]\le 2^{r-1}E[|X|^r+|Y|^r]$ for $r\ge 1$.
	Repeating using the $c_r$-inequality gives
	\begin{align*}
		E[|\lambda' \tilde{\mathbf{Z}}_i\tilde\epsilon_{i,n}I_{i}|^{4}]&=E[|\lambda'\tilde{\mathbf{Z}}_i(Y_i-\tilde\beta_{0,n}-\tilde\beta_{1,n} D_i-\tilde\beta_{2,n} A(X_i))|^{4}I_{i}]\\
		&\le 2^{3c}E[(|\lambda'\tilde{\mathbf{Z}}_i|^{4})(|Y_i|^{4}+|\tilde\beta_{0,n}|^{4}+|\tilde\beta_{1,n}|^{4}D_i+|\tilde\beta_{2,n}|^{4}A(X_i)^{4})I_{i}]\\
		&\le 2^{3c}(|\lambda_1|+|\lambda_2|+|\lambda_3|)^4(E[Y_i^{4}]+\tilde\beta_{0,n}^{4}+\tilde\beta_{1,n}^{4}+\tilde\beta_{2,n}^{4})=2^{3c}O(1)
	\end{align*}
	for some constant $c$, 
	where the last equality is by Assumption \ref{assumption:consistency} \ref{assumption:moment}.
	Thus,
	$
	\sum_{i=1}^n E[|V_{i,n}|^{4}]\rightarrow 0,
	$
	and the conclusion follows from the Lyapunov CLT and the Cram\'{e}r-Wold device.
\end{proof}

We next consider $(A)$. We can write
\begin{align*}
	(A) =&~(\frac{1}{n}\sum_{i=1}^n \mathbf{Z}_i\mathbf{D}_i'I_{i})^{-1}\frac{1}{\sqrt{n}}\sum_{i=1}^n (\mathbf{Z}_iY_iI_{i}-\tilde{\mathbf{Z}}_iY_iI_{i})\\
	&-(\frac{1}{n}\sum_{i=1}^n \mathbf{Z}_i\mathbf{D}_i'I_{i})^{-1}[\frac{1}{\sqrt{n}}\sum_{i=1}^n (\mathbf{Z}_i\mathbf{D}_i'I_{i}-\tilde{\mathbf{Z}}_i\tilde{\mathbf{D}}_i'I_{i})]  (\frac{1}{n}\sum_{i=1}^n \tilde{\mathbf{Z}}_i\tilde{\mathbf{D}}_i'I_{i})^{-1}\frac{1}{n}\sum_{i=1}^n \tilde{\mathbf{Z}}_iY_iI_{i}.
\end{align*}

\begin{step}\label{step:O-delta}
	Let $\{V_i\}_{i=1}^\infty$ be i.i.d. random variables such that $E[|V_i|]<\infty$ and that $E[V_i|X_i]$ is bounded on $N(D^*,\delta')\cap {\cal X}$ for some $\delta'>0$.
	Then, for $l=0,1$, 
	$$
	E[V_ip^{A}(X_i;\delta)^l(p^{A}(X_i;\delta)-A(X_i))1\{p^{A}(X_i;\delta)\in (0,1)\}]=O(\delta).
	$$
\end{step}

\begin{proof}
	For every $x\notin N(D^*,\delta)$, $B(x,\delta)\cap D^*=\emptyset$, so $A$ is continuously differentiable on $B(x,\delta)$.
	By the mean value theorem, for every $x\notin N(D^*,\delta)$ and $a\in B(\bm{0},\delta)$,
	$$
	A(x+a) = A(x)+\nabla A(y(x,a))' a
	$$
	for some point $y(x,a)$ on the line segment connecting $x$ and $x+a$.
	For every $x\notin N(D^*,\delta)$,
	\begin{align*}
		p^{A}(x;\delta)&=\frac{\delta^p\int_{B(\bm{0},1)} A(x+\delta u)du}{\delta^p\int_{B(\bm{0},1)} du}
        =A(x)+\delta\frac{\int_{B(\bm{0},1)} \nabla A(y(x,\delta u))' udu}{\int_{B(\bm{0},1)} du}.
	\end{align*}
	Now, we can write
	\begin{align*}
		&~E[V_ip^{A}(X_i;\delta)^l(p^{A}(X_i;\delta)-A(X_i))1\{p^{A}(X_i;\delta)\in (0,1)\}]\\
		=&~E[V_ip^{A}(X_i;\delta)^l(p^{A}(X_i;\delta)-A(X_i))1\{p^{A}(X_i;\delta)\in (0,1)\}1\{X_i\notin N(D^*,\delta)\}]\\
		&+E[V_ip^{A}(X_i;\delta)^l(p^{A}(X_i;\delta)-A(X_i))1\{p^{A}(X_i;\delta)\in (0,1)\}1\{X_i\in N(D^*,\delta)\}].
	\end{align*}
	For the first term,
	\begin{align*}
		&~|E[V_ip^{A}(X_i;\delta)^l(p^{A}(X_i;\delta)-A(X_i))1\{p^{A}(X_i;\delta)\in (0,1)\}1\{X_i\notin N(D^*,\delta)\}]|\\
		=&~\delta\left\vert E\left[V_ip^{A}(X_i;\delta)^l\frac{\int_{B(\bm{0},1)} \nabla A(y(X_i,\delta u))' udu}{\int_{B(\bm{0},1)} du}1\{p^{A}(X_i;\delta)\in (0,1)\}1\{X_i\notin N(D^*,\delta)\}\right]\right\vert\\
		\le&~\delta E\left[|V_i|p^{A}(X_i;\delta)^l\frac{\int_{B(\bm{0},1)} \sum_{k=1}^p |\frac{\partial A(y(X_i,\delta u))}{\partial x_k}| |u_k|du}{\int_{B(\bm{0},1)} du}1\{p^{A}(X_i;\delta)\in (0,1)\}\times 1\{X_i\notin N(D^*,\delta)\}\right]\\
		\le&~\delta E[|V_i|]\sum_{k=1}^p \sup_{x\in C^*}\left\vert\frac{\partial A(x)}{\partial x_k}\right\vert\frac{\int_{B(\bm{0},1)}  |u_k|du}{\int_{B(\bm{0},1)} du}
		=O(\delta),
	\end{align*}
	where we use the assumption that the partial derivatives of $A$ is bounded on $C^*$.
	For the second term, for sufficiently small $\delta>0$,
	\begin{align*}
		&~|E[V_ip^{A}(X_i;\delta)^l(p^{A}(X_i;\delta)-A(X_i))1\{p^{A}(X_i;\delta)\in (0,1)\}1\{X_i\in N(D^*,\delta)\}]|\\
		\le&~E[|E[V_i|X_i]|1\{X_i\in N(D^*,\delta)\}]
		\le CE[1\{X_i\in N(D^*,\delta)\}]
		=C\Pr(X_i\in N(D^*,\delta))
		=O(\delta),
	\end{align*}
	where $C$ is some constant, the second inequality follows from the assumption that $E[V_i|X_i]$ is bounded on $N(D^*,\delta')\cap {\cal X}$ for some $\delta'>0$, and the last equality follows from Assumption \ref{assumption:asymptotic-normality} \ref{assumption:neighborhood-O-delta}.\end{proof}

\begin{step}\label{step:o_p(1)}
	$\frac{1}{\sqrt{n}}\sum_{i=1}^n (\mathbf{Z}_iY_iI_{i}-\tilde{\mathbf{Z}}_iY_iI_{i})=o_p(1)$ and $\frac{1}{\sqrt{n}}\sum_{i=1}^n (\mathbf{Z}_i\mathbf{D}_i'I_{i}-\tilde{\mathbf{Z}}_i\tilde{\mathbf{D}}_i'I_{i})=o_p(1)$.
\end{step}

\begin{proof}
	We only show that $\frac{1}{\sqrt{n}}\sum_{i=1}^n(p^{A}(X_i;\delta_n)^2-A(X_i)^2)I_{i}=o_p(1)$.
	The proofs for the other elements are similar.
	As for bias,
	\begin{align*}
		&E\left[\frac{1}{\sqrt{n}}\sum_{i=1}^n(p^{A}(X_i;\delta_n)^2-A(X_i)^2)I_{i}\right]
		=\sqrt{n}E[(p^{A}(X_i;\delta_n)^2-A(X_i)^2)I_{i}]\\
		&=\sqrt{n}E[(p^{A}(X_i;\delta_n)+A(X_i))(p^{A}(X_i;\delta_n)-A(X_i))I_{i}]
		=\sqrt{n}O(\delta_n)
		=0,
	\end{align*}
	where the third equality follows from Step \ref{step:O-delta} and the last from the assumption that $n\delta_n^2\rightarrow 0$.
	As for variance, by Lemma \ref{lemma:mean-plim},
	\begin{align*}
		&\Var\left(\frac{1}{\sqrt{n}}\sum_{i=1}^n(p^{A}(X_i;\delta_n)^2-A(X_i)^2)I_{i}\right)
		\le E[(p^{A}(X_i;\delta_n)^2-A(X_i)^2)^2I_{i}]\rightarrow 0.
		\qedhere
	\end{align*}
	\end{proof}

\begin{step}\label{step:sigma-consistency}
	$n\hat{\mathbf{\Sigma}}^c\stackrel{p}{\longrightarrow} (E[\tilde{\mathbf{Z}}_i\tilde{\mathbf{D}}_i'I_i^{A}])^{-1}E[\tilde\epsilon_i^2\tilde{\mathbf{Z}}_i\tilde{\mathbf{Z}}_i'I_i^{A}](E[\tilde{\mathbf{D}}_i\tilde{\mathbf{Z}}_i'I_i^{A}])^{-1}$.
\end{step}

\begin{proof}
	Let $\epsilon_i=Y_i-\mathbf{D}_i'\beta$. We have
\begin{align*}
	&\frac{1}{n}\sum_{i=1}^n \hat \epsilon_i^2\mathbf{Z}_i\mathbf{Z}_i'I_{i}
	=\frac{1}{n}\sum_{i=1}^n(Y_i-\mathbf{D}_i'\hat\beta^c)^2\mathbf{Z}_i\mathbf{Z}_i'I_{i}
	=\frac{1}{n}\sum_{i=1}^n(\epsilon_i-\mathbf{D}_i'(\hat\beta^c-\beta))^2\mathbf{Z}_i\mathbf{Z}_i'I_{i}\\
	&=\frac{1}{n}\sum_{i=1}^n\epsilon_i^2\mathbf{Z}_i\mathbf{Z}_i'I_{i}+\frac{1}{n}\sum_{i=1}^n((\hat\beta_0^c-\beta_0)+D_i(\hat\beta_1^c-\beta_1)+p^{A}(X_i;\delta_n)(\hat\beta_2^c-\beta_2))^2\mathbf{Z}_i\mathbf{Z}_i'I_{i}\\
	&~~~~-\frac{2}{n}\sum_{i=1}^n(Y_i-\mathbf{D}_i'\beta)((\hat\beta_0^c-\beta_0)+D_i(\hat\beta_1^c-\beta_1)+p^{A}(X_i;\delta_n)(\hat\beta_2^c-\beta_2))\mathbf{Z}_i\mathbf{Z}_i'I_{i}\\
	&=\frac{1}{n}\sum_{i=1}^n\epsilon_i^2\mathbf{Z}_i\mathbf{Z}_i'I_{i}+o_p(1)O_p(1),
\end{align*}
where the last equality follows from the result that $\hat\beta^c-\beta=o_p(1)$ and from Lemma \ref{lemma:mean-plim}.
The conclusion is implied by the following consequence of Lemma \ref{lemma:mean-plim}:
\begin{align*}
	\frac{1}{n}\sum_{i=1}^n\epsilon_i^2\mathbf{Z}_i\mathbf{Z}_i'I_{i}&=\frac{1}{n}\sum_{i=1}^n(Y_i^2-2Y_i\mathbf{D}_i'\beta+\beta'\mathbf{D}_i\mathbf{D}_i'\beta)\mathbf{Z}_i\mathbf{Z}_i'I_{i}\\
	&\stackrel{p}{\longrightarrow} E[(Y_i^2-2Y_i\tilde{\mathbf{D}}_i'\beta+\beta'\tilde{\mathbf{D}}_i\tilde{\mathbf{D}}_i'\beta)\tilde{\mathbf{Z}}_i\tilde{\mathbf{Z}}_i'I_i^{A}]=E[\tilde\epsilon_i^2\tilde{\mathbf{Z}}_i\tilde{\mathbf{Z}}_i'I_i^{A}],\\
	\frac{1}{n}\sum_{i=1}^n\mathbf{Z}_i\mathbf{D}_i'I_{i}&\stackrel{p}{\longrightarrow}E[\tilde{\mathbf{Z}}_i\tilde{\mathbf{D}}_i'I_i^{A}]
	\qedhere
\end{align*}

\end{proof}

\begin{step}
	$(\hat\sigma^c)^{-1}(\hat\beta_1^c-\beta_1)\stackrel{d}{\longrightarrow} {\cal N}(0,1)$.
\end{step}

\begin{proof}
	By combining the results from Steps \ref{step:clt}--\ref{step:o_p(1)} and by Lemma \ref{lemma:mean-plim},
	\begin{align*}
		(A)&\stackrel{p}{\longrightarrow} 0,~
		(B)\stackrel{d}{\longrightarrow} {\cal N}(0, (E[\tilde{\mathbf{Z}}_i\tilde{\mathbf{D}}_i'I_i^{A}])^{-1}E[\tilde\epsilon_i^2\tilde{\mathbf{Z}}_i\tilde{\mathbf{Z}}_i'I_i^{A}](E[\tilde{\mathbf{D}}_i\tilde{\mathbf{Z}}_i'I_i^{A}])^{-1}),
	\end{align*}
	and therefore,
	$$
	\sqrt{n}(\hat\beta^c-\tilde\beta_n)\stackrel{d}{\longrightarrow} {\cal N}(0, (E[\tilde{\mathbf{Z}}_i\tilde{\mathbf{D}}_i'I_i^{A}])^{-1}E[\tilde\epsilon_i^2\tilde{\mathbf{Z}}_i\tilde{\mathbf{Z}}_i'I_i^{A}](E[\tilde{\mathbf{D}}_i\tilde{\mathbf{Z}}_i'I_i^{A}])^{-1}).
	$$
	The conclusion then follows from Steps \ref{step:tilde-beta_1} and \ref{step:sigma-consistency}.
\end{proof}

\subsubsection{Proof of Asymptotic Properties of \texorpdfstring{$\hat\beta^s_1$}{hat beta_s} When \texorpdfstring{$\Pr(A(X_i)\in (0,1))>0$}{A is stochastic}}

Let $I_i^s=1\{p^s(X_i;\delta_n)\in (0,1)\}$, $\mathbf{D}_i^s=(1,D_i,p^s(X_i;\delta_n))'$,  $\mathbf{Z}_i^s=(1,Z_i,p^s(X_i;\delta_n))'$,
\begin{align*}
	\hat\beta^{c,s} &= (\sum_{i=1}^n \mathbf{Z}_i^s(\mathbf{D}_i^s)'I_i^s)^{-1}\sum_{i=1}^n \mathbf{Z}_i^sY_iI_i^s, \text{ and} \\
	\hat{\mathbf{\Sigma}}^{c,s}&=(\sum_{i=1}^n \mathbf{Z}_{i}^s(\mathbf{D}_{i}^s)'I_{i}^s)^{-1}(\sum_{i=1}^n (\hat\epsilon_{i}^s)^2\mathbf{Z}_{i}^s(\mathbf{Z}_{i}^s)'I_{i}^s)(\sum_{i=1}^n \mathbf{D}_{i}^s(\mathbf{Z}_{i}^s)'I_{i}^s)^{-1},
\end{align*}
where $\hat\epsilon_i^s=Y_i-(\mathbf{D}_{i}^s)'\hat\beta^{c,s}$.
We only show $\hat\beta_1^{c,s}\stackrel{p}{\longrightarrow} \beta_1$ if $S_n\rightarrow \infty$ and $(\hat\sigma^s)^{-1}(\hat\beta_1^{c,s}-\beta_1)\stackrel{d}{\longrightarrow} {\cal N}(0,1)$ if Assumptions \ref{assumption:asymptotic-normality} and \ref{assumption:simulation} hold and $n\delta_n^2\rightarrow 0$ when $\Var(A(X_i)|I_i^{A}=1)>0$.
For that, it suffices to show that
$	\hat\beta^{c,s}-\hat\beta^c=o_p(1)$
if $S_n\rightarrow \infty$ and that
	$\sqrt{n}(\hat\beta^{c,s}-\hat\beta^c)=o_p(1)$ and 
	$n\hat{\mathbf{\Sigma}}^{c,s}\stackrel{p}{\longrightarrow} (E[\tilde{\mathbf{Z}}_i\tilde{\mathbf{D}}_i'I_i^{A}])^{-1}E[\tilde\epsilon_i^2\tilde{\mathbf{Z}}_i\tilde{\mathbf{Z}}_i'I_i^{A}](E[\tilde{\mathbf{D}}_i\tilde{\mathbf{Z}}_i'I_i^{A}])^{-1}$
if Assumptions \ref{assumption:asymptotic-normality} and \ref{assumption:simulation} hold and $n\delta_n^2\rightarrow 0$.
We have
\begin{align*}
	&\hat\beta^{c,s}-\hat\beta^c=(\frac{1}{n}\sum_{i=1}^n \mathbf{Z}_i^s(\mathbf{D}_i^s)'I_i^s)^{-1}\frac{1}{n}\sum_{i=1}^n \mathbf{Z}_i^sY_iI_i^s-(\frac{1}{n}\sum_{i=1}^n \mathbf{Z}_i\mathbf{D}_i'I_{i})^{-1}\frac{1}{n}\sum_{i=1}^n \mathbf{Z}_iY_iI_{i}\\
	&=(\frac{1}{n}\sum_{i=1}^n \mathbf{Z}_i^s(\mathbf{D}_i^s)'I_i^s)^{-1}(\frac{1}{n}\sum_{i=1}^n \mathbf{Z}_i^sY_iI_i^s-\frac{1}{n}\sum_{i=1}^n\mathbf{Z}_iY_iI_{i})\\
	&~~~-(\frac{1}{n}\sum_{i=1}^n \mathbf{Z}_i^s(\mathbf{D}_i^s)'I_i^s)^{-1}(\frac{1}{n}\sum_{i=1}^n \mathbf{Z}_i^s(\mathbf{D}_i^s)'I_i^s-\frac{1}{n}\sum_{i=1}^n \mathbf{Z}_i\mathbf{D}_i'I_{i})\times (\frac{1}{n}\sum_{i=1}^n \mathbf{Z}_i\mathbf{D}_i'I_{i})^{-1}\frac{1}{n}\sum_{i=1}^n \mathbf{Z}_iY_iI_{i}.
\end{align*}
By Lemma \ref{lemma:mean-simulation-plim}, $\hat\beta^{c,s}-\hat\beta^c=o_p(1)$ if $S_n\rightarrow \infty$, and $\sqrt{n}(\hat\beta^{c,s}-\hat\beta^c)=o_p(1)$ if Assumptions \ref{assumption:asymptotic-normality} and \ref{assumption:simulation} hold and $n\delta_n^2\rightarrow 0$.

By proceeding as in Step \ref{step:sigma-consistency} in Appendix \ref{subsubsection:consistency}, we have
\begin{align*}
	\frac{1}{n}\sum_{i=1}^n (\hat \epsilon_i^s)^2\mathbf{Z}_i^s(\mathbf{Z}_i^s)'I_i^s
	=\frac{1}{n}\sum_{i=1}^n(\epsilon_i^s)^2\mathbf{Z}_i^s(\mathbf{Z}_i^s)'I_i^s+o_p(1),
\end{align*}
where $\epsilon_i^s=Y_i-(\mathbf{D}_i^s)'\beta$.
Then, by Lemma \ref{lemma:mean-simulation-plim},
\begin{align*}
	&\frac{1}{n}\sum_{i=1}^n (\hat \epsilon_i^s)^2\mathbf{Z}_i^s(\mathbf{Z}_i^s)'I_i^s-\frac{1}{n}\sum_{i=1}^n\epsilon_i^2\mathbf{Z}_i\mathbf{Z}_i'I_{i}\\
	&=\frac{1}{n}\sum_{i=1}^n(Y_i^2-2Y_i(\mathbf{D}_i^s)'\beta+\beta'\mathbf{D}_i^s(\mathbf{D}_i^s)'\beta)\mathbf{Z}_i^s(\mathbf{Z}_i^s)'I_i^s\\
	&~~~~-\frac{1}{n}\sum_{i=1}^n(Y_i^2-2Y_i\mathbf{D}_i'\beta+\beta'\mathbf{D}_i\mathbf{D}_i'\beta)\mathbf{Z}_i\mathbf{Z}_i'I_{i}+o_p(1)
	=o_p(1)
\end{align*}
so that
$
\frac{1}{n}\sum_{i=1}^n (\hat \epsilon_i^s)^2\mathbf{Z}_i^s(\mathbf{Z}_i^s)'I_i^s\stackrel{p}{\longrightarrow}E[\tilde\epsilon_i^2\tilde{\mathbf{Z}}_i\tilde{\mathbf{Z}}_i'I_i^{A}].
$
Also, $\frac{1}{n}\sum_{i=1}^n \mathbf{Z}_i^s(\mathbf{D}_i^s)'I_i^s\stackrel{p}{\longrightarrow}E[\tilde{\mathbf{Z}}_i\tilde{\mathbf{D}}_i'I_i^{A}]$
by using Lemma \ref{lemma:mean-simulation-plim}, implying the conclusion.
\qed

\subsubsection{Proof of Asymptotic Properties of \texorpdfstring{$\hat\beta_1$}{hat beta} When \texorpdfstring{$\Pr(A(X_i)\in (0,1))=0$}{A is deterministic}}
\label{subsubsection:consistency-boundary}

Since $\Pr(A(X_i)\in (0,1))=0$, $\mathbf{I}_n=0$ with probability one.
Hence,
$
\hat\beta=(\sum_{i=1}^n \mathbf{Z}_i\mathbf{D}_i'I_{i})^{-1}\sum_{i=1}^n \mathbf{Z}_iY_iI_{i}
$
with probability one.
We use the notation and results provided in Appendix \ref{appendix:notations-lemmas}.
By Lemma \ref{lemma:neighborhood-integral}, under Assumption \ref{assumption:consistency} \ref{assumption:boundary-partition},
there exists $\mu>0$ such that $d_{\Omega^*}^s$ is twice continuously differentiable on $N(\partial\Omega^*,\mu)$ and that
$$
\int_{N(\partial \Omega^*,\delta)}g(x)dx = \int_{-\delta}^{\delta}\int_{\partial \Omega^*}g(u+\lambda \nu_{\Omega^*}(u))J_{p-1}^{\partial\Omega^*}\psi_{\Omega^*}(u,\lambda)d{\cal H}^{p-1}(u)d\lambda
$$
for every $\delta\in (0,\mu)$ and every function $g:\mathbb{R}^p\rightarrow \mathbb{R}$ that is integrable on $N(\partial \Omega^*,\delta)$.

Below we show that $\hat\beta_1\stackrel{p}{\longrightarrow}\beta_1$ if Assumption \ref{assumption:consistency} holds, $\delta_n\rightarrow 0$, and $n\delta_n\rightarrow\infty$, and that $\hat\sigma^{-1}(\hat\beta_1-\beta_1)\stackrel{d}{\longrightarrow} {\cal N}(0,1)$ if $n\delta_n^3\rightarrow 0$ in addition.
The proof proceeds in eight steps.
\setcounter{step}{0}
\begin{step}\label{step:lim_APS}
	There exist $\bar\delta >0$ and a bounded function $r:\partial\Omega^*\cap N({\cal X},\bar\delta)\times (-1,1)\times (0,\bar\delta)\rightarrow \mathbb{R}$ such that
	$$
	p^{A}(u+\delta v \nu_{\Omega^*}(u);\delta)=k(v)+ \delta r(u,v,\delta)
	$$
	for every $(u,v,\delta)\in \partial\Omega^*\cap N({\cal X},\bar\delta)\times(-1,1)\times (0
	,\bar\delta)$,
	where
	\begin{align*}
		k(v) =
		\begin{cases}
			1-\frac{1}{2}I_{(1-v^2)}(\frac{p+1}{2},\frac{1}{2}) & \ \ \ \text{for $v\in [0,1)$}\\
			\frac{1}{2}I_{(1-v^2)}(\frac{p+1}{2},\frac{1}{2}) & \ \ \ \text{for $v\in (-1,0)$}.
		\end{cases}
	\end{align*}
	Here $I_x(\alpha,\beta)$ is the regularized incomplete beta function (the cumulative distribution function of the beta distribution with shape parameters $\alpha$ and $\beta$).
\end{step}

\begin{proof}
	By Assumption \ref{assumption:consistency} \ref{assumption:boundary-measure} \ref{assumption:zero-set}, there exists $\bar\delta \in (0,\frac{\mu}{2})$ such that $A(x)=0$ for almost every $x\in N({\cal X},3\bar\delta)\setminus \Omega^*$.
	By Taylor's theorem, for every $u\in \partial\Omega^*\cap N({\cal X},\bar\delta)$ and $a\in B(\bm{0},2\bar\delta)$,
	\begin{align*}
		d_{\Omega^*}^s(u+a) &= d_{\Omega^*}^s(u)+\nabla d_{\Omega^*}^s(u)' a + a' R(u,a)a= \nu_{\Omega^*}(u)' a + a' R(u,a)a,
	\end{align*}
	where
	$
	R(u,a)=\int_0^1(1-t)D^2d_{\Omega^*}^s(u+ta)dt,
	$
	and the second equality follows since $d_{\Omega^*}^s(u)=0$ and $\nabla d_{\Omega^*}^s(u)=\nu_{\Omega^*}(u)$ for every $u\in \partial\Omega^*\cap N({\cal X},\bar\delta)$ by Lemma \ref{lemma:Crasta2007SDF-C2}.
	Since $D^2d_{\Omega^*}^s$ is continuous and ${\rm cl}(N(\partial\Omega^*,2\bar\delta))$ is bounded and closed, $D^2d_{\Omega^*}^s$ is bounded on ${\rm cl}(N(\partial\Omega^*,2\bar\delta))$.
	Therefore, $R(\cdot,\cdot)$ is bounded on $\partial\Omega^*\cap N({\cal X},\bar\delta)\times  B(\bm{0},2\bar\delta)$.
	
	For $(u,v,\delta)\in \partial\Omega^*\cap N({\cal X},\bar\delta)\times (-1,1)\times(0,\bar\delta)$,
	\begin{align*}
		&p^{A}(u+\delta v \nu_{\Omega^*}(u);\delta)
		=\frac{\delta^p\int_{B(\bm{0},1)}A(u+\delta v \nu_{\Omega^*}(u)+\delta w)dw}{\delta^p\int_{B(\bm{0},1)}dw}\\
		&=\frac{\int_{B(\bm{0},1)}1\{u+\delta v \nu_{\Omega^*}(u)+\delta w\in \Omega^*\}dw}{{\rm Vol}_p}=\frac{\int_{B(\bm{0},1)}1\{d_{\Omega^*}^s(u+\delta (v \nu_{\Omega^*}(u)+w))\ge 0)\}dw}{{\rm Vol}_p}\\
		&=\Bigl[\int_{B(\bm{0},1)}1\{\delta\nu_{\Omega^*}(u)'(v \nu_{\Omega^*}(u)+w)+\\
		&~~~~~~~~~~~~~~~~~~\delta^2 (v \nu_{\Omega^*}(u)+w)'R(u,\delta(v \nu_{\Omega^*}(u)+w))(v \nu_{\Omega^*}(u)+w)\ge 0\}dw\Bigr]/{\rm Vol}_p,
	\end{align*}
	where ${\rm Vol}_p$ denotes the volume of the $p$-dimensional unit ball, and the second equality follows since $u+\delta v \nu_{\Omega^*}(u)+\delta w\in N({\cal X},3\bar\delta)$ and hence $A(u+\delta v \nu_{\Omega^*}(u)+\delta w)=0$ for almost every $w\in B(\bm{0},1)$ such that $u+\delta v \nu_{\Omega^*}(u)+\delta w\notin \Omega^*$.
	Observe that
	\begin{align*}
		&1\{\delta\nu_{\Omega^*}(u)'(v \nu_{\Omega^*}(u)+w)+\delta^2 (v \nu_{\Omega^*}(u)+w)'R(u,\delta(v \nu_{\Omega^*}(u)+w))(v \nu_{\Omega^*}(u)+w)\ge 0\}\\
		&=1\{v +\nu_{\Omega^*}(u)\cdot w+\delta (v \nu_{\Omega^*}(u)+w)'R(u,\delta(v \nu_{\Omega^*}(u)+w))(v \nu_{\Omega^*}(u)+w)\ge 0\}\\
		&=1\{v +\nu_{\Omega^*}(u)\cdot w\ge 0\}-a(u,v,w,\delta)+b(u,v,w,\delta),
	\end{align*}
	where
	\begin{align*}
	    a(u,v,w,\delta)
	    &=1\{v +\nu_{\Omega^*}(u)\cdot w\ge 0,\\
	    &~v +\nu_{\Omega^*}(u)\cdot w+\delta (v \nu_{\Omega^*}(u)+w)'R(u,\delta(v \nu_{\Omega^*}(u)+w))(v \nu_{\Omega^*}(u)+w)< 0\},\\
	    b(u,v,w,\delta)
	    &=1\{v +\nu_{\Omega^*}(u)\cdot w<0, \\
	    &~v +\nu_{\Omega^*}(u)\cdot w+\delta (v \nu_{\Omega^*}(u)+w)'R(u,\delta(v \nu_{\Omega^*}(u)+w))(v \nu_{\Omega^*}(u)+w)\ge 0\}.
	\end{align*}
	$\{w\in B(\bm{0},1):v+\nu_{\Omega^*}(u)\cdot w\ge0\}$ is a region of the $p$-dimensional unit ball cut off by the plane $\{w\in \mathbb{R}^p:v+\nu_{\Omega^*}(u)\cdot w=0\}$.
	The distance from the unit ball's center to the plane is $|v|$.
	By the formula for the volume of a hyperspherical cap \citepappendix{Li2011spherical-cap}, 
	\begin{align*}
		&\int_{B(\bm{0},1)}1\{v+\nu_{\Omega^*}(u)\cdot w\ge 0\}dw
		=
		\begin{cases}
			{\rm Vol}_p(1-\frac{1}{2}I_{(2(1-v)-(1-v)^2)}(\frac{p+1}{2},\frac{1}{2})) & \hspace{-0.5em}\text{if $v\in [0,1)$}\\
			\frac{1}{2}{\rm Vol}_pI_{(2(1+v)-(1+v)^2)}(\frac{p+1}{2},\frac{1}{2})  & \hspace{-0.5em}\text{if $v\in (-1,0)$}.
		\end{cases}
	\end{align*}
	Therefore, for every $(u,v,\delta)\in \partial\Omega^*\cap N({\cal X},\bar\delta)\times(-1,1)\times (0
	,\bar\delta)$,
	$$
	p^{A}(u+\delta v \nu_{\Omega^*}(u);\delta)=k(v)+\frac{\int_{B(\bm{0},1)}(-a(u,v,w,\delta)+ b(u,v,w,\delta))dw}{{\rm Vol}_p}.
	$$
	
	Now let $r(u,v,\delta)=\delta^{-1}(p^{A}(u+\delta v \nu_{\Omega^*}(u);\delta)-k(v))$.
	Since $R(\cdot,\cdot)$ is bounded on $\partial\Omega^*\cap N({\cal X},\bar\delta)\times  B(\bm{0},2\bar\delta)$ and $\|\nu_{\Omega^*}(u)\|=1$, there exists $\bar r>0$ such that
	$$
	|(v \nu_{\Omega^*}(u)+w)'R(u,\delta(v \nu_{\Omega^*}(u)+w))(v \nu_{\Omega^*}(u)+w)|\le \bar r
	$$ for every $(u,v,w,\delta)\in \partial\Omega^*\cap N({\cal X},\bar\delta)\times(-1,1)\times B(\bm{0},1)\times(0,\bar\delta)$.
	Therefore,
	\begin{align*}
	&0\le a(u,v,w,\delta) \le 1\{0\le v +\nu_{\Omega^*}(u)\cdot w<\delta \bar r\},\\
	&0\le b(u,v,w,\delta) \le 1\{-\delta \bar r\le v +\nu_{\Omega^*}(u)\cdot w<0\}.
	\end{align*}
	It then follows that
	\begin{align*}
		-\frac{\int_{B(\bm{0},1)}1\{0\le v +\nu_{\Omega^*}(u)\cdot w<\delta \bar r\}dw}{{\rm Vol}_p}&\le \frac{\int_{B(\bm{0},1)}(-a(u,v,w,\delta)+ b(u,v,w,\delta))dw}{{\rm Vol}_p}\\
		&\le\frac{\int_{B(\bm{0},1)}1\{-\delta \bar r\le v +\nu_{\Omega^*}(u)\cdot w<0\}dw}{{\rm Vol}_p}.
	\end{align*}
	The set $\{w\in B(\bm{0},1):0\le v +\nu_{\Omega^*}(u)\cdot w<\delta \bar r\}$ is a region of the $p$-dimensional unit ball cut off by the two planes $\{w\in \mathbb{R}^p: v +\nu_{\Omega^*}(u)\cdot w=0\}$ and $\{w\in \mathbb{R}^p: v +\nu_{\Omega^*}(u)\cdot w=\delta\bar r\}$.
	Its Lebesgue measure is at most the volume of the ($p-1$)-dimensional unit ball times the distance between the two planes, so
	$$
	-\delta {\rm Vol}_{p-1}\bar r\le -\int_{B(\bm{0},1)}1\{0\le v +\nu_{\Omega^*}(u)\cdot w<\delta \bar r\}dw.
	$$
	Likewise,
	$$
	\int_{B(\bm{0},1)}1\{-\delta \bar r\le v +\nu_{\Omega^*}(u)\cdot w<0\}dw \le \delta {\rm Vol}_{p-1}\bar r.
	$$
	Therefore,
	$$
	-\frac{\delta {\rm Vol}_{p-1}\bar r}{{\rm Vol}_p}\le \frac{\int_{B(\bm{0},1)}(-a(u,v,w,\delta)+ b(u,v,w,\delta))dw}{{\rm Vol}_p}\le \frac{\delta {\rm Vol}_{p-1}\bar r}{{\rm Vol}_p}.
	$$
	It follows that
	\begin{align*}
		r(u,v,\delta) 
		&=\delta^{-1}\frac{\int_{B(\bm{0},1)}(-a(u,v,w,\delta)+ b(u,v,w,\delta))dw}{{\rm Vol}_p}
		\in \left[-\frac{{\rm Vol}_{p-1}\bar r}{{\rm Vol}_p}, \frac{{\rm Vol}_{p-1}\bar r}{{\rm Vol}_p}\right],
	\end{align*}
	and hence $r$ is bounded on $\partial\Omega^*\cap N({\cal X},\bar\delta)\times (-1,1)\times(0,\bar\delta)$.\end{proof}

\begin{step}\label{step:nondegenerateAPS}
	For every $(u,v,\delta)\in \partial\Omega^*\cap N({\cal X},\bar\delta)\times(-1,1)\times (0,\bar\delta)$, $p^{A}(u+\delta v\nu_{\Omega^*}(u);\delta)\in (0,1)$.
\end{step}

\begin{proof}
	Fix $(u,v,\delta)\in \partial\Omega^*\cap N({\cal X},\bar\delta)\times(-1,1)\times (0
	,\bar\delta)$.
	Let $\epsilon\in (0, \delta(1- |v|))$.
	Note that $B(u,\epsilon)\subset B(u+\delta v\nu_{\Omega^*}(u),\delta)$, since for any $x\in B(u,\epsilon)$, $\|u+\delta v\nu_{\Omega^*}(u)-x\|\le \|\delta v\nu_{\Omega^*}(u)\|+\|u-x\|\le \delta|v|+\epsilon<\delta$.
	By Step \ref{step:lim_APS}, $p^{A}(u)=\lim_{\delta'\rightarrow 0}p^{A}(u;\delta')=k(0)=\frac{1}{2}$.
	This implies that
	there exists $\epsilon'\in (0,\epsilon)$ such that $p^{A}(u;\epsilon')\in (0,1)$.
	It then follows that $0<{\cal L}^p(B(u,\epsilon')\cap \Omega^*)\le {\cal L}^p(B(u,\epsilon)\cap \Omega^*)\le {\cal L}^p(B(u+\delta v\nu_{\Omega^*}(u),\delta)\cap \Omega^*)$ and that $0<{\cal L}^p(B(u,\epsilon')\setminus \Omega^*)\le {\cal L}^p(B(u,\epsilon)\setminus \Omega^*)\le {\cal L}^p(B(u+\delta v\nu_{\Omega^*}(u),\delta)\setminus \Omega^*)$.
	Therefore, $p^{A}(u+\delta v\nu_{\Omega^*}(u);\delta)=\frac{{\cal L}^p(B(u+\delta v\nu_{\Omega^*}(u),\delta)\cap \Omega^*)}{{\cal L}^p(B(u+\delta v\nu_{\Omega^*}(u),\delta))}\in (0,1)$.\end{proof}

\begin{step}\label{step:mean-lim-2}
		Let $g:\mathbb{R}^p\rightarrow \mathbb{R}$ be a function that is bounded on $N(\partial\Omega^*,\delta')\cap N({\cal X},\delta')$ for some $\delta'>0$.
		Then, for $l\ge 0$, there exist $\tilde\delta>0$ and constant $C>0$ such that
		\begin{align*}
			|\delta^{-1} E[p^{A}(X_i;\delta)^lg(X_i)1\{p^{A}(X_i;\delta)\in (0,1)\}]| \le C
		\end{align*}
		for every $\delta\in (0,\tilde\delta)$.
		If $g$ is continuous on $N(\partial\Omega^*,\delta')\cap N({\cal X},\delta')$ for some $\delta' >0$, then
		\begin{align*}
			\delta^{-1} E[p^{A}(X_i;\delta)^lg(X_i)1\{p^{A}(X_i;\delta)\in (0,1)\}] &=\int_{-1}^1k(v)^ldv\int_{\partial\Omega^*}g(x)f_X(x)d{\cal H}^{p-1}(x)+o(1)\\
			\delta^{-1} E[Z_ip^{A}(X_i;\delta)^lg(X_i)1\{p^{A}(X_i;\delta)\in (0,1)\}] &=\int_0^1k(v)^ldv\int_{\partial\Omega^*}g(x)f_X(x)d{\cal H}^{p-1}(x)+o(1)
		\end{align*}
		for $l\ge 0$.
		Furthermore, if $g$ is continuously differentiable and $\nabla g$ is bounded on $N(\partial\Omega^*,\delta')\cap N({\cal X},\delta')$ for some $\delta' >0$, then for $l\ge 0$, 
		\begin{align*}
		\delta^{-1} E[p^{A}(X_i;\delta)^lg(X_i)1\{p^{A}(X_i;\delta)\in (0,1)\}] &=\int_{-1}^1k(v)^ldv\int_{\partial\Omega^*}g(x)f_X(x)d{\cal H}^{p-1}(x)+O(\delta)\\
		\delta^{-1} E[Z_ip^{A}(X_i;\delta)^lg(X_i)1\{p^{A}(X_i;\delta)\in (0,1)\}] &=\int_0^1k(v)^ldv\int_{\partial\Omega^*}g(x)f_X(x)d{\cal H}^{p-1}(x)+O(\delta).
		\end{align*}

\end{step}

\begin{proof}
	Let $\bar\delta$ be given in Step \ref{step:lim_APS}.
	Under Assumption \ref{assumption:consistency} \ref{assumption:boundary-continuity}, there exists $\tilde\delta\in (0,\bar\delta)$ such that $f_X$ is bounded, is continuously differentiable, and has bounded partial derivatives on $N(\partial\Omega^*,2\tilde\delta)\cap N({\cal X},2\tilde\delta)$.
	Let $\tilde\delta\in (0,\bar\delta)$ such that both $g$ and $f_X$ are bounded on $N(\partial\Omega^*,2\tilde\delta)\cap N({\cal X},2\tilde\delta)$.
	We first show  $p^{A}(x;\delta)\in \{0,1\}$ for every $x\in {\cal X}\setminus N(\partial\Omega^*,\delta)$ for every $\delta\in (0,\tilde\delta)$.
	Pick $x\in {\cal X}\setminus N(\partial\Omega^*,\delta)$ and $\delta\in (0,\tilde\delta)$.
	Since $B(x,\delta)\cap \partial\Omega^*=\emptyset$,
	either $B(x,\delta)\subset {\rm int}(\Omega^*)$ or $B(x,\delta)\subset{\rm int}(\mathbb{R}^p\setminus \Omega^*)$.
	If $B(x,\delta)\subset {\rm int}(\Omega^*)$, $p^{A}(x;\delta)=1$.
	If $B(x,\delta)\subset{\rm int}(\mathbb{R}^p\setminus \Omega^*)$, $p^{A}(x;\delta)=0$, since $A(x')=0$ for almost every $x'\in B(x,\delta)\subset N({\cal X},3\bar\delta)\setminus\Omega^*$ by the choice of $\bar\delta$.
	Thus $\{x\in{\cal X}:p^{A}(x;\delta)\in (0,1)\}\subset N(\partial\Omega^*,\delta)$ for every $\delta\in (0,\tilde\delta)$.
	By this and Lemma \ref{lemma:neighborhood-integral}, for $\delta\in (0,\tilde\delta)$,
	\begin{align*}
		&\delta^{-1} E[p^{A}(X_i;\delta)^lg(X_i)1\{p^{A}(X_i;\delta)\in (0,1)\}]
		=\delta^{-1}\int p^{A}(x;\delta)^lg(x)1\{p^{A}(x;\delta)\in (0,1)\}f_X(x)dx\\
		&=\delta^{-1}\int_{N(\partial\Omega^*,\delta)} p^{A}(x;\delta)^lg(x)1\{p^{A}(x;\delta)\in (0,1)\}f_X(x)dx\\
		&=\delta^{-1}\int_{-\delta}^{\delta}\int_{\partial\Omega^*}p^{A}(u+\lambda\nu_{\Omega^*}(u);\delta)^lg(u+\lambda\nu_{\Omega^*}(u))1\{p^{A}(u+\lambda\nu_{\Omega^*}(u);\delta)\in (0,1)\}\\
		&~~~~~~~~~~~~~~~\times f_X(u+\lambda\nu_{\Omega^*}(u))J_{p-1}^{\partial\Omega^*}\psi_{\Omega^*}(u,\lambda)d{\cal H}^{p-1}(u)d\lambda.
	\end{align*}
	With change of variables $v = \frac{\lambda}{\delta}$, we have
	\begin{align*}
		&\delta^{-1} E[p^{A}(X_i;\delta)^lg(X_i)1\{p^{A}(X_i;\delta)\in (0,1)\}]\\
		=&\int_{-1}^{1}\int_{\partial\Omega^*} p^{A}(u+\delta v \nu_{\Omega^*}(u);\delta)^l 1\{p^{A}(u+\delta v \nu_{\Omega^*}(u);\delta)\in (0,1)\}\\
		&~~~~~~~~~~~~~~~\times g(u+\delta v \nu_{\Omega^*}(u))f_X(u+\delta v \nu_{\Omega^*}(u))J_{p-1}^{\partial\Omega^*}\psi_{\Omega^*}(u,\delta v)d{\cal H}^{p-1}(u) dv.
	\end{align*}
	For every $(u,v,\delta)\in \partial\Omega^*\setminus N({\cal X},\tilde\delta)\times(-1,1)\times (0
	,\tilde\delta)$, $u+\delta v \nu_{\Omega^*}(u)\notin {\cal X}$, so
	\begin{align*}
		&\delta^{-1} E[p^{A}(X_i;\delta)^lg(X_i)1\{p^{A}(X_i;\delta)\in (0,1)\}]\\
		=&\int_{-1}^{1}\int_{\partial\Omega^*\cap N({\cal X},\tilde\delta)} p^{A}(u+\delta v \nu_{\Omega^*}(u);\delta)^l 1\{p^{A}(u+\delta v \nu_{\Omega^*}(u);\delta)\in (0,1)\}\\
		&~~~~~~~~~~~~~~~\times g(u+\delta v \nu_{\Omega^*}(u))f_X(u+\delta v \nu_{\Omega^*}(u))J_{p-1}^{\partial\Omega^*}\psi_{\Omega^*}(u,\delta v)d{\cal H}^{p-1}(u) dv\\
		=&\int_{-1}^{1}\int_{\partial\Omega^*\cap N({\cal X},\tilde\delta)} (k(v)+\delta r(u,v,\delta))^l\\
		&~~~~~~~~~~~~~~~\times g(u+\delta v \nu_{\Omega^*}(u))f_X(u+\delta v \nu_{\Omega^*}(u))J_{p-1}^{\partial\Omega^*}\psi_{\Omega^*}(u,\delta v)d{\cal H}^{p-1}(u) dv,
	\end{align*}
	where the second equality follows from Steps \ref{step:lim_APS} and \ref{step:nondegenerateAPS}.
	By Lemma \ref{lemma:neighborhood-integral}, $J_{p-1}^{\partial\Omega^*}\psi_{\Omega^*}(\cdot,\cdot)$ is bounded on $\partial\Omega^*\times (-\tilde\delta,\tilde\delta)$.
	Since $r$, $g$ and $f_X$ are also bounded, for some constant $C>0$,
	\begin{align*}
		|\delta^{-1} E[p^{A}(X_i;\delta)^lg(X_i)1\{p^{A}(X_i;\delta)\in (0,1)\}]|
		\le C \int_{-1}^{1}\int_{\partial\Omega^*\cap N({\cal X},\tilde\delta)} d{\cal H}^{p-1}(u) dv,
	\end{align*}
	which is finite by Assumption \ref{assumption:consistency} \ref{assumption:boundary-measure} \ref{assumption:p-1dim}.
	Moreover, if $g$ and $f_X$ are continuous on $N(\partial\Omega^*,2\tilde \delta)\cap N({\cal X},2\tilde \delta)$, by the Dominated Convergence Theorem,
	\begin{align*}
		\delta^{-1} E[p^{A}(X_i;\delta)^lg(X_i)1\{p^{A}(X_i;\delta)\in (0,1)\}] \rightarrow\int_{-1}^1k(v)^ldv\int_{\partial\Omega^*}g(u)f_X(u)d{\cal H}^{p-1}(u),
	\end{align*}
	where we use the fact from Lemma \ref{lemma:neighborhood-integral} that $J_{p-1}^{\partial\Omega^*}\psi_{\Omega^*}(u,\lambda)$ is continuous in $\lambda$ and $J_{p-1}^{\partial\Omega^*}\psi_{\Omega^*}(u,0)=1$.
	
	Note that $A(x)=1$ for every $x\in\Omega^*$ and $A(x)=0$ for almost every $x\in N({\cal X},2\tilde\delta)\setminus\Omega^*$.
	Also, for every $(u,v,\delta)\in\partial\Omega^*\cap N({\cal X},\tilde\delta)\times(-1,1)\times (0
	,\tilde\delta)$, $u+\delta v \nu_{\Omega^*}(u)\in \Omega^*$ if $v\in (0,1)$ and $u+\delta v \nu_{\Omega^*}(u)\in N({\cal X},2\tilde\delta)\setminus\Omega^*$ if $v\in (-1,0]$.
	Therefore,
	\begin{align*}
		&~\delta^{-1} E[Z_ip^{A}(X_i;\delta)^lg(X_i)1\{p^{A}(X_i;\delta)\in (0,1)\}]\\
		=&~\delta^{-1} E[A(X_i)p^{A}(X_i;\delta)^lg(X_i)1\{p^{A}(X_i;\delta)\in (0,1)\}]\\
		=&\int_{-1}^{1}\int_{\partial\Omega^*\cap N({\cal X},\tilde\delta)} A(u+\delta v \nu_{\Omega^*}(u))(k(v)+\delta r(u,v,\delta))^l g(u+\delta v \nu_{\Omega^*}(u))\\
		&~~~~~~~~~~~~~~~\times f_X(u+\delta v \nu_{\Omega^*}(u))J_{p-1}^{\partial\Omega^*}\psi_{\Omega^*}(u,\delta v)d{\cal H}^{p-1}(u) dv\\
		=&\int_{0}^{1}\int_{\partial\Omega^*\cap N({\cal X},\tilde\delta)} (k(v)+\delta r(u,v,\delta))^l g(u+\delta v \nu_{\Omega^*}(u))\\
		&~~~~~~~~~~~~~~~\times f_X(u+\delta v \nu_{\Omega^*}(u))J_{p-1}^{\partial\Omega^*}\psi_{\Omega^*}(u,\delta v)d{\cal H}^{p-1}(u) dv\\
		\rightarrow&\int_{0}^1k(v)^ldv\int_{\partial\Omega^*}g(u)f_X(u)d{\cal H}^{p-1}(u).
	\end{align*}
	
	Now suppose that $g$ and $f_X$ are continuously differentiable on $N(\partial\Omega^*,2\tilde \delta)\cap N({\cal X},2\tilde \delta)$ and that $\nabla g$ and $\nabla f$ are bounded on $N(\partial\Omega^*,2\tilde \delta)\cap N({\cal X},2\tilde \delta)$.
	Using the mean-value theorem, we obtain that, for any $(u,v,\delta)\in\partial\Omega^*\cap N({\cal X},\tilde\delta)\times(-1,1)\times(0,\tilde\delta)$,
	\begin{align*}
		g(u+\delta v \nu_{\Omega^*}(u))&=g(u)+\nabla g(y_g(u,\delta v \nu_{\Omega^*}(u)))'\delta v \nu_{\Omega^*}(u),\\
		f_X(u+\delta v \nu_{\Omega^*}(u))&=f_X(u)+\nabla f_X(y_f(u,\delta v \nu_{\Omega^*}(u)))'\delta v \nu_{\Omega^*}(u)
	\end{align*}
	for some $y_g(u,\delta v\nu_{\Omega^*}(u))$ and $y_f(u,\delta v\nu_{\Omega^*}(u))$ that are on the line segment connecting $u$ and $u+\delta v\nu_{\Omega^*}(u)$.
	In addition,
	\begin{align*}
	J_{p-1}^{\partial\Omega^*}\psi_{\Omega^*}(u,\delta v)&=J_{p-1}^{\partial\Omega^*}\psi_{\Omega^*}(u,0)+\frac{\partial J_{p-1}^{\partial\Omega^*}\psi_{\Omega^*}(u,y_J(u,\delta v))}{\partial \lambda}\delta v=1+\frac{\partial J_{p-1}^{\partial\Omega^*}\psi_{\Omega^*}(u,y_J(u,\delta v))}{\partial \lambda}\delta v
	\end{align*}
	for some $y_J(u,\delta v)$ that is on the line segment connecting $0$ and $\delta v$.
	By Lemma \ref{lemma:neighborhood-integral}, $\tfrac{\partial J_{p-1}^{\partial\Omega^*}\psi_{\Omega^*}(\cdot,\cdot)}{\partial \lambda}$ is bounded on $\partial\Omega^*\times (-\tilde\delta,\tilde\delta)$.
	We then have
	\begin{align*}
		&\delta^{-1}E[p^{A}(X_i;\delta)^lg(X_i)1\{p^{A}(X_i;\delta)\in (0,1)\}]\\
		=&\int_{-1}^{1}\int_{\partial\Omega^*\cap N({\cal X},\tilde\delta)} (k(v)+\delta r(u,v,\delta))^l (g(u)+\nabla g(y_g(u,\delta v \nu_{\Omega^*}(u)))'\delta v \nu_{\Omega^*}(u))\\
		&~~\times (f_X(u)+\nabla f_X(y_f(u,\delta v \nu_{\Omega^*}(u)))'\delta v \nu_{\Omega^*}(u))(1+\tfrac{\partial J_{p-1}^{\partial\Omega^*}\psi_{\Omega^*}(u,y_J(u,\delta v))}{\partial \lambda}\delta v)d{\cal H}^{p-1}(u) dv\\
		=&\int_{-1}^{1}\int_{\partial\Omega^*\cap N({\cal X},\tilde\delta)} (k(v)^lg(u)f_X(u)+\delta h(u,v,\delta))d{\cal H}^{p-1}(u) dv\\
		=&\int_{-1}^{1}k(v)^ldv\int_{\partial\Omega^*} g(u)f_X(u)d{\cal H}^{p-1}(u) +\delta\int_{-1}^{1}\int_{\partial\Omega^*\cap N({\cal X},\tilde\delta)} h(u,v,\delta)d{\cal H}^{p-1}(u) dv
	\end{align*}
	for some function $h$ bounded on $\partial\Omega^*\cap N({\cal X},\tilde\delta)\times(-1,1)\times (0
	,\tilde\delta)$.
	It then follows that
	\begin{align*}
	\delta^{-1} E[p^{A}(X_i;\delta)^lg(X_i)1\{p^{A}(X_i;\delta)\in (0,1)\}] &=\int_{-1}^1k(v)^ldv\int_{\partial\Omega^*}g(u)f_X(u)d{\cal H}^{p-1}(u)+O(\delta).
	\end{align*}
	Similarly,
	\begin{align*}
	&\delta^{-1} E[Z_ip^{A}(X_i;\delta)^lg(X_i)1\{p^{A}(X_i;\delta)\in (0,1)\}]
	=\int_{0}^1k(v)^ldv\int_{\partial\Omega^*}g(u)f_X(u)d{\cal H}^{p-1}(u)+O(\delta).
	\end{align*}
	
\end{proof}

\begin{step}
	Let $S_{\mathbf{D}}=\lim_{\delta\rightarrow 0} \delta^{-1}E[\mathbf{Z}_i\mathbf{D}_i' 1\{p^{A}(X_i;\delta)\in (0,1)\}]$ and $S_Y=\lim_{\delta\rightarrow 0} \delta^{-1}E[\mathbf{Z}_iY_i 1\{p^{A}(X_i;\delta)\in (0,1)\}]$.
	Then the second element of $S_{\mathbf{D}}^{-1}S_Y$ is $\beta_1$.
\end{step}

\begin{proof}
	Note that $D_i=Z_iD_i(1)+(1-Z_i)D_i(0)$ and $Y_i=Z_iY_{1i}+(1-Z_i)Y_{0i}$.
	By Step \ref{step:mean-lim-2},
			\begin{align*}
				S_{\mathbf{D}}=
				\begin{bmatrix}
					2\bar f_X& \int_{\partial\Omega^*}E[D_i(1)+D_i(0)|X_i=x]f_X(x)d{\cal H}^{p-1}(x) & \int_{-1}^1 k(v)dv\bar f_X\\
					\bar f_X & \int_{\partial\Omega^*}E[D_i(1)|X_i=x]f_X(x)d{\cal H}^{p-1}(x) & \int_{0}^1 k(v)dv\bar f_X \\
					\int_{-1}^1 k(v)dv\bar f_X & \begin{array}{l}
						\int_{\partial\Omega^*}(\int_{0}^1 k(v)dvE[D_i(1)|X_i=x] \\
						+ \int_{-1}^0 k(v)dvE[D_i(0)|X_i=x])f_X(x)d{\cal H}^{p-1}(x)
					\end{array} & \int_{-1}^1 k(v)^2dv\bar f_X
				\end{bmatrix},
			\end{align*}
	where $\bar f_X=\int_{\partial\Omega^*}f_X(x)d{\cal H}^{p-1}(x)$, and
	\begin{align*}
	S_Y=
	\begin{bmatrix}
	\int_{\partial\Omega^*}E[Y_{1i}+Y_{0i}|X_i=x]f_X(x)d{\cal H}^{p-1}(x) \\
	\int_{\partial\Omega^*}E[Y_{1i}|X_i=x]f_X(x)d{\cal H}^{p-1}(x)  \\
	\int_{\partial\Omega^*}(\int_{0}^1 k(v)dvE[Y_{1i}|X_i=x] + \int_{-1}^0 k(v)dvE[Y_{0i}|X_i=x])f_X(x)d{\cal H}^{p-1}(x)
	\end{bmatrix}.
	\end{align*}
	After a few lines of algebra, we have
	\begin{align*}
		{\rm det}(S_{\mathbf{D}})
	=&\bar f_X^2\int_{\partial\Omega^*}E[D_i(1)-D_i(0)|X_i=x]f_X(x)d{\cal H}^{p-1}(x)\\
	&\times\left(\int_{-1}^0\left(k(v)-\int_{-1}^0k(s)ds\right)^2dv+\int_{0}^1\left(k(v)-\int_{0}^1k(s)ds\right)^2dv\right).
	\end{align*}
	We verify that ${\rm det}(S_{\mathbf{D}})$ is nonzero.
	Since $\bar f_X>0$ by Assumption \ref{assumption:consistency} \ref{assumption:boundary-measure} \ref{assumption:p-1dim},
	it suffices to show that $\int_{\partial\Omega^*}E[D_i(1)-D_i(0)|X_i=x]f_X(x)d{\cal H}^{p-1}(x)\neq 0$.
	To do so, we first show that $p^A(x)\in\{0,1\}$ for every $x\in {\cal X}\setminus \partial\Omega^*$.
	Pick $x\in {\cal X}\setminus \partial\Omega^*$.
	By definition, either $x\in {\cal X} \cap{\rm int}(\Omega^*)$ or $x\in {\cal X} \cap(\mathbb{R}^p\setminus {\rm cl}(\Omega^*))$.
	If $x\in {\cal X} \cap{\rm int}(\Omega^*)$, then $B(x,\delta)\subset {\rm int}(\Omega^*)$ for any sufficiently small $\delta>0$ so that $p^A(x)=1$.
	If $x\in {\cal X} \cap(\mathbb{R}^p\setminus {\rm cl}(\Omega^*))$, then $B(x,\delta)\subset N({\cal X},\delta')\cap(\mathbb{R}^p\setminus {\rm cl}(\Omega^*))$ for any sufficiently small $\delta>0$, where $\delta'>0$ satisfies Assumption \ref{assumption:consistency} \ref{assumption:boundary-measure} \ref{assumption:zero-set}. Since $A(x')=0$ for almost every $x'\in N({\cal X},\delta')\setminus \Omega^*$, $p^A(x)=0$.
	Note also that $p^A(x)=\lim_{\delta\rightarrow 0}p^A(x;\delta)=k(0)=\frac{1}{2}$ for every $x\in\partial\Omega^*\cap {\cal X}$ by Step \ref{step:lim_APS}.
	It then follows that
	\begin{align*}
	    &~\int_{\partial\Omega^*}E[D_i(1)-D_i(0)|X_i=x]f_X(x)d{\cal H}^{p-1}(x)\\
	    =&~4 \int_{\partial\Omega^*\cap {\cal X}}p^A(x)(1-p^A(x))E[D_i(1)-D_i(0)|X_i=x]f_X(x)d{\cal H}^{p-1}(x)\\
	    =&~4 \int_{{\cal X}}p^A(x)(1-p^A(x))E[D_i(1)-D_i(0)|X_i=x]f_X(x)d{\cal H}^{p-1}(x),
	\end{align*}
	which is nonzero under Assumption \ref{assumption:consistency} \ref{assumption:relevance}.
	
	After another few lines of algebra, we obtain that the second element of $S_{\mathbf{D}}^{-1}S_Y$ is
	$$
	\frac{\int_{\partial\Omega^*}E[(D_i(1)-D_i(0))(Y_i(1)-Y_i(0))|X_i=x]f_X(x)d{\cal H}^{p-1}(x)}{\int_{\partial\Omega^*}E[D_i(1)-D_i(0)|X_i=x]f_X(x)d{\cal H}^{p-1}(x)}.
	$$
	On the other hand, by Step \ref{step:mean-lim-2},
	\begin{align*}
		\beta_1
		&=\lim_{\delta\rightarrow 0}E[\omega_i(\delta)(Y_i(1)-Y_i(0))]\\
		&=\lim_{\delta\rightarrow 0}\frac{\delta^{-1}E[p^{A}(X_i;\delta)(1-p^{A}(X_i;\delta))(D_i(1)-D_i(0))(Y_i(1)-Y_i(0))1\{p^{A}(X_i;\delta)\in (0,1)\}]}{\delta^{-1}E[p^{A}(X_i;\delta)(1-p^{A}(X_i;\delta))(D_i(1)-D_i(0))1\{p^{A}(X_i;\delta)\in (0,1)\}]}\\
		&=\frac{\int_{-1}^1k(v)(1-k(v))dv\int_{\partial\Omega^*}E[(D_i(1)-D_i(0))(Y_i(1)-Y_i(0))|X_i=x]f_X(x)d{\cal H}^{p-1}(x)}{\int_{-1}^1k(v)(1-k(v))dv\int_{\partial\Omega^*}E[D_i(1)-D_i(0)|X_i=x]f_X(x)d{\cal H}^{p-1}(x)}\\
		&=\frac{\int_{\partial\Omega^*}E[(D_i(1)-D_i(0))(Y_i(1)-Y_i(0))|X_i=x]f_X(x)d{\cal H}^{p-1}(x)}{\int_{\partial\Omega^*}E[D_i(1)-D_i(0)|X_i=x]f_X(x)d{\cal H}^{p-1}(x)}.
		\qedhere
	\end{align*}
\end{proof}

\begin{step}\label{step:consistency-2}
	If $\delta_n\rightarrow 0$ and $n\delta_n\rightarrow \infty$ as $n\rightarrow \infty$, then $\hat\beta_1\stackrel{p}{\longrightarrow}\beta_1$.
\end{step}

\begin{proof}
	It suffices to verify the variance of each element of $\frac{1}{n\delta_n}\sum_{i=1}^n\mathbf{Z}_i\mathbf{D}_i' I_{i}$ and $\frac{1}{n\delta_n}\sum_{i=1}^n\mathbf{Z}_iY I_{i}$ is $o(1)$.
	We only verify  $\Var(\frac{1}{n\delta_n}\sum_{i=1}^np^{A}(X_i;\delta_n)Y_iI_{i})=o(1)$.
	Note
	$
	E[Y_i^2|X_i]=E[Z_iY_{1i}^2+(1-Z_i)Y_{0i}^2|X_i]\le E[Y_{1i}^2+Y_{0i}^2|X_i].
	$
	Under Assumption \ref{assumption:consistency} \ref{assumption:boundary-continuity}, there exists $\delta'>0$ such that $E[Y_{1i}^2+Y_{0i}^2|X_i]$ is continuous on $N(\partial\Omega^*,\delta')$.
	Since ${\rm cl}(N(\partial\Omega^*,\frac{1}{2}\delta'))$ is closed and bounded, $E[Y_{1i}^2+Y_{0i}^2|X_i]$ is bounded on ${\rm cl}(N(\partial\Omega^*,\frac{1}{2}\delta'))$.
	For some $C>0$, for any sufficiently large $n$, 
	\begin{align*}
		\Var\left(\frac{1}{n\delta_n}\sum_{i=1}^np^{A}(X_i;\delta_n)Y_iI_{i}\right)
		&\le\frac{1}{n\delta_n}\delta_n^{-1}E[p^{A}(X_i;\delta_n)^2E[Y_i^2|X_i]I_{i}]\le \frac{1}{n\delta_n}C,
	\end{align*}
	where the last inequality holds by Step \ref{step:mean-lim-2}.
	The conclusion follows since $n\delta_n\rightarrow \infty$.\end{proof}

Now let $\beta=(\beta_0,\beta_1,\beta_2)'=S_{\mathbf{D}}^{-1}S_Y$ and let $\epsilon_i=Y_i-\mathbf{D}_i'\beta$.
We can write
\begin{align*}
\sqrt{n\delta_n}(\hat\beta-\beta)
&=(\frac{1}{n\delta_n}\sum_{i=1}^n \mathbf{Z}_i\mathbf{D}_i'I_{i})^{-1}\frac{1}{\sqrt{n\delta_n}}\sum_{i=1}^n \{(\mathbf{Z}_i\epsilon_iI_{i}-E[\mathbf{Z}_i\epsilon_iI_{i}])+E[\mathbf{Z}_i\epsilon_iI_{i}]\}.
\end{align*}

\begin{step}\label{step:clt-2}
	$
	\frac{1}{\sqrt{n\delta_n}}\sum_{i=1}^n (\mathbf{Z}_i\epsilon_iI_{i}-E[\mathbf{Z}_i\epsilon_iI_{i}])\stackrel{d}{\longrightarrow}{\cal N}(0,\mathbf{V}),
	$
	where $\mathbf{V}=\lim_{n\rightarrow\infty}\delta_n^{-1}E[\epsilon_i^2\mathbf{Z}_i\mathbf{Z}_iI_{i}]$.
\end{step}

\begin{proof}
	We use the triangular-array Lyapunov CLT and the Cram\'{e}r-Wold device.
	Pick a nonzero $\lambda\in \mathbb{R}^p$, and let $V_{i,n}=\frac{1}{\sqrt{n\delta_n}}\lambda' (\mathbf{Z}_i\epsilon_iI_{i}-E[\mathbf{Z}_i\epsilon_iI_{i}])$.
	First,
	by Step \ref{step:mean-lim-2-2},
	$
	E[\mathbf{Z}_i\epsilon_iI_{i}]=E[\mathbf{Z}_i(Y_i-\mathbf{D}_i'\beta)I_{i}]=O(\delta_n),
	$
	so
	$
	\delta_n^{-1}E[\mathbf{Z}_i\epsilon_iI_{i}]E[\mathbf{Z}_i'\epsilon_iI_{i}]=o(1).
	$
	We have
	\begin{align*}
		E[\epsilon_i^2\mathbf{Z}_i\mathbf{Z}_i'I_{i}]&=E[(Y_i-\beta_0-\beta_1D_i-\beta_2p^{A}(X_i;\delta_n))^2\mathbf{Z}_i\mathbf{Z}_i'I_{i}]\\
		&=E[Z_i(Y_{1i}-\beta_0-\beta_1D_i(1)-\beta_2p^{A}(X_i;\delta_n))^2\mathbf{Z}_i\mathbf{Z}_i'I_{i}]\\
		&~~~~+E[(1-Z_i)(Y_{0i}-\beta_0-\beta_1D_i(0)-\beta_2p^{A}(X_i;\delta_n))^2\mathbf{Z}_i\mathbf{Z}_i'I_{i}].
	\end{align*}
	Since $E[Y_{1i}|X_i]$, $E[Y_{0i}|X_i]$, $E[D_i(1)|X_i]$, $E[D_i(0)|X_i]$, $E[Y_{1i}^2|X_i]$, $E[Y_{0i}^2|X_i]$, $E[Y_{1i}D_i(1)|X_i]$ and $E[Y_{0i}D_i(0)|X_i]$ are continuous on $N(\partial\Omega^*,\delta')$ for some $\delta'>0$ under Assumption \ref{assumption:consistency} \ref{assumption:boundary-continuity}, $\lim_{n\rightarrow\infty}\delta_n^{-1}E[\epsilon_i^2\mathbf{Z}_i\mathbf{Z}_i'I_{i}]$ exists and finite.
	Therefore,
	$
	\sum_{i=1}^n E[V_{i,n}^2]=\delta_n^{-1}\lambda'(E[\epsilon_i^2\mathbf{Z}_i\mathbf{Z}_i'I_{i}]-E[\mathbf{Z}_i\epsilon_iI_{i}]E[\mathbf{Z}_i'\epsilon_iI_{i}])\lambda\rightarrow \lambda'\mathbf{V}\lambda.
	$
    
    We next verify the Lyapunov condition: for some $t>0$,
	$
	\sum_{i=1}^n E[|V_{i,n}|^{2+t}]\rightarrow 0.
	$
	Note
	\begin{align*}
		\sum_{i=1}^n E[|V_{i,n}|^{4}]
		=\frac{1}{n\delta_n}\delta_n^{-1} E[|\lambda' (\mathbf{Z}_i\epsilon_iI_{i}-E[\mathbf{Z}_i\epsilon_iI_{i}])|^{4}]\le \frac{2^{3}}{n\delta_n} \delta_n^{-1}\{E[|\lambda' \mathbf{Z}_i\epsilon_iI_{i}|^{4}]+|\lambda'E[\mathbf{Z}_i\epsilon_iI_{i}]|^{4}\}
	\end{align*}
	by the $c_r$-inequality.
	Repeating using the $c_r$-inequality gives
	\begin{align*}
		\delta_n^{-1}E[|\lambda' \mathbf{Z}_i\epsilon_iI_{i}|^{4}]&=\delta_n^{-1}E[|\lambda' \mathbf{Z}_i(Y_i-\beta_0-\beta_1 D_i-\beta_2 p^{A}(X_i;\delta_n))|^{4}I_{i}]\\
		&\le 2^{3c}\delta_n^{-1}E[(|\lambda' \mathbf{Z}_i|^{4})(|Y_i|^{4}+|\beta_0|^{4}+|\beta_1|^{4}D_i+|\beta_2|^{4}p^{A}(X_i;\delta_n)^{4})I_{i}]\\
		&\le 2^{3c}(|\lambda_1|+|\lambda_2|+|\lambda_3|)^4\delta_n^{-1}E[(Y_i^{4}+\beta_0^{4}+\beta_1^{4}+\beta_2^{4})I_{i}]= 2^{3c}O(1)
	\end{align*}
	for some finite constant $c$, where the last equality holds by Step \ref{step:mean-lim-2} under Assumption \ref{assumption:consistency} \ref{assumption:boundary-continuity}. 
	Moreover,
	$\delta_n^{-1}|\lambda'E[\mathbf{Z}_i\epsilon_iI_{i}]|^{4}=\delta_n^{3}|\lambda'\delta_n^{-1}E[\mathbf{Z}_i\epsilon_iI_{i}]|^{4}=\delta_n^{3}O(1)=o(1)$.
	Therefore, when $n\delta_n\rightarrow \infty$,
	$
	\sum_{i=1}^n E[|V_{i,n}|^{4}]\rightarrow 0,
	$
	and the conclusion follows from the Lyapunov CLT and the Cram\'{e}r-Wold device.
\end{proof}

\begin{step}\label{step:sigma-consistency-2}
	$n\delta_n\hat{\mathbf{\Sigma}}\stackrel{p}{\longrightarrow}S_{\mathbf{D}}^{-1}\mathbf{V}(S_{\mathbf{D}}')^{-1}$.
\end{step}

\begin{proof}
	Using the result that $\hat\beta-\beta=o_p(1)$ and Step \ref{step:mean-lim-2}, we have
	\begin{align*}
		&\frac{1}{n\delta_n}\sum_{i=1}^n \hat \epsilon_i^2\mathbf{Z}_i\mathbf{Z}_i'I_{i}
		=\frac{1}{n\delta_n}\sum_{i=1}^n(Y_i-\mathbf{D}_i'\hat\beta)^2\mathbf{Z}_i\mathbf{Z}_i'I_{i}
		=\frac{1}{n\delta_n}\sum_{i=1}^n(\epsilon_i-\mathbf{D}_i'(\hat\beta-\beta))^2\mathbf{Z}_i\mathbf{Z}_i'I_{i}\\
		&=\frac{1}{n\delta_n}\sum_{i=1}^n\epsilon_i^2\mathbf{Z}_i\mathbf{Z}_i'I_{i}+\frac{1}{n\delta_n}\sum_{i=1}^n((\hat\beta_0-\beta_0)+D_i(\hat\beta_1-\beta_1)+p^{A}(X_i;\delta_n)(\hat\beta_2-\beta_2))^2\mathbf{Z}_i\mathbf{Z}_i'I_{i}\\
		&~~~~-\frac{2}{n\delta_n}\sum_{i=1}^n(Y_i-\mathbf{D}_i'\beta)((\hat\beta_0-\beta_0)+D_i(\hat\beta_1-\beta_1)+p^{A}(X_i;\delta_n)(\hat\beta_2-\beta_2))\mathbf{Z}_i\mathbf{Z}_i'I_{i}\\
		&=\frac{1}{n\delta_n}\sum_{i=1}^n\epsilon_i^2\mathbf{Z}_i\mathbf{Z}_i'I_{i}+o_p(1)O_p(1).
	\end{align*}
	
	To show $\frac{1}{n\delta_n}\sum_{i=1}^n\epsilon_i^2\mathbf{Z}_i\mathbf{Z}_i'I_{i}\stackrel{p}{\longrightarrow}\mathbf{V}$, it suffices to verify that the variance of each element of $\frac{1}{n\delta_n}\sum_{i=1}^n\epsilon_i^2\mathbf{Z}_i\mathbf{Z}_i'I_{i}$ is $o(1)$. We only verify that $\Var(\frac{1}{n\delta_n}\sum_{i=1}^n\epsilon_i^2p^{A}(X_i;\delta_n)^2I_{i})=o(1)$.
	Using the $c_r$-inequality, we have that for some constant $c$,
	\begin{align*}
		&\Var\left(\frac{1}{n\delta_n}\sum_{i=1}^n\epsilon_i^2p^{A}(X_i;\delta_n)^2I_{i}\right)\le \frac{1}{n\delta_n}\delta_n^{-1}E[\epsilon_i^4I_{i}]=\frac{1}{n\delta_n}\delta_n^{-1}E[(Y_i-\beta_0-\beta_1D_i-\beta_2p^{A}(X_i))^4I_{i}]\\
		&\le\frac{1}{n\delta_n}2^{3c}\delta_n^{-1}E[(Y_i^4+\beta_0^4+\beta_1^4D_i+\beta_2^4p^{A}(X_i)^4)I_{i}]
        =\frac{1}{n\delta_n}2^{3c}O(1)=o(1),
	\end{align*}
	where the second last equality holds by Step \ref{step:mean-lim-2} under Assumption \ref{assumption:consistency} \ref{assumption:boundary-continuity}.
	Thus
	$
	\frac{1}{n\delta_n}\sum_{i=1}^n\hat \epsilon_i^2\mathbf{Z}_i\mathbf{Z}_i'I_{i}\stackrel{p}{\longrightarrow}\mathbf{V}.
	$
	It follows that
	$$
	n\delta_n\hat{\mathbf{\Sigma}}=(\frac{1}{n\delta_n}\sum_{i=1}^n\mathbf{Z}_i\mathbf{D}_i' I_{i})^{-1}(\frac{1}{n\delta_n}\sum_{i=1}^n \hat \epsilon_i^2\mathbf{Z}_i\mathbf{Z}_i'I_{i})(\frac{1}{n\delta_n}\sum_{i=1}^n\mathbf{D}_i\mathbf{Z}_i' I_{i})^{-1}\stackrel{p}{\longrightarrow}S_{\mathbf{D}}^{-1}\mathbf{V}(S_{\mathbf{D}}')^{-1}.\qedhere
	$$
	\end{proof}

\begin{step}\label{step:asymptotic-normal}
	$\hat\sigma^{-1}(\hat\beta_1-\beta_1)\stackrel{d}{\longrightarrow} {\cal N}(0,1)$.
\end{step}

\begin{proof}
	Let $\beta_n=S_{\mathbf{D}}^{-1}\delta_n^{-1}E[\mathbf{Z}_iY_iI_{i}]$.
	Since $\epsilon_i=Y_i-\mathbf{D}'\beta=Y_i-\mathbf{D}_i'\beta_n+\mathbf{D}_i'(\beta_n-\beta)$,
	\begin{align*}
		&\frac{1}{\sqrt{n\delta_n}}\sum_{i=1}^nE[\mathbf{Z}_i\epsilon_iI_{i}]=\sqrt{n\delta_n}\delta_n^{-1}E[\mathbf{Z}_i(Y_i-\mathbf{D}'\beta)I_{i}]\\
		&=\sqrt{n\delta_n}\delta_n^{-1}\{E[\mathbf{Z}_iY_iI_{i}]-E[\mathbf{Z}_i\mathbf{D}_i'I_{i}]\beta_n+E[\mathbf{Z}_i\mathbf{D}_i'I_{i}](\beta_n-\beta)\}\\
		&=\sqrt{n\delta_n}\{(S_{\mathbf{D}}-\delta_n^{-1}E[\mathbf{Z}_i\mathbf{D}_i'I_{i}])S_{\mathbf{D}}^{-1}\delta_n^{-1}E[\mathbf{Z}_iY_iI_{i}]+\delta_n^{-1}E[\mathbf{Z}_i\mathbf{D}_i'I_{i}]S_{\mathbf{D}}^{-1}(\delta_n^{-1}E[\mathbf{Z}_iY_iI_{i}]-S_Y)\}\\
		&=\sqrt{n\delta_n}(O(\delta_n)O(1)+O(1)O(\delta_n))=O(\sqrt{n\delta_n}\delta_n),
	\end{align*}
	where we use Step \ref{step:mean-lim-2} for the second last equality.
	Thus, when $n\delta_n^3\rightarrow 0$,
	\begin{align*}
		\sqrt{n\delta_n}(\hat\beta-\beta)
		&=(\frac{1}{n\delta_n}\sum_{i=1}^n \mathbf{Z}_i\mathbf{D}_i'I_{i})^{-1}\frac{1}{\sqrt{n\delta_n}}\sum_{i=1}^n \{(\mathbf{Z}_i\epsilon_iI_{i}-E[\mathbf{Z}_i\epsilon_iI_{i}])+E[\mathbf{Z}_i\epsilon_iI_{i}]\}\\
		&\stackrel{d}{\longrightarrow} {\cal N}(0,S_{\mathbf{D}}^{-1}\mathbf{V}(S_{\mathbf{D}}')^{-1}).
	\end{align*}
	The conclusion then follows from Step \ref{step:sigma-consistency-2}.
\end{proof}

\setcounter{secnumdepth}{4}

\paragraph{Asymptotic Distribution When $n\delta_n^3\rightarrow C\in[0,\infty)$}\label{appendix:asymptotic-bias}

    ~\\
    To derive the asymptotic bias when $n\delta_n^3$ goes to some nonzero constant,
    note that from Step \ref{step:asymptotic-normal},
	\begin{align*}
		&\frac{1}{\sqrt{n\delta_n}}\sum_{i=1}^nE[\mathbf{Z}_i\epsilon_iI_{i}]\\
		&=\sqrt{n\delta_n}\{(S_{\mathbf{D}}-\delta_n^{-1}E[\mathbf{Z}_i\mathbf{D}_i'I_{i}])S_{\mathbf{D}}^{-1}\delta_n^{-1}E[\mathbf{Z}_iY_iI_{i}]+\delta_n^{-1}E[\mathbf{Z}_i\mathbf{D}_i'I_{i}]S_{\mathbf{D}}^{-1}(\delta_n^{-1}E[\mathbf{Z}_iY_iI_{i}]-S_Y)\}\\
        &=\sqrt{n\delta_n}\delta_n\{(T_{\mathbf{D}}+o(1))S_{\mathbf{D}}^{-1}(S_Y+o(1))+(S_{\mathbf{D}}+o(1))S_{\mathbf{D}}^{-1}(T_Y+o(1))\}\\
		&=\sqrt{n\delta_n^3}(T_{\mathbf{D}}S_{\mathbf{D}}^{-1}S_Y+T_Y+o(1)),
	\end{align*}
    provided $T_{\mathbf{D}}$ and $T_Y$ exist,
    where $T_{\mathbf{D}}=\lim_{\delta\rightarrow 0}\delta^{-1}(S_{\mathbf{D}}-\delta^{-1}E[\mathbf{Z}_i\mathbf{D}_i'1\{p^{A}(X_i;\delta)\in (0,1)\}])$ and $T_Y=\lim_{\delta\rightarrow 0}\delta^{-1}(\delta^{-1}E[\mathbf{Z}_iY_i1\{p^{A}(X_i;\delta)\in (0,1)\}]-S_Y)$.
    Thus, if $\delta_n\rightarrow 0$, $n\delta_n\rightarrow \infty$, and $n\delta_n^3\rightarrow C\in [0,\infty)$, then
	\begin{align*}
		\sqrt{n\delta_n}(\hat\beta-\beta)
		&=(\frac{1}{n\delta_n}\sum_{i=1}^n \mathbf{Z}_i\mathbf{D}_i'I_{i})^{-1}\frac{1}{\sqrt{n\delta_n}}\sum_{i=1}^n \{(\mathbf{Z}_i\epsilon_iI_{i}-E[\mathbf{Z}_i\epsilon_iI_{i}])+E[\mathbf{Z}_i\epsilon_iI_{i}]\}\\
		&\stackrel{d}{\longrightarrow} {\cal N}(\sqrt{C}S_{\mathbf{D}}^{-1}(T_{\mathbf{D}}S_{\mathbf{D}}^{-1}S_Y+T_Y),S_{\mathbf{D}}^{-1}\mathbf{V}(S_{\mathbf{D}}')^{-1}).
	\end{align*}
    The asymptotic bias can be nonzero if $C>0$.

    Now, consider the terms $T_{\mathbf{D}}=\lim_{\delta\rightarrow 0}\delta^{-1}(S_{\mathbf{D}}-\delta^{-1}E[\mathbf{Z}_i\mathbf{D}_i'1\{p^{A}(X_i;\delta)\in (0,1)\}])$ and $T_Y=\lim_{\delta\rightarrow 0}\delta^{-1}(\delta^{-1}E[\mathbf{Z}_iY_i1\{p^{A}(X_i;\delta)\in (0,1)\}]-S_Y)$.
    To guarantee the existence of these limits, we assume that for every $(u,v)\in \partial\Omega^*\cap N({\cal X},\bar\delta)\times(-1,1)$,
    $\lim_{\delta\rightarrow 0}r(u,v,\delta)$ exists, where the function $r$ is defined in Step \ref{step:lim_APS}.
    For example, if the boundary is a hyperplane, $r(u,v,\delta)=0$ for any $\delta>0$, so this condition holds with $\lim_{\delta\rightarrow 0}r(u,v,\delta)=0$.    
    Under this condition, the limits $T_{\mathbf{D}}$ and $T_Y$ exist, and all of their elements can be obtained by using the results below.
    
    Let $\bar r(u,v)=\lim_{\delta\rightarrow 0}r(u,v,\delta)$.
    Let $g:\mathbb{R}^p\rightarrow \mathbb{R}$ be a continuously differentiable function such that $\nabla g$ is bounded on $N(\partial\Omega^*,\delta')\cap N({\cal X},\delta')$ for some $\delta' >0$.
    Following Step \ref{step:mean-lim-2}, we can show that
	\begin{align*}
		&\delta^{-1}E[p^{A}(X_i;\delta)g(X_i)1\{p^{A}(X_i;\delta)\in (0,1)\}]\\
		=&\int_{-1}^{1}\int_{\partial\Omega^*\cap N({\cal X},\tilde\delta)} (k(v)+\delta r(u,v,\delta)) (g(u)+\nabla g(y_g(u,\delta v \nu_{\Omega^*}(u)))'\delta v \nu_{\Omega^*}(u))\\
		&~~\times (f_X(u)+\nabla f_X(y_f(u,\delta v \nu_{\Omega^*}(u)))'\delta v \nu_{\Omega^*}(u))(1+\tfrac{\partial J_{p-1}^{\partial\Omega^*}\psi_{\Omega^*}(u,y_J(u,\delta v))}{\partial \lambda}\delta v)d{\cal H}^{p-1}(u) dv\\
        =&\int_{-1}^{1}k(v)dv\int_{\partial\Omega^*} g(u)f_X(u)d{\cal H}^{p-1}(u)+\delta \int_{-1}^{1}\int_{\partial\Omega^*\cap N({\cal X},\tilde\delta)} \Biggl(k(v)g(u)f_X(u)\tfrac{\partial J_{p-1}^{\partial\Omega^*}\psi_{\Omega^*}(u,y_J(u,\delta v))}{\partial \lambda}v\\
        &~~~~~~~~+k(v)g(u)\nabla f_X(y_f(u,\delta v \nu_{\Omega^*}(u)))' v \nu_{\Omega^*}(u)+k(v)\nabla g(y_g(u,\delta v \nu_{\Omega^*}(u)))' v \nu_{\Omega^*}(u) f_X(u)\\
        &~~~~~~~~+r(u,v,\delta)g(u)f_X(u)+\delta h(u,v,\delta)\Biggr)d{\cal H}^{p-1}(u) dv
	\end{align*}
	for some function $h$ bounded on $\partial\Omega^*\cap N({\cal X},\tilde\delta)\times(-1,1)\times (0
	,\tilde\delta)$.
    By the Dominated Convergence Theorem, as $\delta\rightarrow 0$,
	\begin{align*}
		&\delta^{-1}\left(\delta^{-1}E[p^{A}(X_i;\delta)g(X_i)1\{p^{A}(X_i;\delta)\in (0,1)\}]-\int_{-1}^{1}k(v)dv\int_{\partial\Omega^*} g(u)f_X(u)d{\cal H}^{p-1}(u)\right)\\
        &\rightarrow \int_{-1}^{1}\int_{\partial\Omega^*\cap N({\cal X},\tilde\delta)} (k(v)g(u)f_X(u)\tfrac{\partial J_{p-1}^{\partial\Omega^*}\psi_{\Omega^*}(u,0)}{\partial \lambda}v+k(v)g(u)\nabla f_X(u)' v \nu_{\Omega^*}(u)\\
        &~~~~~~~~+k(v)\nabla g(u)' v \nu_{\Omega^*}(u)f_X(u)+\bar r(u,v)g(u)f_X(u))d{\cal H}^{p-1}(u) dv\\
        &=\int_{-1}^{1}\int_{\partial\Omega^*} (k(v)g(u)f_X(u)\tfrac{\partial J_{p-1}^{\partial\Omega^*}\psi_{\Omega^*}(u,0)}{\partial \lambda}v+k(v)g(u)\nabla f_X(u)' v \nu_{\Omega^*}(u)\\
        &~~~~~~~~+k(v)\nabla g(u)' v \nu_{\Omega^*}(u)f_X(u)+\bar r(u,v)g(u)f_X(u))d{\cal H}^{p-1}(u) dv,
	\end{align*}
    where the equality holds since $f_X(u)=0$ and $\nabla f_X(u)=\bm{0}$ for all $u\notin {\cal X}$.
    Likewise,
	\begin{align*}
		&\delta^{-1}\left(\delta^{-1}E[Z_ip^{A}(X_i;\delta)g(X_i)1\{p^{A}(X_i;\delta)\in (0,1)\}]-\int_{0}^{1}k(v)dv\int_{\partial\Omega^*} g(u)f_X(u)d{\cal H}^{p-1}(u)\right)\\
        &\rightarrow \int_{0}^{1}\int_{\partial\Omega^*} (k(v)g(u)f_X(u)\tfrac{\partial J_{p-1}^{\partial\Omega^*}\psi_{\Omega^*}(u,0)}{\partial \lambda}v+k(v)g(u)\nabla f_X(u)' v \nu_{\Omega^*}(u)\\
        &~~~~~~~~+k(v)\nabla g(u)' v \nu_{\Omega^*}(u)f_X(u)+\bar r(u,v)g(u)f_X(u))d{\cal H}^{p-1}(u) dv,\\
		&\delta^{-1}\left(\delta^{-1}E[g(X_i)1\{p^{A}(X_i;\delta)\in (0,1)\}]-2\int_{\partial\Omega^*} g(u)f_X(u)d{\cal H}^{p-1}(u)\right)\\
        &\rightarrow \int_{-1}^{1}\int_{\partial\Omega^*} (g(u)f_X(u)\tfrac{\partial J_{p-1}^{\partial\Omega^*}\psi_{\Omega^*}(u,0)}{\partial \lambda}v+g(u)\nabla f_X(u)' v \nu_{\Omega^*}(u)\\
        &~~~~~~~~+\nabla g(u)' v \nu_{\Omega^*}(u)f_X(u))d{\cal H}^{p-1}(u) dv=0,
	\end{align*}
    where the equality holds since $\int_{-1}^{1}vdv=0$,
 	\begin{align*}
		&\delta^{-1}\left(\delta^{-1}E[Z_ig(X_i)1\{p^{A}(X_i;\delta)\in (0,1)\}]-\int_{\partial\Omega^*} g(u)f_X(u)d{\cal H}^{p-1}(u)\right)\\
        &\rightarrow \int_{0}^{1}\int_{\partial\Omega^*} (g(u)f_X(u)\tfrac{\partial J_{p-1}^{\partial\Omega^*}\psi_{\Omega^*}(u,0)}{\partial \lambda}v+g(u)\nabla f_X(u)' v \nu_{\Omega^*}(u)+\nabla g(u)' v f_X(u))d{\cal H}^{p-1}(u) dv\\
        &=\frac{1}{2}\int_{\partial\Omega^*} (g(u)f_X(u)\tfrac{\partial J_{p-1}^{\partial\Omega^*}\psi_{\Omega^*}(u,0)}{\partial \lambda}+g(u)\nabla f_X(u)' \nu_{\Omega^*}(u)+\nabla g(u)'\nu_{\Omega^*}(u)f_X(u))d{\cal H}^{p-1}(u),
	\end{align*}
 where the equality holds since $\int_{0}^{1}vdv=\frac{1}{2}$, and
    \begin{align*}
 		&\delta^{-1}\left(\delta^{-1}E[p^{A}(X_i;\delta)^21\{p^{A}(X_i;\delta)\in (0,1)\}]-\int_{-1}^{1}k(v)^2dv\int_{\partial\Omega^*} f_X(u)d{\cal H}^{p-1}(u)\right)\\
        &\rightarrow \int_{-1}^{1}\int_{\partial\Omega^*} (k(v)^2f_X(u)\tfrac{\partial J_{p-1}^{\partial\Omega^*}\psi_{\Omega^*}(u,0)}{\partial \lambda}v+k(v)^2\nabla f_X(u)' v \nu_{\Omega^*}(u)\\
        &~~~~~~~~+2k(v)\bar r(u,v)f_X(u))d{\cal H}^{p-1}(u) dv.
    \end{align*}
\setcounter{secnumdepth}{3}

\subsubsection{Proof of Asymptotic Properties of \texorpdfstring{$\hat\beta^s_1$}{hat beta_s} 
When \texorpdfstring{$\Pr(A(X_i)\in (0,1))=0$}{A is deterministic}}
\label{subsubsection:consistency-simulation-boundary}
Let $I_i^s=1\{p^s(X_i;\delta_n)\in (0,1)\}$, $\mathbf{D}_i^s=(1,D_i,p^s(X_i;\delta_n))'$ and $\mathbf{Z}_i^s=(1,Z_i,p^s(X_i;\delta_n))'$.
$\hat\beta^{s}$ and $\hat{\mathbf{\Sigma}}^{s}$ are given by
$	\hat\beta^{s} = (\sum_{i=1}^n \mathbf{Z}_i^s(\mathbf{D}_i^s)'I_i^s)^{-1}\sum_{i=1}^n \mathbf{Z}_i^sY_iI_i^s
$
and
$
\hat{\mathbf{\Sigma}}^{s}=(\sum_{i=1}^n \mathbf{Z}_{i}^s(\mathbf{D}_{i}^s)'I_{i}^s)^{-1}(\sum_{i=1}^n (\hat\epsilon_{i}^s)^2\mathbf{Z}_{i}^s(\mathbf{Z}_{i}^s)'I_{i}^s)(\sum_{i=1}^n \mathbf{D}_{i}^s(\mathbf{Z}_{i}^s)'I_{i}^s)^{-1},
$
where $\hat\epsilon_i^s=Y_i-(\mathbf{D}_{i}^s)'\hat\beta^{s}$.
It is sufficient to show that
	$\hat\beta^s-\hat\beta=o_p(1)$ 
if $S_n\rightarrow \infty$, and that
$\sqrt{n\delta_n}(\hat\beta^{s}-\hat\beta)=o_p(1)$ and
$n\delta_n\hat{\mathbf{\Sigma}}^{s}\stackrel{p}{\longrightarrow} S_{\mathbf{D}}^{-1}\mathbf{V}(S_{\mathbf{D}}')^{-1}$
if Assumption \ref{assumption:simulation} holds.
\setcounter{step}{0}
\begin{step}\label{step:simulation-mean-boundary}
	Let $\{V_i\}_{i=1}^\infty$ be i.i.d. random variables.
	If $E[V_i|X_i]$ and $E[V_i^2|X_i]$ are bounded on $N(\partial\Omega^*,\delta')\cap N({\cal X},\delta')$ for some $\delta' >0$, and $S_n\rightarrow \infty$, then
	$$
	\frac{1}{n\delta_n}\sum_{i=1}^n V_i p^s(X_i;\delta_n)^lI_i^s-\frac{1}{n\delta_n}\sum_{i=1}^n V_i p^{A}(X_i;\delta_n)^l I_{i}=o_p(1)
	$$
	for $l=0,1,2,3,4$.
	If, in addition, Assumption \ref{assumption:simulation} holds, then for $l=0,1,2$, 
	$$
	\frac{1}{\sqrt{n\delta_n}}\sum_{i=1}^n V_i p^s(X_i;\delta_n)^lI_i^s-\frac{1}{\sqrt{n\delta_n}}\sum_{i=1}^n V_i p^{A}(X_i;\delta_n)^l I_{i}=o_p(1).
	$$
\end{step}

\begin{proof}
	We have
	\begin{align*}
	&\frac{1}{n\delta_n}\sum_{i=1}^n V_i p^s(X_i;\delta_n)^lI_i^s-\frac{1}{n\delta_n}\sum_{i=1}^n V_i p^{A}(X_i;\delta_n)^l I_{i}\\
	=&\frac{1}{n\delta_n}\sum_{i=1}^n V_i p^s(X_i;\delta_n)^l (I_i^s-I_{i})+\frac{1}{n\delta_n}\sum_{i=1}^n V_i( p^s(X_i;\delta_n)^l-p^{A}(X_i;\delta_n)^l)I_{i}.
	\end{align*}
	We first consider $\frac{1}{n\delta_n}\sum_{i=1}^n V_i( p^s(X_i;\delta_n)^l-p^{A}(X_i;\delta_n)^l)I_{i}$.
	By using the argument in the proof of Step \ref{step:mean-lim-2} in Appendix \ref{subsubsection:consistency-boundary}, we have
	\begin{align*}
	&~\left\vert E\left[\frac{1}{n\delta_n}\sum_{i=1}^n V_i( p^s(X_i;\delta_n)^l-p^{A}(X_i;\delta_n)^l)I_{i}\right]\right\vert\\
	\le&~\delta_n^{-1}E[|E[V_i|X_i]||E[p^s(X_i;\delta_n)^l-p^{A}(X_i;\delta_n)^l|X_i]|I_{i}]\\
	=&~\int_{-1}^{1}\int_{\partial\Omega^*\cap N({\cal X},\tilde\delta)} |E[V_i|X_i=u+\delta_n v \nu_{\Omega^*}(u)]|\{|E[p^s(u+\delta_n v \nu_{\Omega^*}(u);\delta_n)^l\\
	&~~~~~~~~~~~~~~~-p^{A}(u+\delta_n v \nu_{\Omega^*}(u);\delta_n)^l]|\} f_X(u+\delta_n v \nu_{\Omega^*}(u))J_{p-1}^{\partial\Omega^*}\psi_{\Omega^*}(u,\delta_n v)d{\cal H}^{p-1}(u) dv,
	\end{align*}
	where the choice of $\tilde\delta$ is as in the proof of Step \ref{step:mean-lim-2}.
	By Lemma \ref{lemma:simulation-error}, 
	\begin{align*}
	&~\left\vert E\left[\frac{1}{n\delta_n}\sum_{i=1}^n V_i( p^s(X_i;\delta_n)^l-p^{A}(X_i;\delta_n)^l)I_{i}\right]\right\vert\\
	\le&~ \frac{1}{S_n}\int_{-1}^{1}\int_{\partial\Omega^*\cap N({\cal X},\tilde\delta)} |E[V_i|X_i=u+\delta_n v \nu_{\Omega^*}(u)]|\\
	&~~~~~~~~~~~~~~~\times f_X(u+\delta_n v \nu_{\Omega^*}(u))J_{p-1}^{\partial\Omega^*}\psi_{\Omega^*}(u,\delta_n v)d{\cal H}^{p-1}(u) dv
	=O(S_n^{-1})
	\end{align*}
	for $l=0,1,2$. Also, by Lemma \ref{lemma:simulation-error},
	\begin{align*}
	&~\left\vert E\left[\frac{1}{n\delta_n}\sum_{i=1}^n V_i(p^s(X_i;\delta_n)^3-p^{A}(X_i;\delta_n)^3)I_{i}\right]\right\vert\\
	\le&~\delta_n^{-1}E[|E[V_i|X_i]||E[(p^s(X_i;\delta_n)-p^{A}(X_i;\delta_n))\\
	&~~~~~~~~~~~~~~~\times (p^s(X_i;\delta_n)^2+p^s(X_i;\delta_n)p^{A}(X_i;\delta_n)+p^{A}(X_i;\delta_n)^2)|X_i]|I_{i}]\\
	\le&~3\delta_n^{-1}E[|E[V_i|X_i]|E[|p^s(X_i;\delta_n)-p^{A}(X_i;\delta_n)||X_i]I_{i}]\\
	=&~3\int_{-1}^{1}\int_{\partial\Omega^*\cap N({\cal X},\tilde\delta)} |E[V_i|X_i=u+\delta_n v \nu_{\Omega^*}(u)]|\\
    &~~~~~~~~~~~~~~~\times E[|p^s(u+\delta_n v \nu_{\Omega^*}(u);\delta_n)-p^{A}(u+\delta_n v \nu_{\Omega^*}(u);\delta_n)|]\\
	&~~~~~~~~~~~~~~~\times f_X(u+\delta_n v \nu_{\Omega^*}(u))J_{p-1}^{\partial\Omega^*}\psi_{\Omega^*}(u,\delta_n v)d{\cal H}^{p-1}(u) dv
	\le (\frac{1}{S_n\epsilon^2}+\epsilon)O(1)
	\end{align*}
	for every $\epsilon>0$.
	We can make the right-hand side arbitrarily close to zero by taking sufficiently small $\epsilon>0$ and sufficiently large $S_n$, which implies that $|E[\frac{1}{n\delta_n}\sum_{i=1}^n V_i(p^s(X_i;\delta_n)^3-p^{A}(X_i;\delta_n)^3)I_{i}]|=o(1)$ if $S_n\rightarrow \infty$.
	Likewise, we can show $|E[\frac{1}{n\delta_n}\sum_{i=1}^n V_i(p^s(X_i;\delta_n)^4-p^{A}(X_i;\delta_n)^4)I_{i}]|=o(1)$.

	As for variance, for $l=0,1,2$,
	\begin{align*}
	&\Var\left(\frac{1}{n\delta_n}\sum_{i=1}^n V_i(p^s(X_i;\delta_n)^l-p^{A}(X_i;\delta_n)^l)I_{i}\right)
	\le\frac{1}{n\delta_n}\delta_n^{-1}E[V_i^2(p^s(X_i;\delta_n)^l-p^{A}(X_i;\delta_n)^l)^2I_{i}]\\
	&=\frac{1}{n\delta_n}\delta_n^{-1}E[E[V_i^2|X_i]E[(p^s(X_i;\delta_n)^l-p^{A}(X_i;\delta_n)^l)^2|X_i]I_{i}]\\
	&\le \frac{4}{n\delta_nS_n}\delta_n^{-1}E[E[V_i^2|X_i]I_{i}]
	=O((n\delta_nS_n)^{-1}),
 \end{align*}
    and for $l=3,4$,
    \begin{align*}
	&\Var\left(\frac{1}{n\delta_n}\sum_{i=1}^n V_i(p^s(X_i;\delta_n)^l-p^{A}(X_i;\delta_n)^l)I_{i}\right)
	\le \frac{1}{n\delta_n}\delta_n^{-1}E[V_i^2 I_{i}]
	=o(1).
	\end{align*}
	Therefore, $\frac{1}{n\delta_n}\sum_{i=1}^n V_i(p^s(X_i;\delta_n)^l-p^{A}(X_i;\delta_n)^l)I_{i}=o_p(1)$ if $S_n\rightarrow \infty$ for $l=0,1,2,3,4$, and $\frac{1}{\sqrt{n\delta_n}}\sum_{i=1}^n V_i(p^s(X_i;\delta_n)^l-p^{A}(X_i;\delta_n)^l)I_{i}=o_p(1)$ if $(n\delta_n)^{-1/2}S_n\rightarrow \infty$ for $l=0,1,2$.
	
	We next show $\frac{1}{n\delta_n}\sum_{i=1}^n V_i p^s(X_i;\delta_n)^l (I_{i}^s-I_{i})=o_p(1)$ if $S_n\rightarrow \infty$ for $l\ge 0$.
	Note
	\begin{align*}
	&\left\vert E\left[\frac{1}{n\delta_n}\sum_{i=1}^n V_i p^s(X_i;\delta_n)^l (I_{i}^s-I_{i})\right]\right\vert
    \le \delta_n^{-1}E[|E[V_i|X_i]|E[|I_{i}^s-I_{i}||X_i]].
	\end{align*}
	Since $I_{i}^s-I_{i}\le 0$ with strict inequality only if $I_{i}=1$,
	$$
	E[|I_{i}^s-I_{i}||X_i]=-E[I_{i}^s-I_{i}|X_i]I_{i}=(1-E[I_{i}^s|X_i])I_{i}=\Pr(p^s(X_i;\delta_n)\in \{0,1\}|X_i)I_{i}.
	$$
	We then have
	\begin{align*}
	&\left\vert E\left[\frac{1}{n\delta_n}\sum_{i=1}^n V_i p^s(X_i;\delta_n)^l (I_{i}^s-I_{i})\right]\right\vert
	\le\delta_n^{-1}E[|E[V_i|X_i]|\Pr(p^s(X_i;\delta_n)\in \{0,1\}|X_i)I_{i}]\\
	&\le\delta_n^{-1}E[|E[V_i|X_i]|((1-p^{A}(X_i;\delta_n))^{S_n}+p^{A}(X_i;\delta_n)^{S_n})I_{i}]\\
	&\le\int_{-1}^{1}\int_{\partial\Omega^*\cap N({\cal X},\tilde\delta)} |E[V_i|X_i=u+\delta_n v \nu_{\Omega^*}(u)]|\{(1-p^{A}(u+\delta_n v \nu_{\Omega^*}(u);\delta_n))^{S_n}\\
	&~~~~~~~~~~~~+p^{A}(u+\delta_n v \nu_{\Omega^*}(u);\delta_n)^{S_n}\}f_X(u+\delta_n v \nu_{\Omega^*}(u))J_{p-1}^{\partial\Omega^*}\psi_{\Omega^*}(u,\delta_n v)d{\cal H}^{p-1}(u) dv,
	\end{align*}
	where the second inequality follows from Lemma \ref{lemma:simulation-error}.
	For every $(u,v)\in \partial\Omega^*\cap N({\cal X},\tilde\delta)\times (-1,1)$, $\lim_{\delta\rightarrow 0}p^{A}(u+\delta v \nu_{\Omega^*}(u);\delta)=k(v)\in (0,1)$ by Step \ref{step:lim_APS}. 
	Since $E[V_i|X_i]$, $f_X$ and $J_{p-1}^{\partial\Omega^*}\psi_{\Omega^*}$ are bounded, by the Bounded Convergence Theorem,
	$
	|E[\frac{1}{n\delta_n}\sum_{i=1}^n V_i p^s(X_i;\delta_n)^l (I_{i}^s-I_{i})]|=o(1)
	$
	if $S_n\rightarrow \infty$. 
	As for variance,
	\begin{align*}
	&\Var\left(\frac{1}{n\delta_n}\sum_{i=1}^n V_i p^s(X_i;\delta_n)^l (I_{i}^s-I_{i})\right)
	\le\frac{1}{n\delta_n}\delta_n^{-1}E[E[V_i^2|X_i] E[|I_{i}^s-I_{i}||X_i]]=o(1).
	\end{align*}

	Lastly, we show that, for $l\ge 0$, $\frac{1}{\sqrt{n\delta_n}}\sum_{i=1}^n V_i p^s(X_i;\delta_n)^l (I_{i}^s-I_{i})=o_p(1)$ if Assumption \ref{assumption:simulation} holds.
	Let $\eta_n=\gamma\frac{\log n}{S_n}$, where $\gamma>1/2$. 
	We have
	\begin{align*}
	&~\left\vert E\left[\frac{1}{\sqrt{n\delta_n}}\sum_{i=1}^n V_i p^s(X_i;\delta_n)^l (I_i^s-I_{i})\right]\right\vert\\
	\le&\sqrt{n\delta_n^{-1}}E[|E[V_i|X_i]|((1-p^{A}(X_i;\delta_n))^{S_n}+p^{A}(X_i;\delta_n)^{S_n})I_{i}]\\
	=&\sqrt{n\delta_n^{-1}}E[|E[V_i|X_i]|((1-p^{A}(X_i;\delta_n))^{S_n}+p^{A}(X_i;\delta_n)^{S_n})1\{p^{A}(X_i;\delta_n)\in (0,\eta_n)\cup (1-\eta_n,1)\}]\\
	&+\sqrt{n\delta_n^{-1}}E[|E[V_i|X_i]|((1-p^{A}(X_i;\delta_n))^{S_n}+p^{A}(X_i;\delta_n)^{S_n})1\{p^{A}(X_i;\delta_n)\in (\eta_n, 1-\eta_n)\}]\\
	\le&~(\sup_{x\in N(\partial\Omega^*,2\tilde \delta)\cap N({\cal X},2\tilde \delta)}|E[V_i|X_i=x]|)\Big(\sqrt{n\delta_n^{-1}}\Pr(p^{A}(X_i;\delta_n)\in (0,\eta_n)\cup (1-\eta_n,1))\\
	&~~~~~~~~~~+2\sqrt{n\delta_n}(1-\eta_n)^{S_n}\delta_n^{-1}E[1\{p^{A}(X_i;\delta_n)\in (\eta_n, 1-\eta_n)\}]\Big).
	\end{align*}
	By Assumption \ref{assumption:simulation}, $\sqrt{n\delta_n^{-1}}\Pr(p^{A}(X_i;\delta_n)\in (0,\eta_n)\cup (1-\eta_n,1))=o(1)$.
	For the second term,
	\begin{align*}
		2\sqrt{n\delta_n}(1-\eta_n)^{S_n}\delta_n^{-1}E[1\{p^{A}(X_i;\delta_n)\in (\eta_n, 1-\eta_n)\}]&\le 2\sqrt{n\delta_n}(1-\eta_n)^{S_n}\delta_n^{-1}E[I_{i}]\\
		&=2\sqrt{n\delta_n}(1-\eta_n)^{S_n} O(1).
	\end{align*}
	Using the fact that $e^t\ge 1+t$ for every $t\in\mathbb{R}$, we have
	\begin{align*}
	&\sqrt{n\delta_n}(1-\eta_n)^{S_n}\le \sqrt{n\delta_n}(e^{-\eta_n})^{S_n}=\sqrt{n\delta_n}e^{-\gamma\log n}=\sqrt{n\delta_n}n^{-\gamma}=n^{1/2-\gamma}\delta_n^{1/2}\rightarrow 0,
	\end{align*}
	since $\gamma>1/2$.
	As for variance,
	\begin{align*}
	&\Var\left(\frac{1}{\sqrt{n\delta_n}}\sum_{i=1}^n V_i p^s(X_i;\delta_n)^l (I_i^s-I_{i})\right)
	\le \delta_n^{-1}E[E[V_i^2|X_i]E[|I_i^s-I_{i}||X_i]]\\
    &\le \delta_n^{-1}E[|E[V_i^2|X_i]|\Pr(p^s(X_i;\delta_n)\in \{0,1\}|X_i)I_{i}]=o(1).
	\qedhere
	\end{align*}\end{proof}

We have
\begin{align*}
\hat\beta^{s}-\hat\beta
&=(\frac{1}{n\delta_n}\sum_{i=1}^n \mathbf{Z}_i^s(\mathbf{D}_i^s)'I_i^s)^{-1}\frac{1}{n\delta_n}\sum_{i=1}^n \mathbf{Z}_i^sY_iI_i^s-(\frac{1}{n\delta_n}\sum_{i=1}^n \mathbf{Z}_i\mathbf{D}_i'I_{i})^{-1}\frac{1}{n\delta_n}\sum_{i=1}^n \mathbf{Z}_iY_iI_{i}\\
&=(\frac{1}{n\delta_n}\sum_{i=1}^n \mathbf{Z}_i^s(\mathbf{D}_i^s)'I_i^s)^{-1}(\frac{1}{n\delta_n}\sum_{i=1}^n \mathbf{Z}_i^sY_iI_i^s-\frac{1}{n\delta_n}\sum_{i=1}^n\mathbf{Z}_iY_iI_{i})-(\frac{1}{n\delta_n}\sum_{i=1}^n \mathbf{Z}_i^s(\mathbf{D}_i^s)'I_i^s)^{-1}\\
&~~~~~\times (\frac{1}{n\delta_n}\sum_{i=1}^n \mathbf{Z}_i^s(\mathbf{D}_i^s)'I_i^s-\frac{1}{n\delta_n}\sum_{i=1}^n \mathbf{Z}_i\mathbf{D}_i'I_{i})(\frac{1}{n\delta_n}\sum_{i=1}^n \mathbf{Z}_i\mathbf{D}_i'I_{i})^{-1}\frac{1}{n\delta_n}\sum_{i=1}^n \mathbf{Z}_iY_iI_{i}.
\end{align*}
By Step \ref{step:simulation-mean-boundary}, $\hat\beta^{s}-\hat\beta=o_p(1)$ if $S_n\rightarrow \infty$, and $\sqrt{n\delta_n}(\hat\beta^{s}-\hat\beta)=o_p(1)$ if Assumption \ref{assumption:simulation} holds. 
By proceeding as in Step \ref{step:sigma-consistency-2} in Appendix \ref{subsubsection:consistency-boundary}, we have
\begin{align*}
\frac{1}{n\delta_n}\sum_{i=1}^n (\hat \epsilon_i^s)^2\mathbf{Z}_i^s(\mathbf{Z}_i^s)'I_i^s
=\frac{1}{n\delta_n}\sum_{i=1}^n(\epsilon_i^s)^2\mathbf{Z}_i^s(\mathbf{Z}_i^s)'I_i^s+o_p(1),
\end{align*}
where $\epsilon_i^s=Y_i-(\mathbf{D}_i^s)'\beta$.
Then, by Step \ref{step:simulation-mean-boundary},
\begin{align*}
&~\frac{1}{n\delta_n}\sum_{i=1}^n (\hat \epsilon_i^s)^2\mathbf{Z}_i^s(\mathbf{Z}_i^s)'I_i^s-\frac{1}{n\delta_n}\sum_{i=1}^n\epsilon_i^2\mathbf{Z}_i\mathbf{Z}_i'I_{i}\\
=&~\frac{1}{n\delta_n}\sum_{i=1}^n(Y_i-(\mathbf{D}_i^s)'\beta)^2\mathbf{Z}_i^s(\mathbf{Z}_i^s)'I_i^s
-\frac{1}{n\delta_n}\sum_{i=1}^n(Y_i-\mathbf{D}_i'\beta)^2\mathbf{Z}_i\mathbf{Z}_i'I_{i}+o_p(1)
=o_p(1)
\end{align*}
so that
$
\frac{1}{n\delta_n}\sum_{i=1}^n (\hat \epsilon_i^s)^2\mathbf{Z}_i^s(\mathbf{Z}_i^s)'I_i^s\stackrel{p}{\longrightarrow}\mathbf{V}.
$
Also, $\frac{1}{n\delta_n}\sum_{i=1}^n \mathbf{Z}_i^s(\mathbf{D}_i^s)'I_i^s\stackrel{p}{\longrightarrow} S_{\mathbf{D}}$
by using Step \ref{step:simulation-mean-boundary}, implying the conclusion.\qed

\section{Proofs of Lemmas}

\subsection{Proof of Lemma \ref{lemma:Crasta2007SDF-C2}}\label{proof:lemma:Crasta2007SDF-C2}

	We apply results from \citeappendix{Crasta2007SDF}.
	Let $K=\{x\in\mathbb{R}^p:\|x\|\le 1\}$.
	$K$ is nonempty, compact, convex subset of $\mathbb{R}^p$ with the origin as an interior point.
	The polar body of $K$, defined as $K_0=\{y\in \mathbb{R}^p:y\cdot x\le 1\text{ for all $x\in K$}\}$, is $K$ itself.
	The gauge functions $\rho_K, \rho_{K_0}:\mathbb{R}^p\rightarrow [0, \infty]$ of $K$ and $K_0$ are given by
	\begin{align*}
	\rho_K(x) &\equiv {\rm inf}\{t\ge 0:x\in tK\} =\|x\|,~~
	\rho_{K_0}(x) \equiv {\rm inf}\{t\ge 0:x\in tK_0\} =\|x\|.
	\end{align*}
	Given $\rho_{K_0}$, the Minkowski distance from a set $S\subset\mathbb{R}^p$ is defined as
	$$
	\delta_S(x)\equiv\inf_{y\in S}\rho_{K_0}(x-y),~~~x\in\mathbb{R}^p.
	$$
	Note that we can write
	\begin{align*}
		d_S^s(x) = \begin{cases}
			\delta_{\partial S}(x)& \ \ \ \text{if $x\in {\rm cl}(S)$}\\
			-\delta_{\partial S}(x)& \ \ \ \text{if $x\in \mathbb{R}^p\setminus {\rm cl}(S)$}.
		\end{cases}
	\end{align*}
	It then follows from Theorem 4.16 of \citeappendix{Crasta2007SDF} that $d_S^s$ is twice continuously differentiable on $N(\partial S,\mu)$ for some $\mu>0$, and for every $x_0\in \partial S$,
	\begin{align*}
		\nabla d_S^s(x_0)&=\frac{\nu_S(x_0)}{\rho_K(\nu_S(x_0))}=\frac{\nu_S(x_0)}{\|\nu_S(x_0)\|}=\nu_S(x_0),
	\end{align*}
	where the last equality follows since $\nu_S(x_0)$ is a unit vector.
	It then follows that $\|\nabla d_S^s(x_0)\|=\|\nu_S(x_0)\|=1$ for every $x_0\in \partial S$.
	Also, it is obvious that, for every $x_0\in \partial S$, $\Pi_{\partial S}(x_0)=\{x_0\}$ and $x_0=x_0+d_S^s(x_0)\nu_S(x_0)$, since $d_S^s(x_0)=0$.
	In addition, as stated in the proof of Theorem 4.16 of \citeappendix{Crasta2007SDF}, $\mu$ is chosen so that (4.7) in Proposition 4.6 of \citeappendix{Crasta2007SDF} holds for every $x_0\in\partial S$ and every $t\in (-\mu,\mu)$.
	That is, $\Pi_{\partial S}(x_0+t\nabla \rho_K(\nu_S(x_0)))=\{x_0\}$ for every $x_0\in\partial S$ and every $t\in (-\mu,\mu)$.
	Since $\nabla \rho_K(\nu_S(x_0))=\frac{\nu_S(x_0)}{\|\nu_S(x_0)\|}=\nu_S(x_0)$, $\Pi_{\partial S}(x_0+t\nu_S(x_0))=\{x_0\}$ for every $x_0\in\partial S$ and every $t\in (-\mu,\mu)$.
	
	Furthermore, for every $x\in N(\partial S,\mu)\setminus \partial S$, $\Pi_{\partial S}(x)$ is a singleton as shown in the proof of Theorem 4.16 of \citeappendix{Crasta2007SDF}.
	Let $\pi_{\partial S}(x)$ be the unique element in $\Pi_{\partial S}(x)$.
	By Lemma 4.3 of \citeappendix{Crasta2007SDF}, for every $x\in N(\partial S,\mu)\setminus \partial S$,
	\begin{align*}
		\nabla d_S^s(x)&=\frac{\nu_S(\pi_{\partial S}(x))}{\rho_K(\nu_S(\pi_{\partial S}(x)))}=\frac{\nu_S(\pi_{\partial S}(x))}{\|\nu_S(\pi_{\partial S}(x))\|}=\nu_S(\pi_{\partial S}(x)),
	\end{align*}
	where the last equality follows since $\nu_S(\pi_{\partial S}(x))$ is a unit vector.
	It then follows that $\|\nabla d_S^s(x)\|=\|\nu_S(\pi_{\partial S}(x))\|=1$ for every $x\in N(\partial S,\mu)\setminus \partial S$.
	Lastly, note that
	\begin{align*}
		\delta_{\partial S}(x) &= \begin{cases}
			d_S^s(x) & \ \ \ \text{if $x\in N(\partial S,\mu)\cap {\rm int}(S)$}\\
			-d_S^s(x) & \ \ \ \text{if $x\in N(\partial S,\mu)\setminus {\rm cl}(S)$},
		\end{cases}\\
				\nabla \delta_{\partial S}(x) &= \begin{cases}
			\nabla d_S^s(x) & \ \ \ \text{if $x\in N(\partial S,\mu)\cap {\rm int}(S)$}\\
			-\nabla d_S^s(x) & \ \ \ \text{if $x\in N(\partial S,\mu)\setminus {\rm cl}(S)$},
		\end{cases}
	\end{align*}
	so $\delta_{\partial S}(x)\nabla \delta_{\partial S}(x)= d_S^s(x)\nabla d_S^s(x)=d_S^s(x)\nu_S(\pi_{\partial S}(x))$ for every $x\in N(\partial S,\mu)\setminus \partial S$.
	By Proposition 3.3 (i) of \citeappendix{Crasta2007SDF}, for every $x\in N(\partial S,\mu)\setminus \partial S$,
	$\nabla \rho_K(\nabla \delta_{\partial S}(x))=\frac{x-\pi_{\partial S}(x)}{\delta_{\partial S}(x)}$,
	which implies that
	\begin{align*}
		x&=\pi_{\partial S}(x)+\delta_{\partial S}(x)\nabla \rho_K(\nabla \delta_{\partial S}(x))\\
		&=\pi_{\partial S}(x)+\delta_{\partial S}(x)\frac{\nabla \delta_{\partial S}(x)}{\|\nabla \delta_{\partial S}(x)\|}=\pi_{\partial S}(x)+d_S^s(x)\nu_S(\pi_{\partial S}(x)).\tag*{\qed}
	\end{align*}

\subsection{Proof of Lemma \ref{lemma:C1-submanifold}}\label{proof:lemma:C1-submanifold}

	Fix any $x^*\in \partial S$.
	By Lemma \ref{lemma:Crasta2007SDF-C2}, $\nabla d_S^s(x^*)$ is nonzero.
	Without loss of generality, let $\frac{\partial d_S^s(x^*)}{\partial x_p}\neq 0$.
	Let $\psi: \mathbb{R}^p\rightarrow \mathbb{R}^p$ be the function such that $\psi(x)=(x_1,...,x_{p-1},d_S^s(x))$.
	$\psi$ is continuously differentiable, and 
	the Jacobian matrix of $\psi$ at $x^*$ is given by
	\begin{align*}
		J\psi(x^*)=
		\begin{pmatrix}
			\frac{\partial \psi_1}{\partial x_1}(x^*) & \cdots& \frac{\partial \psi_1}{\partial x_p}(x^*)\\
			\vdots &\ddots &\vdots\\
			\frac{\partial \psi_p}{\partial x_1}(x^*) & \cdots& \frac{\partial \psi_p}{\partial x_p}(x^*)
		\end{pmatrix}
		=
		\begin{pmatrix}
			&  & & 0\\
			& I_{p-1} &  &\vdots\\
			&  & &0\\
			\frac{\partial d_S^s(x^*)}{\partial x_1} & \cdots &\frac{\partial d_S^s(x^*)}{\partial x_{p-1}}& \frac{\partial d_S^s(x^*)}{\partial x_p}
		\end{pmatrix}.
	\end{align*}
	Since $\frac{\partial d_S^s(x^*)}{\partial x_p}\neq 0$, the Jacobian matrix is invertible.
	By the Inverse Function Theorem, there exist an open set $V$ containing $x^*$ and an open set $W$ containing $\psi(x^*)$ such that $\psi: V\rightarrow W$ has an inverse function $\psi^{-1}: W\rightarrow V$ that is continuously differentiable.
	We make $V$ small enough so that $\frac{\partial d_S^s(x)}{\partial x_p}\neq 0$ for every $x\in V$.
	The Jacobian matrix of $\psi^{-1}$ is given by $J\psi^{-1}(y)=J\psi(\psi^{-1}(y))^{-1}$ for all $y\in W$.
	
	Now note that $\psi(x)=(x_1,...,x_{p-1},0)$ for all $x\in V\cap \partial S$ by the definition of $d_S^s$.
	Let $U=\{(x_1,...,x_{p-1})\in \mathbb{R}^{p-1}: x\in V\cap \partial S\}$ and $\phi:U\rightarrow \mathbb{R}^p$ be a function such that $\phi(u)=\psi^{-1}((u,0))$ for all $u\in U$.
	Below we verify that $\phi$ is one-to-one and continuously differentiable, that $J\phi(u)$ is of rank $p-1$ for all $u\in U$, that $\phi(U)=V\cap \partial S$, and that $U$ is open.
	
	First, $\phi$ is one-to-one, since $\psi^{-1}$ is one-to-one, and $(u,0)\neq (u',0)$ if $u\neq u'$.
	Second, $\phi$ is continuously differentiable, since $\psi^{-1}$ is so.
	The Jacobian matrix of $\phi$ at $u\in U$ is by definition
	\begin{align*}
		J\phi(u)=
		\begin{pmatrix}
			\frac{\partial \psi_1^{-1}}{\partial y_1}((u,0)) & \cdots& \frac{\partial \psi_1^{-1}}{\partial y_{p-1}}((u,0))\\
			\vdots &\ddots &\vdots\\
			\frac{\partial \psi_p^{-1}}{\partial y_1}((u,0)) & \cdots& \frac{\partial \psi_p^{-1}}{\partial y_{p-1}}((u,0))
		\end{pmatrix}.
	\end{align*}
	Note that this is the left $p\times (p-1)$ submatrix of $J\psi^{-1}((u,0))$.
	Since $J\psi^{-1}((u,0))$ has full rank, $J\phi(u)$ is of rank $p-1$.
	Moreover,
	\begin{align*}
		&\phi(U)=\{\psi^{-1}((u,0)):u\in U\}
		=\{\psi^{-1}((x_1,...,x_{p-1},0)): x\in V\cap \partial S\}\\
		&=\{\psi^{-1}(\psi(x)): x\in V\cap \partial S\}=V\cap \partial S.
	\end{align*}
	
	Lastly, we show that $U$ is open.
	Pick any $\bar u\in U$.
	Then, there exists $\bar x_p\in\mathbb{R}$ such that $(\bar u,\bar x_p)\in V\cap \partial S$.
	As $(\bar u,\bar x_p)\in V\cap \partial S$, $d_S^s((\bar u,\bar x_p))=0$.
	Since $\frac{\partial d_S^s((\bar u,\bar x_p))}{\partial x_p}\neq 0$, it follows by the Implicit Function Theorem that there exist an open set $T\subset \mathbb{R}^{p-1}$ containing $\bar u$ and a continuously differentiable function $g: T\rightarrow \mathbb{R}$ such that $g(\bar u)=\bar x_p$ and $d_S^s(u,g(u))=0$ for all $u\in T$.
	Since $g$ is continuous, $(\bar u, g(\bar u))\in V$, and $V$ is open, there exists an open set $T'\subset T$ containing $\bar u$ such that $(u,g(u))\in V$ for all $u\in T'$.
	By the definition of $d_S^s$, $d_S^s(x)=0$ if and only if $x\in \partial S$.
	Therefore, if $u\in T'$, $(u,g(u))$ must be contained by $\partial S$, for otherwise $d_S^s(u,g(u))\neq 0$, which is a contradiction.
	Thus, $(u,g(u))\in V\cap \partial S$ and hence $u\in U$ for all $u\in T'$.
	This implies that $T'$ is an open subset of $U$ containing $\bar u$, which proves that $U$ is open.
	\qed

\subsection{Proof of Lemma \ref{lemma:neighborhood-integral}}\label{proof:lemma:neighborhood-integral}

We first introduce the coarea formula and the area formula, which we will use to prove Lemma \ref{lemma:neighborhood-integral}.

\begin{lemma}[Coarea Formula, Lemma 5.1.4 and Corollary 5.2.6 of \citeappendix{Krantz2008book}]\label{lemma:coarea}
	If $f:\mathbb{R}^m\rightarrow \mathbb{R}^n$ is a Lipschitz function and $m\ge n$, then
	$$
	\int_Sg(x)J_nf(x)d{\cal L}^m(x) = \int_{\mathbb{R}^n}\int_{\{x'\in S:f(x')=y\}}g(x)d{\cal H}^{m-n}(x)d{\cal L}^n(y)
	$$
	for every Lebesgue measurable subset $S$ of $\mathbb{R}^m$ and every ${\cal L}^m$-measurable function $g:S\rightarrow \mathbb{R}$, where for each $x\in \mathbb{R}^m$ at which $f$ is differentiable,
	$
	J_n f(x) = \sqrt{{\rm det}((Jf(x))(Jf(x))')}.
	$
\end{lemma}

\begin{lemma}[Area Formula, Lemma 5.3.5 and Theorem 5.3.7 of \citeappendix{Krantz2008book}]\label{lemma:area}
	Suppose $m\le \nu$ and $f:\mathbb{R}^n\rightarrow \mathbb{R}^\nu$ is Lipschitz.
	If $S$ is an $m$-dimensional $C^1$ submanifold of $\mathbb{R}^n$, then
	$$
	\int_Sg(x)J_m^Sf(x)d{\cal H}^m(x) = \int_{\mathbb{R}^\nu}\sum_{x\in S:f(x)=y}g(x)d{\cal H}^m(y)
	$$
	for every ${\cal H}^m$-measurable function $g:S\rightarrow \mathbb{R}$, where for each $x\in \mathbb{R}^n$ at which $f$ is differentiable,
		$J_m^Sf(x)=\frac{{\cal H}^m(\{Jf(x)y:y\in P\})}{{\cal H}^m(P)}$
	for an arbitrary $m$-dimensional parallelepiped $P$ contained in $T_S(x)$.
\end{lemma}

	Let $\bar\mu=\frac{1}{2}\min_{m,m'\in\{1,...,M\},m\neq m'}{\rm dist}(\Omega_m,\Omega_{m'})$ so that $\{N(\partial\Omega_m,\bar\mu)\}_{m=1}^M$ is a partition of $N(\partial\Omega,\bar\mu)$.
	Note that for every $m\in\{1,...,M\}$, $d_{\Omega}^s(x)=d_{\Omega_m}^s(x)$ for every $x\in N(\partial\Omega_m,\bar\mu)$.
	By Lemma \ref{lemma:Crasta2007SDF-C2}, for every $m\in\{1,...,M\}$, there exists $\bar\mu_m>0$ such that $d_{\Omega_m}^s$ is twice continuously differentiable on $N(\partial\Omega_m,\bar\mu_m)$.
	Letting $\mu\in(0,\min\{\bar\mu,\bar\mu_1,...,\bar\mu_M\})$, we have that $d_{\Omega}^s$ is twice continuously differentiable on $N(\partial\Omega,\mu)$.
	This implies that $d_{\Omega}^s$ is Lipschitz on $N(\partial\Omega,\mu)$.
	For every $\delta\in (0,\mu)$ and every function $g:\mathbb{R}^p\rightarrow \mathbb{R}$ that is integrable on $N(\partial \Omega,\delta)$,
	\begin{align}
		\int_{N(\partial \Omega,\delta)}g(x)dx &= \int_{\{x'\in\mathbb{R}^p:d_{\Omega}^s(x')\in (-\delta,\delta)\}}g(x)\sqrt{{\rm det}(\|\nabla d_{\Omega}^s(x)\|)}dx \nonumber\\
		&= \int_{\{x'\in\mathbb{R}^p:d_{\Omega}^s(x')\in (-\delta,\delta)\}}g(x)\sqrt{{\rm det}(\nabla d_{\Omega}^s(x)'\nabla d_{\Omega}^s(x))}dx \nonumber\\
		&= \int_{\{x'\in\mathbb{R}^p:d_{\Omega}^s(x')\in (-\delta,\delta)\}}g(x)\sqrt{{\rm det}((J d_{\Omega}^s(x))(J d_{\Omega}^s(x))')}dx \nonumber\\
		&= \int_{\mathbb{R}}\int_{\{x'\in \mathbb{R}^p:d_{\Omega}^s(x')\in (-\delta,\delta), d_{\Omega}^s(x')=\lambda\}}g(x)d{\cal H}^{p-1}(x)d\lambda \nonumber\\
		&= \int_{-\delta}^{\delta}\int_{\{x'\in \mathbb{R}^p: d_{\Omega}^s(x')=\lambda\}}g(x)d{\cal H}^{p-1}(x)d\lambda, \label{eq:coarea}
	\end{align}
	where the first equality follows since $\|\nabla d_{\Omega}^s(x)\|=1$ for every $x\in N(\partial \Omega,\delta)$ by Lemma \ref{lemma:Crasta2007SDF-C2}, the third equality follows from the definition of the Jacobian matrix, and the fourth equality follows from Lemma \ref{lemma:coarea}.
	
	Let $\Gamma(\lambda)=\{x\in \mathbb{R}^p: d_{\Omega}^s(x)=\lambda\}$ for each $\lambda\in (-\mu,\mu)$.
	Since $\nabla d_{\Omega}^s$ is differentiable on $N(\partial\Omega,\mu)$, $\psi_\Omega(x,\lambda)$ is defined on $N(\partial\Omega,\mu)\times \mathbb{R}$.
	We show that $\{\psi_\Omega(x_0,\lambda):x_0\in\partial\Omega\}\subset\Gamma(\lambda)$ for every $\lambda\in (-\mu,\mu)$.
	By Lemma \ref{lemma:Crasta2007SDF-C2}, for every $x_0\in\partial\Omega$, $\psi_\Omega(x_0,\lambda)=x_0+\lambda \nu_\Omega(x_0)$ and
		$\Pi_{\partial \Omega}(\psi_\Omega(x_0,\lambda))
		=\Pi_{\partial \Omega}(x_0+\lambda \nu_\Omega(x_0))
	    =\{x_0\}$.
	Hence,
	\begin{align*}
		d(\psi_\Omega(x_0,\lambda), \partial \Omega)&=\|\psi_\Omega(x_0,\lambda)-x_0\|
		=\|\lambda \nu_\Omega(x_0)\|
		=|\lambda|.
	\end{align*}
	Since $\nu_\Omega(x_0)$ is an inward normal vector, $\psi_\Omega(x_0,\lambda)\in {\rm cl}(\Omega)$ if $0\le \lambda<\mu$, and $\psi_\Omega(x_0,\lambda)\in \mathbb{R}^p\setminus {\rm cl}(\Omega)$ if $-\mu<\lambda< 0$.
	It follows that
	\begin{align*}
		d_\Omega^s(\psi_\Omega(x_0,\lambda)) &= \begin{cases}
			|\lambda|& \ \ \ \text{if $0\le \lambda <\mu$}\\
			-|\lambda| & \ \ \ \text{if $-\mu<\lambda<0$}
		\end{cases}~~=\lambda,
	\end{align*}
	so $\{\psi_\Omega(x_0,\lambda):x_0\in\partial\Omega\}\subset\Gamma(\lambda)$.
	It also holds that
	$\Gamma(\lambda)\subset \{\psi_\Omega(x_0,\lambda):x_0\in\partial\Omega\}$, since by Lemma \ref{lemma:Crasta2007SDF-C2}, for every $x\in \Gamma(\lambda)$,
	\begin{align*}
		\psi_\Omega(\pi_{\partial \Omega}(x),\lambda)&=\pi_{\partial \Omega}(x)+\lambda\nabla d_{\Omega}^s(\pi_{\partial \Omega}(x))=\pi_{\partial \Omega}(x)+d_{\Omega}^s(x) \nu_\Omega(\pi_{\partial \Omega}(x))=x,
	\end{align*}
	where $\pi_{\partial \Omega}(x)$ is the unique element in $\Pi_{\partial \Omega}(x)$.
	Thus, $\{\psi_\Omega(x_0,\lambda):x_0\in\partial\Omega\}=\Gamma(\lambda)$.
	
	Now note that $\{\partial\Omega_m\}_{m=1}^M$ is a partition of $\partial\Omega$, since ${\rm dist}(\Omega_m,\Omega_{m'})>0$ for any $m,m'\in\{1,...,M\}$ such that $m\neq m'$.
	By Lemma \ref{lemma:C1-submanifold}, $\partial\Omega_m$ is a $(p-1)$-dimensional $C^1$ submanifold of $\mathbb{R}^p$ for every $m\in\{1,...,M\}$, and hence $\partial\Omega$ is a $(p-1)$-dimensional $C^1$ submanifold of $\mathbb{R}^p$.
	Furthermore, since $\nabla d_{\Omega}^s$ is continuously differentiable on $N(\partial\Omega,\mu)$, $\psi_\Omega(\cdot,\lambda)$ is continuously differentiable on $N(\partial\Omega,\mu)$, which implies that $\psi_\Omega(\cdot,\lambda)$ is Lipschitz on $N(\partial\Omega,\mu)$ for every $\lambda\in\mathbb{R}$.
	Applying Lemma \ref{lemma:area}, we have that for every $\lambda\in (-\mu,\mu)$,
	\begin{align}
		\int_{\partial \Omega}g(u+\lambda \nu_\Omega(u))J_{p-1}^{\partial\Omega}\psi_\Omega(u,\lambda)d{\cal H}^{p-1}(u)&=
		\int_{\partial\Omega}g(\psi_\Omega(u,\lambda))J_{p-1}^{\partial\Omega}\psi_\Omega(u,\lambda)d{\cal H}^{p-1}(u) \nonumber\\
		& =\int_{\mathbb{R}^p}\sum_{u\in\partial \Omega:\psi_\Omega(u,\lambda)=x}g(\psi_\Omega(u,\lambda))d{\cal H}^{p-1}(x).\label{eq:area}
	\end{align}
	If $x\notin \{\psi_\Omega(u,\lambda):u\in\partial\Omega\}$, $\{u\in\partial \Omega:\psi_\Omega(u,\lambda)=x\}=\emptyset$.
	If $x\in \{\psi_\Omega(u,\lambda):u\in\partial\Omega\}$, there exists $u\in\partial\Omega$ such that $x=\psi_\Omega(u,\lambda)$.
	Since $\Pi_{\partial\Omega}(x)=\Pi_{\partial\Omega}(u+\lambda \nabla d_\Omega^s(u))=\Pi_{\partial\Omega}(u+\lambda \nu_\Omega(u))=\{u\}$ by Lemma \ref{lemma:Crasta2007SDF-C2}, such $u$ is unique, and hence $\{u\in\partial \Omega:\psi_\Omega(u,\lambda)=x\}$ is a singleton.
	It follow that
	\begin{align}
		\int_{\mathbb{R}^p}\sum_{u\in\partial \Omega:\psi_\Omega(u,\lambda)=x}g(\psi_\Omega(u,\lambda))d{\cal H}^{p-1}(x)
		&= \int_{\{\psi_\Omega(u,\lambda):u\in\partial\Omega\}}g(x)d{\cal H}^{p-1}(x) = \int_{\Gamma(\lambda)} g(x)d{\cal H}^{p-1}(x), \label{eq:integral-level-set}
	\end{align}
	where the last equality holds since $\{\psi_\Omega(u,\lambda):u\in\partial\Omega\}=\Gamma(\lambda)$.
	Combining (\ref{eq:coarea}), (\ref{eq:area}) and (\ref{eq:integral-level-set}), we obtain
	$$
	\int_{N(\partial \Omega,\delta)}g(x)dx = \int_{-\delta}^{\delta}\int_{\partial \Omega}g(u+\lambda \nu_\Omega(u))J_{p-1}^{\partial\Omega}\psi_\Omega(u,\lambda)d{\cal H}^{p-1}(u)d\lambda.
	$$
	
	We next show that $J_{p-1}^{\partial\Omega}\psi_\Omega(x,\cdot)$ is continuously differentiable in $\lambda$ and $J_{p-1}^{\partial\Omega}\psi_\Omega(x,0)=1$ for every $x\in\partial\Omega$.
	Fix an $x\in\partial\Omega$, and let $V_\Omega(x)$ be an arbitrary $p\times (p-1)$ matrix whose columns $v_1(x),...,v_{p-1}(x)\in\mathbb{R}^p$ form an orthonormal basis of $T_{\partial\Omega}(x)$.
	Let $P(x)\subset T_{\partial\Omega}(x)$ be a parallelepiped determined by $v_1(x),...,v_{p-1}(x)$, that is, let $P(x)=\{\sum_{k=1}^{p-1}c_k v_k(x): 0\le c_k\le 1\text{ for $k=1,...,p-1$}\}$.
	Since $v_1(x),...,v_{p-1}(x)$ are linearly independent, $P(x)$ is a $(p-1)$-dimensional parallelepiped.
	It follows that for each fixed $\lambda\in \mathbb{R}$,
	\begin{align*}
		\{J\psi_\Omega(x,\lambda) y:y\in P(x)\}&=\{J\psi_\Omega(x,\lambda) \sum_{k=1}^{p-1}c_k v_k(x): 0\le c_k\le 1\text{ for $k=1,...,p-1$}\}\\
		&=\{\sum_{k=1}^{p-1}c_k J\psi_\Omega(x,\lambda) v_k(x): 0\le c_k\le 1\text{ for $k=1,...,p-1$}\}\\
		&=\{\sum_{k=1}^{p-1}c_k w_k(x,\lambda): 0\le c_k\le 1\text{ for $k=1,...,p-1$}\},
	\end{align*}
	where $w_k(x,\lambda)=J\psi_\Omega(x,\lambda)v_k(x)$ for $k=1,...,p-1$.
	Since $J\psi_\Omega(x,\lambda)v_k(x)$ is the $k$-th column of $J\psi_\Omega(x,\lambda)V_\Omega(x)$, $\{J\psi_\Omega(x,\lambda) y:y\in P(x)\}$ is the parallelepiped determined by the columns of $J\psi_\Omega(x,\lambda)V_\Omega(x)$.
	By Proposition 5.1.2 of \citeappendix{Krantz2008book}, we have that
	\begin{align*}
		J_{p-1}^{\partial\Omega}\psi_\Omega(x,\lambda)&=\frac{{\cal H}^{p-1}(\{\sum_{k=1}^{p-1}c_k w_k(x,\lambda): 0\le c_k\le 1\text{ for $k=1,...,p-1$}\})}{{\cal H}^{p-1}(P(x))}\\
		&=\frac{\sqrt{{\rm det}((J\psi_\Omega(x,\lambda)V_\Omega(x))'(J\psi_\Omega(x,\lambda) V_\Omega(x)))}}{\sqrt{{\rm det}(V_\Omega(x)'V_\Omega(x))}}\\
		&=\frac{\sqrt{{\rm det}((V_\Omega(x)+\lambda D^2d_\Omega^s(x)V_\Omega(x))'(V_\Omega(x)+\lambda D^2d_\Omega^s(x)V_\Omega(x)))}}{\sqrt{{\rm det}(I_{p-1})}}\\
		&=\sqrt{{\rm det}(V_\Omega(x)'V_\Omega(x)+2V_\Omega(x)' \lambda D^2d_\Omega^s(x)V_\Omega(x)+V_\Omega(x)'(\lambda D^2d_\Omega^s(x))^2V_\Omega(x))}\\
		&=\sqrt{{\rm det}(I_{p-1}+\lambda V_\Omega(x)' (2 D^2d_\Omega^s(x)+\lambda (D^2d_\Omega^s(x))^2)V_\Omega(x)))}\\
		&=\sqrt{{\rm det}(I_p+\lambda V_\Omega(x)V_\Omega(x)' (2 D^2d_\Omega^s(x)+\lambda (D^2d_\Omega^s(x))^2))},
	\end{align*}
	where we use the fact that $V_\Omega(x)'V_\Omega(x)=I_{p-1}$ and the fact that ${\rm det}(I_m+AB)={\rm det}(I_n+BA)$ for an $m\times n$ matrix $A$ and an $n\times m$ matrix $B$ (the Weinstein-Aronszajn identity).
	For every $x\in\partial\Omega$, $J_{p-1}^{\partial\Omega}\psi_\Omega(x,\cdot)$ is continuously differentiable in $\lambda$, and
	$J_{p-1}^{\partial\Omega}\psi_\Omega(x,0)=\sqrt{{\rm det}(I_p)}=1$.
	
	Lastly, we show that $J_{p-1}^{\partial\Omega}\psi_\Omega(\cdot,\cdot)$ and $\tfrac{\partial J_{p-1}^{\partial\Omega}\psi_\Omega(\cdot,\cdot)}{\partial\lambda}$ are bounded on $\partial\Omega\times (-\mu^*,\mu^*)$ for some $\mu^*\in (0,\mu)$.
    Let $f:\mathbb{R}\times \mathbb{R}^{p \times (p-1)}\times \mathbb{R}^{p\times p}\rightarrow \mathbb{R}$ be a function such that
    $
    f(\lambda,V,D)={\rm det}(I_p+2\lambda VV'D+\lambda^2VV'D^2).
    $
    Note that $J_{p-1}^{\partial\Omega}\psi_\Omega(x,\lambda)=\sqrt{f(\lambda,V_\Omega(x),D^2d_\Omega^s(x))}$.

    Let $S=\{(V,D^2d_\Omega^s(x))\in\mathbb{R}^{p \times (p-1)}\times\mathbb{R}^{p\times p}:\|v_k\|=1 \text{ for $k=1,...,p-1$}, x\in\partial\Omega\}$, where $v_k$ denotes the $k$th column of $V$.
    Since $D^2d_\Omega^s(\cdot)$ is continuous on $\partial\Omega$, and $\partial\Omega$ is closed and bounded, $S$ is closed and bounded.
    Observe that 
    \begin{align*}
    	\frac{\partial f(\lambda,V,D)}{\partial\lambda}=\sum_{i,j}\frac{\partial {\rm det}(I_p+2\lambda VV'D+\lambda^2VV'D^2)}{\partial b_{ij}}(2(VV'D)_{ij}+2\lambda (VV'D^2)_{ij}),
    \end{align*}
    where $\tfrac{\partial {\rm det}(B)}{\partial b_{ij}}$ denotes the partial derivative of the function ${\rm det}:\mathbb{R}^{p\times p}\rightarrow \mathbb{R}$ with respect to the $(i,j)$ entry of $B$, which is continuous.
    Since the right-hand side is continuous in $(\lambda,V,D)$, there exists $\bar M>0$ such that $|\frac{\partial f(\lambda,V,D)}{\partial\lambda}|\le  \bar M$ for all $(\lambda,V,D)\in[-\mu,\mu]\times S$.

    By the mean value theorem, for every $(\lambda,V,D)\in[-\mu,\mu]\times S$,
    \begin{align*}
    f(\lambda,V,D)&=f(0,V,D)+\frac{\partial f(\tilde \lambda,V,D)}{\partial\lambda} \lambda\in [1-\bar M|\lambda|,1+\bar M|\lambda|],
    \end{align*}
    where $\tilde\lambda$ lies on the line segment connecting $0$ and $\lambda$ and the second line holds since $f(0,V,D)=1$ by construction.
    Pick $\mu^* \in (0,\bar\mu]$ such that $1-\bar M\mu^*>0$.
    Since $\{(V_\Omega(x),D^2d_\Omega^s(x)):x\in\partial\Omega\}\subset S$, it follows that $J_{p-1}^{\partial\Omega}\psi_\Omega(x,\lambda)=\sqrt{f(\lambda,V_\Omega(x),D^2d_\Omega^s(x))}$ is bounded on $\partial\Omega\times (-\mu^*,\mu^*)$.
    Moreover, for every $(x,\lambda)\in\partial\Omega\times (-\mu^*,\mu^*)$,
    \begin{align*}
    \frac{\partial J_{p-1}^{\partial\Omega}\psi_\Omega(x,\lambda)}{\partial\lambda}&=\frac{1}{2\sqrt{f(\lambda,V_\Omega(x),D^2d_\Omega^s(x))}}\frac{\partial f(\lambda,V_\Omega(x),D^2d_\Omega^s(x))}{\partial\lambda}\\
    &\in \left(-\frac{\bar M}{2\sqrt{1-\bar M\mu^*}},\frac{\bar M}{2\sqrt{1-\bar M\mu^*}}\right).
    \end{align*}
    Thus, $\frac{\partial J_{p-1}^{\partial\Omega}\psi_\Omega(x,\lambda)}{\partial\lambda}$ is bounded on $\partial\Omega\times (-\mu^*,\mu^*)$.
    \qed

\subsection{Proof of Lemma \ref{lemma:mean-plim}}\label{proof:lemma:mean-plim}

		We show that
		$E[V_i p^{A}(X_i;\delta)^l 1\{p^{A}(X_i;\delta)\in (0,1)\}^m]\rightarrow E[V_i A(X_i)^l 1\{A(X_i)\in (0,1)\}^m]$
		for $l\ge 0$ and $m=0,1$ as $\delta\rightarrow 0$, and that
		$\Var(\frac{1}{n}\sum_{i=1}^n V_i p^{A}(X_i;\delta_n)^l I_{i,n})\rightarrow 0$
		for $l\ge 0$ as $n\rightarrow \infty$.
		We first prove the first part.
		Suppose $A$ is continuous at $x$ and $A(x)\in (0,1)$. Then $\lim_{\delta\rightarrow 0}p^{A}(x;\delta)=A(x)$ by Part \ref{LPSexistence:continuity} of Corollary \ref{corollary:LPSexistence}, and hence $p^{A}(x;\delta)\in (0,1)$ for sufficiently small $\delta>0$.
		It follows that $1\{p^{A}(x;\delta)\in (0,1)\}\rightarrow 1 = 1\{A(x)\in (0,1)\}$ as $\delta\rightarrow 0$.
		Suppose $x\in {\rm int}({\cal X}_0)\cup {\rm int}({\cal X}_1)$. Then $B(x,\delta)\subset {\cal X}_0$ or $B(x,\delta)\subset {\cal X}_1$ for sufficiently small $\delta>0$ by the fact that ${\rm int}({\cal X}_0)$ and ${\rm int}({\cal X}_1)$ are open, and hence $1\{p^{A}(x;\delta)\in (0,1)\}\rightarrow 0 = 1\{A(x)\in (0,1)\}$ as $\delta\rightarrow 0$.
		Therefore, $\lim_{\delta\rightarrow 0}p^{A}(x;\delta)=A(x)$ and $\lim_{\delta\rightarrow 0}1\{p^{A}(x;\delta)\in (0,1)\}= 1\{A(x)\in (0,1)\}$ for almost every $x\in{\cal X}$, since $A$ is continuous at $x$ for almost every $x\in{\cal X}$ by Assumption \ref{assumption:A} \ref{assumption:A-continuity}, and either $A(x)\in (0,1)$ or $x\in {\rm int}({\cal X}_0)\cup {\rm int}({\cal X}_1)$ for almost every $x\in {\cal X}$ by Assumption \ref{assumption:A} \ref{assumption:A-measure-zero-boundary}.
		By the Dominated Convergence Theorem,
		\begin{align*}
			E[V_i p^{A}(X_i;\delta)^l 1\{p^{A}(X_i;\delta)\in (0,1)\}^m] 
			&\rightarrow E[V_iA(X_i)^l 1\{A(X_i)\in (0,1)\}^m]
		\end{align*}
		as $\delta\rightarrow 0$.
		As for variance, as $n\rightarrow \infty$, 
		\begin{align*}
			\Var\left(\frac{1}{n}\sum_{i=1}^n V_i p^{A}(X_i;\delta_n)^l I_{i,n}\right) &\le \frac{1}{n}E[V_i^2 p^{A}(X_i;\delta_n)^{2l} (I_{i,n})^2]\le \frac{1}{n}E[V_i^2]\rightarrow 0.\tag*{\qed}
		\end{align*}

\subsection{Proof of Lemma \ref{lemma:simulation-error}}\label{proof:lemma:simulation-error}
	By construction, $E[A(X_s^*)]=p^{A}(x;\delta_n)$, so
	\begin{align*}
	&E[p^s(x;\delta_n)-p^{A}(x;\delta_n)]=E\left[\frac{1}{S_n}\sum_{s=1}^{S_n}A(X_s^*)\right]-p^{A}(x;\delta_n)=0,\\
	&E[(p^s(x;\delta_n)-p^{A}(x;\delta_n))^2]=E[p^s(x;\delta_n)^2-p^{A}(x;\delta_n)^2]\\
    &=\Var(p^s(x;\delta_n))=\Var\left(\frac{1}{S_n}\sum_{s=1}^{S_n}A(X_s^*)\right)=\frac{1}{S_n}\Var(A(X_s^*))\le\frac{1}{S_n}E[A(X_s^*)^2]\le\frac{1}{S_n},\\
	&E[(p^s(x;\delta_n)^2-p^{A}(x;\delta_n)^2)^2]
	=E[(p^s(x;\delta_n)+p^{A}(x;\delta_n))^2(p^s(x;\delta_n)-p^{A}(x;\delta_n))^2]\\
	&\le4E[(p^s(x;\delta_n)-p^{A}(x;\delta_n))^2]
	\le\frac{4}{S_n}.
	\end{align*}
	
	We have the following bounds on $\Pr(A(X_s^*)=0)$ and $\Pr(A(X_s^*)=1)$:
	\begin{align*}
	0\le \Pr(A(X_s^*)=0) \le 1-p^{A}(x;\delta_n),~~
	&0\le \Pr(A(X_s^*)=1) \le p^{A}(x;\delta_n).
	\end{align*}
	It follows that
	\begin{align*}
	\Pr(p^s(x;\delta_n)\in\{0,1\})&=\Pr(A(X_s^*)=0)^{S_n}+\Pr(A(X_s^*)=1)^{S_n}\\
	&\le (1-p^{A}(x;\delta_n))^{S_n}+p^{A}(x;\delta_n)^{S_n}.
	\end{align*}
	
	Lastly, for any $\epsilon>0$,
	\begin{align*}
	&~E[|p^s(x;\delta_n)-p^{A}(x;\delta_n)|]\\
	=&~E[|p^s(x;\delta_n)-p^{A}(x;\delta_n)|||p^s(x;\delta_n)-p^{A}(x;\delta_n)|\ge \epsilon]\Pr(|p^s(x;\delta_n)-p^{A}(x;\delta_n)|\ge \epsilon)\\
	&+E[|p^s(x;\delta_n)-p^{A}(x;\delta_n)|||p^s(x;\delta_n)-p^{A}(x;\delta_n)|< \epsilon]\Pr(|p^s(x;\delta_n)-p^{A}(x;\delta_n)|< \epsilon)\\
	<&~1\cdot\frac{\Var(p^s(x;\delta_n))}{\epsilon^2}+\epsilon\cdot 1
	\le \frac{1}{S_n\epsilon^2}+\epsilon,
	\end{align*}
	where we use Chebyshev's inequality for the first inequality.
	We can make $E[|p^s(x;\delta_n)-p^{A}(x;\delta_n)|]$ arbitrarily close to zero by taking sufficiently small $\epsilon>0$ and sufficiently large $S_n$, which implies that $E[|p^s(x;\delta_n)-p^{A}(x;\delta_n)|]=o(1)$ if $S_n\rightarrow \infty$.
	\qed

	
\subsection{Proof of Lemma \ref{lemma:mean-simulation-plim}}\label{proof:lemma:mean-simulation-plim}

	We have
	\begin{align*}
		&\frac{1}{n}\sum_{i=1}^n V_i p^s(X_i;\delta_n)^lI_{i,n}^s-\frac{1}{n}\sum_{i=1}^n V_i p^{A}(X_i;\delta_n)^l I_{i,n}\\
		=&\frac{1}{n}\sum_{i=1}^n V_i p^s(X_i;\delta_n)^l (I_{i,n}^s-I_{i,n})+\frac{1}{n}\sum_{i=1}^n V_i( p^s(X_i;\delta_n)^l-p^{A}(X_i;\delta_n)^l)I_{i,n}.
	\end{align*}
	We first show that $\frac{1}{n}\sum_{i=1}^n V_i p^s(X_i;\delta_n)^l (I_{i,n}^s-I_{i,n})=o_p(1)$ if 
    $\delta_n\rightarrow 0$ for $l\ge 0$.
	We have
	\begin{align*}
		&\left\vert E\left[\frac{1}{n}\sum_{i=1}^n V_i p^s(X_i;\delta_n)^l (I_{i,n}^s-I_{i,n})\right]\right\vert
		=|E[V_i p^s(X_i;\delta_n)^l (I_{i,n}^s-I_{i,n})]|\\
		&\le E[|E[V_i|X_i]||E[p^s(X_i;\delta_n)^l (I_{i,n}^s-I_{i,n})|X_i]|]\le E[|E[V_i|X_i]|E[|I_{i,n}^s-I_{i,n}||X_i]].
	\end{align*}
	Note that by construction, $1\{p^s(X_i;\delta_n)\in (0,1)\}\le 1\{p^{A}(X_i;\delta_n)\in (0,1)\}$ with probability one conditional on $X_i=x$, so that
	$
	E[|I_{i,n}^s-I_{i,n}||X_i=x]=-E[I_{i,n}^s-I_{i,n}|X_i=x].
	$
	Suppose $A$ is continuous at $x$ and $A(x)\in (0,1)$. Then $A(x^*)\in (0,1)$ for all $x^*\in B(x,\delta_n)$ and hence $p^{A}(x;\delta_n)\in (0,1)$ and $p^s(x;\delta_n)\in (0,1)$ for sufficiently small $\delta_n>0$, so that $E[I_{i,n}^s-I_{i,n}|X_i=x]\rightarrow 0$ as $n\rightarrow \infty$.
	Suppose $x\in {\rm int}({\cal X}_0)\cup {\rm int}({\cal X}_1)$. Then $B(x,\delta_n)\subset {\cal X}_0$ or $B(x,\delta_n)\subset {\cal X}_1$ for sufficiently small $\delta_n>0$ by the fact that ${\rm int}({\cal X}_0)$ and ${\rm int}({\cal X}_1)$ are open, and hence $p^{A}(x;\delta_n)\in \{0,1\}$ and $p^s(x;\delta_n)\in \{0,1\}$ for sufficiently small $\delta_n>0$, so that $E[I_{i,n}^s-I_{i,n}|X_i=x]\rightarrow 0$ as $n\rightarrow \infty$.
	Therefore, $E[I_{i,n}^s-I_{i,n}|X_i=x]\rightarrow 0$ for almost every $x\in{\cal X}$, since $A$ is continuous at $x$ for almost every $x\in{\cal X}$ by Assumption \ref{assumption:A} \ref{assumption:A-continuity}, and either $A(x)\in (0,1)$ or $x\in {\rm int}({\cal X}_0)\cup {\rm int}({\cal X}_1)$ for almost every $x\in {\cal X}$ by Assumption \ref{assumption:A} \ref{assumption:A-measure-zero-boundary}.
	By the Dominated Convergence Theorem,
	$
	-E[|E[V_i|X_i]|E[I_{i,n}^s-I_{i,n}|X_i]] \rightarrow 0
$
	as $n\rightarrow \infty$.
		As for variance,
	\begin{align*}
		\Var\left(\frac{1}{n}\sum_{i=1}^n V_i p^s(X_i;\delta_n)^l (I_{i,n}^s-I_{i,n})\right)
		&\le\frac{1}{n}E[V_i^2 p^s(X_i;\delta_n)^{2l} (I_{i,n}^s-I_{i,n})^2]\le\frac{1}{n}E[V_i^2]\rightarrow 0.
	\end{align*}
	
	Next, we show that, for $l\ge 0$, $\frac{1}{\sqrt{n}}\sum_{i=1}^n V_i p^s(X_i;\delta_n)^l (I_{i,n}^s-I_{i,n})=o_p(1)$ if Assumption \ref{assumption:simulation} holds and $E[V_i|X_i]$ is bounded.
	Let $\eta_n=\gamma\frac{\log n}{S_n}$, where $\gamma>1/2$.
	We have
	\begin{align*}
		&\left\vert E\left[\frac{1}{\sqrt{n}}\sum_{i=1}^n V_i p^s(X_i;\delta_n)^l (I_{i,n}^s-I_{i,n})\right]\right\vert
		\le\sqrt{n}E[|E[V_i|X_i]|E[|I_{i,n}^s-I_{i,n}||X_i]]\\
		&=\sqrt{n}E[|E[V_i|X_i]|E[|I_{i,n}^s-1||X_i] I_{i,n}]=\sqrt{n}E[|E[V_i|X_i]|\Pr(p^s(X_i;\delta_n)\in \{0,1\}|X_i) I_{i,n}]\\
        &\le\sqrt{n}E[|E[V_i|X_i]|((1-p^{A}(X_i;\delta_n))^{S_n}+p^{A}(X_i;\delta_n)^{S_n})I_{i,n}]\\
		&=\sqrt{n}E[|E[V_i|X_i]|((1-p^{A}(X_i;\delta_n))^{S_n}+p^{A}(X_i;\delta_n)^{S_n})1\{p^{A}(X_i;\delta_n)\in (0,\eta_n)\cup (1-\eta_n,1)\}]\\
		&~~~~+\sqrt{n}E[|E[V_i|X_i]|((1-p^{A}(X_i;\delta_n))^{S_n}+p^{A}(X_i;\delta_n)^{S_n})1\{p^{A}(X_i;\delta_n)\in [\eta_n, 1-\eta_n]\}]\\
		&\le(\sup_{x\in{\cal X}}|E[V_i|X_i=x]|)(\sqrt{n}\Pr(p^{A}(X_i;\delta_n)\in (0,\eta_n)\cup (1-\eta_n,1))+2\sqrt{n}(1-\eta_n)^{S_n}),
	\end{align*}
	where the first equality follows from the fact that $I_{i,n}^s\le I_{i,n}$ with strict inequality only if $I_{i,n}=1$, and the second inequality follows from Lemma \ref{lemma:simulation-error}.
	By Assumption \ref{assumption:simulation}, $\sqrt{n}\Pr(p^{A}(X_i;\delta_n)\in (0,\eta_n)\cup (1-\eta_n,1))=o(1)$.
	As for $\sqrt{n}(1-\eta_n)^{S_n}$, 
	using the fact that $e^t\ge 1+t$ for every $t\in\mathbb{R}$, we have
	\begin{align*}
		&\sqrt{n}(1-\eta_n)^{S_n}\le \sqrt{n}(e^{-\eta_n})^{S_n}=\sqrt{n}e^{-\eta_n S_n}=\sqrt{n}e^{-\gamma\log n}=\sqrt{n}n^{-\gamma}=n^{1/2-\gamma}\rightarrow 0,
	\end{align*}
	since $\gamma>1/2$.
	As for variance,
	\begin{align*}
		&\Var\left(\frac{1}{\sqrt{n}}\sum_{i=1}^n V_i p^s(X_i;\delta_n)^l (I_{i,n}^s-I_{i,n})\right)
		\le E[V_i^2 p^s(X_i;\delta_n)^{2l} (I_{i,n}^s-I_{i,n})^2]\\
		&\le E[V_i^2 |I_{i,n}^s-I_{i,n}|]= E[E[V_i^2|X_i]E[|I_{i,n}^s-I_{i,n}||X_i]]=o(1).
	\end{align*}
 
	Now, we show that $\frac{1}{n}\sum_{i=1}^n V_i( p^s(X_i;\delta_n)^l-p^{A}(X_i;\delta_n)^l)I_{i,n}=o_p(1)$ if $S_n\rightarrow \infty$ for $l=0,1,2,3,4$.
    For $l=0,1,2,3,4$, we can write
    \begin{align*}
	\left\vert E\left[\frac{1}{n}\sum_{i=1}^n V_i(p^s(X_i;\delta_n)^l-p^{A}(X_i;\delta_n)^l)I_{i,n}\right]\right\vert\le C E[|E[V_i|X_i]|E[|p^s(X_i;\delta_n)-p^{A}(X_i;\delta_n)||X_i]]
	\end{align*}
    for some constant $C>0$. For example, 
	\begin{align*}
	&~\left\vert E\left[\frac{1}{n}\sum_{i=1}^n V_i(p^s(X_i;\delta_n)^4-p^{A}(X_i;\delta_n)^4)I_{i,n}\right]\right\vert\\
	=&~|E[V_i(p^s(X_i;\delta_n)^2+p^{A}(X_i;\delta_n)^2)(p^s(X_i;\delta_n)+p^{A}(X_i;\delta_n))(p^s(X_i;\delta_n)-p^{A}(X_i;\delta_n))I_{i,n}]|\\
	\le&~ 4E[|E[V_i|X_i]|E[|p^s(X_i;\delta_n)-p^{A}(X_i;\delta_n)||X_i]].
	\end{align*}
    By Lemma \ref{lemma:simulation-error}, $E[|E[V_i|X_i]|E[|p^s(X_i;\delta_n)-p^{A}(X_i;\delta_n)||X_i]I_{i,n}]=o(1)$.
	As for variance, for $l\ge 0$,
	\begin{align*}
		\Var\left(\frac{1}{n}\sum_{i=1}^n V_i(p^s(X_i;\delta_n)^l-p^{A}(X_i;\delta_n)^l)I_{i,n}\right)
		&\le\frac{1}{n}E[V_i^2(p^s(X_i;\delta_n)^l-p^{A}(X_i;\delta_n)^l)^2 I_{i,n}]=o(1).
	\end{align*}

    Lastly, we show that, for $l=0,1,2$, $\frac{1}{\sqrt{n}}\sum_{i=1}^n V_i(p^s(X_i;\delta_n)^l-p^{A}(X_i;\delta_n)^l)I_{i,n}=o_p(1)$ if Assumptions \ref{assumption:asymptotic-normality} and \ref{assumption:simulation} hold, $n\delta_n^2\rightarrow 0$, and $E[V_i|X_i]$ is bounded.
    Let $C^*$ and $D^*$ be defined as in Appendix \ref{appendix:assumptions}.
    We first obtain a bound on 
    $E[p^s(x;\delta_n)^2-p^{A}(x;\delta_n)^2]$ 
    that holds for every $x\notin N(D^*,\delta_n)$ and $n$.
    Fix $x\in N(D^*,\delta_n)$, and let $X_1^*,...,X_{S_n}^*$ be $S_n$ independent draws from the uniform distribution on $B(x,\delta_n)$ so that
	$
	p^s(x;\delta_n)=\frac{1}{S_n}\sum_{s=1}^{S_n}A(X_s^*).
	$
    We have
    \begin{align*}
        E[p^s(x;\delta_n)^2-p^{A}(x;\delta_n)^2]=\Var(p^s(x;\delta_n))=\frac{1}{S_n}\Var(A(X_s^*)).
    \end{align*}
    We compute a bound on $\Var(A(X_s^*))$.
    Since $x\notin N(D^*,\delta_n)$, $B(x,\delta_n)\cap D^*=\emptyset$, so $A$ is continuously differentiable on $B(x,\delta_n)$.
	By the mean value theorem, for every $a\in B(\bm{0},\delta_n)$,
	$$
	A(x+a) = A(x)+\nabla A(y(x,a))' a
	$$
	for some point $y(x,a)$ on the line segment connecting $x$ and $x+a$.
    Hence,
	\begin{align*}
		p^{A}(x;\delta_n)&=\frac{\delta_n^p\int_{B(\bm{0},1)} A(x+\delta_n u)du}{\delta_n^p\int_{B(\bm{0},1)} du}
        =A(x)+\delta_n\frac{\int_{B(\bm{0},1)} \nabla A(y(x,\delta_n u))' udu}{\int_{B(\bm{0},1)} du}.
	\end{align*}
    For every $v\in B(\bm{0},1)$,
    \begin{align*}
        &|A(x+\delta_n v)-p^{A}(x;\delta_n)|=\delta_n\left\vert\nabla A(y(x,\delta_n v))' v-\frac{\int_{B(\bm{0},1)} \nabla A(y(x,\delta_n u))' udu}{\int_{B(\bm{0},1)} du}\right\vert\\
        &\le \delta_n\left(\sum_{k=1}^p \left\vert\frac{\partial A(y(x,\delta_n v))}{\partial x_k}\right\vert |v_k|+\frac{\int_{B(\bm{0},1)} \sum_{k=1}^p |\frac{\partial A(y(x,\delta_n u))}{\partial x_k}| |u_k|du}{\int_{B(\bm{0},1)} du}\right)\\
        &\le \delta_n\left(\sum_{k=1}^p\sup_{x^*\in C^*}\left\vert\frac{\partial A(x^*)}{\partial x_k}\right\vert+\sum_{k=1}^p \sup_{x^*\in C^*}\left\vert\frac{\partial A(x^*)}{\partial x_k}\right\vert\frac{\int_{B(\bm{0},1)}  |u_k|du}{\int_{B(\bm{0},1)} du}\right)= C\delta_n,
    \end{align*}
    where $C=\sum_{k=1}^p\sup_{x^*\in C^*}\left\vert\frac{\partial A(x^*)}{\partial x_k}\right\vert \left(1+\frac{\int_{B(\bm{0},1)}  |u_k|du}{\int_{B(\bm{0},1)} du}\right)$ is finite under Assumption \ref{assumption:asymptotic-normality}.
    Note that $C$ is independent of $x$ and $n$.
    It follows that
    \begin{align*}
        \Var(A(X_s^*))=E[(A(X_s^*)-p^A(x;\delta_n))^2]=\frac{\delta_n^p\int_{B(\bm{0},1)} (A(x+\delta_n u)-p^A(x;\delta_n))^2du}{\delta_n^p\int_{B(\bm{0},1)} du}\le C^2\delta_n^2,
    \end{align*}
    so that $E[p^s(x;\delta_n)^2-p^{A}(x;\delta_n)^2]\le C^2\frac{\delta_n^2}{S_n}$.
    
	Now, for $l=0,1,2$,
	\begin{align*}
        &E\left[\frac{1}{\sqrt{n}}\sum_{i=1}^n V_i(p^s(X_i;\delta_n)^l-p^{A}(X_i;\delta_n)^l)I_{i,n}\right]=\sqrt{n}E[V_i(p^s(X_i;\delta_n)^l-p^{A}(X_i;\delta_n)^l)I_{i,n}]\\
		&=\sqrt{n}E[V_i(p^s(X_i;\delta_n)^l-p^{A}(X_i;\delta_n)^l)I_{i,n}1\{X_i\notin N(D^*,\delta_n)\}]\\
		&~~~~~+\sqrt{n}E[V_i(p^s(X_i;\delta_n)^l-p^{A}(X_i;\delta_n)^l)I_{i,n}1\{X_i\in N(D^*,\delta_n)\}]\\
        &\le \sqrt{n}E[|E[V_i|X_i]||E[(p^s(X_i;\delta_n)^l-p^{A}(X_i;\delta_n)^l)|X_i]|1\{X_i\notin N(D^*,\delta_n)\}]\\
        &~~~~+\sqrt{n}E[|E[V_i|X_i]|1\{X_i\in N(D^*,\delta_n)\}]\\
        &\le \sqrt{n}(\sup_{x\in{\cal X}}|E[V_i|X_i=x]|)\left(C^2\frac{\delta_n^2}{S_n}+\Pr(X_i\in N(D^*,\delta_n))\right)=O\left(\frac{\sqrt{n}\delta_n^2}{S_n}\right)+O(\sqrt{n}\delta_n),
	\end{align*}
    where the last equality follows from Assumption \ref{assumption:asymptotic-normality} \ref{assumption:neighborhood-O-delta}.
    Since $(n\delta_n)^{-1/2}S_n\rightarrow\infty$ and $n\delta_n^2\rightarrow 0$, $O\left(\frac{\sqrt{n}\delta_n^2}{S_n}\right)+O(\sqrt{n}\delta_n)=o(1)$.

    As for variance, by Lemma \ref{lemma:simulation-error}, for $l=0,1,2$,
	\begin{align*}
		&\Var\left(\frac{1}{\sqrt{n}}\sum_{i=1}^n V_i(p^s(X_i;\delta_n)^l-p^{A}(X_i;\delta_n)^l)I_{i,n}\right)
		\le E[V_i^2(p^s(X_i;\delta_n)^l-p^{A}(X_i;\delta_n)^l)^2I_{i,n}]\\
		&\le E[E[V_i^2|X_i]E[(p^s(X_i;\delta_n)^l-p^{A}(X_i;\delta_n)^l)^2|X_i]I_{i,n}]
		\le \frac{4}{S_n}E[E[V_i^2|X_i]I_{i,n}]=o(1). \tag*{\qed}
	\end{align*}

\section{Extensions and Discussions}

\subsection{Existence of the Approximate Propensity Score}\label{section:existence}

Proposition \ref{proposition:id} assumes that APS exists, but is it fair to assume so? In general, APS may fail to exist.
Nevertheless, APS exists for almost every $x$, as shown in the following proposition.


\begin{proposition}\label{proposition:APSexistence-a.e.}
	$p^{A}(x)$ exists and is equal to $A(x)$ for almost every $x\in {\cal X}$ (with respect to the Lebesgue measure).
\end{proposition}

\begin{proof}
	Since $A$ is a ${\cal L}^p$-measurable and bounded function, $A$ is locally integrable with respect to the Lebesgue measure, i.e., for every ball $B\subset \mathbb{R}^p$, $\int_B A(x)dx$ exists. An application of the Lebesgue differentiation theorem (see e.g. Theorem 1.4 in Chapter 3 of \citeappendix{Stein2005book}) to the function $A$ shows that for almost every $x\in \mathbb{R}^p$,
\begin{align*}
\lim_{\delta\rightarrow 0}\frac{\int_{B(x,\delta)}A(x^*)dx^*}{\int_{B(x,\delta)}dx^*}&=A(x).
\qedhere
\end{align*}

\end{proof}

Does APS exist at a specific point $x$? What is the value of APS at $x$ if it is not equal to $A(x)$?
We show that APS exists and is of a particular form for most covariate points and typical algorithms.
For each $x\in {\cal X}$ and each $q\in {\rm Supp}(A(X_i))$, define
\begin{align*}
	{\cal U}_{x,q}\equiv \{u\in B(\bm{0},1): \lim_{\delta\rightarrow 0}A(x+\delta u)=q\}.
\end{align*}
${\cal U}_{x,q}$ is the set of vectors in $B(\bm{0},1)$ such that the value of $A$ approaches $q$ as we approach $x$ from the direction of the vector. 
With this notation, we obtain a sufficient condition for the existence of APS at a point $x$. 

\begin{proposition}\label{proposition:LPSexistence}
	Take any $x\in{\cal X}$. 
	If there exists a countable set $Q\subset {\rm Supp}(A(X_i))$ such that ${\cal L}^p(\cup_{q\in Q}{\cal U}_{x,q})={\cal L}^p(B(\bm{0},1))$ and ${\cal U}_{x,q}$ is ${\cal L}^p$-measurable for all $q\in Q$, then $p^{A}(x)$ exists and is given by
		\begin{align*}
		p^{A}(x)=\frac{\sum_{q\in Q}q{\cal L}^p({\cal U}_{x,q})}{{\cal L}^p(B(\bm{0},1))}.
		\end{align*}
\end{proposition}

\begin{proof}
	With change of variables $u=\frac{x^*-x}{\delta}$, we have
\begin{align*}
	&p^{A}(x;\delta)=\frac{\int_{B(x,\delta)} A(x^*)dx^*}{\int_{B(x,\delta)} dx^*}=\frac{\delta^p\int_{B(\bm{0},1)} A(x+\delta u)du}{\delta^p\int_{B(\bm{0},1)} du}\\
	&=\frac{\int_{\cup_{q\in Q}{\cal U}_{x,q}}A(x+\delta u)du+\int_{B(\bm{0},1)\setminus\cup_{q\in Q}{\cal U}_{x,q}}A(x+\delta u)du}{\int_{B(\bm{0},1)}du}=\frac{\sum_{q\in Q}\int_{{\cal U}_{x,q}}A(x+\delta u)du}{\int_{B(\bm{0},1)}du},
\end{align*}
where the last equality follows from the assumption that ${\cal L}^p(\cup_{q\in Q}{\cal U}_{x,q})={\cal L}^p(B(\bm{0},1))$.
By the definition of ${\cal U}_{x,q}$, for each $q\in Q$, $\lim_{\delta\rightarrow 0}A(x+\delta u)=q$ for any $u\in{\cal U}_{x,q}$.
By the Dominated Convergence Theorem,
\begin{align*}
	p^{A}(x)&=\lim_{\delta\rightarrow 0} p^{A}(x;\delta)=\frac{\sum_{q\in Q}q{\cal L}^p({\cal U}_{x,q})}{{\cal L}^p(B(\bm{0},1))}.
\end{align*}
The numerator exists, since $q\le 1$ for all $q\in Q$ and $\sum_{q\in Q}{\cal L}^p({\cal U}_{x,q})={\cal L}^p(B(\bm{0},1))$.
\end{proof}

If almost every point in $B(\bm{0},1)$ is contained by one of countably many ${\cal U}_{x,q}$'s, therefore, APS exists and is equal to the weighted average of the values of $q$ with the weight proportional to the hypervolume of ${\cal U}_{x,q}$.
This result implies that APS exists in practically important cases. 

\begin{corollary}\label{corollary:LPSexistence}
~
\begin{enumerate}
 \item{\rm (Continuity points)} \label{LPSexistence:continuity} If $A$ is continuous at $x\in{\cal X}$, then $p^{A}(x)$ exists and $p^{A}(x)=A(x)$.
 \item{\rm (Interior points)} \label{LPSexistence:interior} Let ${\cal X}_q=\{x\in {\cal X}:A(x)=q\}$ for some $q\in [0,1]$.
 Then, for any interior point $x\in {\rm int}({\cal X}_q)$, $p^{A}(x)$ exists and $p^{A}(x)=q$.
 \item{\rm (Smooth boundary points)} Suppose that $\{x\in{\cal X}:A(x)=q_1\}=\{x\in{\cal X}:f(x)\ge 0\}$ and $\{x\in{\cal X}:A(x)=q_2\}=\{x\in{\cal X}:f(x)< 0\}$ for some $q_1,q_2\in [0,1]$, where $f:\mathbb{R}^p\rightarrow \mathbb{R}$.
 Let $x\in {\cal X}$ be a boundary point such that $f(x)=0$, and suppose that $f$ is continuously differentiable in a neighborhood of $x$ with $\nabla f(x)\neq \bm{0}$.
 In this case, $p^{A}(x)$ exists and $p^{A}(x)=\frac{1}{2}(q_1+q_2)$.
\item{\rm (Intersection points under CART and random forests)} \label{LPSexistence:cart} Let $p=2$, and suppose that $\{x\in{\cal X}:A(x)=q_1\}=\{(x_1,x_2)'\in {\cal X}: x_1\le 0 \text{ or } x_2\le 0\}$, $\{x\in{\cal X}:A(x)=q_2\}=\{(x_1,x_2)'\in {\cal X}: x_1> 0, x_2> 0\}$, and $\bm{0}=(0,0)'\in {\cal X}$.
This is an example in which tree-based algorithms such as Classification And Regression Tree (CART) and random forests are used to create $A$.
In this case, $p^{A}(\bm{0})$ exists and $p^{A}(\bm{0})=\frac{3}{4}q_1+\frac{1}{4}q_2$.
\end{enumerate}
\end{corollary}

\subsection{Discrete Covariates}\label{section:discrete-X}
	In this section, we provide the definition of APS and identification and asymptotic normality results when $X_i$ includes discrete covariates.
	Suppose that $X_i=(X_{di},X_{ci})$, where $X_{di}\in \mathbb{R}^{p_d}$ is a vector of discrete covariates, and $X_{ci}\in \mathbb{R}^{p_c}$ is a vector of continuous covariates.
	Let ${\cal X}_d$ denote the support of $X_{di}$ and be assumed to be finite.
	We also assume that $X_{ci}$ is continuously distributed conditional on $X_{di}$, and let ${\cal X}_c(x_d)$ denote the support of $X_{ci}$ conditional on $X_{di}=x_d$ for each $x_d\in {\cal X}_d$.
	Let ${\cal X}_{c,0}(x_d)=\{x_c\in {\cal X}_c(x_d):A(x_d,x_c)=0\}$ and ${\cal X}_{c,1}(x_d)=\{x_c\in {\cal X}_c(x_d):A(x_d,x_c)=1\}$.
	
	Define APS as follows: for each $x=(x_d,x_c)\in{\cal X}$,
	\begin{align*}
		p^{A}(x;\delta)&\equiv\frac{\int_{B(x_c,\delta)}A(x_d,x_c^{*})dx_c^*}{\int_{B(x_c,\delta)}dx_c^*},~~
		p^{A}(x)\equiv \lim_{\delta\rightarrow 0}p^{A}(x;\delta), 
	\end{align*}
	where $B(x_c, \delta)=\{x_c^*\in\mathbb{R}^{p_c}:\|x_c-x_c^*\|\le\delta\}$ is the $\delta$-ball around $x_c\in \mathbb{R}^{p_c}$.
	That is, we take the average of the $A(x_d,x_c^*)$ values when $x_c^*$ is uniformly distributed on $B(x_c,\delta)$ holding $x_d$ fixed, and let $\delta\rightarrow 0$.
	Below, we assume Assumptions \ref{assumption:A}, \ref{continuity}, \ref{assumption:consistency} and \ref{assumption:asymptotic-normality} hold conditional on $X_{di}$.
	
	\setcounterref{assumption}{assumption:A}
	\addtocounter{assumption}{-1}
	\begin{assumption}[Almost Everywhere Continuity of $A$]\label{assumption:A-discreteX}
		~
		\begin{enumerate}[label=(\alph*)]
			\item For every $x_d\in {\cal X}_d$, $A(x_d,\cdot)$ is continuous almost everywhere with respect to the Lebesgue measure ${\cal L}^{p_c}$.
			\item For every $x_d\in {\cal X}_d$, ${\cal L}^{p_c}({\cal X}_{c,k}(x_d))={\cal L}^{p_c}({\rm int}({\cal X}_{c,k}(x_d)))$ for $k=0,1$.
		\end{enumerate}
	\end{assumption}

\subsubsection{Identification}
	\setcounterref{assumption}{continuity}
	\addtocounter{assumption}{-1}
	\begin{assumption}[Local Mean Continuity]\label{continuity-discreteX}
		For every $x_d\in {\cal X}_d$ and $z\in\{0,1\}$, the conditional expectation functions $E[Y_{zi}|X_i=(x_d,x_c)]$ and $E[D_i(z)|X_i=(x_d,x_c)]$ are continuous in $x_c$ at any point $x_c\in {\cal X}_c(x_d)$ such that $p^{A}(x_d,x_c)\in (0,1)$ and $A(x_d,x_c)\in \{0,1\}$. 
	\end{assumption}
	
	Let ${\rm int}_c({\cal X})=\{(x_d,x_c)\in{\cal X}:x_c\in{\rm int}({\cal X}_c(x_d))\}$.
	We say that a set $S\subset \mathbb{R}^p$ \textit{is open relative to} ${\cal X}$ if there exists an open set $U\subset\mathbb{R}^p$ such that $S=U\cap {\cal X}$.
	For a set $S\subset \mathbb{R}^p$, let ${\cal X}_d^S=\{x_d\in {\cal X}_d:(x_d,x_c)\in S \text{ for some } x_c\in \mathbb{R}^{p_c}\}$ and ${\cal X}_c^S(x_d)=\{x_c\in {\cal X}_c: (x_d,x_c)\in S\}$ for each $x_d\in {\cal X}_d^S$.
	
	\begin{proposition}\label{proposition:id-discreteX}
	Under Assumptions \ref{assumption:A-discreteX} and \ref{continuity-discreteX}:
	\begin{enumerate}[label=(\alph*)]
		\item $E[Y_{1i}-Y_{0i}| X_i=x]$ and $E[D_i(1)-D_i(0)| X_i=x]$ are identified for every $x\in {\rm int}_c({\cal X})$ such that $p^{A}(x)\in (0,1)$.
		\item Let $S$ be any subset of ${\cal X}$ open relative to ${\cal X}$ such that $p^{A}(x)$ exists for all $x\in S$.
		Then either $E[Y_{1i}-Y_{0i}| X_i \in S]$ or $E[D_i(1)-D_i(0)| X_i \in S]$, or both are identified only if $p^{A}(x)\in (0,1)$ for almost every $x_c\in {\cal X}_c^S(x_d)$ for every $x_d\in {\cal X}_d^S$.
	\end{enumerate}
	\end{proposition}
	
	\begin{proof}
				We can prove Part (a) using the same argument in the proof of Proposition \ref{proposition:id} (a).
		For Part (b), suppose to the contrary that there exists $x_d\in {\cal X}_d^S$ such that ${\cal L}^{p_c}(\{x_c\in {\cal X}_c^S(x_d):p^{A}(x_d,x_c)\in \{0,1\}\})>0$.
		Without loss of generality, assume ${\cal L}^{p_c}(\{x_c\in {\cal X}_c^S(x_d):p^{A}(x_d,x_c)=1\})>0$.
		The proof proceeds in five steps.
		\setcounter{step-subsection}{0}
		\begin{step-subsection}\label{step:necessary-nonzero-discrete}
			${\cal L}^{p_c}({\cal X}_c^S(x_d)\cap {\cal X}_{c,1}(x_d))>0$.
		\end{step-subsection}
		\begin{step-subsection}\label{step:necessary-nonempty-discrete}
			${\cal X}_c^S(x_d)\cap {\rm int}({\cal X}_{c,1}(x_d))\neq \emptyset$.
		\end{step-subsection}
		\begin{step-subsection}\label{step:necessary-interior-discrete}
			$p^{A}(x_d,x_c)=1$ for any $x_c\in {\rm int}({\cal X}_{c,1}(x_d))$.
		\end{step-subsection}
		\begin{step-subsection}\label{step:discrete-X}
			For every $x_c^*\in {\cal X}_c^S(x_d)\cap {\rm int}({\cal X}_{c,1}(x_d))$, there exists $\delta>0$ such that $B(x_c^*,\delta)\subset {\cal X}_c^S(x_d)\cap {\rm int}({\cal X}_{c,1}(x_d))$.
		\end{step-subsection}
		\begin{step-subsection}\label{step:necessary-nonID-discrete}
			$E[Y_{1i}-Y_{0i}|X_i\in S]$ is not identified.
		\end{step-subsection}
		
		Following the argument in the proof of Proposition \ref{proposition:id} (b), we can prove Steps \ref{step:necessary-nonzero-discrete}--\ref{step:necessary-interior-discrete}.
		Once Step \ref{step:discrete-X} is established, we prove Step \ref{step:necessary-nonID-discrete} by following the proof of Step \ref{step:necessary-nonID} in Appendix \ref{proof:id} with $B(x_c^*,\delta)$ and $B(x_c^*,\epsilon)$ in place of $B(x^*, \delta)$ and $B(x^*, \epsilon)$, respectively, using the fact that $\Pr(X_{ci}\in B(x_c^*,\epsilon)|X_{di}=x_d)>0$ by the definition of support.
		
		Below, we provide the proof of Step \ref{step:discrete-X}.
		Pick an $x_c^*\in {\cal X}_c^S(x_d)\cap {\rm int}({\cal X}_{c,1})$.
		Then, $x^*=(x_d,x_c^*)\in S$.
		Since $S$ is open relative to ${\cal X}$, there exists an open set $U\in \mathbb{R}^p$ such that $S=U\cap {\cal X}$.
		This implies that for any sufficiently small $\delta>0$, $B(x^*,\delta)\cap {\cal X}\subset U\cap {\cal X}=S$.
		It then follows that $\{x_c\in \mathbb{R}^{p_c}: (x_d,x_c)\in B(x^*,\delta)\cap {\cal X}\}\subset \{x_c\in \mathbb{R}^{p_c}: (x_d,x_c)\in S\}$, equivalently, $B(x_c^*,\delta)\cap {\cal X}_c(x_d)\subset {\cal X}_c^S(x_d)$.
		By choosing a sufficiently small $\delta>0$ so that $B(x_c^*,\delta)\subset {\rm int}({\cal X}_{c,1}(x_d))\subset {\cal X}_c(x_d)$, we have $B(x_c^*,\delta)\subset {\cal X}_c^S(x_d) \cap {\rm int}({\cal X}_{c,1}(x_d))$.
	\end{proof}

\subsubsection{Estimation}
	For each $x_d\in{\cal X}_d$, let $\Omega^*(x_d)=\{x_c\in \mathbb{R}^{p_c}: A(x_d,x_c)=1\}$.
	Also, let ${\cal X}_d^*=\{x_d\in{\cal X}_d:\Var(A(X_i)|X_{di}=x_d)>0\}$, and let $f_{X_c|X_d}$ denote the probability density function of $X_{ci}$ conditional on $X_{di}$.
	In addition, for each $x_d\in{\cal X}_d$, let
	$$C^*(x_d)=\{x_c\in \mathbb{R}^{p_c}: \text{$A(x_d,\cdot)$ is continuously differentiable at $x_c$}\},
	$$
	and let $D^*(x_d)=\mathbb{R}^{p_c}\setminus C^*(x_d)$.
	
	\setcounterref{assumption}{assumption:consistency}
	\addtocounter{assumption}{-1}
	\begin{assumption}\label{assumption:consistency-discreteX}
		~
		\begin{enumerate}
			\renewcommand{\theenumi}{(\alph{enumi})}
			\renewcommand{\labelenumi}{(\alph{enumi})}
			\item{\rm (Finite Moments)}\label{assumption:moment-discreteX} $E[Y_i^{4}]<\infty$.
			\item{\rm (Nonzero First Stage)}\label{assumption:relevance-discreteX}
			$\int_{{\cal X}}p^A(x)(1-p^A(x))E[D_i(1)-D_i(0)|X_i=x]f_X(x)d\mu(x)\neq 0$, where $\mu$ is the Lebesgue measure ${\cal L}^p$ when $\Pr(A(X_i)\in (0,1))>0$ and is the $(p-1)$-dimensional Hausdorff measure ${\cal H}^{p-1}$ when $\Pr(A(X_i)\in (0,1))=0$.
		\end{enumerate}
		\vspace{.5em}
		\noindent
		If $\Pr(A(X_i)\in (0,1))=0$, then the following conditions \ref{assumption:nondegeneracy-discreteX}--\ref{assumption:boundary-continuity-discreteX} hold.
		\begin{enumerate}
			\renewcommand{\theenumi}{(\alph{enumi})}
			\renewcommand{\labelenumi}{(\alph{enumi})}
			\setcounter{enumi}{2}
			\item {\rm (Nonzero Variance)}\label{assumption:nondegeneracy-discreteX}
			${\cal X}_d^*\neq\emptyset$.
			\item {\rm ($C^2$ Boundary of $\Omega^*(x_d)$)}\label{assumption:boundary-partition-discreteX}
			For each $x_d\in{\cal X}_d^*$, there exists a partition $\{\Omega^*_{1}(x_d),...,\Omega^*_{M}(x_d)\}$ of $\Omega^*(x_d)$ such that
			\begin{enumerate}[label=(\roman*)]
				\item ${\rm dist}(\Omega^*_m(x_d),\Omega^*_{m'}(x_d))>0$ for any $m,m'\in\{1,...,M\}$ such that $m\neq m'$;
				\item $\Omega^*_m(x_d)$ is nonempty, bounded, open, connected, and twice continuously differentiable for each $m\in \{1,...,M\}$.
			\end{enumerate}
			
			\item {\rm (Regularity of Deterministic $A$)} \label{assumption:boundary-measure-discreteX}
		    For each $x_d\in{\cal X}_d^*$, the following holds.	\begin{enumerate}[label=(\roman*)]
				\item \label{assumption:p-1dim-discreteX}
			
			${\cal H}^{p_c-1}(\partial\Omega^*(x_d))<\infty$, and $\int_{\partial\Omega^*(x_d)} f_{X_c|X_d}(x_c|x_d) d{\cal H}^{p_c-1}(x_c)>0$;
			\item\label{assumption:zero-set-discreteX} There exists $\delta>0$ such that $A(x_d,x_c)=0$ for almost every $x_c\in N({\cal X}_c(x_d), \delta)\setminus \Omega^*(x_d)$.
			\end{enumerate}
			\item {\rm (Conditional Means and Density near $\partial\Omega^*(x_d)$)} \label{assumption:boundary-continuity-discreteX}
			For each $x_d\in{\cal X}_d^*$,
			there exists $\delta>0$ such that
			\begin{enumerate}[label=(\roman*)]
				\item$E[Y_{1i}|X_i=(x_d,\cdot)]$, $E[Y_{0i}|X_i=(x_d,\cdot)]$, $E[D_i(1)|X_i=(x_d,\cdot)]$, $E[D_i(0)|X_i=(x_d,\cdot)]$ and $f_{X_c|X_d}(\cdot|x_d)$ are continuously differentiable and have bounded partial derivatives on $N(\partial\Omega^*(x_d),\delta)$;
				\item $E[Y_{1i}^2|X_i=(x_d,\cdot)]$, $E[Y_{0i}^2|X_i=(x_d,\cdot)]$, $E[Y_{1i}D_i(1)|X_i=(x_d,\cdot)]$ and $E[Y_{0i}D_i(0)|X_i=(x_d,\cdot)]$ are continuous on $N(\partial\Omega^*(x_d),\delta)$;
				\item $E[Y_i^{4}|X_i=(x_d,\cdot)]$ is bounded on $N(\partial\Omega^*(x_d),\delta)$.
			\end{enumerate}
		\end{enumerate}
	\end{assumption}
	
	\begin{assumption}\label{assumption:asymptotic-normality-discreteX}
		If $\Pr(A(X_i)\in (0,1))>0$, then the following conditions \ref{assumption:neighborhood-O-delta-discreteX}--\ref{assumption:bounded-mean-discreteX} hold.
		\begin{enumerate}
			\renewcommand{\theenumi}{(\alph{enumi})}
			\renewcommand{\labelenumi}{(\alph{enumi})}
			\item {\rm (Probability of Neighborhood of $D^*(x_d)$)}\label{assumption:neighborhood-O-delta-discreteX}
			For each $x_d\in{\cal X}_d^*$, $\Pr(X_i\in N(D^*(x_d),\delta))=O(\delta)$.
			\item {\rm (Bounded Partial Derivatives of $A$)}\label{assumption:bounded-derivatives-A-discreteX}
			For each $x_d\in{\cal X}_d^*$, the partial derivatives of $A(x_d,\cdot)$ are bounded on $C^*(x_d)$.
			\item {\rm (Bounded Conditional Mean)}\label{assumption:bounded-mean-discreteX}
			For each $x_d\in{\cal X}_d^*$, $E[Y_i|X_i=(x_d,\cdot)]$ is bounded on ${\cal X}_c(x_d)$.
			
		\end{enumerate}
		
	\end{assumption}
	
	\begin{theorem}\label{theorem:2sls-consistency-discreteX}
		Suppose that Assumptions \ref{assumption:A-discreteX} and \ref{assumption:consistency-discreteX} hold and $\delta_n\rightarrow 0$, $n\delta_n\rightarrow \infty$, and $S_n\rightarrow \infty$ as $n\rightarrow \infty$.
		Then the 2SLS estimators $\hat \beta_1$ and $\hat\beta_1^s$ converge in probability to
		$$
		\beta_1\equiv \lim_{\delta \rightarrow 0} E[\omega_i(\delta)(Y_i(1)-Y_i(0))],
		$$
		where
		$$
		\omega_i(\delta)=\frac{p^{A}(X_i;\delta)(1-p^{A}(X_i;\delta))(D_i(1)-D_i(0))}{E[p^{A}(X_i;\delta)(1-p^{A}(X_i;\delta))(D_i(1)-D_i(0))]}.
		$$
		Suppose, in addition, that Assumptions \ref{assumption:asymptotic-normality-discreteX} and \ref{assumption:simulation} hold and $n\delta_n^2\rightarrow 0$ as $n\rightarrow \infty$.
		Then
		\begin{align*}
			\hat\sigma^{-1}_n(\hat\beta_1-\beta_1)&\stackrel{d}{\longrightarrow} {\cal N}(0,1),~~
			(\hat\sigma^{s}_n)^{-1}(\hat\beta_1^s-\beta_1)\stackrel{d}{\longrightarrow} {\cal N}(0,1).
		\end{align*}

	\end{theorem}
	
	\begin{proof}
		The proof is analogous to the proof of Theorem \ref{theorem:2sls-consistency}.
The only difference is that, when we prove the convergence of expectations, we show the convergence of the expectations conditional on $X_{di}$, and then take the expectations over $X_{di}$.
	\end{proof}

\section{Other Examples}\label{section:examples}

Here we give other algorithm examples and discuss the applicability of our framework. 

\begin{example}[Bandit Algorithms]\label{example:rf}
We are constantly exposed to digital information (movie, music, news, search results, advertisements, and recommendations) through a variety of devices and platforms. 
Tech companies allocate these pieces of content by using bandit algorithms. 
Our method is applicable to bandit algorithms. 
For simplicity, assume a perfect-compliance scenario where the company perfectly controls the treatment assignment $(D_i=Z_i)$.
The algorithms below first use past data and supervised learning to estimate the conditional means and variances of potential outcomes, $E[Y_i(z)|X_i]$ and $\Var(Y_i(z)|X_i)$, for each $z\in \{0, 1\}$.
Let $\mu_z$ and $\sigma^2_z$ denote the estimated functions.
The algorithms use $\mu_z(X_i)$ and $\sigma^2_z(X_i)$ to determine the treatment assignment for individual $i$.

\begin{enumerate}[label=(\alph*)]
	\item (Thompson Sampling Using Gaussian Priors)
The algorithm first samples potential outcomes from the normal distribution with mean $(\mu_0(X_i), \mu_1(X_i))$ and variance ${\rm diag}(\sigma^2_0(X_i), \sigma^2_1(X_i))$. 
It then chooses the treatment with the highest sampled potential outcome:
	$$
	Z^{TS}_i \equiv \argmax_{z\in \{0, 1\}}y(z), ~~A^{TS}(X_i)= E[\argmax_{z\in \{0, 1\}}y(z)|X_i],
	$$
	where $y(z)\sim {\cal N}(\mu_z(X_i), \sigma^2_z(X_i))$ independently across $z$. 
	This algorithm often induces quasi-experimental variation in treatment assignment, as a strand of the computer science literature has observed \citepappendix{precup2000eligibility, li2010contextual, narita2019efficient, saito2021open, narita2022general}. 
	If the functions $\mu_0(\cdot)$, $\mu_1(\cdot)$, $\sigma^2_0(\cdot)$ and $\sigma^2_1(\cdot)$ are continuous, 
	the function $A$ and APS have an analytical expression:
	$$A^{TS}(x)=p^{TS}(x)=1-\Phi\left(\dfrac{\mu_0(x)-\mu_1(x)}{\sqrt{\sigma^2_0(x)+\sigma^2_1(x)}}\right),$$
	where $\Phi$ is the standard normal cumulative distribution function.
	This APS is nondegenerate, meaning that the data from the algorithm allow for causal-effect identification.
	Furthermore, if the functions $\mu_0(\cdot)$, $\mu_1(\cdot)$, $\sigma^2_0(\cdot)$ and $\sigma^2_1(\cdot)$ are continuously differentiable, this algorithm satisfies Assumption \ref{assumption:asymptotic-normality} \ref{assumption:neighborhood-O-delta}, which is required for asymptotic normality when $\Pr(A(X_i)\in (0,1))>0$.

	\item (Upper Confidence Bound, UCB)
	Unlike the above stochastic algorithm, the UCB algorithm is a deterministic algorithm, producing a less obvious example of our framework. 
	This algorithm chooses the treatment with the highest upper confidence bound for the potential outcome:
	\begin{align*}
		Z^{UCB}_i &\equiv \argmax_{z=0, 1}	\{\mu_z(X_i)+\alpha \sigma_z(X_i)\},\\
        A^{UCB}(x) &=\argmax_{z=0, 1}	\{\mu_z(x)+\alpha \sigma_z(x)\},
	\end{align*}
	where $\alpha$ is chosen so that $|\mu_z(x)-E[Y_i(z)|X_i=x]|\le \alpha \sigma_z(x)$ at least with some probability, for example, $0.95$, for every $x$.
		Suppose that the function $g=\mu_1-\mu_0+\alpha (\sigma_1-\sigma_0)$ is continuous on ${\cal X}$ and is continuously differentiable in a neighborhood of $x$ with $\nabla g(x)\neq \bm{0}$ for any $x\in {\cal X}$ such that $g(x)=0$.
APS for this case is given by
	$$p^{UCB}(x)=\begin{cases}
	0 & \ \ \ \text{if $\mu_1(x)+\alpha\sigma_1(x)<\mu_0(x)+\alpha\sigma_0(x)$}\\
	0.5 & \ \ \ \text{if $\mu_1(x)+\alpha\sigma_1(x)=\mu_0(x)+\alpha\sigma_0(x)$}\\
	1 & \ \ \ \text{if $\mu_1(x)+\alpha\sigma_1(x)>\mu_0(x)+\alpha\sigma_0(x)$}.
	\end{cases}$$
	This means that the UCB algorithm produces complicated quasi-experimental variation along the boundary in the covariate space where the algorithm's treatment recommendation changes.
	If, in addition, $g$ is twice continuously differentiable along the boundary, this algorithm satisfies Assumption \ref{assumption:consistency} \ref{assumption:boundary-partition}, which is required for consistency and asymptotic normality when $\Pr(A(X_i)\in (0,1))=0$. 
	It is possible to identify and estimate causal effects across the boundary. 
\end{enumerate}

\end{example}

\begin{example}[Unsupervised Learning]\label{example:unsupervised}
	Customer segmentation is a core marketing practice that divides a company's customers into groups based on their characteristics and behavior so that the company can effectively target marketing activities at each group.
	Many businesses today use unsupervised learning algorithms, clustering algorithms in particular, to perform customer segmentation.
	Using our notation, assume that a company decides whether it targets a campaign at customer $i$ ($Z_i=1$) or not ($Z_i=0$).
	The company first uses a clustering algorithm such as $K$-means clustering or Gaussian mixture model clustering to divide customers into $K$ groups, making a partition $\{S_1,...,S_K\}$ of the covariate space $\mathbb{R}^p$.
	The company then conducts the campaign targeted at some of the groups:
	\begin{align*}
		Z^{CL}_i&\equiv1\{X_i\in \cup_{k\in T} S_k\},~~
		A^{CL}(x)= 1\{x\in \cup_{k\in T} S_k\}, 
	\end{align*}
	where $T\subset \{1,..,K\}$ is the set of the indices of the target groups.
	
	For example, suppose that the company uses $K$-means clustering, which creates a partition in which a covariate value $x$ belongs to the group with the nearest centroid.
	Let $c_1,...,c_K$ be the centroids of the $K$ groups. Define a set-valued function $C:\mathbb{R}^p\rightarrow 2^{\{1,...,K\}}$, where $2^{\{1,...,K\}}$ is the power set of $\{1,...,K\}$, as
	$C(x)\equiv\argmin_{k\in \{1,...,K\}}\|x-c_k\|$.
	If $C(x)$ is a singleton, $x$ belongs to the unique group in $C(x)$.
	If $C(x)$ contains more than one indices, the group to which $x$ belongs is arbitrarily determined. 
	APS for this case is given by
	$$p^{CL}(x)=\begin{cases}
	0 & \ \ \ \text{if $C(x)\cap T= \emptyset$}\\
	0.5 & \ \ \ \text{if $|C(x)|=2$, $x\in \partial(\cup_{k\in T} S_k)$}\\
	1 & \ \ \ \text{if $C(x)\subset T$}
	\end{cases}$$
	and $p^{CL}(x)\in (0,1)$ if $|C(x)|\ge 3$ and $x\in \partial(\cup_{k\in T} S_k)$,
	where $|C(x)|$ is the number of elements in $C(x)$.
	\footnote{
	If $|C(x)|=2$ and $x\in \partial(\cup_{k\in T} S_k)$, $x$ is on a linear boundary between one target group and one non-target group, and hence APS is $0.5$.
	If $|C(x)|\ge 3$ and $x\in \partial(\cup_{k\in T} S_k)$, $x$ is a common endpoint of several group boundaries, and APS is determined by the angles at which the boundaries intersect.}
	Thus, it is possible to identify causal effects across the boundary $\partial(\cup_{k\in T} S_k)$.
	Assumption \ref{assumption:consistency} \ref{assumption:boundary-partition} approximately holds in that the target group $\cup_{k\in T} S_k$ is arbitrarily well approximated by a set that satisfies the differentiability condition.
\end{example}

\begin{example}[Mechanism Design: Matching and Auction]\label{example:md} 
Centralized economic mechanisms such as matching and auction are also suitable examples, as summarized below \citep{abdulkadirouglu2017research, abdulkadiroglu2019breaking, atila2013,kawai2020using, narita2021theory, narita2018toward}: 
	\begin{center}
		\begin{tabular}{c | c | c c c}
			\hline
			& Matching (e.g., School Choice) & Auction  \\
			\hline\hline
			$i$ & Student & Bidder  \\
			\hline
			$X_i$ & Preference/Priority/Tie-breaker & Bid  \\
			\hline
			$Z_i$ & \begin{tabular}{@{}c@{}}Whether student $i$ is \\ assigned treatment school\end{tabular} & \begin{tabular}{@{}c@{}}Whether bidder $i$ \\ wins the good\end{tabular} \\
			\hline
			$D_i$ & \begin{tabular}{@{}c@{}}Whether student $i$ \\ attends treatment school\end{tabular} & Same as $Z_i$ \\
			\hline
			$Y_i$ & \begin{tabular}{@{}c@{}}Student $i$'s \\ future test score \end{tabular} & \begin{tabular}{@{}c@{}}Bidder $i$'s future \\ economic performance\end{tabular} \\
			\hline
		\end{tabular}
	\end{center}
	In mechanism design and other algorithms with capacity constraints, the treatment recommendation for individual $i$ may depend not only on $X_i$ but also on the characteristics of others.
	These interactive situations can be accommodated by our framework if we consider the following large market setting.\footnote{The approach proposed by \citeappendix{Borusyak2020nonramdom} is applicable to finite-sample settings if the treatment recommendation probability, which may depend on all individuals' characteristics, is nondegenerate for multiple individuals.}
	Suppose that there is a continuum of individuals $i\in [0,1]$ and that the recommendation probability for individual $i$ with covariate $X_i$ is determined by a function $M$ as follows:
	$$
	\Pr(Z_i=1|X_i;F_{X_{-i}}) = M(X_i;F_{X_{-i}}).
	$$
	Here $F_{X_{-i}}=\Pr(\{j\in [0,1]\setminus\{i\}: X_j\le x\})$ is the distribution of $X$ among all individuals $j\in [0,1]\setminus\{i\}$. 
	The function $M:\mathbb{R}^p\times {\cal F}\rightarrow [0,1]$, where ${\cal F}$ is a set of distributions on $\mathbb{R}^p$, gives the recommendation probability for each individual in the market. 
	With a continuum of individuals, for any $i\in [0,1]$, $F_{X_{-i}}$ is the same as the distribution of $X$ in the whole market, denoted by $F_X$.
	Therefore, the data generated by the mechanism $M$ are equivalent to the data generated by the algorithm $A:\mathbb{R}^p\rightarrow [0,1]$ such that $A(x)\equiv M(x;F_X)$ for all $x\in \mathbb{R}^p$.
	Our framework is applicable to this large-market interactive setting.
	\end{example}

\noindent The above discussions can be summarized as follows. 

\begin{corollary}
 In all the above examples, there exists $x\in{\rm int}({\cal X})$ such that $p^{A}(x)\in (0,1)$.
 Therefore, a causal effect is identified under Assumptions \ref{assumption:A} and \ref{continuity}.
\end{corollary}

\section{Monte Carlo Simulation: Details}\label{appendix:simulation}
{\bf Parameter Choice.}
For the variance-covariance matrix $\mathbf{\Sigma}$ of $X_i$, we first create a $100\times 100$ symmetric matrix $\mathbf{V}$ such that the diagonal elements are one, $\mathbf{V}_{ij}$ is nonzero and equal to $\mathbf{V}_{ji}$ for $(i,j)\in \{2,3,4,5,6\}\times\{35,66,78\}$, and everything else is zero.
We draw values from ${\rm Unif}(-0.5,0.5)$ independently for the nonzero off-diagonal elements of $\mathbf{V}$.
We then create matrix $\mathbf{\Sigma} = \mathbf{V}  \times \mathbf{V}$, which is a positive semidefinite matrix. 
For $\alpha_0$ and $\alpha_1$, we first draw $\tilde \alpha_{0j}$, $j=51,...,100$, from ${\rm Unif}(-100,100)$ independently across $j$, and draw $\tilde \alpha_{1j}$, $j=1,...,100$, from ${\rm Unif}(-150,200)$ independently across $j$.
We then set $\tilde \alpha_{0j}=\tilde \alpha_{1j}$ for $j=1,...,50$, and calculate $\alpha_0$ and $\alpha_1$ by normalizing $\tilde\alpha_0$ and $\tilde\alpha_1$ so that $\Var(X_i'\alpha_0)=\Var(X_i'\alpha_1)=1$.

{\bf Training of Prediction Model.}
We construct $\tau_{pred}$ using an independent sample $\{(\tilde Y_i, \tilde X_i, \tilde D_i, \tilde Z_i)\}_{i=1}^{\tilde n}$ of size $\tilde n=2,000$. The distribution of $(\tilde Y_i, \tilde X_i, \tilde D_i, \tilde Z_i)$ is the same as that of $(Y_i,X_i,D_i,Z_i)$ except (1) that $\tilde Y_i(1)$ is generated as $\tilde Y_i(1)=\tilde Y_i(0)+0.5 \tilde X_i'\alpha_1+0.5 \epsilon_{1i}$, where $\epsilon_{1i}\sim {\cal N}(0,1)$ and (2) that $\tilde Z_i\sim {\rm Bernoulli}(0.5)$.
This can be viewed as data from a past randomized experiment conducted to construct the algorithm.
We then use random forests separately for the subsamples with $\tilde Z_i=1$ and $\tilde Z_i=0$ to predict $\tilde Y_i$ from $\tilde X_i$.
Let $\mu_z(x)$ be the trained prediction model. Set $\tau_{pred}(x)=\mu_1(x)-\mu_0(x)$.
We generate the sample $\{(\tilde Y_i, \tilde X_i, \tilde D_i, \tilde Z_i)\}_{i=1}^{\tilde n}$ and construct $\tau_{pred}$ only once, and we use it for all of the $1,000$ simulation samples.
The distribution of the sample $\{(Y_i,X_i,D_i,Z_i)\}_{i=1}^n$ is held fixed for all simulations.

When training $\mu_z$, we first randomly split the sample $\{(\tilde Y_i, \tilde X_i, \tilde D_i, \tilde Z_i)\}_{i=1}^{\tilde n}$ into train (80\%) and test datasets (20\%). We use random forests on the training sample to obtain the prediction model $\mu_z$ and validate its performance on the test sample. The trained algorithm has an accuracy of 80.5\% on the test data.

\section{Empirical Policy Application: Details}\label{appendix:empirics}
\subsection{Hospital Cost Data}
We use publicly available Healthcare Cost Report Information System (HCRIS) data, to project funding eligibility and amounts for all hospitals in the dataset. 
This data set contains information on various hospital characteristics including utilization, number of employees, medicare cost data and financial statement data. 
We use the methodology detailed in the \href{https://www.hhs.gov/coronavirus/cares-act-provider-relief-fund/general-information/index.html}{CARES Act website} to project 
funding based on 2018 financial year cost reports.

The data is available from financial year 1996 to 2019. As the coverage is higher for 2018 (compared to 2019), we utilize the data corresponding to the 2018 financial year. Hospitals are uniquely identified in a financial year by their CMS (Center for Medicaid and Medicare Services) Certification Number. We have data for 4,705 providers for the 2018 financial year. We focus on 4,648 acute care and critical access hospitals that are either located in one of the 50 states or Washington DC.  

\textbf{Disproportionate patient percentage.} 
Disproportionate patient percentage is equal to the percentage of Medicare inpatient days attributable to patients eligible for both Medicare Part A and Supplemental Security Income (SSI) summed with the percentage of total inpatient days attributable to patients eligible for Medicaid but not Medicare Part A.\footnote{For the precise definition, see \url{https://www.cms.gov/Medicare/Medicare-Fee-for-Service-Payment/AcuteInpatientPPS/dsh}.} In the data, this variable is missing for 1560 hospitals. We impute the disproportionate patient percentage to 0 when it is missing. 

\textbf{Uncompensated care per bed.} 
Cost of uncompensated care refers to the care provided by the hospital for which no compensation was received from the patient or the insurer. It is the sum of a hospital’s bad debt and the financial assistance it provides.\footnote{The precise definition can be found at \url{https://www.aha.org/fact-sheets/2020-01-06-fact-sheet-uncompensated-hospital-care-cost}.} The cost of uncompensated care is missing for 86 hospitals, which we impute to 0. We divide the cost of uncompensated care by the number of beds in the hospital to obtain the cost per bed. The data on bed count is missing for 15 hospitals, which we drop  from the analysis, leaving us with 4,633 hospitals in 2,473 counties.

\textbf{Profit Margin.} 
Hospital profit margins are indicative of the financial health of the hospitals. 
We calculate profit margins as the ratio of net income to total revenue where total revenue is the sum of net patient revenue and total other income. After the calculation, profit margins are missing for 92 hospitals, which we impute to 0. 

\textbf{Funding.} 
We calculate the projected funding using the formula on the \href{https://www.hhs.gov/coronavirus/cares-act-provider-relief-fund/general-information/index.html#targeted.}{CARES ACT website}. Hospitals that do not qualify on any of the three dimensions are not given any funding. Each eligible hospital is assigned an individual facility score, which is calculated as the product of disproportionate patient percentage and number of beds in that hospital. We calculate cumulative facility score  as the sum of all individual facility scores in the dataset. Each hospital receives a share of \$10 billion, where the share is determined by the ratio of the individual facility score of that hospital to the cumulative facility score. The amount of funding received by hospitals is bounded below at \$5 million and capped above at \$50 million.

\subsection{Hospital Utilization Data}
We use the publicly available COVID-19 Reported Patient Impact and Hospital Capacity by Facility dataset for our outcome variables. This provides facility level data on hospital utilization aggregated on a weekly basis, from July 31st onwards. These reports are derived from two main sources -- (1) HHS TeleTracking and (2) reporting provided directly to HHS Protect by state/territorial health departments on behalf of health care facilities.\footnote{Source:
\url{https://healthdata.gov/Hospital/COVID-19-Reported-Patient-Impact-and-Hospital-Capa/uqq2-txqb}.}

The hospitals are uniquely identified for a given collection week (which goes from Friday to Thursday) by their CMS Certification number. All hospitals that are registered with CMS by June 1st 2020 are included in the population. We merge the hospital cost report data with the utilization data using the CMS certification number. According to the terms and conditions of the CARES Health Care Act, the recipients may use the relief funds only to ``prevent, prepare for, and respond to coronavirus'' and for ``health care related expenses or lost revenues that are attributable to coronavirus''. Therefore, for our analysis we focus on 4 outcomes that were directly affected by COVID-19, for the week spanning July 31st to August 6th 2020. The outcome measures are described below.\footnote{
We conduct sanity checks and  impute observations to missing if they fail our checks. For example, we impute the value \# Confirmed/ Suspected COVID Patients and \# Confirmed COVID Patients to missing when the latter is greater than the former. \# Confirmed/ Suspected COVID Patients should be greater than or equal to \# Confirmed COVID Patients as the former includes the latter. Similarly, we impute \# Confirmed/ Suspected COVID Patients in ICU and \# Confirmed COVID Patients in ICU to be missing when the latter is greater than the former.}
\begin{enumerate}
    \item Total reports of patients currently hospitalized in an adult inpatient bed who have laboratory-confirmed or suspected COVID-19, including those in observation beds reported during the 7-day period. 
    \item Total reports of patients currently hospitalized in an adult inpatient bed who have laboratory-confirmed COVID-19 or influenza, including those in observation beds. Including patients who have both laboratory-confirmed COVID-19 and laboratory confirmed influenza during the 7-day period.
    \item Total reports of patients currently hospitalized in a designated adult ICU bed who have suspected or laboratory-confirmed COVID-19.
    \item Total reports of patients currently hospitalized in a designated adult ICU bed who have laboratory-confirmed COVID-19 or influenza, including patients who have both laboratory-confirmed COVID-19 and laboratory-confirmed influenza.\footnote{In the dataset, when the values of the 7 day sum are reported to be less than 4, they are replaced with -999,999. We recode these values to be missing. The results in Table 4 remain almost the same even if we impute the suppressed values (coded as -999,999) with 0s. Results are available upon request.}
\end{enumerate}

\subsection{Additional Empirical Results}\label{appendix:additional_empirics}





\paragraph{Persistence and Heterogeneity}

\begin{figure*}[!t]
        \centering
        \caption[]{\small Dynamic Effects of Funding on Weekly Hospital Outcomes} 
        \begin{subfigure}[b]{0.49\textwidth}   
            \centering 
            \includegraphics[width=\textwidth]{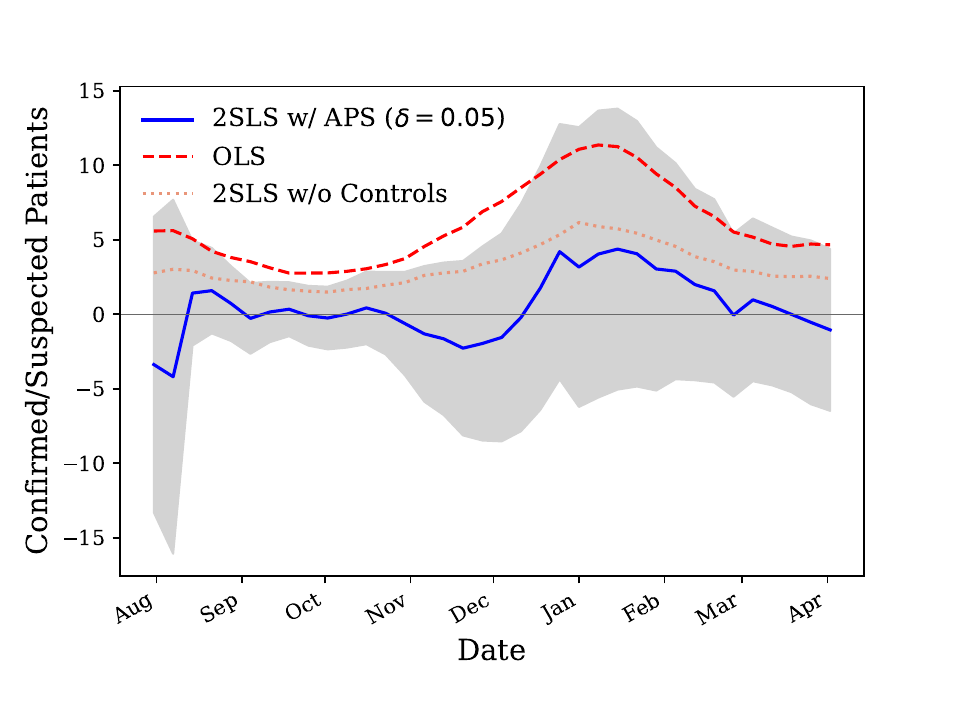}
            \caption[]%
            {{\footnotesize \# Confirmed/Suspected COVID Patients}}    
            \label{fig:mean and std of net34}
        \end{subfigure}
        \begin{subfigure}[b]{0.49\textwidth}   
            \centering 
            \includegraphics[width=\textwidth]{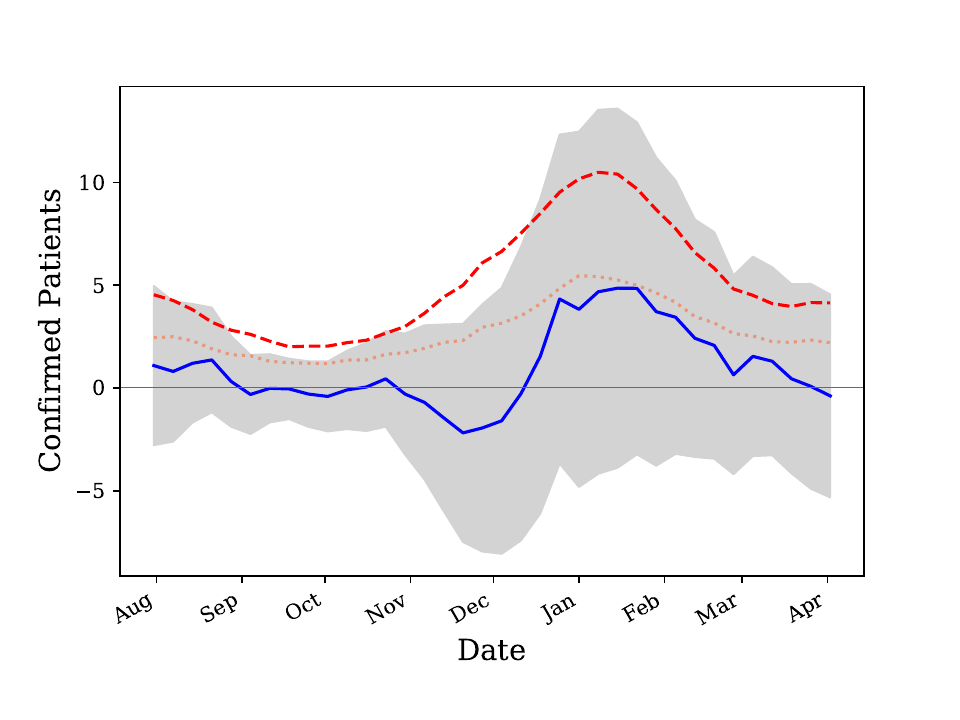}
            \caption[]%
            {{\footnotesize \# Confirmed COVID Patients}} 
            \label{fig:mean and std of net44}
        \end{subfigure}
        \vskip0.1\baselineskip
        \begin{subfigure}[b]{0.49\textwidth}
            \centering
            \includegraphics[width=\textwidth]{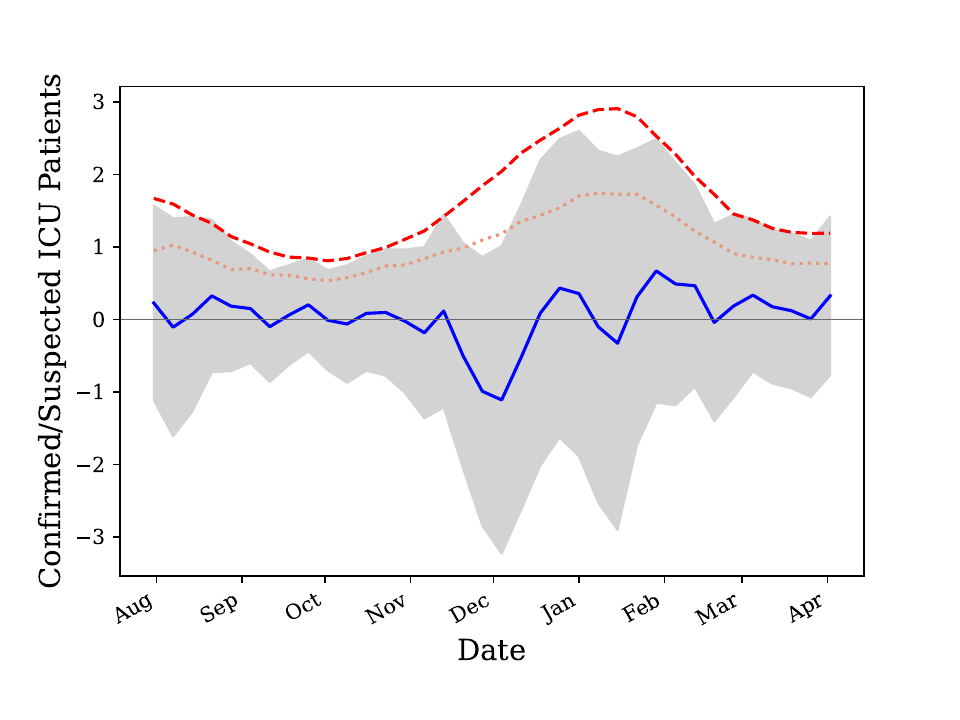}
            \caption[]%
            {{\footnotesize \# Confirmed/Suspected COVID Patients in ICU}}
            \label{fig:mean and std of net14}
        \end{subfigure}
        \hfill
        \begin{subfigure}[b]{0.49\textwidth}  
            \centering 
            \includegraphics[width=\textwidth]{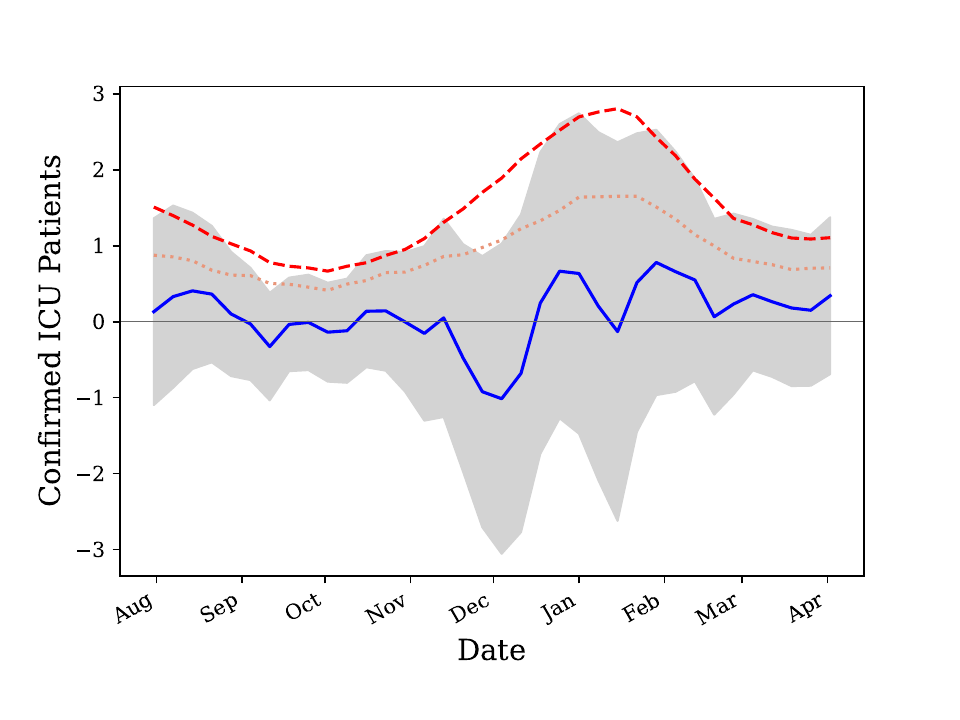}
            \caption[]%
            {{\footnotesize \# Confirmed COVID Patients in ICU}}
            \label{fig:mean and std of net24}
        \end{subfigure}
        \caption*{\scriptsize{\it Note}: The figure shows the results of estimating our main 2SLS specification about the effect of \$1mm of relief funding on weekly hospital outcomes from 07/31/2020 to 04/02/2021. The outcomes record the 7-day sum of the number of hospitalized patients with the specified condition.
        We compute the Approximate Propensity Score with $S = 10,000$ and $\delta = 0.05$. The estimates from the uncontrolled OLS, uncontrolled 2SLS, and 2SLS with the Approximate Propensity Score controls are plotted on the y-axis. Grey areas are 95\% confidence intervals.}
        \label{fig:main_ts}
\end{figure*}

The analysis in the main body looks at the immediate effects of relief funding. However, the effects of relief funding might kick in after a time lag, given that expansion in capacity and staff takes time. 
To investigate the relevance of this concern, we measure the evolving effects of relief funding. We estimate our main 2SLS specification on the 7-day average of each hospital outcome for each week from July 31st, 2020 to April 2nd, 2021. We plot the estimated dynamic effects in Figure \ref{fig:main_ts}. The estimated dynamic effects are similar to the initial null effects in Table \ref{table:2sls_safetynet}, even several months after the distribution of relief funding. This dynamic analysis suggests that funding has no substantial effect even in the long run. 

\begin{figure*}[!ht]
        \centering
        \caption[]{Heterogeneous Effects of Hospital Funding by Hospital Characteristics}
        \vspace{-1em}
        \begin{subfigure}[b]{0.49\textwidth}   
            \centering 
            \includegraphics[width=0.95\textwidth]{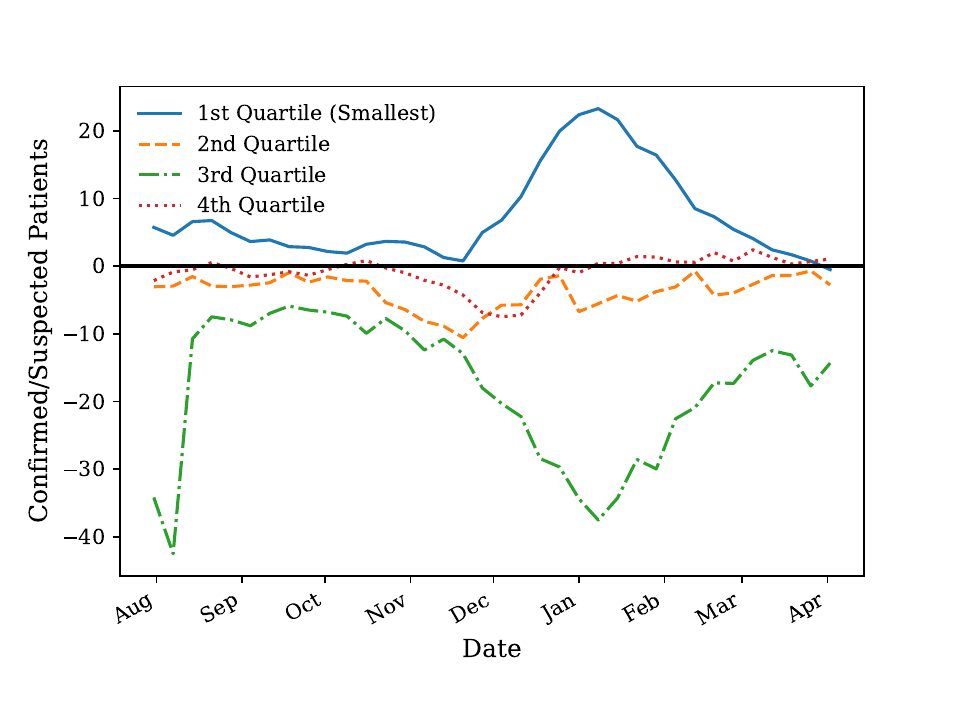}
            \caption[]%
            {{\small Net Income}}    
            \label{fig:het_condense_income}
        \end{subfigure}
        \begin{subfigure}[b]{0.49\textwidth}   
            \centering 
            \includegraphics[width=0.95\textwidth]{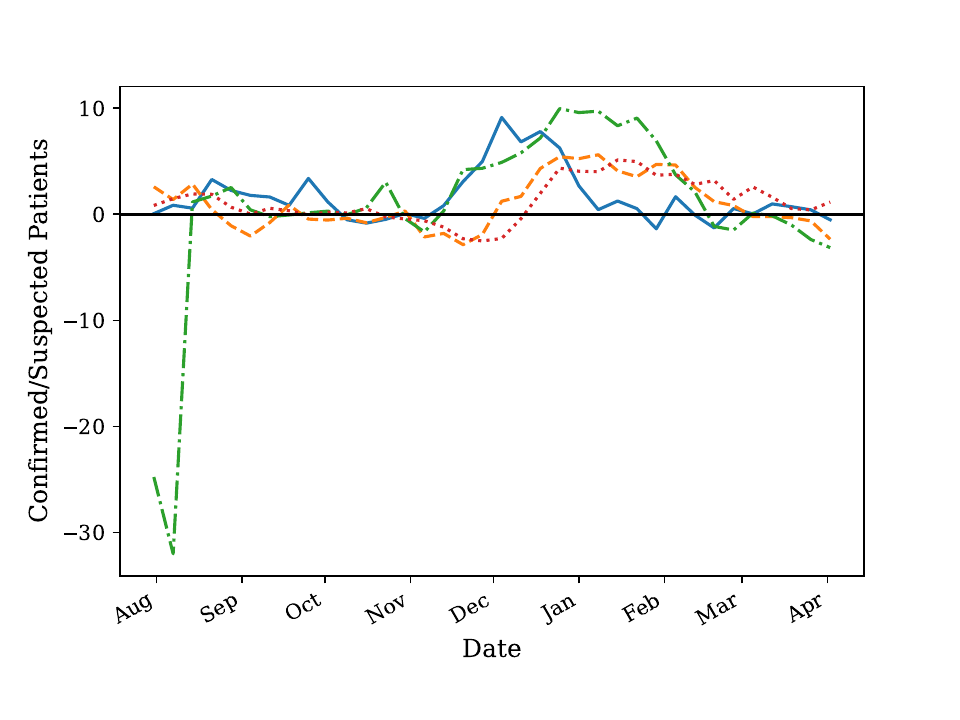}
            \caption[]%
            {{\small Total Full-Time Employees}}
        \end{subfigure}
        \vskip0.1\baselineskip
        \begin{subfigure}[b]{0.49\textwidth}
            \centering
            \includegraphics[width=0.95\textwidth]{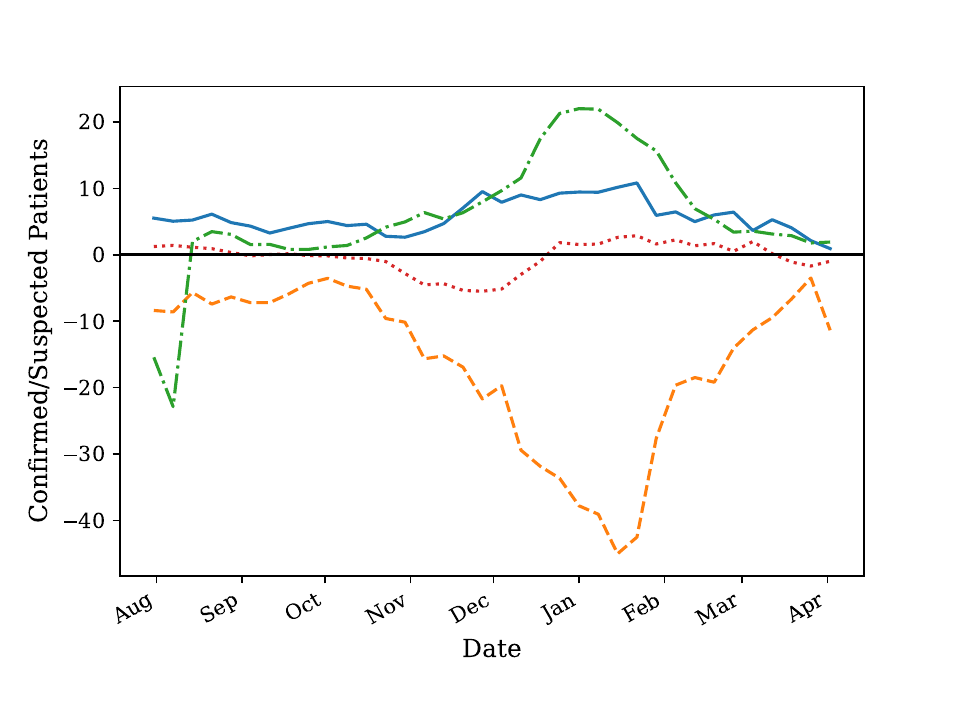}
            \caption[]%
            {{\small Total Beds}}  
            \end{subfigure}
       \hfill
        \begin{subfigure}[b]{0.49\textwidth}
            \centering
            \includegraphics[width=0.95\textwidth]{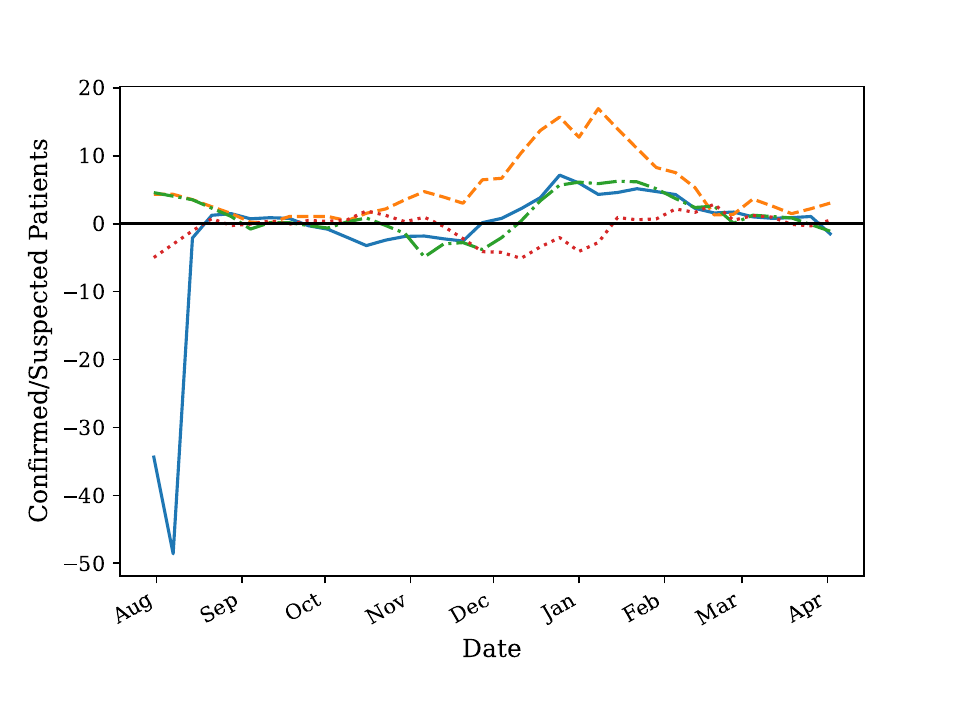}
            \caption[]%
            {{\small Inpatient Length of Stay}}
        \end{subfigure} 
        \vskip0.1\baselineskip
        \begin{subfigure}[b]{0.49\textwidth}   
            \centering 
            \includegraphics[width=0.95\textwidth]{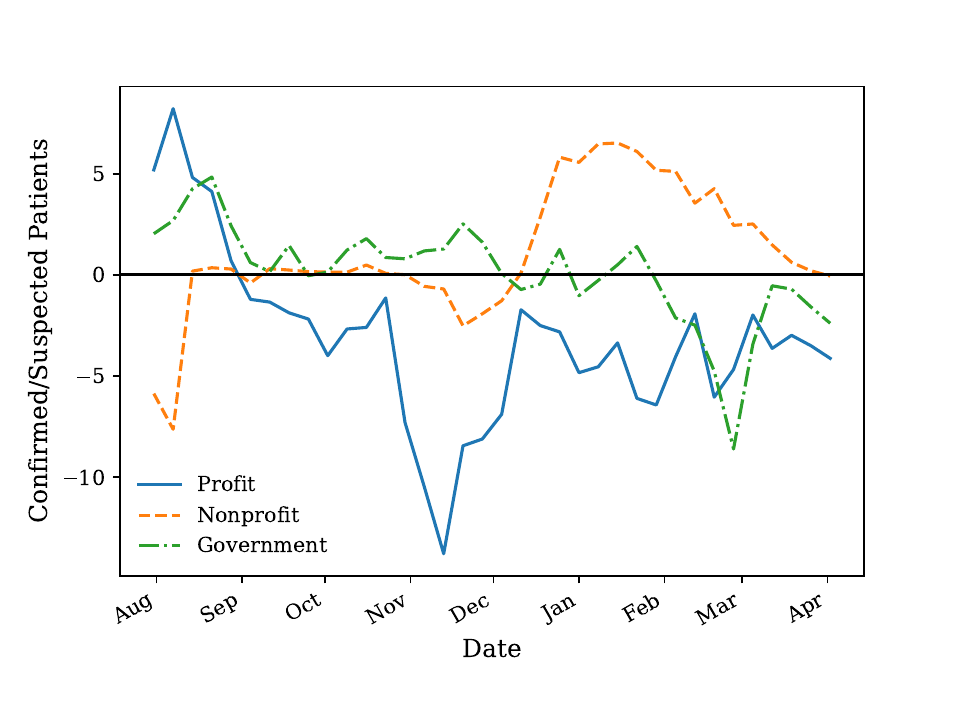}
            \caption[]%
            {{\small Ownership Type}}  
        \end{subfigure}
\caption*{\scriptsize{\it Note}: The figure shows the results of estimating our main 2SLS specification of the effect of \$1mm of relief funding on weekly confirmed/suspected COVID-19 patients from 07/31/2020 to 04/12/2021, where the sample is stratified by quartiles of different hospital characteristics, or ownership type. No effect estimates are significantly different from each other at the 5\% level. 
We compute APS with $S = 10,000$ and $\delta = 0.05$.}
\label{fig:het_condense}
\end{figure*}

We further extend this analysis by estimating the heterogeneous effects of funding for different types of hospitals. Figure \ref{fig:het_condense} plots the estimates by repeating the same dynamic analysis as in Figure \ref{fig:main_ts}, but for different groups of hospitals defined by hospital size and ownership type. 
Overall, hospitals of different types sometimes face different trends of funding effects, but none of the differences is statistically significant at the 5\% level. 
We do not find any strong evidence of heterogeneity in the funding effects at any point in time.  

Having said that, there is some suggestive indication of potential heterogeneity. In Figure \ref{fig:het_condense_income}, for example, the estimated funding effect spiked among the hospitals in the lowest quartile of revenue from December 2020 to February 2021. This trend may suggest that the funding was able to alleviate the financial burden faced by struggling hospitals in this strata and allowed them to take on new patients during the winter surge.
There is also a sizable dip in the funding effect of for-profit hospitals around the same period. This could be due to regional differences in the distribution of hospital ownership. Nonprofits and government-managed hospitals tend to be in rural areas, which both received more funding and experienced a worse surge during the winter. On the other hand, the for-profits that received funding tend to be in urban areas and experienced a less extreme winter wave. 

\begin{figure}[!ht]
    \centering
    \caption{Funding Distribution for Eligible Hospitals}
    \includegraphics[height=2.6in]{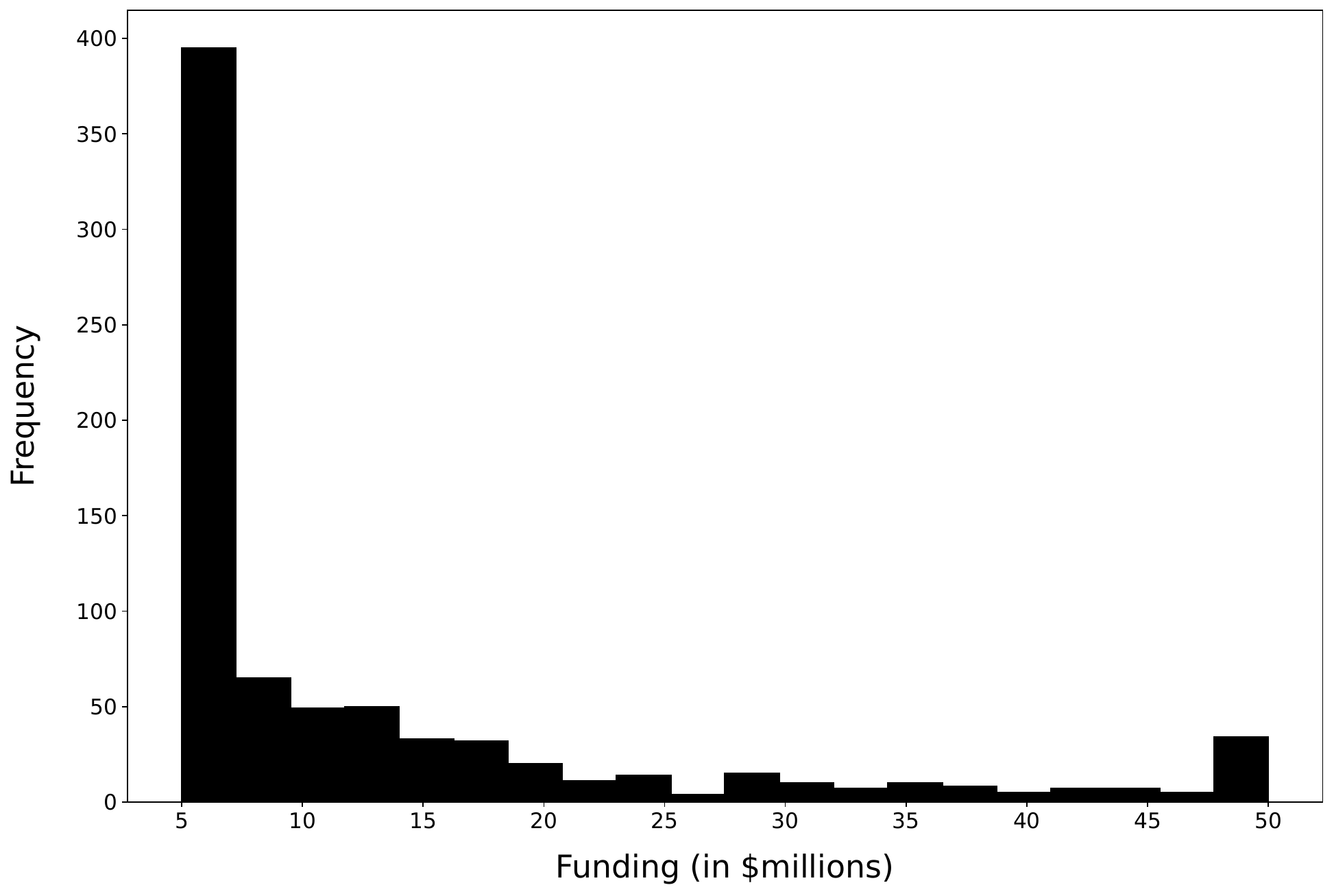}
    \caption*{\scriptsize{\it Note}: The figure shows the distribution of funding amounts for eligible hospitals. Each eligible hospital is assigned an individual facility score, which is the product of Disproportionate Patient Percentage and number of beds in the hospital. The share of \$10 billion received by an eligible hospital is determined by the ratio of the individual facility score of that hospital to the sum of facility scores across all eligible hospitals. The amount of funding that can be received by an eligible hospital is calculated as the product of this ratio and \$10 billion, and is bounded below at \$5 million and bounded above at \$50 million. }
    \label{fig:dist_funding}
\end{figure}

\begin{figure}[!htbp]
    \centering
    \caption{Fixed-bandwidth APS Estimation with Varying Simulations $S$}
    \includegraphics[height=2.8in]{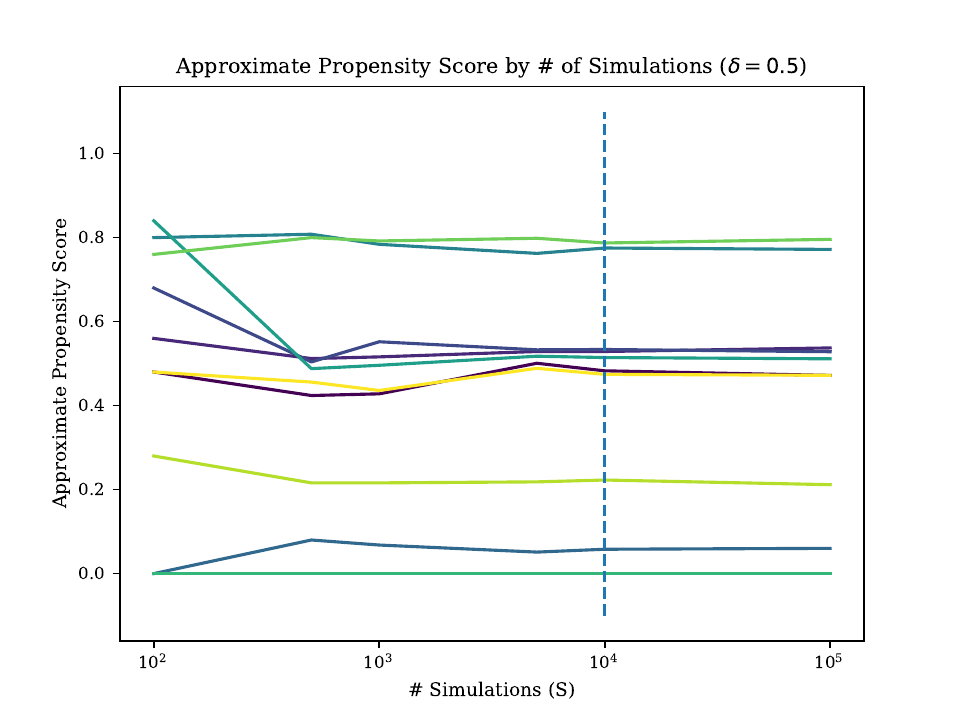}
    \caption*{\scriptsize{\it Note}: The above figure plots the fixed-bandwidth APS estimates for 10 randomly selected hospitals along the eligibility margin for varying numbers of simulations $S$. Each line represents a different hospital. The dotted line at $10^4$ indicates the number of simulations we use for our main analysis. }
    \label{fig:aps_sim}
\end{figure}

\begin{table}[!ht] \centering
\newcolumntype{C}{>{\centering\arraybackslash}X}

\caption{Differential Attrition}
{\scriptsize
\begin{tabularx}{\textwidth}{lCCCCCCCCCC}

\toprule
{}&{}&{}&\multicolumn{7}{c}{Our Method with Approximate Propensity Score Controls}\\
{}&{Ineligible}&{No}&\\\cline{4-10}
{}&{Hospitals}&{Controls}&
{ $\delta = 0.01$}&{ $\delta = 0.025$}&{ $\delta = 0.05$}&{ $\delta = 0.075$}&{ $\delta = 0.1$}&{ $\delta = 0.25$}&{ $\delta = 0.5$} \tabularnewline
{}&{(1)}&{(2)}&{(3)}&{(4)}&{(5)}&{(6)}&{(7)}&{(8)}&{(9)}\tabularnewline
\midrule\addlinespace[1.5ex]
\#Confirmed/Suspected&.745&0.11&0.23&$-$0.05&0.03&0.04&0.05&0.07&0.10 \tabularnewline
Covid Patients&&(0.01)&(0.14)&(0.10)&(0.07)&(0.06)&(0.05)&(0.03)&(0.02) \tabularnewline
&&N=4633&N=91&N=239&N=494&N=684&N=905&N=1751&N=2397 \tabularnewline
\#Confirmed Covid Patients&.754&0.08&0.13&$-$0.09&$-$0.02&0.00&0.02&0.07&0.10 \tabularnewline
&&(0.01)&(0.17)&(0.11)&(0.07)&(0.06)&(0.05)&(0.03)&(0.03) \tabularnewline
&&N=4633&N=91&N=239&N=494&N=684&N=905&N=1751&N=2397 \tabularnewline
\#Confirmed/Suspected&.728&0.09&0.16&$-$0.05&$-$0.11&$-$0.07&$-$0.03&0.05&0.09 \tabularnewline
Covid Patients in ICU&&(0.02)&(0.14)&(0.10)&(0.07)&(0.06)&(0.05)&(0.03)&(0.03) \tabularnewline
&&N=4633&N=91&N=239&N=494&N=684&N=905&N=1751&N=2397 \tabularnewline
\#Confirmed Covid Patients&.744&0.07&0.24&0.11&0.01&$-$0.04&0.00&0.03&0.07 \tabularnewline
in ICU&&(0.02)&(0.18)&(0.12)&(0.08)&(0.06)&(0.05)&(0.04)&(0.03) \tabularnewline
&&N=4633&N=91&N=239&N=494&N=684&N=905&N=1751&N=2397 \tabularnewline
\bottomrule 

\end{tabularx}
\caption*{
\scriptsize {\it {Note}}: This table reports differential safety net eligibility effects on the availability of outcome data at the hospital level. Column 1 presents the average of the  availability indicators of the outcome variables for the ineligible hospitals. In column 2, we regress the availability indicator on dummy for safety net eligibility without any controls. In columns 3-9, we run this regression controlling for the Approximate Propensity Score with different values of bandwidth $\delta$ on the sample with nondegenerate Approximate Propensity Score. All Approximate Propensity Scores are computed by averaging 10,000 simulation draws. 
The outcome variables are the 7 day totals for the week spanning July 31st, 2020 to August 6th, 2020. Confirmed or Suspected COVID patients refer to the sum of patients in inpatient beds with lab-confirmed/suspected COVID-19. Confirmed COVID patients refer to the sum of patients in inpatient beds with lab-confirmed COVID-19, including those with both lab-confirmed COVID-19 and influenza. Inpatient bed totals also include observation beds. Similarly, Confirmed/Suspected COVID patients in ICU refer to the sum of patients in ICU beds with lab-confirmed or suspected COVID-19. Confirmed COVID patients in ICU refers to the sum of patients in ICU beds with lab-confirmed COVID-19, including those with both lab-confirmed COVID-19 and influenza. Robust standard errors are reported in parenthesis.
}
\label{table:differential_attrition}
}

\end{table}

\clearpage

\bibliographystyleappendix{ecca}
\bibliographyappendix{reference}

\end{document}